\documentclass{jfm}

\usepackage{graphicx}
\usepackage{newtxtext}
\usepackage{natbib}
\usepackage[hidelinks]{hyperref}
\usepackage{placeins}
\usepackage{natbib}

\usepackage{amsmath,mathtools}
\usepackage{color} 

\usepackage{tabularx}
\usepackage{longtable}
\usepackage{listings}
\usepackage{booktabs}
\usepackage{float}
\usepackage{makecell}
\usepackage{epstopdf}
\usepackage{pifont} 
\usepackage{wasysym}
\usepackage{tikz}
\usepackage{upgreek}
\usepackage[justification=justified,singlelinecheck=false]{caption}

\usepackage[normalem]{ulem}

\usepackage{flafter}  
\usepackage{ulem}
\usepackage{slashbox}

\definecolor{darkorange}{rgb}{0.8500 0.3250 0.0980}

\definecolor{darkgreen}{rgb}{0.4660 0.6740 0.1880}

\def\mline{\vrule width4pt height2.5pt depth -2pt}
\def\dashed{\mline\hskip3.5pt\mline\thinspace}

\newcommand{\RomanNumeralCaps}[1]
\linenumbers

\usepackage{lmodern,pdftexcmds,amsmath}

\DeclareMathAlphabet{\mathsfbi}{OT1}{\sfdefault}{bx}{sl}
\DeclareMathVersion{sfletters}
\SetSymbolFont{letters}{sfletters}{OML}{ntxsfmi}{b}{it}

\makeatletter
\newcommand{\mathbfsbilow}[1]{%
  \text{\mathversion{sfletters}$\m@th#1$}%
}
\makeatother

\usepackage{etoolbox}

\title{Staircase solutions and stability in vertically confined salt-finger convection}

\author{Chang Liu\aff{1}
  \corresp{\email{chang\_liu@berkeley.edu}}, Keith Julien\aff{2} \and 
  Edgar Knobloch\aff{1}
 }
\affiliation{\aff{1}Department of Physics, University of California, Berkeley, CA 94720, USA
\aff{2}Department of Applied Mathematics, University of Colorado, Boulder, CO 80309, USA
}

\begin{document}
\maketitle

\begin{abstract}
Bifurcation analysis of confined salt-finger convection using single-mode equations obtained from a severely truncated Fourier expansion in the horizontal is performed. Strongly nonlinear staircase-like solutions having, respectively, one (S1), two (S2) and three (S3) regions of mixed salinity in the vertical direction are computed using numerical continuation, and their stability properties are determined. Near onset, the one-layer S1 solution is stable and corresponds to maximum salinity transport among the three solutions. The S2 and S3 solutions are unstable but exert an influence on the statistics observed in direct numerical simulations (DNS) in larger two-dimensional (2D) domains. Secondary bifurcations of S1 lead either to tilted-finger (TF1) or to traveling wave (TW1) solutions, both accompanied by the spontaneous generation of large-scale shear, a process favored for lower density ratios and Prandtl numbers ($Pr$). These states at low $Pr$ are associated, respectively, with two-layer and three-layer staircase-like salinity profiles in the mean. States breaking reflection symmetry in the midplane are also computed. In 2D and for low $Pr$ the DNS results favor direction-reversing tilted fingers (RTF) resembling the pulsating wave state observed in other systems. Two-layer and three-layer mean salinity profiles corresponding to RTF and TW1 are observed in 2D DNS averaged over time. The single-mode solutions close to the high wavenumber onset are in an excellent agreement with 2D DNS in small horizontal domains and compare well with 3D DNS.

\end{abstract}


\section{Introduction}
\label{sec:intro}

Oceanographic measurements have widely reported staircase-like structures with regions of nearly constant density in the vertical direction separated by interfaces with a sharp density gradient \citep{tait1968some,tait1971thermohaline,schmitt1987c,schmitt2005enhanced,padman1989thermal,muench1990temperature,zodiatis1996thermohaline,morell2006thermohaline,timmermans2008ice,fer2010seismic,spear2012thermohaline}. Such structures are typical of tropical and subtropical regions where warm salty water often overlies cold fresh water and have been ascribed to the presence of salt-finger convection. Indeed, staircases have been observed in the western tropical Atlantic \citep{schmitt1987c,schmitt2005enhanced}, Tyrrhenian Sea \citep{zodiatis1996thermohaline}, and the Mediterranean outflow in the Northeast Atlantic \citep{tait1968some,tait1971thermohaline}, where conditions for salt-finger convection prevail. These staircases are typically characterized by a large coherence length in the horizontal, much larger than the step height in the vertical. For example, the Caribbean-Sheets and Layers Transect (C-SALT) field programs show that the well-mixed layers were 5–30 m thick and laterally coherent over scales of 200–400 km \citep{schmitt1987c}. The presence of staircases also enhances the tracer mixing rate. For example, the North Atlantic Tracer Release Experiments (NATRE) \citep{schmitt2005enhanced} reveal a mixing rate in the western tropical Atlantic five times larger than that in the eastern subtropical Atlantic. Spontaneous formation of staircases has also been observed in idealized laboratory experiments \citep{linden1978formation,krishnamurti2003double,krishnamurti2009heat} and direct numerical simulations (DNS) \citep{piacsek1980nonlinear,radko2003mechanism,radko2005determines,stellmach2011dynamics,yang2020multiple}.

Understanding the origin of spontaneous staircase formation is important for parameterization and accurate modeling of oceanographic processes. Various mechanisms for staircase formation have been proposed \citep[\S 8]{radko2013double}. The collective instability mechanism \citep{stern1969collective,holyer1981collective} identifies conditions for the formation of a staircase based on the instability of the salt-finger field to internal gravity waves, but requires a closure model for the Reynolds stress and the temperature and salinity fluxes based on laboratory measurement and ocean observation. The onset of staircase formation is also predicted by models with negative density diffusion \citep{phillips1972turbulence,posmentier1977generation}. The related $\gamma$-instability of \citet{radko2003mechanism} requires a parameterization of the Nusselt number and the flux ratio but shows good agreement with 2D and 3D direct numerical simulation of salt-finger convection \citep{radko2003mechanism,stellmach2011dynamics}. However, these predictions generally only focus on the onset of a large-scale instability suggesting the appearance of a staircase, but do not provide a detailed profile of the final staircase or its parameter dependence. In this connection the model of \citet{balmforth1998dynamics} is of particular interest. The model parametrizes the coupling between buoyancy flux and local turbulent kinetic energy but succeeds in generating robust staircases. Nevertheless, the use of closure models employed in all these predictions results in uncertainty in the applicable parameter regime. 

This work aims to discuss an alternative mechanism for spontaneous staircase formation from a bifurcation theory point of view, focusing on the computation of strongly nonlinear staircase-like solutions and analyzing their stability. Bifurcation analysis has been widely employed to provide insight into pattern formation in fluid dynamics \citep{crawford1991symmetry}. For example, a secondary bifurcation of steady convection rolls in Rayleigh-B\'enard convection to tilted rolls was shown to be accompanied by the generation of large-scale shear \citep{howard1986large,rucklidge1996analysis}, resembling both experimental observation \citep{krishnamurti1981large} and direct numerical simulations \citep{goluskin2014convectively}. A sequence of local and global bifurcations of such tilted convection cells in magnetoconvection was shown to lead to a pulsating wave characterized by periodic reversals in the direction of the tilt and the accompanying large-scale shear \citep{matthews1993pulsating,proctor1993symmetries,proctor1994nonlinear,rucklidge1996analysis}. In the diffusive configuration in which cold fresh water overlies warm salty water, a similar analysis found stable traveling waves near onset \citep{knobloch1986doubly} and provided insight into the transition to chaos \citep{knobloch1992heteroclinic}. Such chaotic or even fully developed turbulent states generally visit neighborhoods of (unstable) steady, periodic, or traveling wave solutions, and these visits leave an imprint on the flow statistics; see, e.g., \citet{kawahara2001periodic,van2006periodic} and the reviews by \citet{kawahara2012significance,graham2021exact}. 

This work focuses on vertically confined salt-finger convection in order to understand the interior between two well-mixed layers. While stress-free velocity boundary conditions are suitable for understanding oceanographic scenarios, no-slip boundary conditions are more relevant to laboratory experiments \citep{hage2010high}. A wide range of bifurcation analyses of related problems have been performed on vertically confined systems with different boundary conditions including Rayleigh-B\'enard convection. In the salt-finger case the fluxes generated in a vertically confined system agree quantitatively with those obtained in a vertically periodic domain when normalized by the bulk conductive fluxes \citep{li2022wall} despite the elimination of the elevator mode that is present in vertically periodic domains; see, e.g., \citet[\S 2.1]{stern1969collective,holyer1984stability,radko2013double}.

In order to facilitate bifurcation analysis, we focus here on the single-mode equations obtained from a severely truncated Fourier expansion in the horizontal. Such single-mode equations reduce the two (or three) spatial dimensions in the primitive equations into one vertical dimension, with the dependence on the horizontal direction parameterized by a single assumed horizontal wavenumber. Such a truncation may provide insight into salt-finger convection as the underlying flow structures are dominated by well-organized columnar structures; see, e.g., the visualization of confined salt-finger convection by \citet[figure 7]{yang2016scaling}. This single-mode formulation also explicitly isolates and describes the interaction between horizontally averaged modes (corresponding to the staircase or large-scale shear) and the horizontal harmonics (corresponding to fingers), in the spirit of mean-field theory; see, e.g, \citep[\S 3.2.1]{garaud2018double}. Moreover, the single-mode equations do not require any closure assumptions for the Reynolds stress and the temperature and salinity fluxes to parametrize the feedback between the fluctuations and horizontally averaged quantities, while preserving the nonlinear interaction between them; see figure \ref{fig:illustration}(b) below.

Single-mode equations have been widely employed to provide insight into related problems. For example, single-mode solutions (also called `single-$\alpha$ mean-field theory') for Rayleigh-B\'enard convection with stress-free boundary conditions have been shown to reproduce the mean temperature profile expected at a high Rayleigh number \citep{herring1963investigation}. For no-slip boundaries at high Prandtl numbers, the Nusselt number predicted from this framework is within 20\% of the experimental value, and the root-mean-square velocity and temperature fluctuations also resemble the profiles seen in experimental measurements \citep{herring1964investigation}. The single-mode theory has since been extended to the time-dependent problem \citep{elder1969temporal} and to more general planforms (e.g., hexagonal) \citep{gough1975modal,toomre1977numerical}, showing qualitative agreement with experimental results. The vertical vorticity mode was included within the single-mode equations with hexagon planform by \citep{lopez1983time,murphy1984influence,massaguer1988instability,massaguer1990nonlinear} and is excited beyond a secondary bifurcation of convection rolls. Such single-mode equations have also proved to provide insight into many other problems including rotating convection \citep{baker1975modal}, plane Poiseuille flow \citep{zahn1974nonlinear} and double-diffusive convection \citep{gough1982single,paparella1999sheared,paparella2002shear}. In particular, DNS of single-mode solutions of salt-finger convection show that fingers can tilt leading to the spontaneous formation of large-scale shear, which may be steady or oscillate, together with a staircase-like profile in the horizontally averaged salinity \citep{paparella1999sheared,paparella1997few}. Moreover, in certain asymptotic regimes single-mode solutions may be exact; see, e.g., the high wavenumber (tall and thin) limit of Rayleigh-B\'enard convection \citep{blennerhassett1994nonlinear}, convection in a porous medium \citep{lewis1997high} or salt-finger convection \citep{proctor1986planform}. Such high wavenumber regimes are naturally achieved in convection with strong restraints such as rapid rotation \citep{julien1997fully} or strong magnetic field \citep{julien1999strongly,calkins2016convection,plumley2018self} and the resulting {\it exact} single-mode equations have proved immensely useful for understanding these systems; see the review by \citet{julien2007reduced}. 
 
This work performs bifurcation analysis of vertically confined salt-finger convection using single-mode equations. The resulting equations are solved for the vertical structure of the solutions as a function of the density ratio, the Prandtl number and the assumed horizontal wavenumber. We fix the diffusivity ratio and thermal Rayleigh number and focus almost exclusively on no-slip velocity boundary conditions. We found staircase-like solutions with one (S1), two (S2) and three (S3) steps in the mean salinity profile in the vertical direction, all of which bifurcate from the trivial solution. In each case salinity gradients are expelled from regions of closed streamlines, leading to a mixed layer. Secondary bifurcations of S1 lead to either tilted fingers (TF1) or traveling waves (TW1), both of which break the horizontal reflection symmetry via the spontaneous formation of large-scale shear. Secondary bifurcations of S2 and S3 lead to asymmetric solutions (A2 and A3) that spontaneously break midplane reflection symmetry. 

The stability and accuracy of the single-mode solutions are further analyzed with the assistance of 2D DNS. Near onset, the one-layer solution S1 is stable and corresponds to maximum salinity transport among all known solutions, a fact that is consistent with the prediction of the `relative stability' criterion \citep{malkus1958finite}. The associated Sherwood number (or salinity Nusselt number) near the high wavenumber onset is in excellent agreement with DNS in small horizontal domains. In large domains the DNS reveals a tendency to revert to the characteristic finger scale but the wavenumber of the final state exhibits a strong dependence on initial conditions (if the final state is steady) or the state may remain chaotic. The S2, S3, A2, and A3 solutions are all unstable within the explored parameter regimes, but may be imprinted on the statistics of the chaotic state. At lower density ratios, the S1, S2 and S3 salinity profiles all sharpen and the large-scale shear and tilt angle of the TF1 state increases. The TW1 solutions are also present in this regime, while low Prandtl numbers favor direction-reversing tilted fingers (RTF), a state in which the tilt direction and associated large-scale shear reverse periodically as in the pulsating wave.

The remainder of this paper is organized as follows. \S \ref{sec:single_mode} describes the problem setup and the formulation of the single-mode equations. The equations are solved using numerical continuation and their stability properties are established for Prandtl number $Pr=7$ in \S \ref{sec:result_Pr_7}. The tilted finger and traveling wave states are analyzed for $Pr=0.05$ in \S \ref{sec:results_Pr_0p05}. The paper concludes with a brief summary and discussion in \S \ref{sec:conclusion}.  

\section{Single-mode equations for confined salt-finger convection}
\label{sec:single_mode}

\begin{figure}
(a) \hspace{0.49\textwidth} (b)

    \centering
    
    \includegraphics[width=0.54\textwidth]{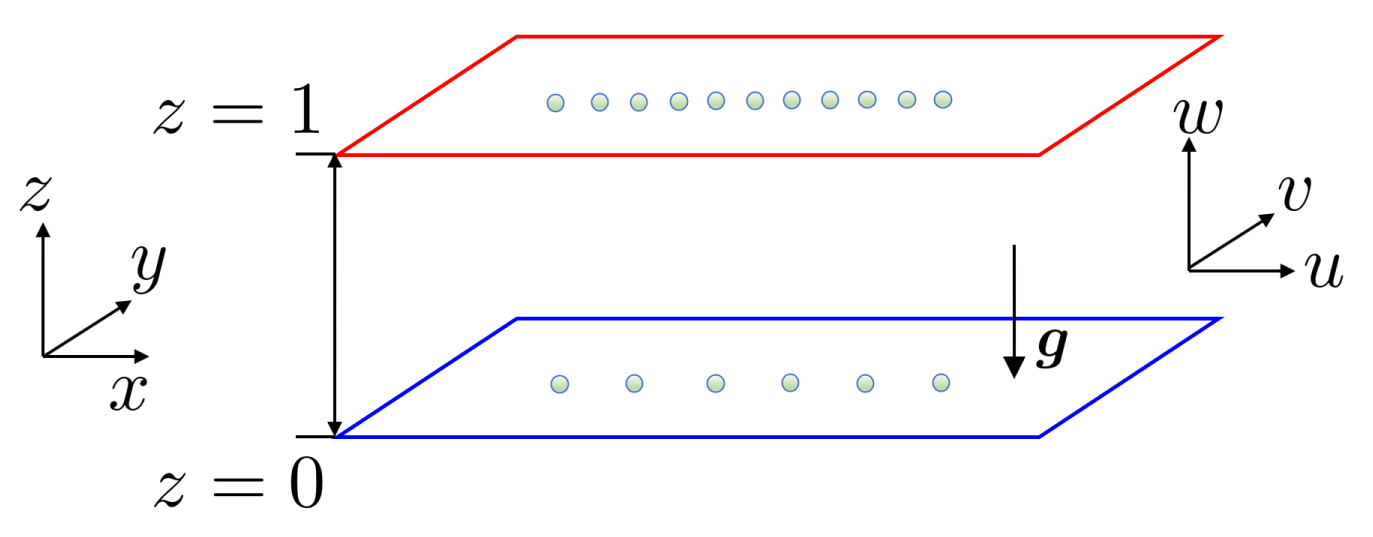}
    \includegraphics[width=0.44\textwidth,trim=-0 -0.3in 0 0]{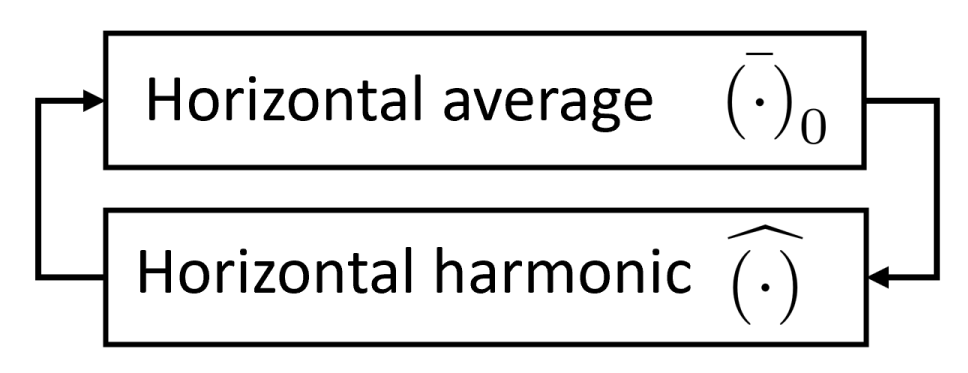}
    \caption{(a) The salt-finger convection setup. The red color indicates a hot plate while the blue color suggests a cold plate. The number of circles on top and bottom plates suggests the salinity at the top is higher than at the bottom. (b) Nonlinear interaction within the single-mode equations \eqref{eq:single_mode}: horizontal averages $\bar{(\cdot)}_0$ influence the horizontal harmonics $\widehat{(\cdot)}$ through \eqref{eq:single_mode_a}-\eqref{eq:single_mode_d}, while the horizontal harmonics $\widehat{(\cdot)}$ contribute to the horizontal averages $\bar{(\cdot)}_0$ through \eqref{eq:single_mode_e}-\eqref{eq:single_mode_g}.}
    \label{fig:illustration}
\end{figure}
We consider a fluid between two infinitely long parallel plates separated by a distance $h$. The temperature and salinity at these two plates are maintained at constant values with the top plate maintained at a higher temperature and salinity, as illustrated in figure \ref{fig:illustration}(a). The equation of state $(\rho_*-\rho_{r*})/\rho_{r*}=-\alpha (T_*-T_{r*})+\beta (S_*-S_{r*})$ is linear, with constant expansion/contraction coefficients $\alpha$, $\beta$ and reference density, temperature, and salinity $\rho_{r*},T_{r*},S_{r*}$, respectively. The subscript $_*$ denotes a dimensional variable. In the following we nondimensionalize the temperature $T_*$ by the temperature difference between the top and bottom layer, $T=T_*/\Delta T$ ($\Delta T>0$), and likewise for the salinity $S_*$, $S=S_*/\Delta S$ ($\Delta S>0$). Spatial variables are nondimensionalized by the height $h$ of the layer while time and velocity are nondimensionalized using the thermal diffusion time $h^2/\kappa_T$ and the corresponding speed $\kappa_T/h$, respectively. Here $\kappa_T$ is the thermal diffusivity. We decompose the temperature and salinity into a linear base state and deviation,
\begin{subequations}
\label{eq:total_T_S}
\begin{align}
    T=z+{\tilde T},\;\;S=z+{\tilde S}, \tag{\theequation a-b}
\end{align}
\end{subequations}
and introduce the velocity field $\boldsymbol{u}:=(u,v,w)$ in Cartesian coordinates $(x,y,z)$ with $z$ in the upward vertical direction. Dropping the tildes and assuming the Boussinesq approximation the system is governed by
\begin{subequations}
\label{eq:convection_double_diffusive}
\begin{align}
    \partial_t\boldsymbol{u}+\boldsymbol{u}\cdot \boldsymbol{\nabla}\boldsymbol{u}=&Pr\nabla^2\boldsymbol{u}-\boldsymbol{\nabla} p+Pr Ra_T\left( T-R_\rho^{-1} S\right)\boldsymbol{e}_z,\label{eq:convection_double_diffusive_mom}\\
    \boldsymbol{\nabla}\cdot \boldsymbol{u}=&0,\label{eq:convection_double_diffusive_mass}\\
    \partial_t T+\boldsymbol{u}\cdot \boldsymbol{\nabla}T+w=&\nabla^2 T,\label{eq:convection_double_diffusive_T}\\
    \partial_t S+\boldsymbol{u}\cdot \boldsymbol{\nabla}S+w=&\tau \nabla^2 S.
    \label{eq:convection_double_diffusive_S}
\end{align}
\end{subequations}
Here $p$ is the dimensionless pressure $p=p_*h^2/\kappa_T^2 \rho_{r*}$ while $\boldsymbol{e}_z$ in \eqref{eq:convection_double_diffusive_mom} is the unit vector in the vertical direction associated with buoyancy. The governing parameters include the Prandtl number, the diffusivity ratio, the density ratio, and the thermal Rayleigh number defined by:
\begin{subequations}
\begin{align}
    Pr:=\frac{\nu}{\kappa_T},\;\;\tau:=\frac{\kappa_S}{\kappa_T},\;\;R_\rho:=\frac{\alpha \Delta T}{\beta \Delta S},\;\; Ra_T:=\frac{ g \alpha \Delta T h^3}{\kappa_T\nu }, \tag{\theequation a-d}
\end{align} 
\end{subequations}
where $\nu$ is the viscosity and $\kappa_S$ is the salinity diffusivity. 

We impose constant temperature and salinity as boundary conditions at the top and the bottom plates:
\begin{subequations}
\label{eq:BC}
\begin{align}
    &T(x,y,z=0,t)=T(x,y,z=1,t)=0,\label{eq:BC_T_S_bom}\\
    &S(x,y,z=0,t)=S(x,y,z=1,t)=0, \label{eq:BC_T_S_top}
\end{align}
\end{subequations}
while for the velocity we adopt no-slip boundary conditions:
\begin{align}
        \boldsymbol{u}(x,y,z=0,t)=\boldsymbol{u}(x,y,z=1,t)=0.\label{eq:BC_no_slip}
\end{align}
Periodic boundary conditions in the horizontal are imposed on all variables.

We next formulate single-mode equations following the procedure in \citet{herring1963investigation,herring1964investigation,gough1975modal,gough1982single,paparella1999sheared,paparella2002shear}. The single-mode ansatz is
\begin{subequations}
\label{eq:normal_mode}
\begin{align}
    S(x,y,z,t)=&\bar{S}_0(z,t)+\widehat{S}(z,t)\,e^{\text{i}(k_x x+k_y y)}+c.c.,\label{eq:normal_mode_S}\\
    T(x,y,z,t)=&\bar{T}_0(z,t)+\widehat{T}(z,t)\,e^{\text{i}(k_x x+k_y y)}+c.c.,\label{eq:normal_mode_T}\\
    \boldsymbol{u}(x,y,z,t)=&\bar{U}_0(z,t)\boldsymbol{e}_x+\widehat{\boldsymbol{u}}(z,t)\,e^{\text{i}(k_x x+k_y y)}+c.c.,\label{eq:normal_mode_u}\\
    p(x,y,z,t)=&\bar{P}_0(z,t)+\widehat{p}(z,t)\,e^{\text{i}(k_x x+k_y y)}+c.c.,\label{eq:normal_mode_p}
\end{align}
\end{subequations}
where c.c. denotes the complex conjugate. Equation \eqref{eq:normal_mode_S} decomposes the departure of the salinity from the linear profile into a horizontally averaged quantity $\bar{S}_0(z,t)$ and a single harmonic in the horizontal direction associated with the wavenumber pair $(k_x,k_y)$ and characterized by the complex amplitude $\widehat{S}(z,t)$. The temperature is decomposed similarly. To allow for mean flow in the horizontal we decompose the velocity into a large-scale shear $\bar{U}_0(z,t)\boldsymbol{e}_x$ and a harmonic associated with the same wavenumber pair $(k_x,k_y)$ as in equation \eqref{eq:normal_mode_u}, assuming that the large-scale shear $\bar{U}_0$ is generated in the $x$-direction. This is appropriate in a 2D configuration, but in 3D the large-scale shear can be oriented in principle in any horizontal direction, a possibility that is left for future study. The horizontally averaged vertical velocity is zero based on the continuity equation \eqref{eq:convection_double_diffusive_mass} and the boundary conditions $w(z=0)=w(z=1)=0$. The horizontally averaged quantities $\bar{(\cdot)}_0$ are real and the amplitudes of the horizontal harmonics $\widehat{(\cdot)}$ are in general complex. Equations \eqref{eq:normal_mode} assume a horizontal planform in the form of square cells, an assumption that also includes 2D rolls when $k_y=0$. Other planforms such as hexagons generate additional self-interaction terms in the single-mode equations \citep{gough1975modal} and we leave this extension to future work.

We now substitute \eqref{eq:normal_mode} into the governing equations \eqref{eq:convection_double_diffusive}, dropping all harmonics beyond the first, and balance separately the horizontally averaged components and the harmonic components. We eliminate the horizontally averaged pressure via $-\partial_z \bar{P}_0+PrRa_T(\bar{T}_0-R_\rho^{-1}\bar{S}_0)=0$ and eliminate the harmonic component of the pressure using the continuity equation \eqref{eq:convection_double_diffusive_mass}. The resulting single-mode equations can then be expressed in terms of the vertical velocity $w$ and vertical vorticity $\zeta:=\partial_y u-\partial_x v$:
\begin{subequations}
\label{eq:single_mode}
\begin{align}
\partial_t\widehat{\nabla}^2\widehat{w}+\text{i}k_x\bar{U}_0\widehat{\nabla}^2\widehat{w}-\text{i}k_x\bar{U}_0''\widehat{w}=&Pr\widehat{\nabla}^4\widehat{w}+Pr\widehat{\nabla}_\perp^2 Ra_T\left(\widehat{T}-R_\rho^{-1}\widehat{S}\right),\label{eq:single_mode_a}\\
\partial_t \widehat{\zeta}+\text{i}k_x \bar{U}_0 \widehat{\zeta}+\text{i}k_y \bar{U}_0'\widehat{w}=&Pr\widehat{\nabla}^2 \widehat{\zeta},\label{eq:single_mode_b}\\
\partial_t\widehat{T}+\text{i}k_x\bar{U}_0\widehat{T}+\widehat{w}\partial_z \bar{T}_0+\widehat{w}=&\widehat{\nabla}^2\widehat{T},\label{eq:single_mode_c}\\
\partial_t\widehat{S}+\text{i}k_x\bar{U}_0\widehat{S}+\widehat{w}\partial_z \bar{S}_0+\widehat{w}=&\tau \widehat{\nabla}^2\widehat{S},\label{eq:single_mode_d}\\
\partial_t \bar{U}_0+\partial_z\left(\widehat{w}^*\widehat{u}+\widehat{w}\widehat{u}^*\right)=&Pr\partial_z^2\bar{U}_0,\label{eq:single_mode_e}\\
\partial_t \bar{T}_0+\partial_z\left(\widehat{w}^*\widehat{T}+\widehat{w}\widehat{T}^*\right)=&\partial_z^2 \bar{T}_0,\label{eq:single_mode_f}\\
\partial_t \bar{S}_0+\partial_z\left(\widehat{w}^*\widehat{S}+\widehat{w}\widehat{S}^*\right)=&\tau \partial_z^2 \bar{S}_0,\label{eq:single_mode_g}\\
\widehat{u}=&\frac{\text{i}k_x\partial_z \widehat{w}}{k_x^2+k_y^2}-\frac{\text{i}k_y}{k_x^2+k_y^2}\widehat{\zeta},\label{eq:single_mode_h}
\end{align}
\end{subequations}
where the superscript $^*$ denotes a complex conjugate and
$\widehat{\nabla}^2:=\partial_z^2-k_x^2-k_y^2$, $\widehat{\nabla}^2_\perp:=-k_x^2-k_y^2$, $\widehat{\nabla}^4:=\partial_z^4-2(k_x^2+k_y^2)\partial_z^2+(k_x^2+k_y^2)^2$, $\bar{U}_0':=\partial_z \bar{U}_0$ and $\bar{U}_0'':=\partial_z^2 \bar{U}_0$. The corresponding boundary conditions for the salinity and temperature are:
\begin{subequations}
\label{eq:BC_T_S_single_mode}
\begin{align}
    &\widehat{S}(z=0,t)=\widehat{S}(z=1,t)=\widehat{T}(z=0,t)=\widehat{T}(z=1,t)\\
    =\,&\bar{S}_0(z=0,t)=\bar{S}_0(z=1,t)=\bar{T}_0(z=0,t)=\bar{T}_0(z=1,t)\\
    =\,&0,
\end{align}
\end{subequations}
while the no-slip boundary conditions in \eqref{eq:BC_no_slip} correspond to
\begin{subequations}
\label{eq:BC_no_slip_single_mode}
\begin{align}
    &\widehat{w}(z=0,t)=\widehat{w}(z=1,t)=\partial_z\widehat{w}(z=0,t)=\partial_z\widehat{w}(z=1,t)\\
    =\,&\widehat{\zeta}(z=0,t)=\widehat{\zeta}(z=1,t)=\bar{U}_0(z=0,t)=\bar{U}_0(z=1,t)\\
    =\,&0.
\end{align}
\end{subequations}

\begin{table}
    \centering
    \begin{tabular}{ccc}
        \hline
\backslashbox{Symmetry name}{Formulation}         & Primitive equations in \eqref{eq:convection_double_diffusive} & Single-mode in \eqref{eq:single_mode} \\
         \hline
        Midplane reflection  &  \thead{$z\rightarrow 1-z$\\ $(w,T,S)\rightarrow-(w,T,S)$ } & \thead{$z\rightarrow 1-z$ \\
        $(\widehat{w},\widehat{T},\bar{T}_0,\widehat{S},\bar{S}_0)\rightarrow -(\widehat{w},\widehat{T},\bar{T}_0,\widehat{S},\bar{S}_0)$} \\ 
        Horizontal reflection &\thead{$x\rightarrow-x$\\ 
        $u\rightarrow -u$} & \thead{$k_x\rightarrow -k_x$ \\ $(\widehat{u}, \bar{U}_0,\widehat{\zeta})\rightarrow -(\widehat{u}, \bar{U}_0,\widehat{\zeta})$} \\
        Horizontal translation & $x\rightarrow x+\delta x$ & $\widehat{(\cdot)}\rightarrow \widehat{(\cdot)}e^{\text{i}k_x \delta x}$ \\
        \hline
    \end{tabular}
    \caption{Symmetry properties of the primitive equations in \eqref{eq:convection_double_diffusive} and the single-mode equations in \eqref{eq:single_mode}.}
    \label{tab:symmetry}
\end{table}

Equations \eqref{eq:single_mode_a}-\eqref{eq:single_mode_d} are the governing equation for the first harmonic in the presence of the large-scale fields $\bar{U}_0$, $\bar{T}_0$ and $\bar{S}_0$. In particular, equation \eqref{eq:single_mode_a} is the Orr-Sommerfeld equation modified by an additional buoyancy term, while \eqref{eq:single_mode_b} is known as the Squire equation for the vertical vorticity \citep{schmid2012stability}. Note that the large-scale shear $\bar{U}_0$ is self-induced (i.e., a variable to be solved for) instead of an imposed background shear. Thus the terms involving $\bar{U}_0$ in equations \eqref{eq:single_mode_a}-\eqref{eq:single_mode_d} are also \emph{nonlinear}. The remaining equations \eqref{eq:single_mode_e}-\eqref{eq:single_mode_g} are the governing equations for the horizontally averaged quantities, distorted by the harmonic fluctuations. The resulting nonlinear interaction between horizontally averaged quantities and horizontal harmonics is summarized schematically in figure \ref{fig:illustration}(b). Equation \eqref{eq:single_mode_h} is used to obtain the horizontal velocity $\widehat{u}$ from the $\widehat{w}$ and $\widehat{\zeta}$ required for the computation of the Reynolds stress in \eqref{eq:single_mode_e}. 

Similar single-mode equations were previously used to study double-diffusive convection focusing on the diffusive regime; see, e.g., \citet[equations (3.7)-(3.11)]{gough1982single}, but did not include the coupling to the large large-scale shear $\bar{U}_0$. A numerical simulation of the same single-mode equation in a 2D configuration (i.e., with $k_y=0$) and stress-free boundary conditions was employed to study double-diffusive convection in both the fingering regime \citep{paparella1999sheared} and the diffusive regime \citep{paparella2002shear}.

Here we focus on the bifurcation properties of the single-mode equations \eqref{eq:single_mode}. These reflect the symmetries of the primitive equations \eqref{eq:convection_double_diffusive}, including midplane reflection, horizontal reflection, and horizontal translation, as summarized in table \ref{tab:symmetry}. In the following we use the numerical software pde2path \citep{uecker2014pde2path,uecker2021numerical} to compute strongly nonlinear solutions of the above problem as a function of the system parameters and analyze their stability. The vertical direction is discretized using the Chebyshev collocation method with derivatives calculated using the Chebyshev differentiation matrix \citep{weideman2000matlab} implemented following \citet{uecker2021pde2path}. The number of grid points used, including the boundary, ranges from $N_z=65$ for results at $R_\rho=40$ to $N_z=257$ for results at $R_\rho=2$; our continuation in $R_\rho$ also uses $N_z=257$ grid points. The solutions are obtained by arclength continuation including prediction and Newton-correction steps from a given solution profile at a nearby parameter; see \citet[\S 2.1]{uecker2014pde2path} and \citet[\S 3.1]{uecker2021numerical}. The tolerance of the maximal absolute value of the residue at each vertical location ($L_\infty$ norm) is set to $10^{-6}$.

The presence of horizontal translation symmetry within the single-mode equations requires a phase condition whenever $k_x\neq 0$ in order to fix the solution phase and obtain a unique solution. The implementation  of this condition following \citet{rademacher2017symmetries} requires the predictor $\boldsymbol{\psi}(z,t)$ from a solution $\boldsymbol{\psi}_{\text{old}}(z,t)$ to be orthogonal to $\text{i}\boldsymbol{\psi}_{\text{old}}(z,t)$:
\begin{align}
     \int_{0}^1 \text{i}\boldsymbol{\psi}_{\text{old}}(z,t)[\boldsymbol{\psi}(z,t)-\boldsymbol{\psi}_{\text{old}}(z,t)]^*  dz=0,\label{eq:phase_old}
\end{align}
where  
\begin{align}
\boldsymbol{\psi}(z,t):=\begin{bmatrix}\widehat{w}(z,t),\widehat{\zeta}(z,t),\widehat{T}(z,t),\widehat{S}(z,t)\end{bmatrix}^{\text{T}}.
\end{align}
The horizontally averaged modes are not involved in setting the phase.

To compute a steady nonlinear wave traveling in the $x$ direction with speed $c$ we write equations \eqref{eq:single_mode_a}-\eqref{eq:single_mode_d} in the comoving frame,
\begingroup
\allowdisplaybreaks
\begin{subequations}
\label{eq:single_mode_phase}
\begin{align}
    \partial_t\widehat{\nabla}^2\widehat{w}+\text{i}k_x\bar{U}_0\widehat{\nabla}^2\widehat{w}-\text{i}ck_x \widehat{\nabla}^2 \widehat{w}-\text{i}k_x\bar{U}_0''\widehat{w}=&Pr\widehat{\nabla}^4\widehat{w}+Pr\widehat{\nabla}_\perp^2 Ra_T\left(\widehat{T}-R_\rho^{-1}\widehat{S}\right),\label{eq:single_mode_a_phase}\\ \partial_t \widehat{\zeta}+\text{i}k_x \bar{U}_0 \widehat{\zeta}-\text{i}ck_x\widehat{\zeta}+\text{i}k_y \bar{U}_0'\widehat{w}=&Pr\widehat{\nabla}^2 \widehat{\zeta},\label{eq:single_mode_b_phase}\\
\partial_t\widehat{T}+\text{i}k_x\bar{U}_0\widehat{T}-\text{i}ck_x\widehat{T}+\widehat{w}\partial_z \bar{T}_0+\widehat{w}=&\widehat{\nabla}^2\widehat{T},\label{eq:single_mode_c_phase}\\
\partial_t\widehat{S}+\text{i}k_x\bar{U}_0\widehat{S}-\text{i}ck_x \widehat{S}+\widehat{w}\partial_z \bar{S}_0+\widehat{w}=&\tau \widehat{\nabla}^2\widehat{S},\label{eq:single_mode_d_phase}
\end{align}
\end{subequations}
\endgroup
and set the time derivatives in these equations and in \eqref{eq:single_mode_e}-\eqref{eq:single_mode_g} to zero. With the phase condition \eqref{eq:phase_old} the resulting problem has a unique nonlinear eigenvalue $c$ and associated solution profile. Both are updated at each step of the continuation procedure. Steady solutions have $c=0$, typically within the order of machine precision.

The stability of each solution is examined via the eigenvalues of the associated Jacobian matrix. The eigenvalue computation uses the \texttt{eigs} command in MATLAB to compute a subset of the eigenvalues. We compute 40 eigenvalues near 0 to identify bifurcation points and also compute 1 eigenvalue near 100 to help identify unstable eigenvalues; see discussion in \citet[Remark 3.12(b)]{uecker2021numerical}. For validation we reproduced the results for single-mode equations for Rayleigh-B\'enard convection \citep{herring1963investigation,herring1964investigation,toomre1977numerical}, as well as the high wavenumber asymptotic single-mode equations for this problem \citep[section 3]{bassom1994strongly} and for porous medium convection \citep[section 3]{lewis1997high}. Selected solution profiles and eigenvalues obtained from pde2path were also validated against the results obtained from the nonlinear boundary value problem (NLBVP) and eigenvalue problem (EVP) solvers in Dedalus \citep{burns2020dedalus}, where the EVP solver is chosen to return the full set of eigenvalues. Finally, selected predictions from the bifurcation diagram and stability properties (figure \ref{fig:bif_diag_low_Ra_S2T_low_Pr}(a)) were also validated against direct numerical simulations (IVP solver) of the single-mode equations \eqref{eq:single_mode} using Dedalus \citep{burns2020dedalus}. 

We also performed 2D direct numerical simulations (DNS) of the primitive equations \eqref{eq:convection_double_diffusive} using Dedalus \citep{burns2020dedalus} to further analyze the accuracy and stability of the solutions obtained from the bifurcation analysis of the single-mode equations. We focus on 2D domains with periodic boundary conditions in the horizontal and the no-slip boundary conditions \eqref{eq:BC_no_slip} in the vertical direction. We use a Chebyshev spectral method in the vertical direction with $N_z=128$ grid points with a Fourier spectral method in the horizontal direction with $N_x=128$ and a dealiasing scaling factor 3/2. To check the accuracy we doubled the number of grid points in both $x$ and $z$ directions for selected results and confirmed that the resulting Sherwood number $Sh$ in figure \ref{fig:bif_diag_R_rho_T2S_tau_0p01} (steady rolls resembling the S1 solution) does not change up to eight decimal places. Time is advanced using a 3rd-order 4-stage diagonal implicit Runge-Kutta (DIRK) scheme coupled with a 4-stage explicit Runge-Kutta (ERK) scheme (RK443) \citep[section 2.8]{ascher1997implicit}. We mention that we do not expect the DNS results to coincide in all cases with the single-mode results. This is because the former include, in principle, all spatial harmonics of the fundamental wavenumber; it is precisely these that are omitted from the single-mode theory. We are interested in identifying parameter regimes in which the single-mode theory provides quantitatively accurate results as well as regimes in which it fails.  In all cases, the horizontal domain is selected based on the wavenumber employed in constructing the single-mode solutions, as described in detail in the next section.

The single-mode equations are parametrized by the horizontal wavenumbers $k_x$ and $k_y$, in addition to the physical parameters, and the main challenge of the approach is therefore the correct choice of these wavenumbers, cf.~\citet{toomre1977numerical}. In fact, salt fingers have a reasonably well-defined horizontal scale $d$ that depends on both the thermal and salinity Rayleigh numbers, and hence on the density ratio, as seen in both 3D DNS \citep{yang2016scaling} and in experiments \citep{hage2010high}. Our approach permits us to examine the properties of fingers of different widths through the choice of the wavenumber $k_x$. In the following, we find that if $k_x$ (and $k_y$) is close to the onset wavenumber the resulting single-mode solution represents an accurate description of the system, in the sense that DNS in a domain of width $L_x=2\pi/k_x$ returns solutions with the same properties. However, wider fingers in DNS with $L_x=2\pi/k_x$ generally break up into fingers with wavenumber $k_x\sim 2\pi/d$. In the following we use the wavenumber $k_x$ as a proxy for the domain size and note when such states are in fact unstable to perturbations with a higher wavenumber, resulting in several fingers in the original $2\pi/k_x$ domain; in single-mode theory such instabilities are of course excluded.

\section{Single-mode solutions and stability at $Pr=7$}

\label{sec:result_Pr_7}

\begin{table}
    \centering
    \begin{tabular}{cccccc}
    \hline
    Abbreviation & Description & Bifurcate from & Stability & $\bar{U}_0(z)$ & Color \\
    \hline
       S1  & Symmetric one-layer solutions  & Trivial & S/U & $=0$&  Black ({\color{black}$\mline\mline\mline$})\\
        S2 & Symmetric two-layer solutions & Trivial & U & $=0$&  Red ({\color{red}$\mline\mline\mline$})\\
        S3 & Symmetric three-layer solutions & Trivial & U & $=0$ & Blue ({\color{blue}$\mline\mline\mline$})\\
        TF1 & Tilted fingers & S1 & S/U & $\neq 0$& Magenta ({\color{magenta}$\mline\mline\mline$})\\
        TW1 & Traveling waves & S1 & S/U & $\neq 0$ & Green ({\color{green}$\mline\mline\mline$})\\
        A2 & Asymmetric two-layer solutions & S2 & U & $=0$ & Cyan ({\color{cyan}$\mline\mline\mline$})\\
        A3 & Asymmetric three-layer solutions & S3  & U & $=0$ & Brown ({\color{darkorange}$\mline\mline\mline$})\\
        \hline
    \end{tabular}
    \caption{Summary of solution features, stability, induced large-scale shear $\bar{U}_0(z)$ and line color employed in the bifurcation diagrams (figures \ref{fig:bif_diag_low_Ra_S2T}, \ref{fig:bif_diag_low_Ra_S2T_stress_free}, \ref{fig:bif_diag_mid_Ra_S2T}, \ref{fig:bif_diag_R_rho_T2S_tau_0p01} and \ref{fig:bif_diag_low_Ra_S2T_low_Pr}). Symmetry is with respect to mid-plane reflection. S/U indicates both stable and unstable solutions exist along this solution branch and U indicates that only unstable solutions were found. }
    \label{tab:summary_case}
\end{table}

We start with the results for $Pr=7$, a Prandtl number value appropriate to oceanographic applications, and fix $\tau=0.01$, corresponding to the diffusivity ratio between salinity and temperature. The thermal Rayleigh number is fixed at $Ra_T=10^5$ for a direct comparison with DNS results \citep{yang2015salinity,yang2016scaling}. Table \ref{tab:summary_case} provides a summary of the solutions, all of which are described in more detail below. All the solutions in the table are steady-state solutions, including the traveling waves (TW1) which are computed as steady states in a frame moving with the phase speed $c$ of the wave. This speed solves a nonlinear eigenvalue problem. The bifurcation diagrams presented in this work (figures \ref{fig:bif_diag_low_Ra_S2T}, \ref{fig:bif_diag_low_Ra_S2T_stress_free}, \ref{fig:bif_diag_mid_Ra_S2T}, \ref{fig:bif_diag_R_rho_T2S_tau_0p01} and \ref{fig:bif_diag_low_Ra_S2T_low_Pr}) all show the time-averaged Sherwood number $Sh$ as a function of the horizontal domain size $L_x$ or equivalently the fundamental wavenumber $k_x=2\pi/L_x$, where
\begin{align}
    Sh:=\langle \partial_z S |_{z=0} \rangle_{h,t}+1
\end{align}
with $\langle \cdot \rangle_{h,t}$ indicating average in both the horizontal direction and over time. Without confusion, $\langle \cdot \rangle_{h,t}$ is also referred to as a \emph{mean}. In particular, for steady-state single-mode solutions, we obtain
\begin{align}
    Sh=\langle \partial_z \bar{S}_0 |_{z=0}\rangle_{t}+1=\partial_z\bar{S}_0|_{z=0}+1,
\end{align}
where $\langle \cdot \rangle_t$ denotes time-averaging. We also use $\langle \cdot \rangle_h$ to denote averaging in the horizontal. For time-dependent solutions such as those obtained from DNS, these two types of averaging yield distinct results.

In the bifurcation diagrams reported below, we use thick lines to indicate stable solutions while thin lines represent unstable solutions. When the maximum amplitude of the large-scale shear mode $\underset{z}{\text{max}}\;|\bar{U}_0(z)|$ along a branch is of the order of machine precision or smaller we report it as zero (see table \ref{tab:summary_case}). In all other cases $\bar{U}_0(z)\neq 0$. When the large-scale shear is not generated, $\bar{U}_0=0$, the harmonic quantities $\widehat{w}(z,t)$, $\widetilde{u}(z,t):=-\text{i}\widehat{u}(z,t)$, $\widehat{S}(z,t)$ and $\widehat{T}(z,t)$ are real for a suitably chosen phase. Here, the definition of $\widetilde{u}(z,t)$ is motivated by the continuity equation $\text{i}k_x \widehat{u}+\partial_z\widehat{w}=0$ indicating that for a suitable choice of phase both $\widetilde{u}(z,t)$ and $\widehat{w}(z,t)$ may be simultaneously real-valued. This is the case whenever $\bar{U}_0=0$. However, for solutions associated with an induced large-scale shear ($\bar{U}_0\neq0$) this is no longer so, and we denote the real and imaginary parts of the harmonic profile by $\mathcal{R}e[\cdot]$ and $\mathcal{I}m[\cdot]$, respectively. We refer to the base state in which all variables within the single-mode theory vanish as the trivial state. This state corresponds to $Sh=1$. Instabilities of this state generate the primary solution branches referred to below as S1, S2 and S3. These branches are continued from their respective origin on the trivial branch, and their stability and secondary bifurcation points are examined. These give rise to the TF1, TW1, A2 and A3 solution branches which are also continued as detailed in table \ref{tab:summary_case}. Except for the results with stress-free velocity boundary conditions in subsection \ref{subsec:results_Pr_7_stress_free}, all our results are computed for no-slip boundary conditions.

\subsection{Solution profiles and comparison with DNS at $R_\rho=40$}
\label{subsec:results_Pr_7_R_rho_40}

We start from the dynamics near onset when $R_\rho=40$, i.e., for $R_\rho\tau\sim O(1)$. This is the parameter regime close to the onset of the salt-finger instability, which occurs at $R_{\rho,{\rm crit}}=1/\tau$ in vertically periodic domains \citep[equation (2.4)]{radko2013double} previously studied by \citet{radko1999salt,radko2000finite} and \citet{xie2017reduced,xie2019jet}. We focus on the bifurcation diagram as a function of $k_x$ measuring the finger width, and compare the results with DNS in domains of length $L_x=2\pi/k_x$. This approach provides insight into the effects of the domain size on a system with a characteristic scale, here $d$, much as in Rayleigh-B\'enard convection \citep{van2012flow,wagner2013aspect,goluskin2014convectively}.

\begin{figure}
    \centering

(a) 2D: $k_y=0$ \hspace{0.35\textwidth} (b) 3D: $k_y=k_x$

    \includegraphics[width=0.49\textwidth]{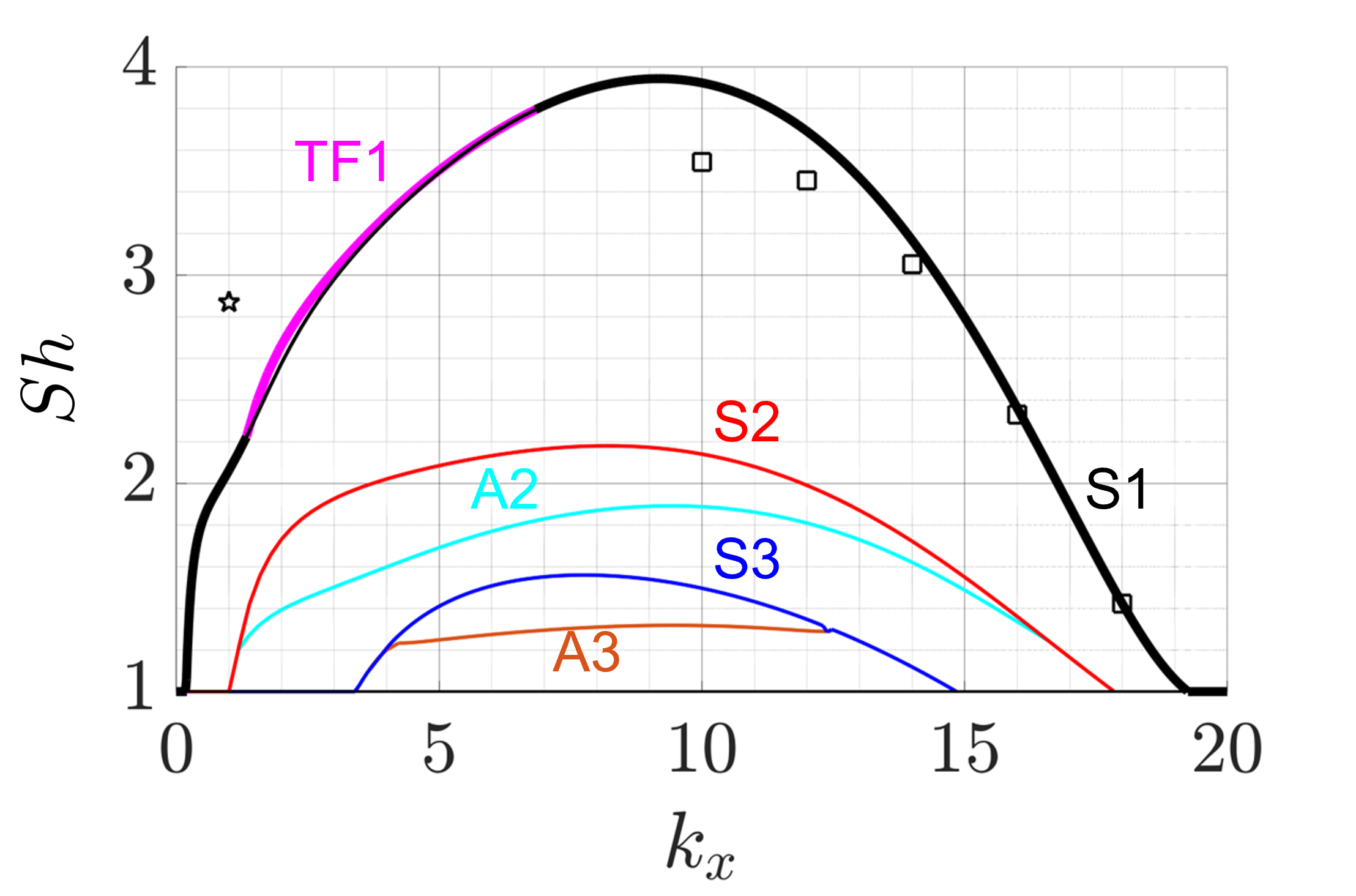}
    \includegraphics[width=0.49\textwidth]{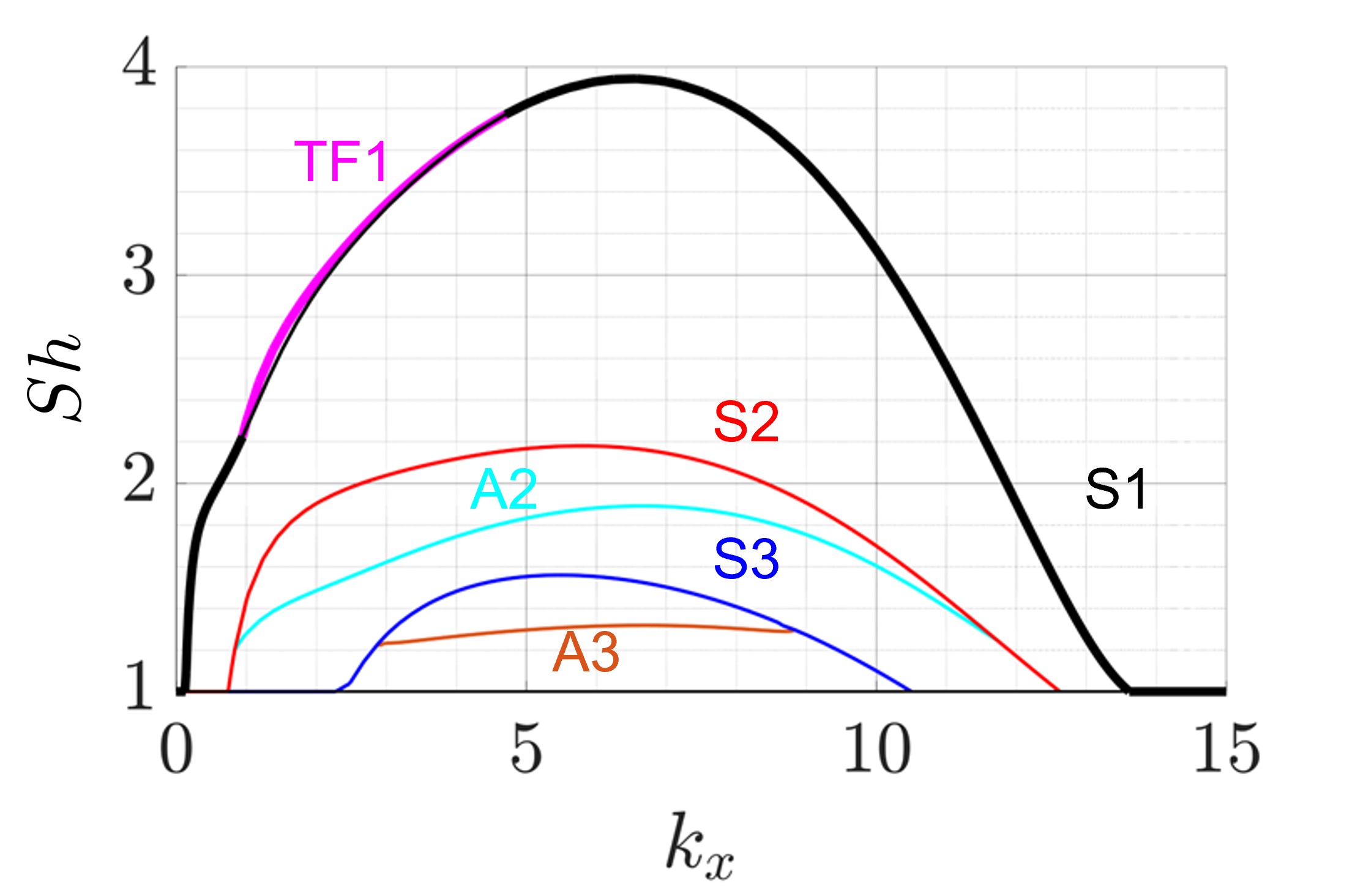}
    
    (c) \hspace{0.49\textwidth} (d)
    \includegraphics[width=0.49\textwidth]{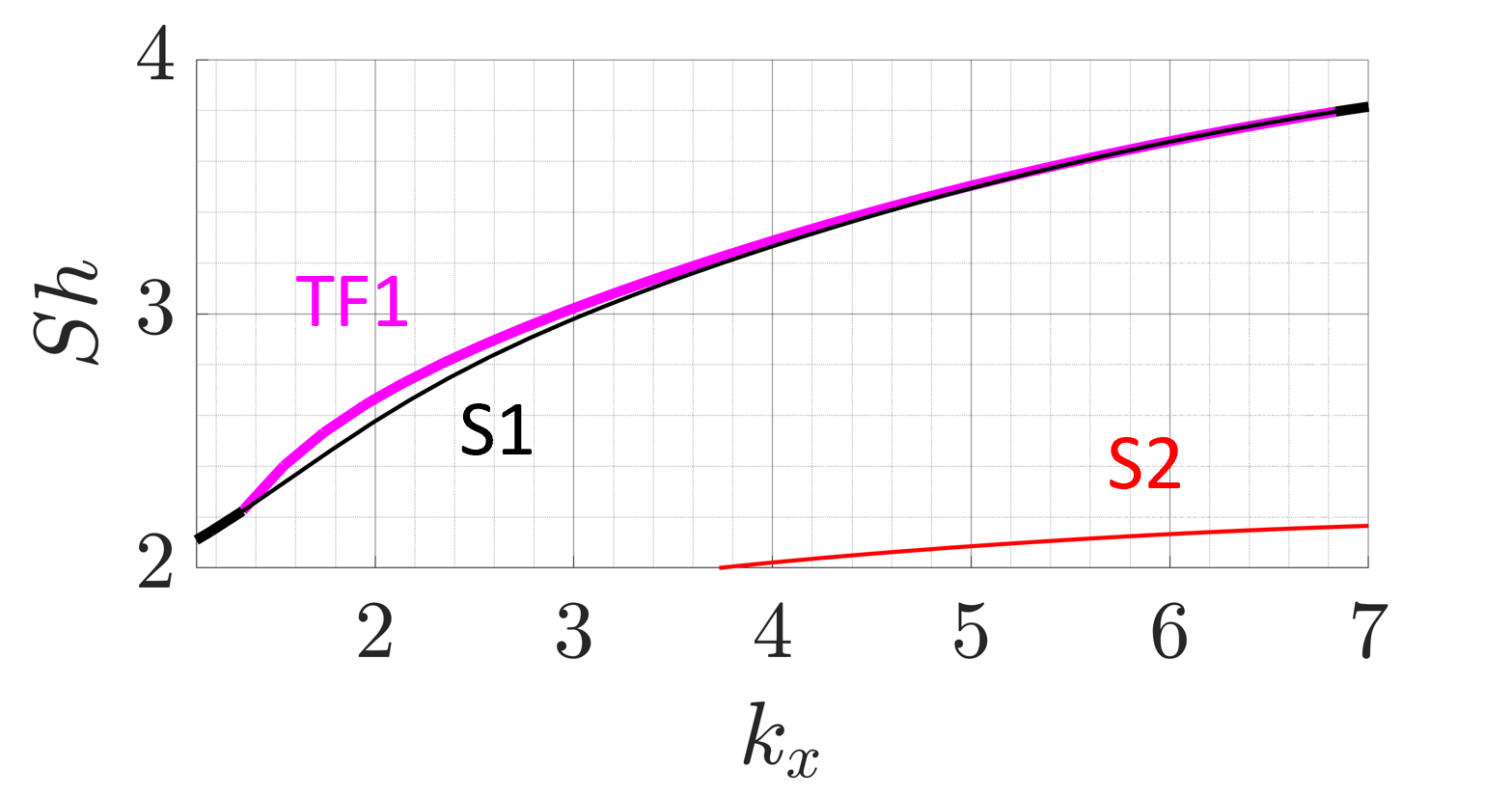}
    \includegraphics[width=0.49\textwidth]{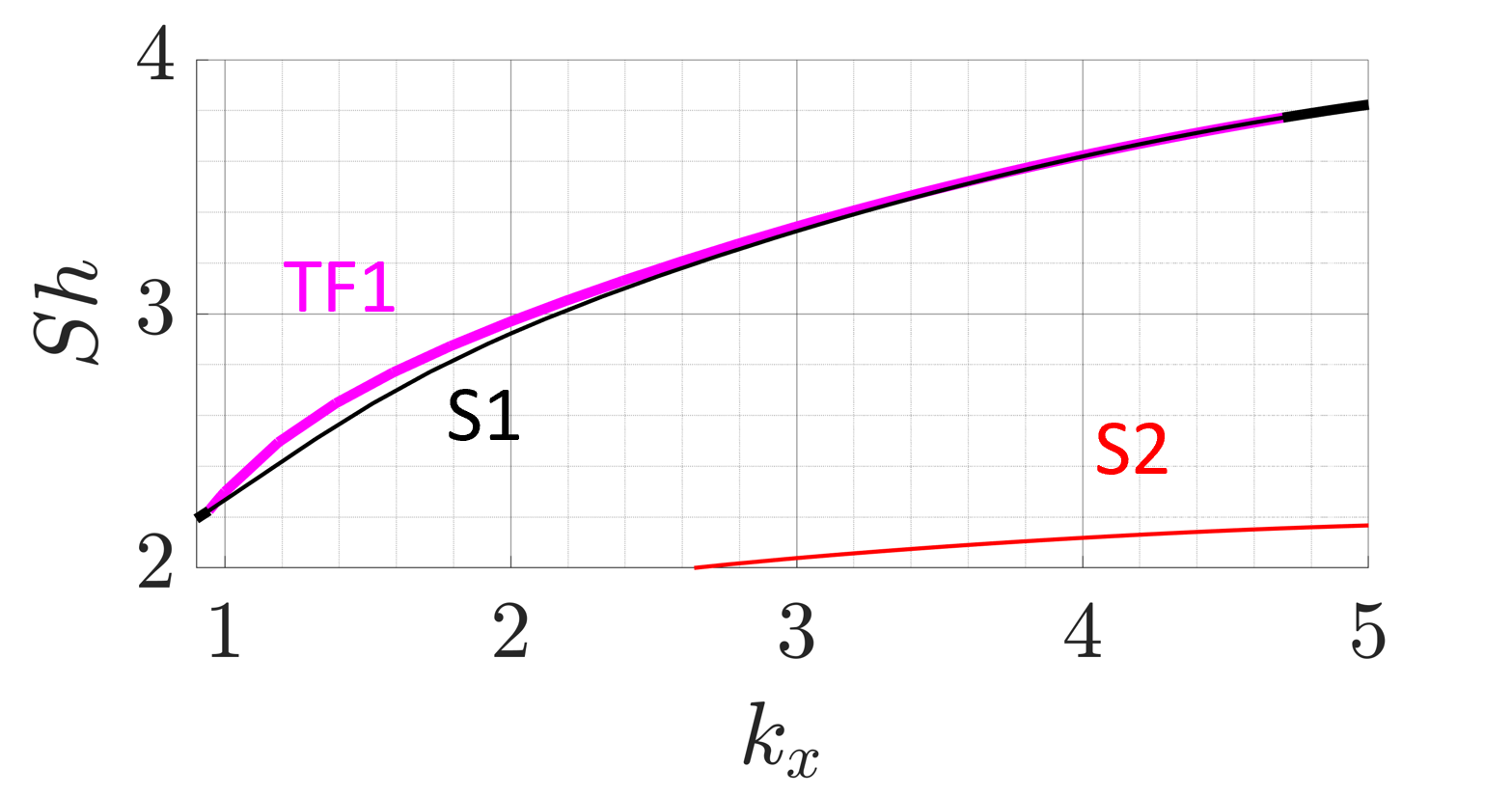}
    \caption{Bifurcation diagrams as a function of the fundamental wavenumber $k_x$ from the single-mode equations \eqref{eq:single_mode} for (a) 2D: $k_y=0$, (b) 3D: $k_y=k_x$, and the parameters $R_\rho=40$, $Pr=7$, $\tau=0.01$, $Ra_T=10^5$. The black squares show the $Sh$ of the steady state reached using 2D DNS in domains of size $L_x=2\pi /k_x$ (table \ref{tab:DNS_transition_low_Ra_S2T_Pr_7}); the black pentagram corresponds to DNS with $L_x=2\pi$ displaying persistent chaotic behavior. Panels (c) and (d) show zooms of the 2D and 3D results near the TF1 branch, respectively.}
    \label{fig:bif_diag_low_Ra_S2T}
\end{figure}

Figure \ref{fig:bif_diag_low_Ra_S2T} shows the bifurcation diagram at $R_\rho=40$ with no-slip boundary conditions. We consider both 2D results with $k_y=0$ in figure \ref{fig:bif_diag_low_Ra_S2T}(a) and 3D results with $k_y=k_x$ in figure \ref{fig:bif_diag_low_Ra_S2T}(b) indicating the same aspect ratio in the $x$ and $y$ horizontal directions. Here, the bifurcation diagram for the 3D states with no large-scale flow $\bar{U}_0$ can be transformed into the diagram for the 2D states upon defining an equivalent 2D horizontal wavenumber
\begin{align}
    k_{x,2D}=\sqrt{k_x^2+k_y^2}.
    \label{eq:kx_2D}
\end{align}
This is because the horizontal wavenumbers in the single-mode equations \eqref{eq:single_mode} for these states always appear in the combination $k_x^2+k_y^2$. This is not the case when $\bar{U}_0\neq 0$ and in this case the TF1 states are indeed no longer identical as elaborated in figures \ref{fig:bif_diag_mid_Ra_S2T} and \ref{fig:bif_diag_low_Ra_S2T_low_Pr} below.

\begin{figure}
\centering
    (a) \hspace{0.2\textwidth} (b) \hspace{0.2\textwidth} (c) \hspace{0.2\textwidth} (d)
    
    \includegraphics[width=0.24\textwidth]{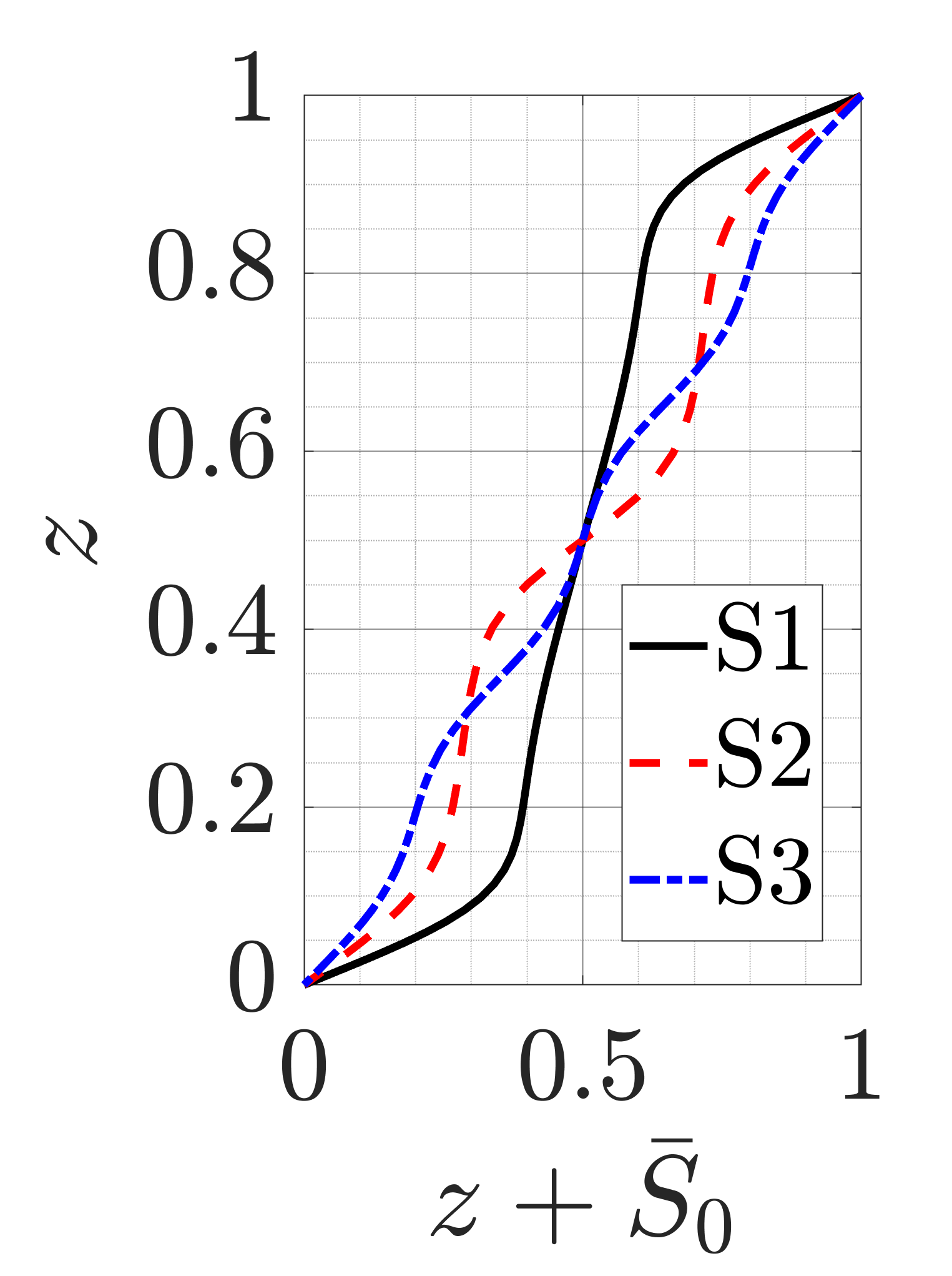}
    \includegraphics[width=0.24\textwidth]{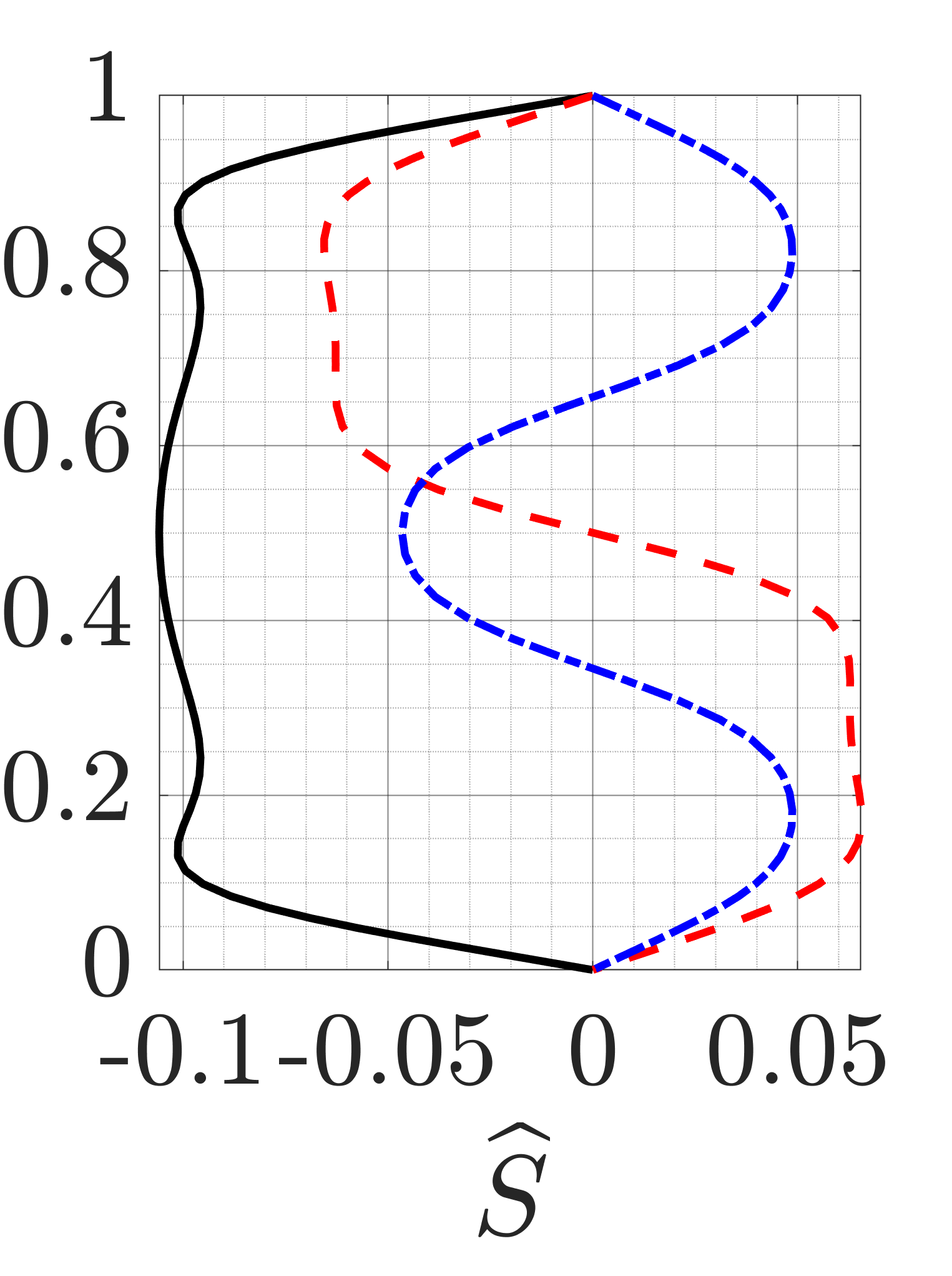}
    \includegraphics[width=0.24\textwidth]{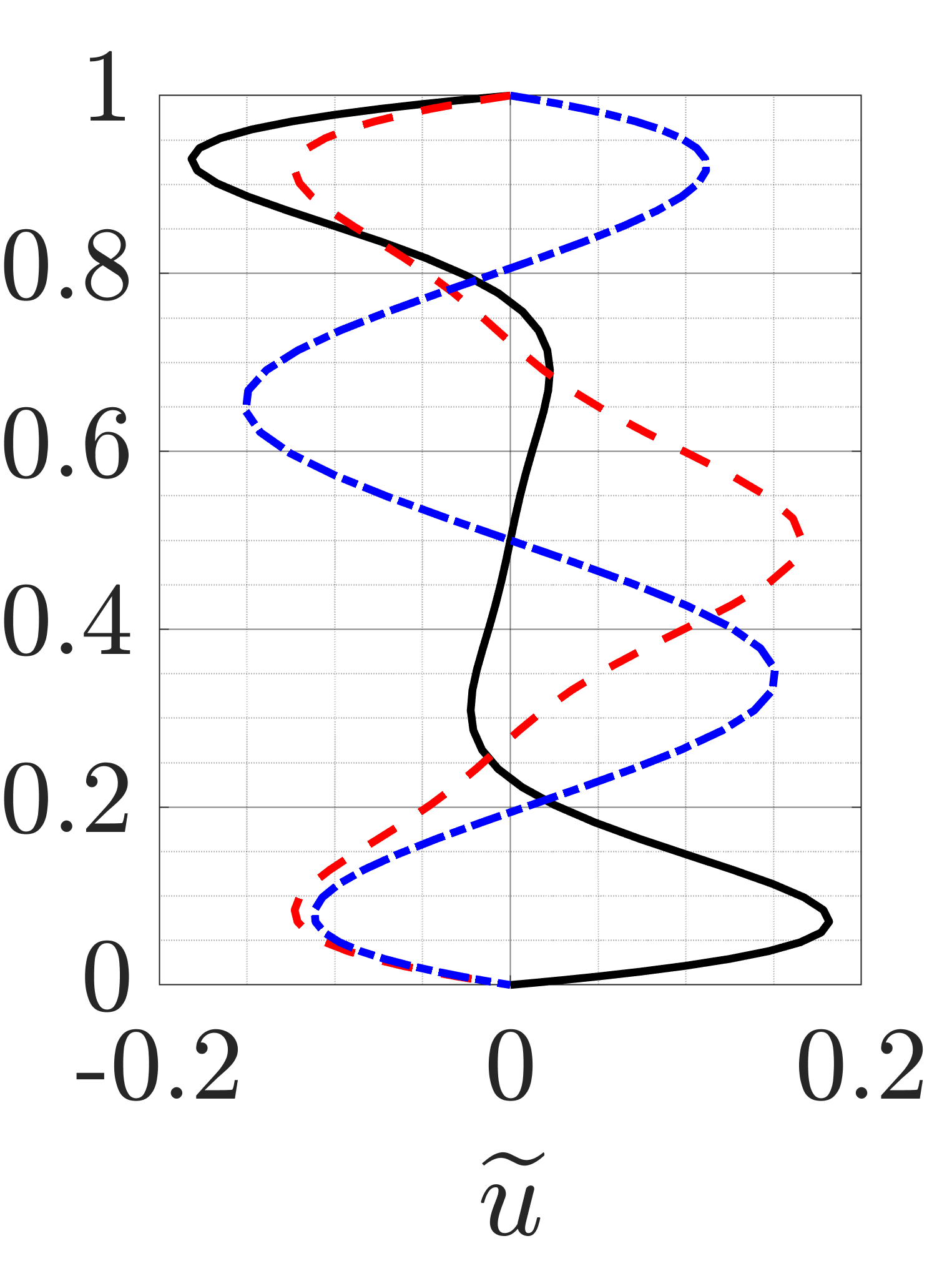}
    \includegraphics[width=0.24\textwidth]{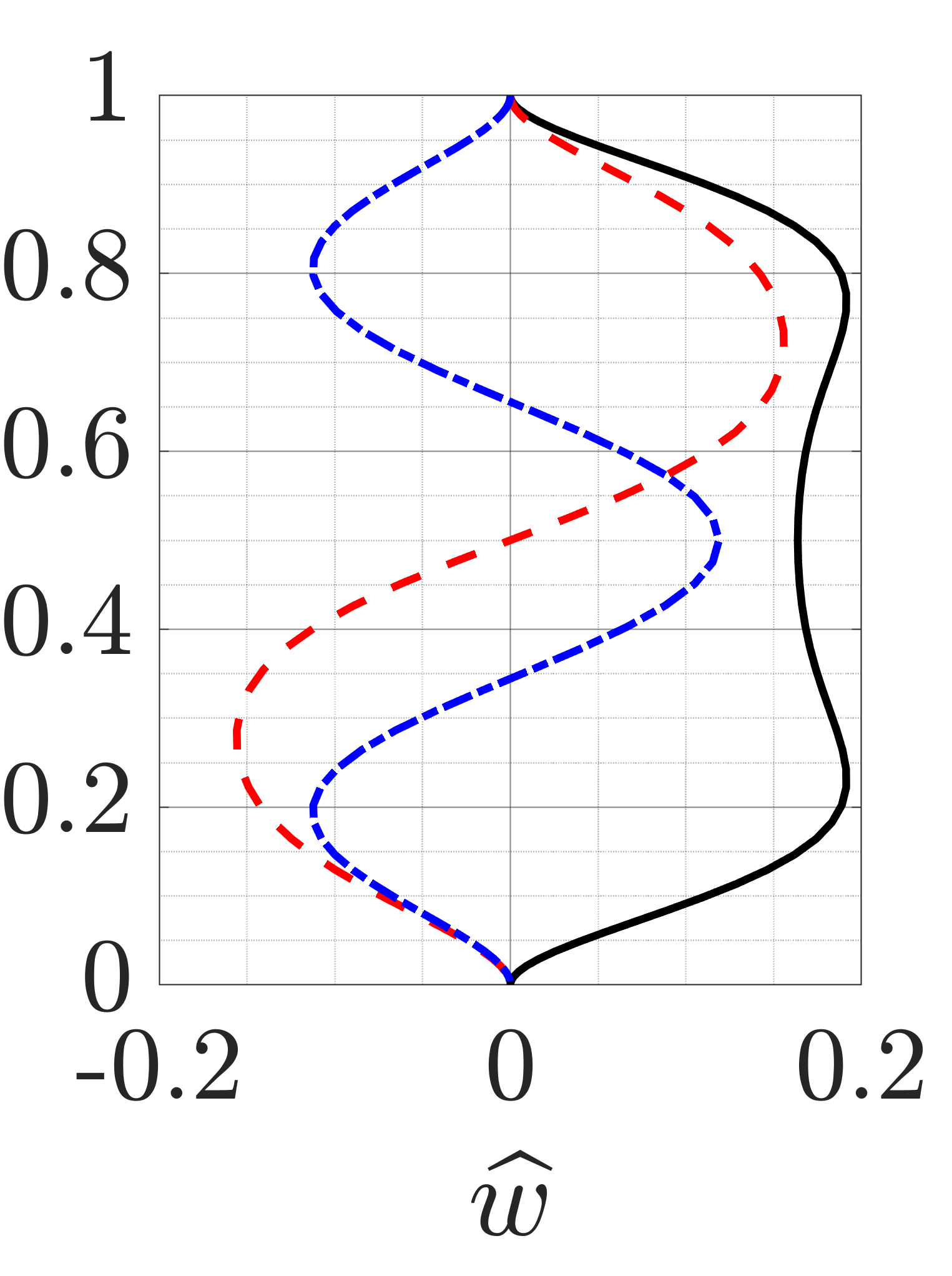}

    (e) S1 \hspace{0.24\textwidth} (f) S2 \hspace{0.24\textwidth} (g) S3
        
    \includegraphics[width=0.31\textwidth]{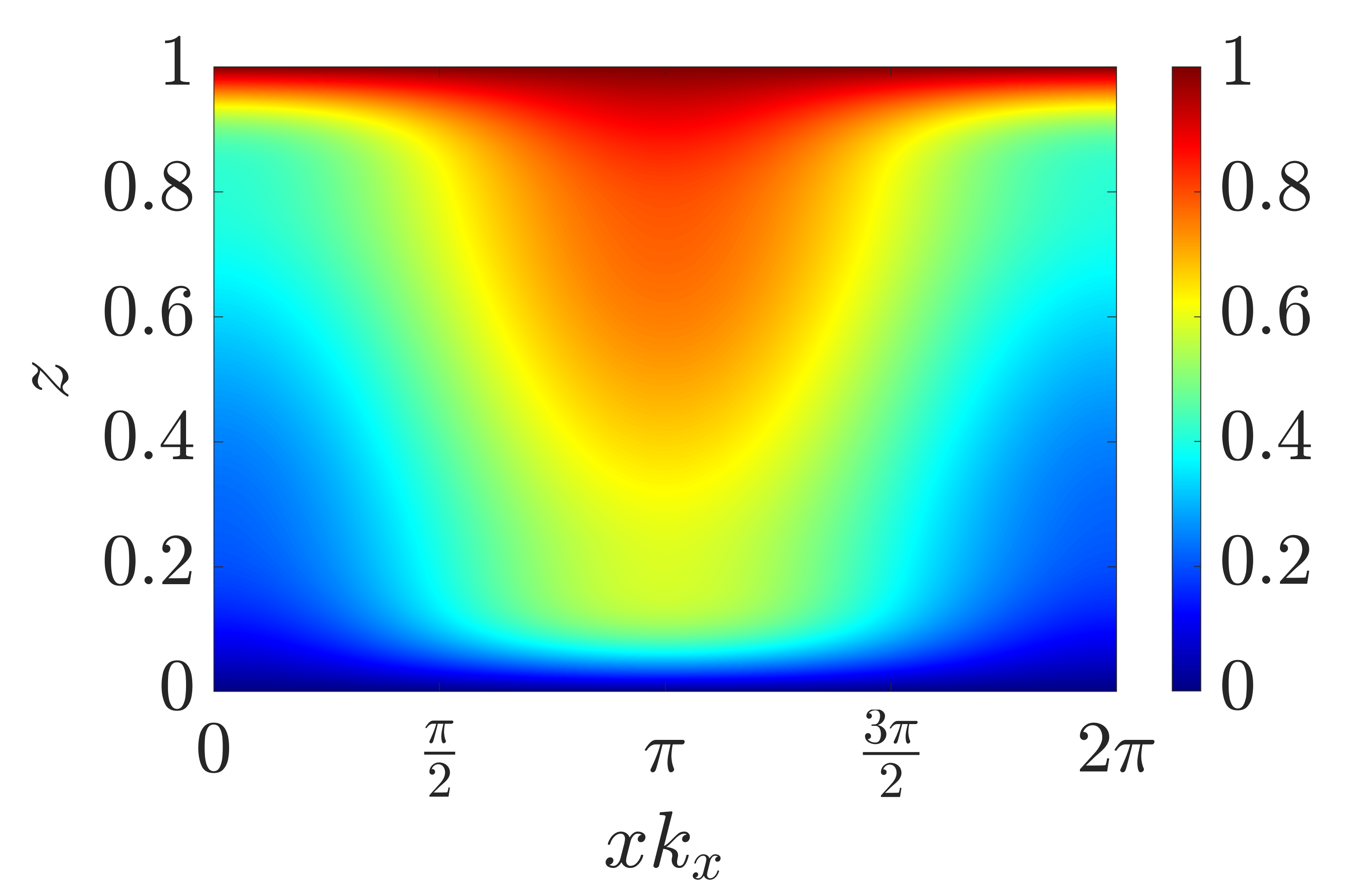}
    \includegraphics[width=0.31\textwidth]{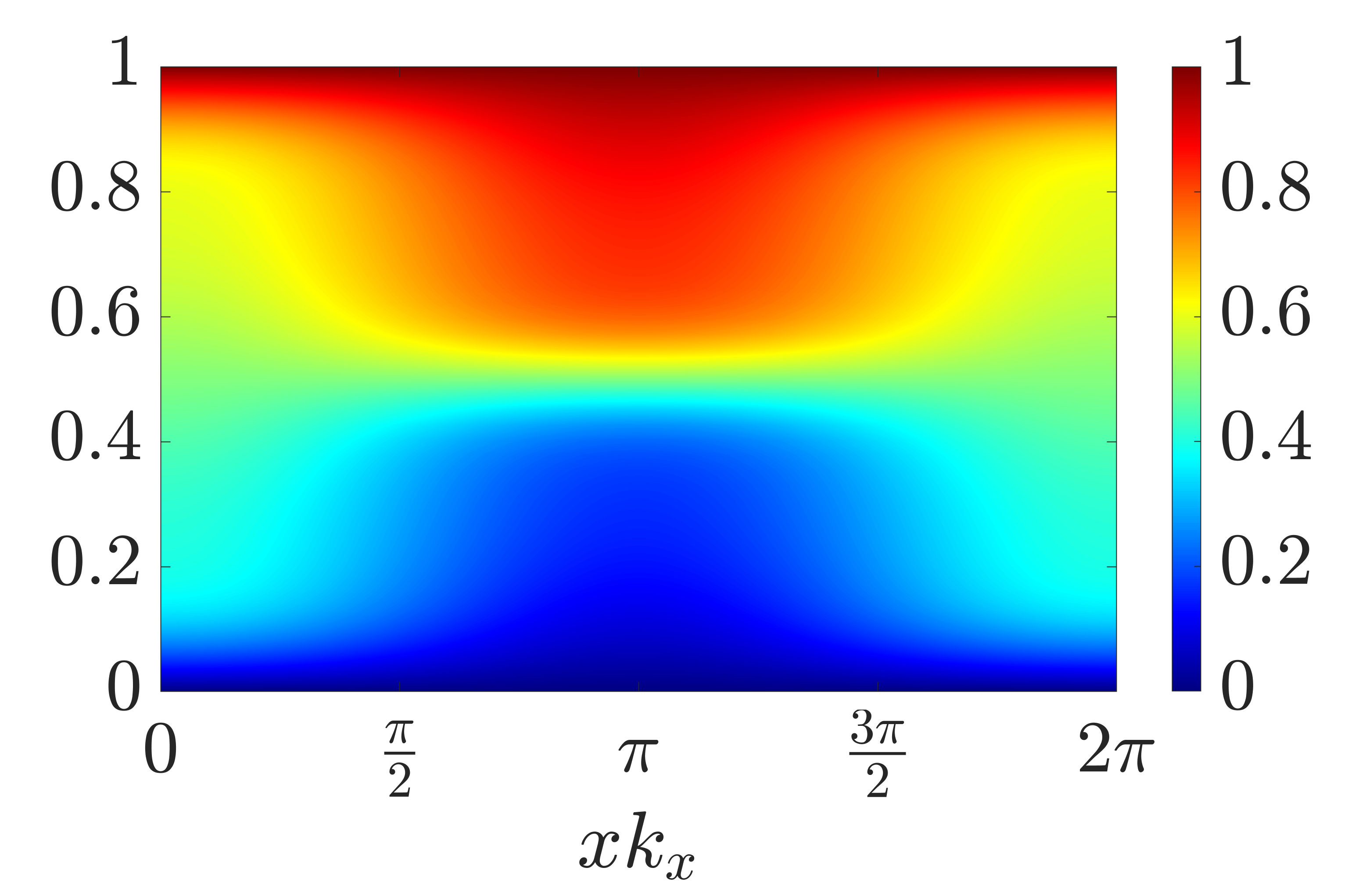}
    \includegraphics[width=0.31\textwidth]{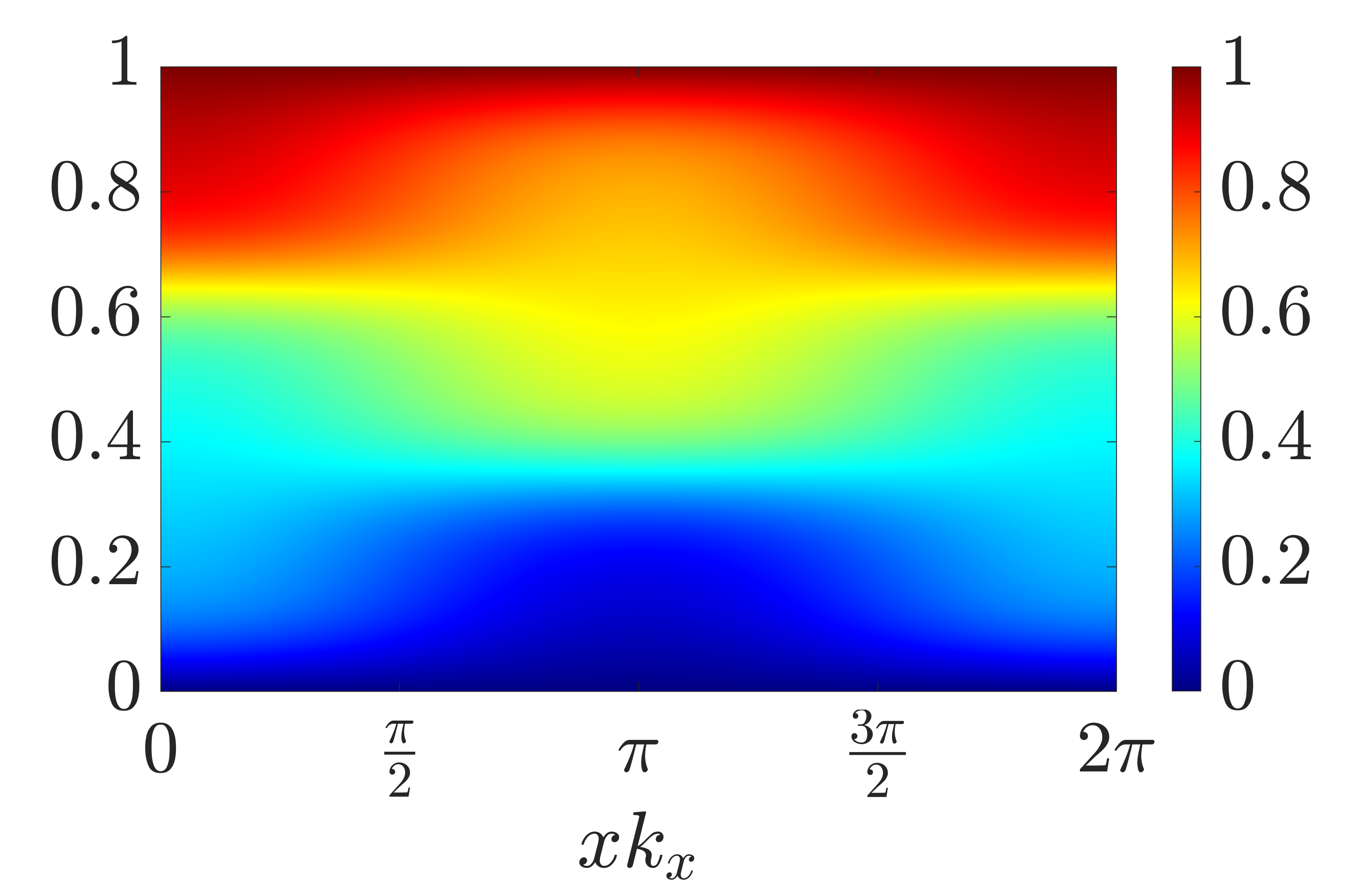}

(h) S1 \hspace{0.24\textwidth} (i) S2 \hspace{0.24\textwidth} (j) S3
         
          \includegraphics[width=0.31\textwidth]{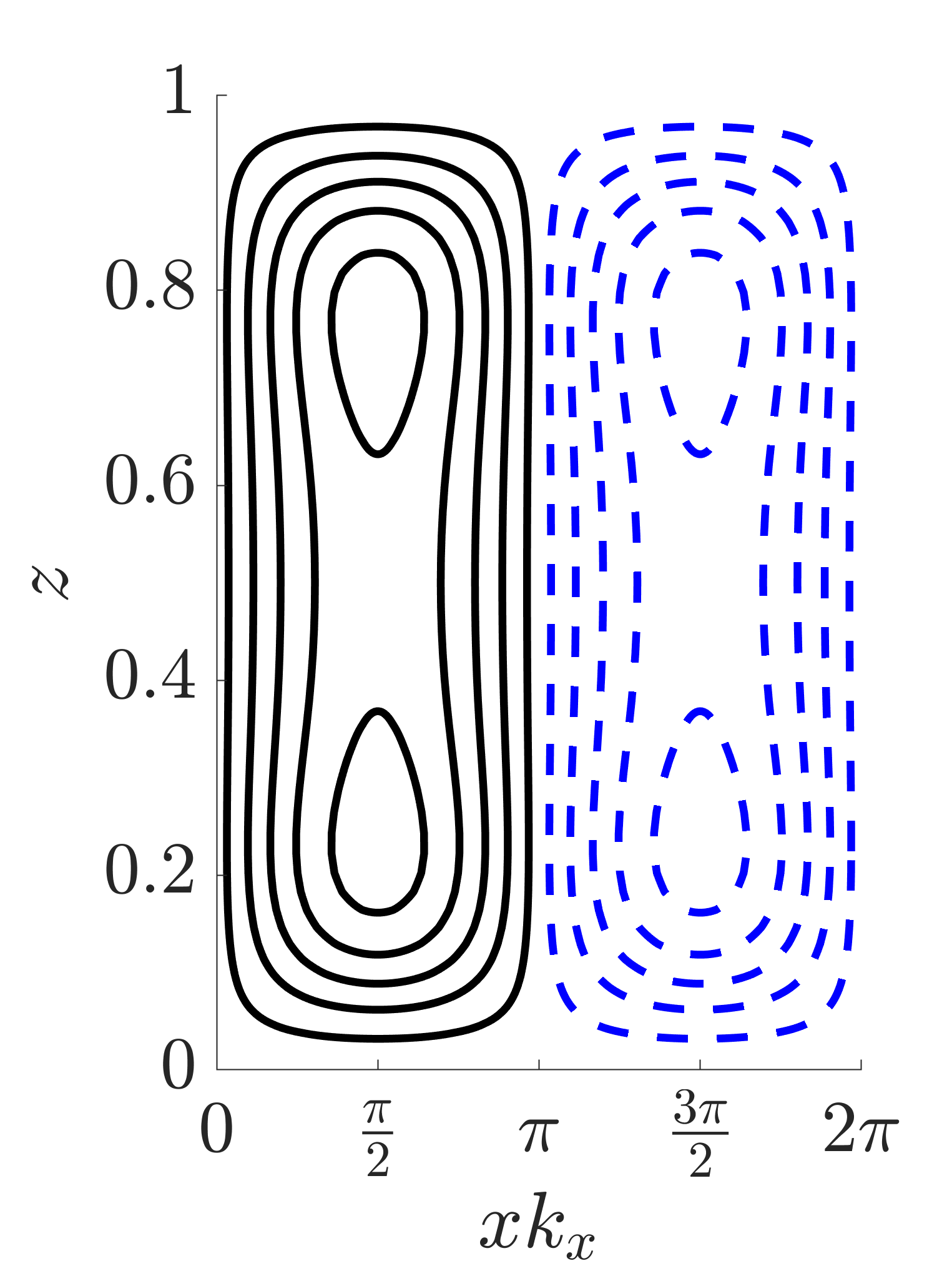}
    \includegraphics[width=0.31\textwidth]{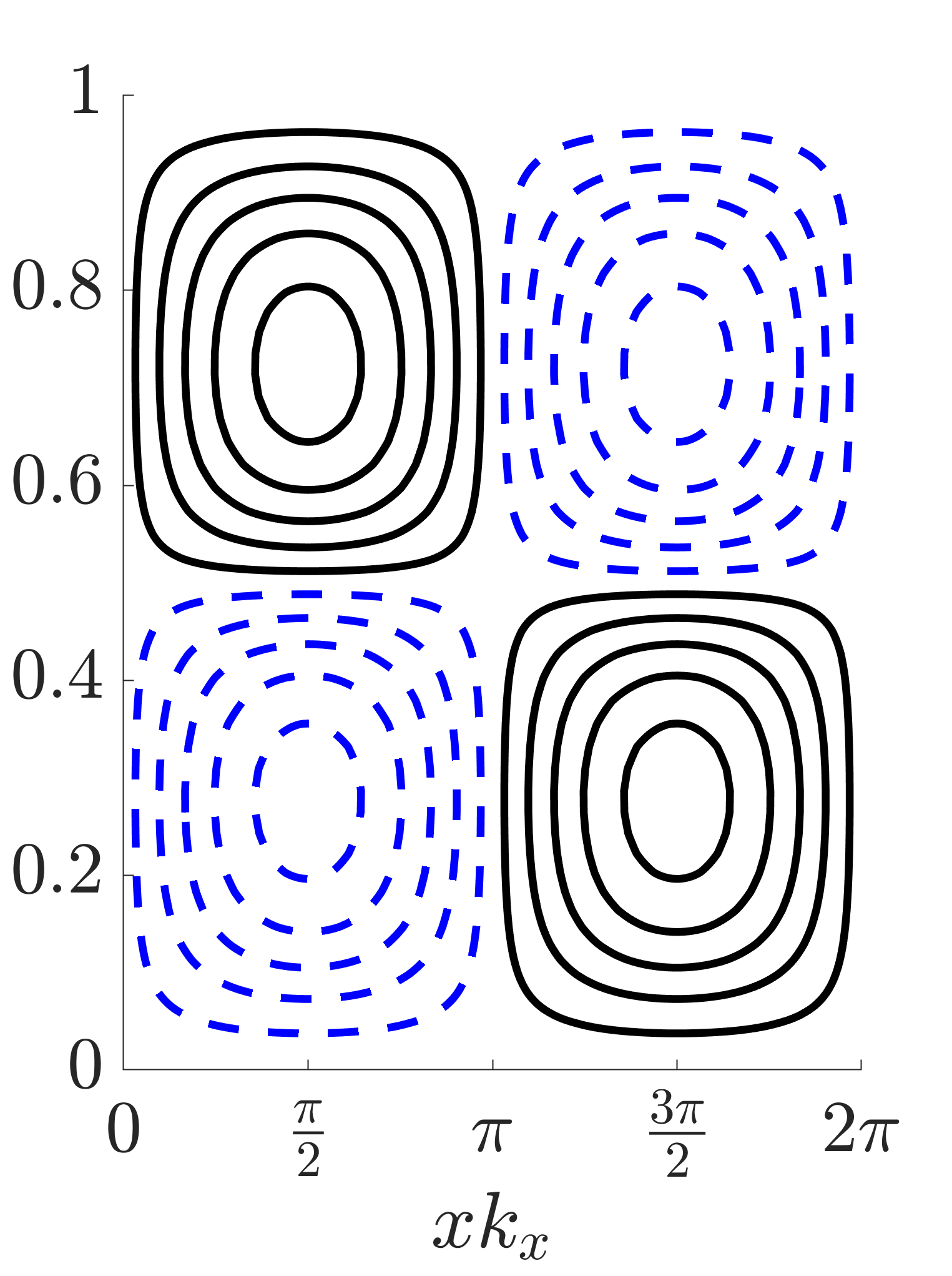}
    \includegraphics[width=0.31\textwidth]{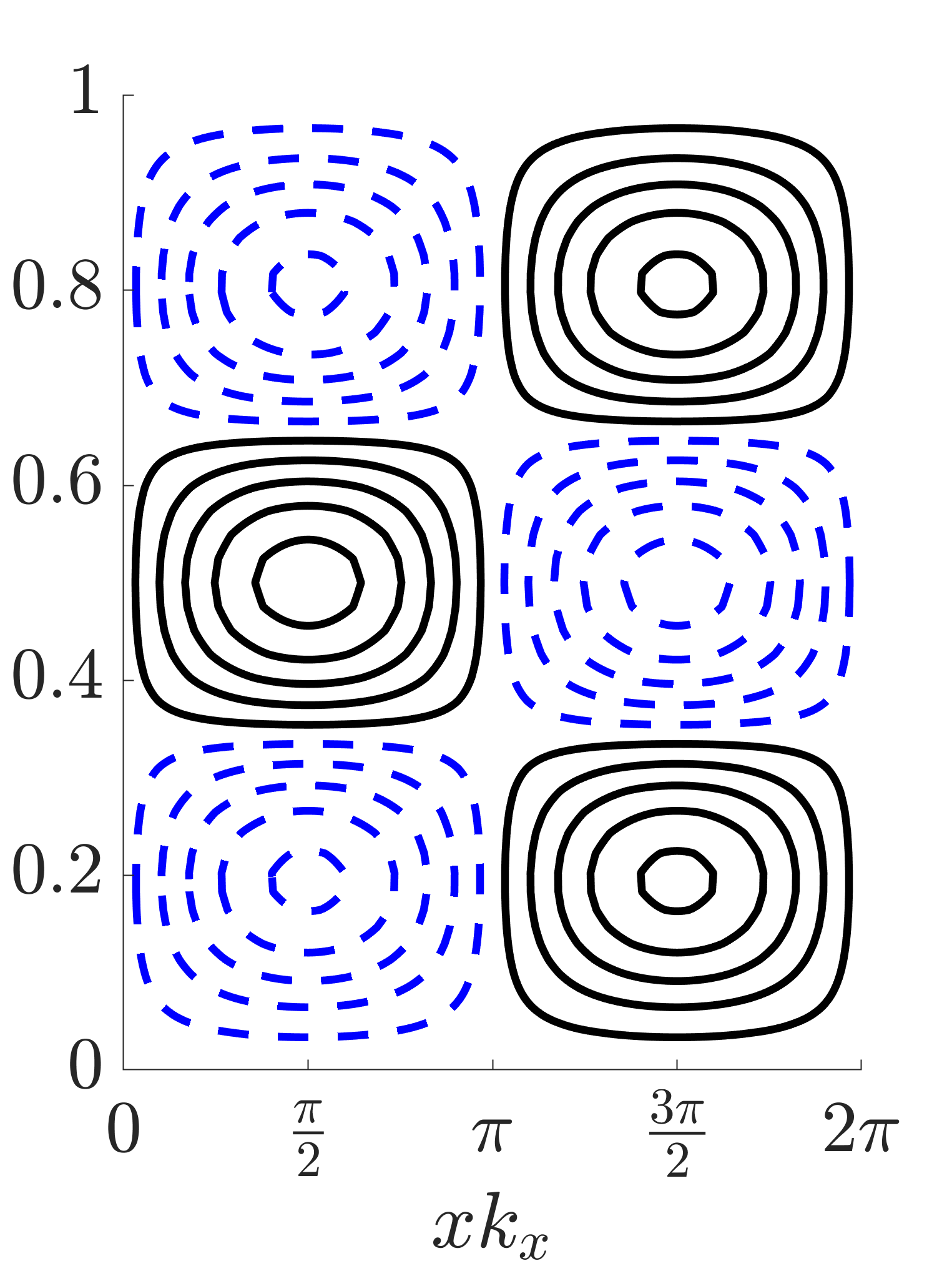}

    \caption{Solution profiles from the single-mode equations \eqref{eq:single_mode} in the form of S1, S2 and S3 when $k_x=8$, $k_y=0$,  $R_\rho=40$, $Pr=7$, $\tau=0.01$, $Ra_T=10^5$. The first row shows the profiles of (a) $z+\bar{S}_0$, (b) $\widehat{S}$, (c) $\widetilde{u}$, and (d) $\widehat{w}$. The second row shows the reconstructed total salinity using \eqref{eq:normal_mode_S} and \eqref{eq:total_T_S} for (e) S1, (f) S2 and (g) S3 solutions. The third row shows isocontours of the streamfunction for (h) S1, (i) S2 and (j) S3 solutions.  }
    \label{fig:profile_R_rho_T2S_40_tau_0p01}
\end{figure}

The solutions S1, S2 and S3 all bifurcate from the trivial solution. The S1, S2 and S3 solution profiles with $k_x=8$, $k_y=0$ are shown in figure \ref{fig:profile_R_rho_T2S_40_tau_0p01}. The horizontally averaged total salinity profiles $z+\bar{S}_0(z)$ in figure \ref{fig:profile_R_rho_T2S_40_tau_0p01}(a) show that these solutions are, respectively, associated with one, two and three mixed layers with reduced vertical gradient, resembling the staircase structures observed in field measurements, e.g., \citet{schmitt1987c}. Staircase-like solutions are also shown in some snapshots from 2D direct numerical simulations \citep[figure 2]{piacsek1980nonlinear} and reproduced by \citet[figure 3]{zhang2018numerical}. Recent 3D DNS results show the coexistence of multiple states with one, two, or three mixed regions using different initial conditions \citep[figure 2]{yang2020multiple}. In all three cases the mixed regions correspond to large values of the salinity amplitude $\widehat{S}$ as shown in figure \ref{fig:profile_R_rho_T2S_40_tau_0p01}(b), as well as a large vertical velocity $\widehat{w}$ as shown in figure \ref{fig:profile_R_rho_T2S_40_tau_0p01}(d). In contrast, the horizontal velocity peaks outside of the mixed region in each solution. See \S\ref{subsec:results_Pr_7_R_rho_dependence} for further discussion.

The second row of figure \ref{fig:profile_R_rho_T2S_40_tau_0p01} reconstructs the total salinity using equations \eqref{eq:normal_mode_S} and \eqref{eq:total_T_S}, while the third row of figure \ref{fig:profile_R_rho_T2S_40_tau_0p01} shows the isocontours of the 2D streamfunction computed as $\psi(x,z)=\bar{\psi}_0(z)+\widehat{\psi}(z)e^{\text{i}k_x x}+c.c.$ with $\bar{\psi}_0(z)=-\int_0^z\bar{U}_0(\xi)d \xi$ and $\widehat{\psi}=\widehat{w}/(\text{i}k_x)$. The isocontours are equispaced between $\pm 0.9$ of the maximum value. The line ($\mline\mline$) is used for positive (clockwise) streamlines while ({\color{blue}$\dashed$}, blue) indicates negative (counterclockwise) streamlines throughout this work. These results suggest that each mixed region is associated with one downward and one upward moving plume. Since salinity tends to be homogenized in regions of closed steady streamlines \citep{rhines1983rapidly} while salinity gradients are expelled from these regions, the resulting mean salinity exhibits an overall staircase-like profile.

Previous analysis of salt-finger convection in the spirit of single-mode solutions \citep{radko2000finite,proctor1986planform} also found the S1 type of solution, but S2 and S3 solutions were not found. This is likely because these studies \citep{radko2000finite,proctor1986planform} focused on the asymptotic behavior close to the onset of instability but did not go beyond the first bifurcation point from the trivial solution.

\begin{figure}
    \centering
    TF1
    
 (a) \hspace{0.15\textwidth} (b) \hspace{0.23\textwidth} (c) \hspace{0.23\textwidth} (d)

    \includegraphics[width=0.19\textwidth]{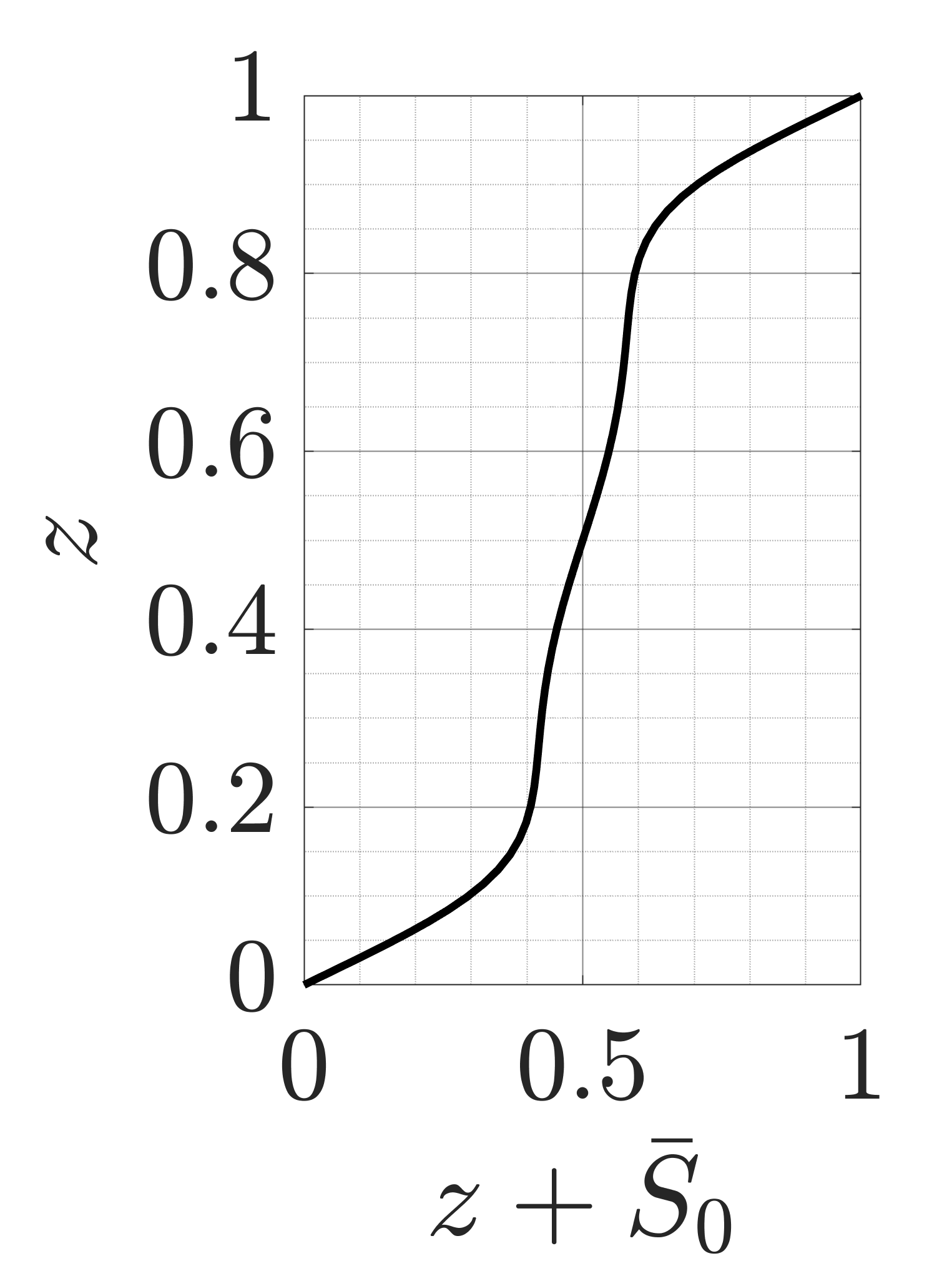}
    \includegraphics[width=0.19\textwidth]{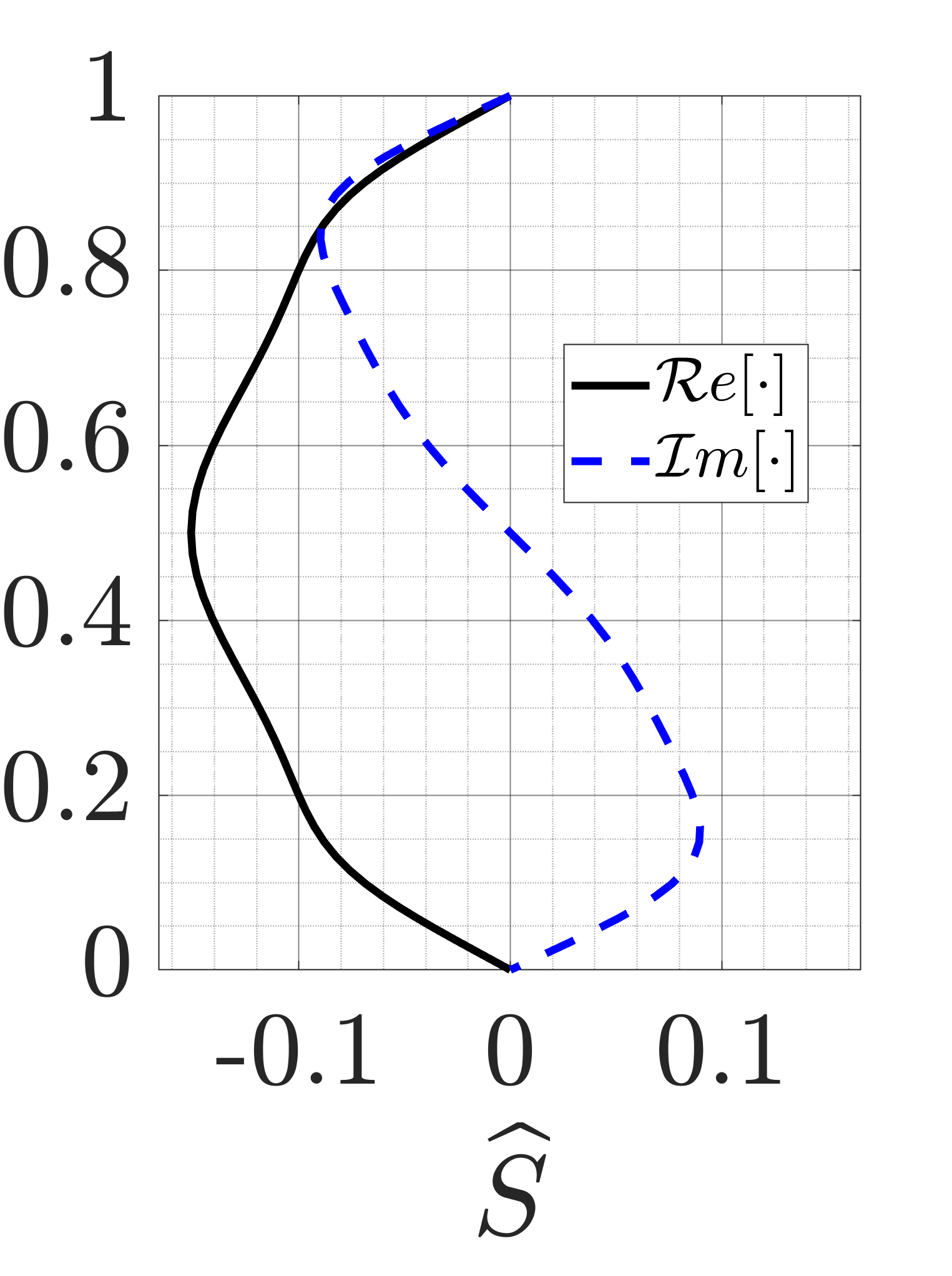}
    \includegraphics[width=0.38\textwidth,trim=-0 0in 0 0]{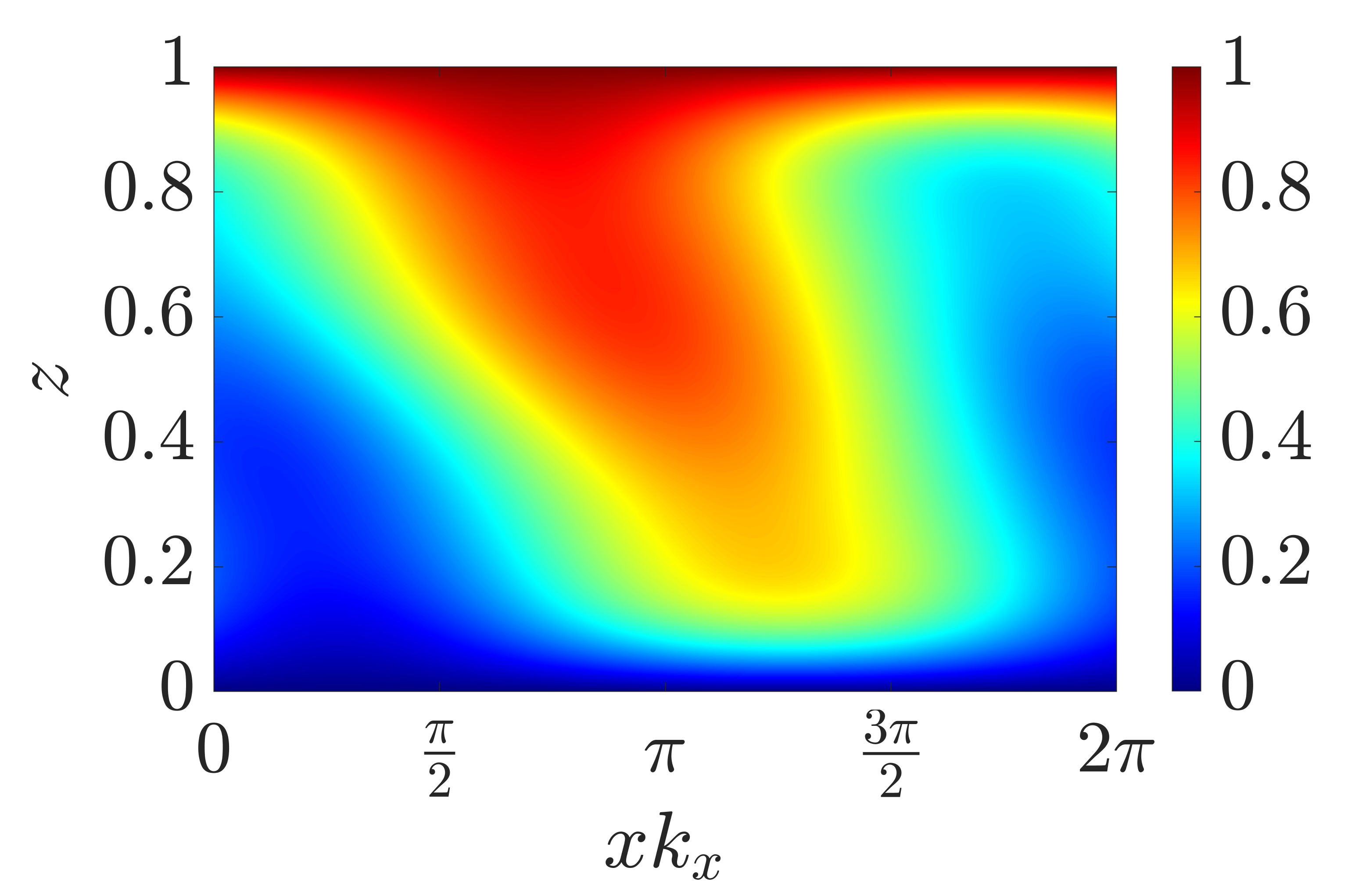}
    \includegraphics[width=0.175\textwidth,trim=-0 -0.4in 0 0]{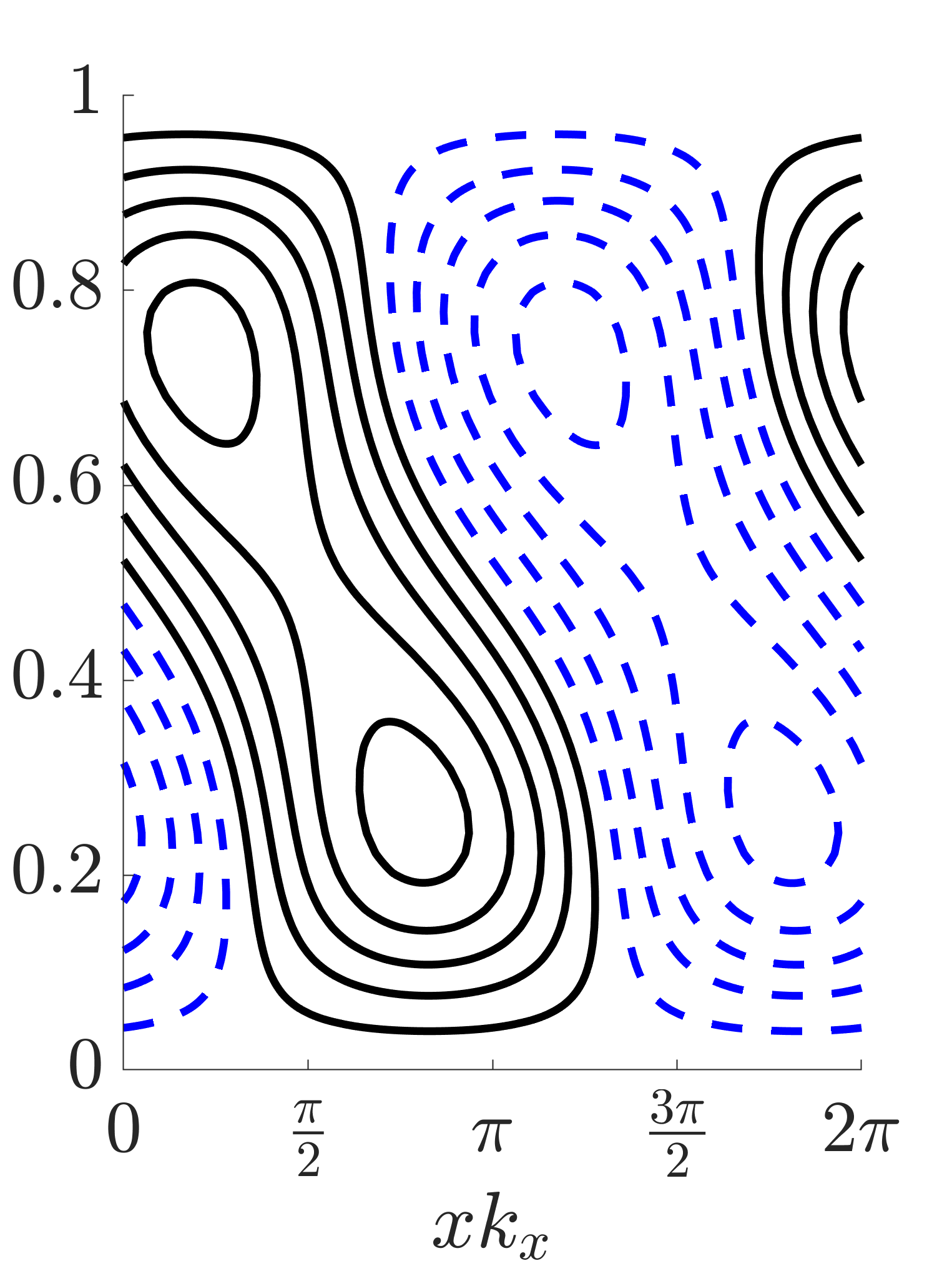}
    \caption{Solution profiles of the TF1 state from the single-mode equations \eqref{eq:single_mode} with $k_x=4$, $k_y=0$, $R_\rho=40$, $Pr=7$, $\tau=0.01$, $Ra_T=10^5$. Panels (a) and (b) show the profiles of $z+\bar{S}_0$ and $\widehat{S}$, respectively, while (c) and (d) show the reconstructed total salinity using \eqref{eq:normal_mode_S} and \eqref{eq:total_T_S} and the isocontours of the streamfunction.}
    \label{fig:profile_R_rho_T2S_40_tau_0p01_Pr_7_TF1}
\end{figure}

\begin{figure}
    \centering
    
    (a) \hspace{0.2\textwidth} (b) \hspace{0.2\textwidth} (c) \hspace{0.2\textwidth} (d)

    \includegraphics[width=0.24\textwidth]{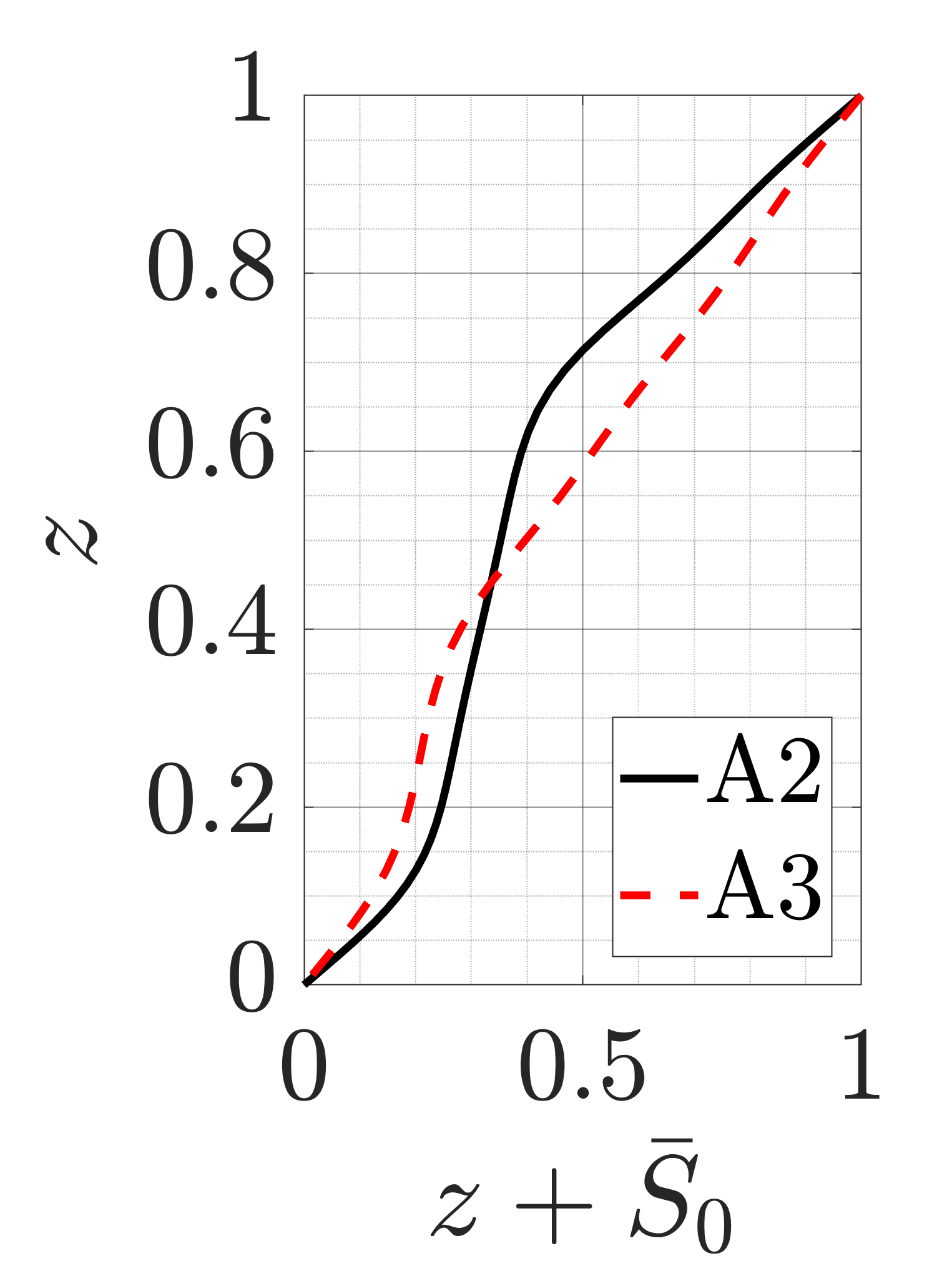}
    \includegraphics[width=0.24\textwidth]{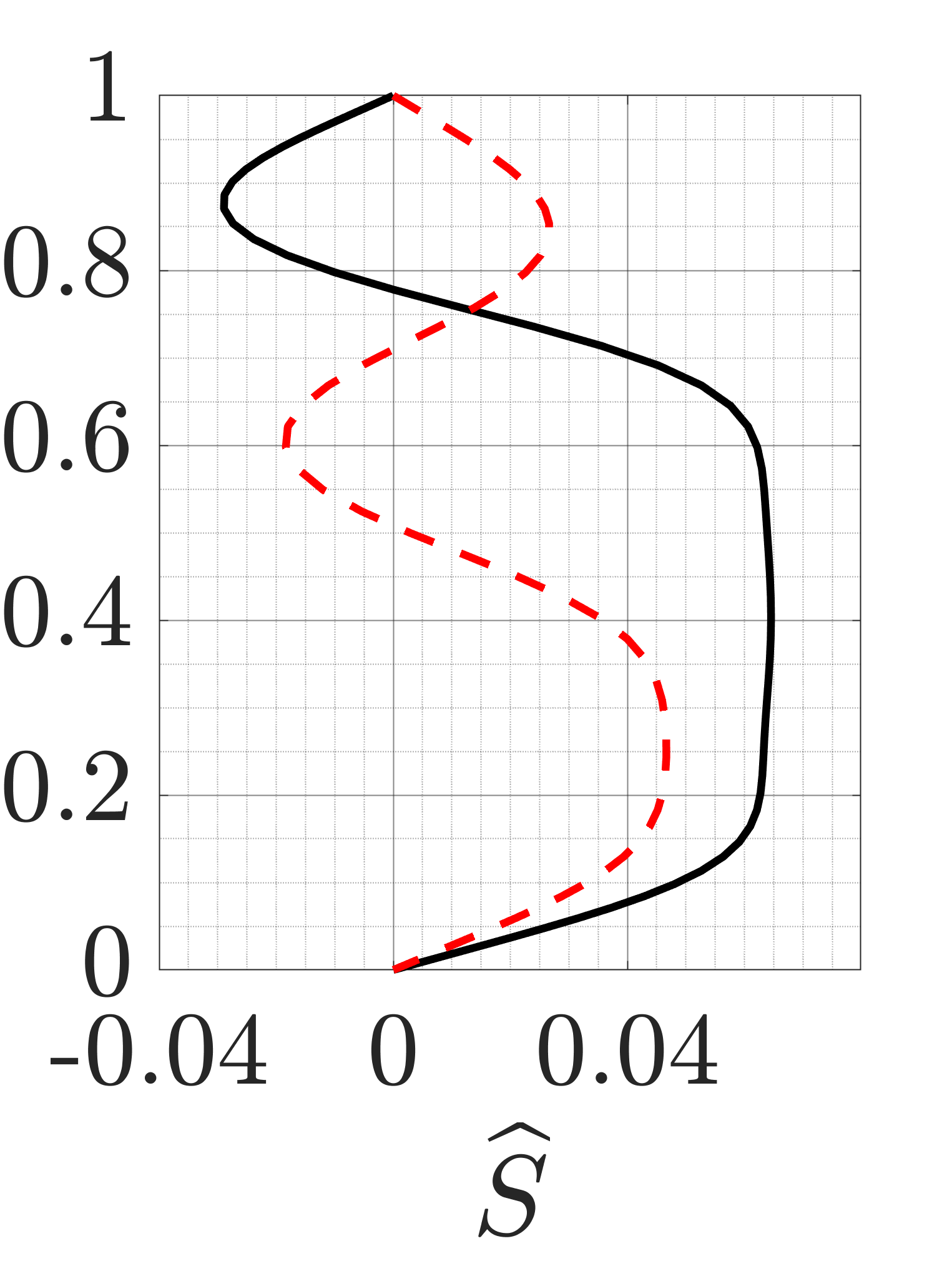}
    \includegraphics[width=0.24\textwidth]{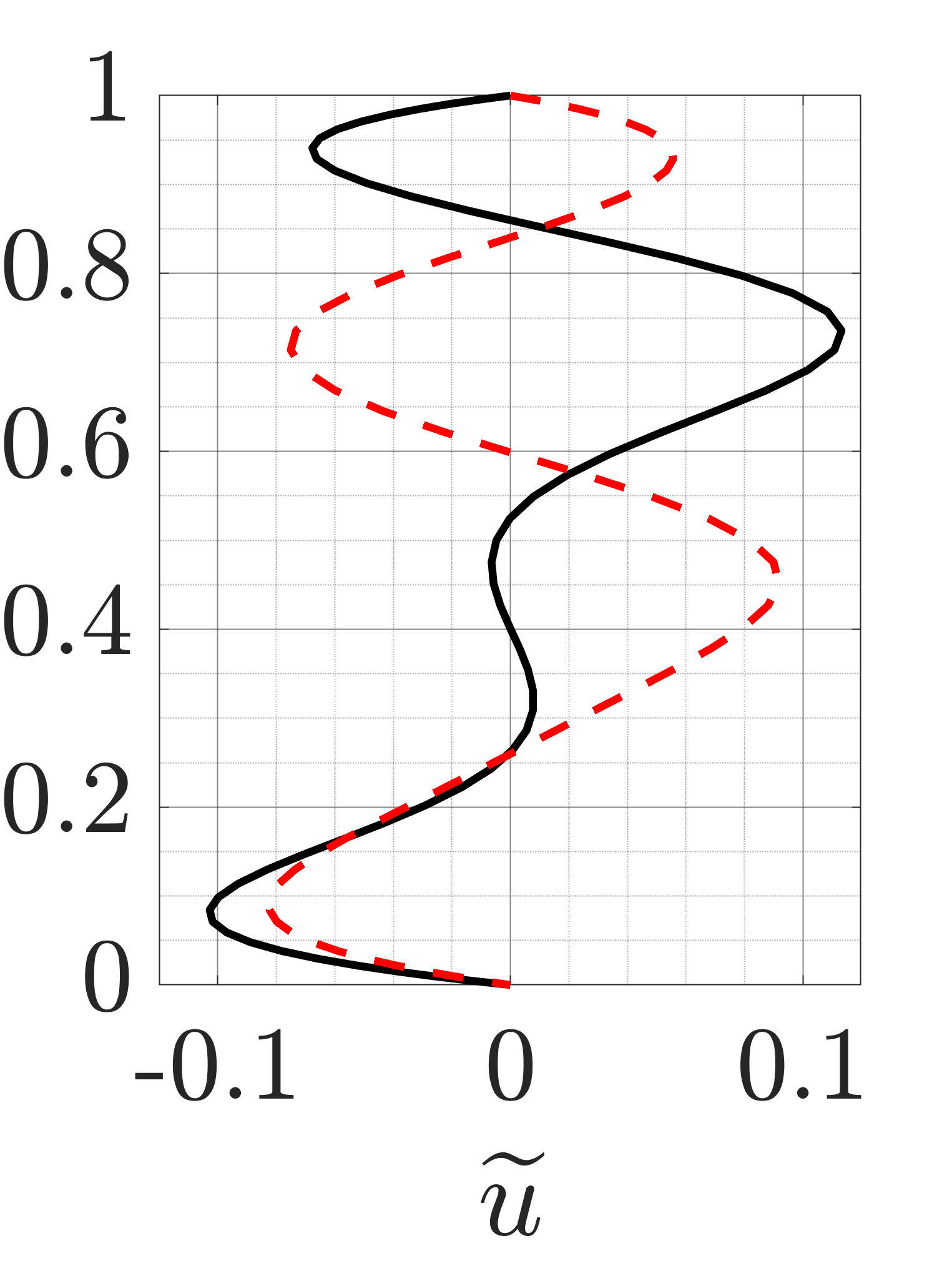}
    \includegraphics[width=0.24\textwidth]{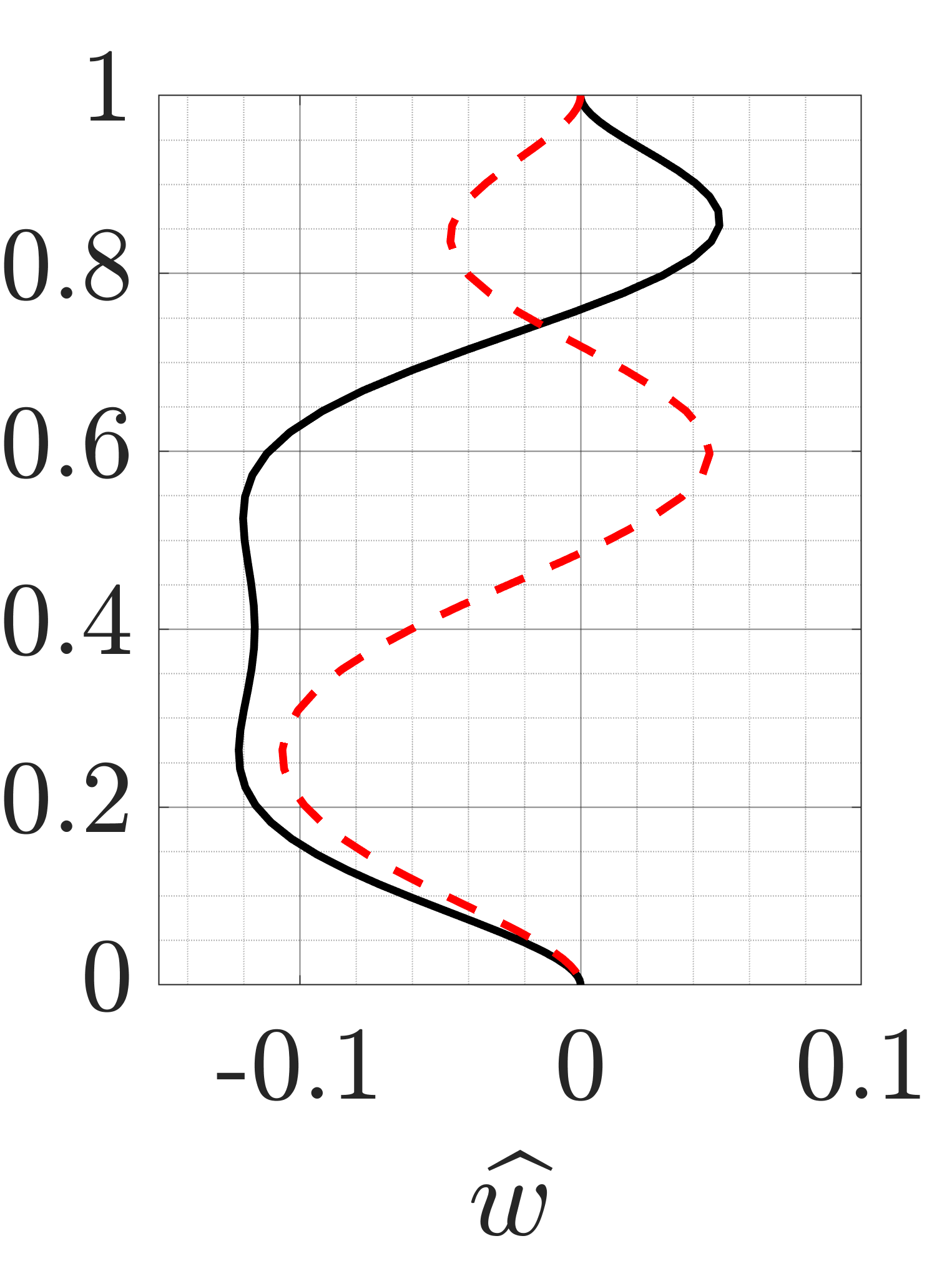}

(e)A2 \hspace{0.4\textwidth} (f)A3

    \includegraphics[width=0.33\textwidth]{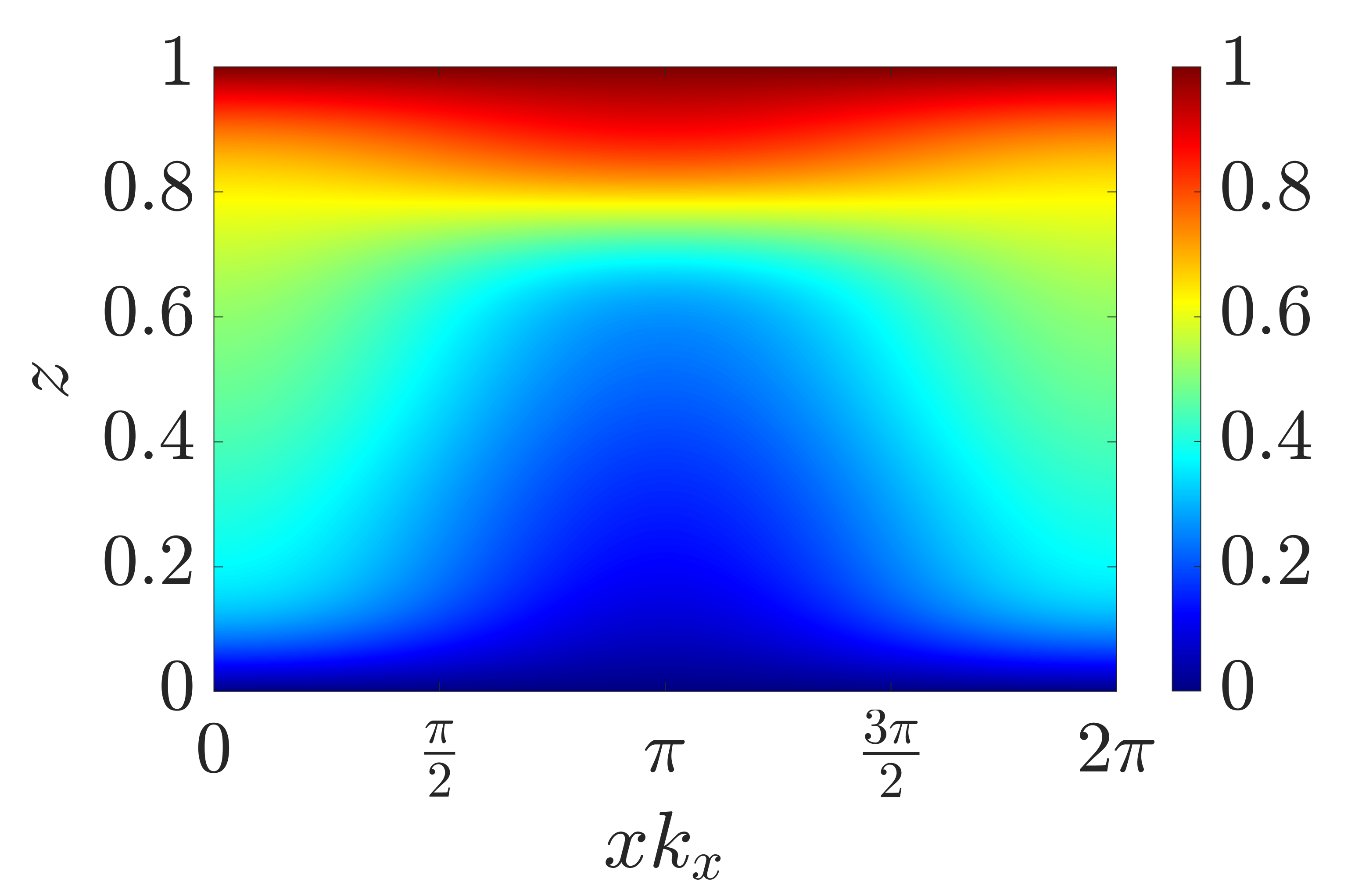}
    \includegraphics[width=0.15\textwidth,trim=-0 -0.7in 0 0]{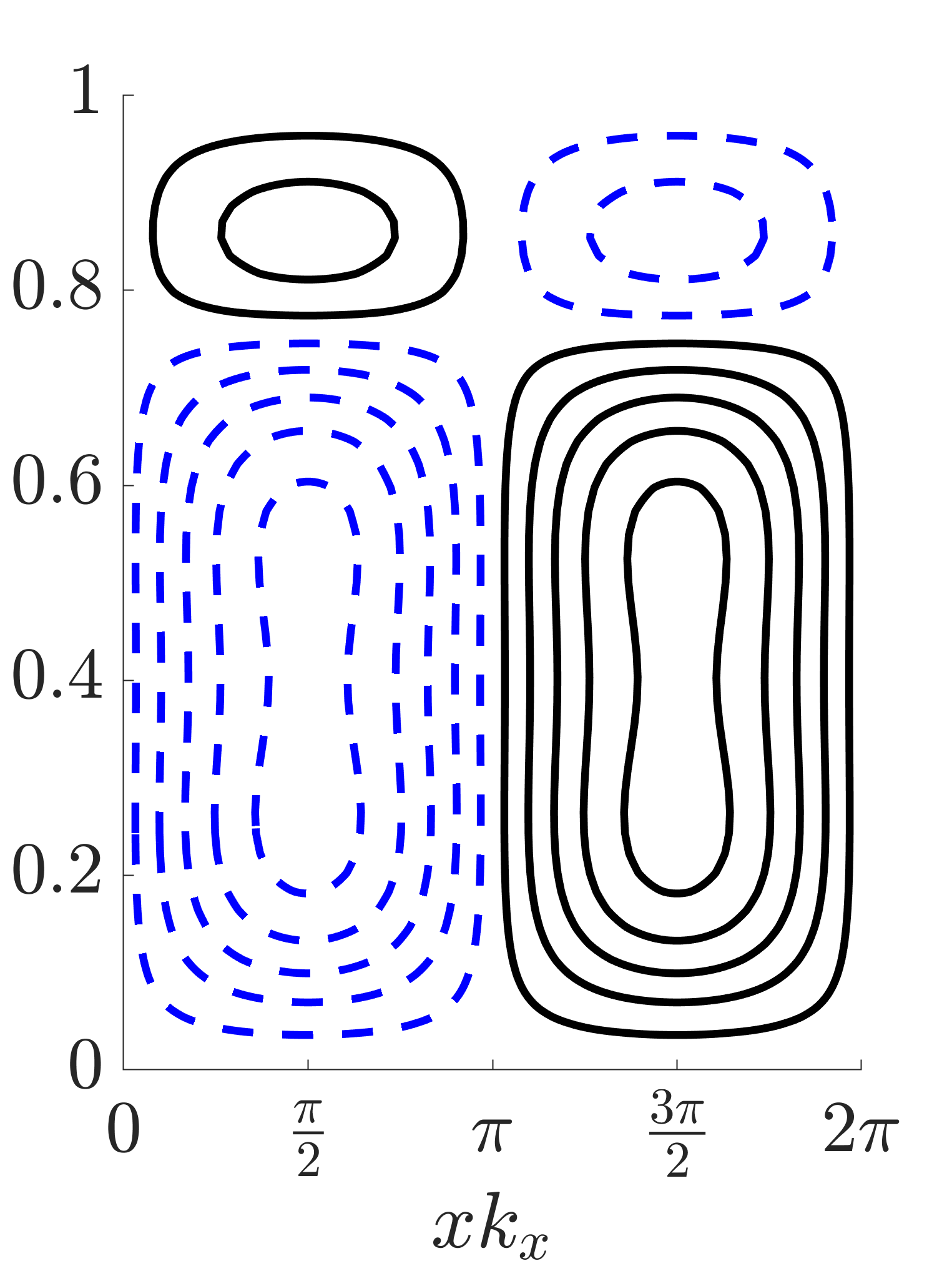}
    \includegraphics[width=0.33\textwidth]{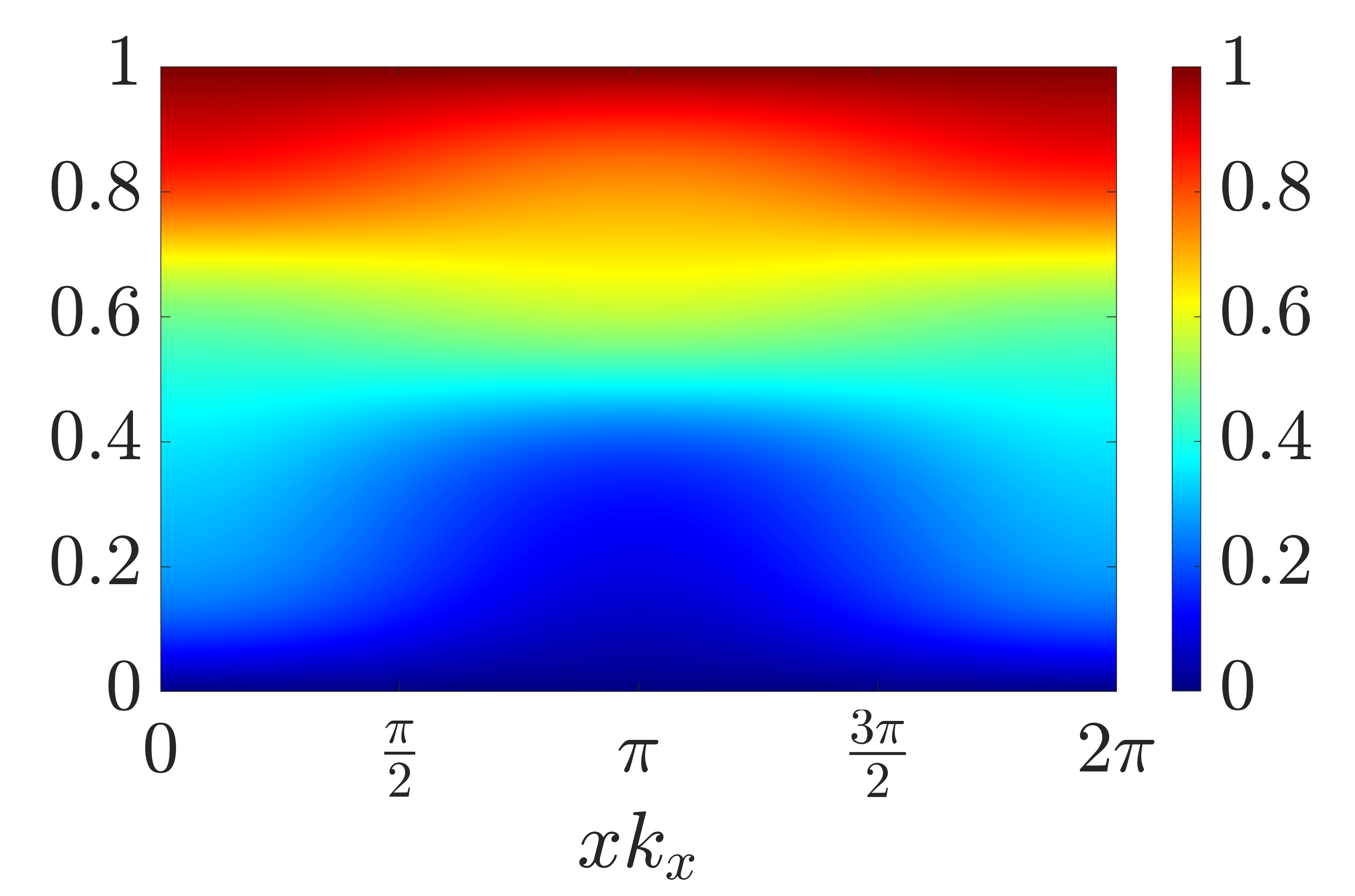}
    \includegraphics[width=0.15\textwidth,trim=-0 -0.7in 0 0]{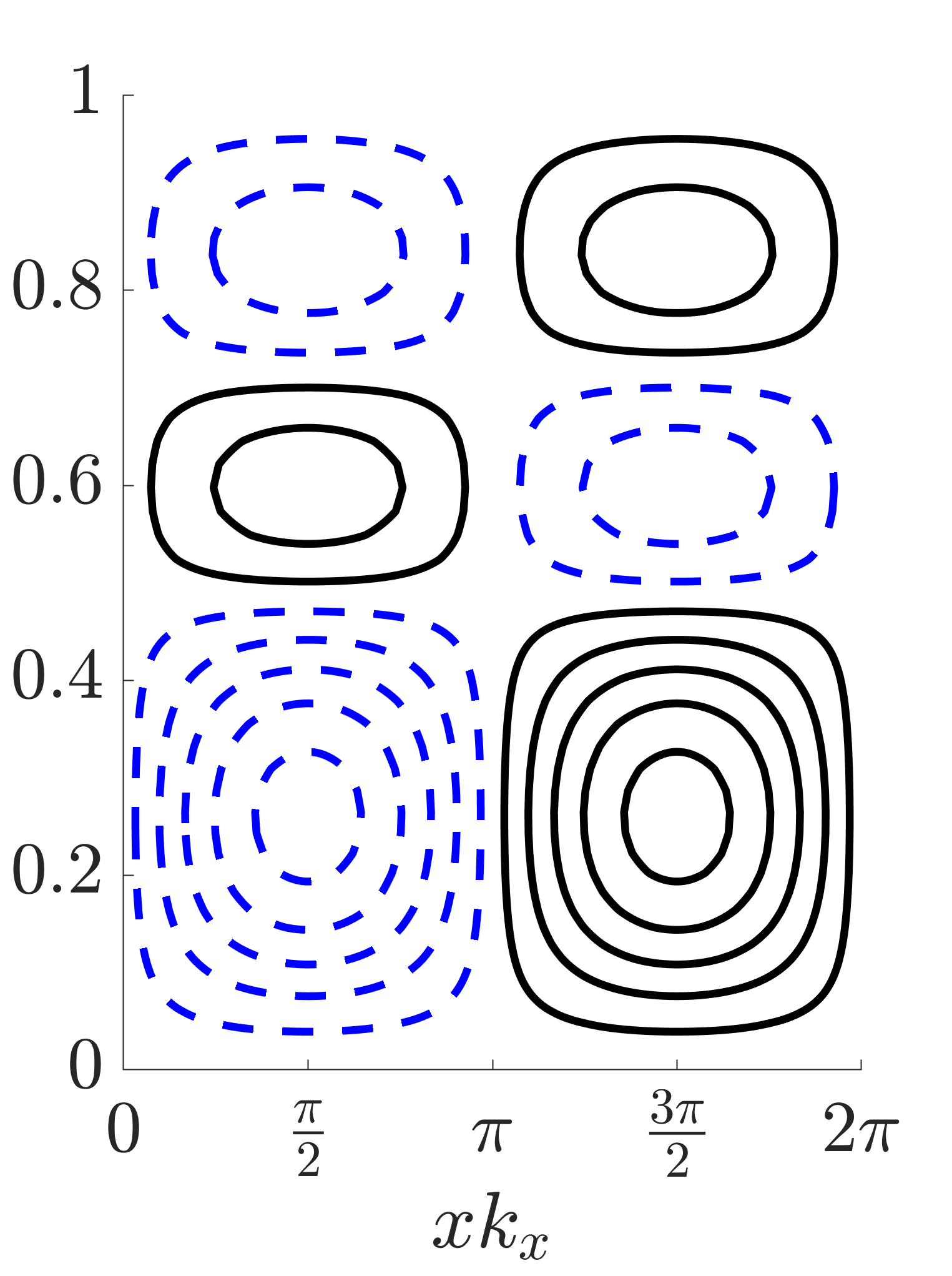}

        \caption{Solution profiles of the A2 and A3 states from the single-mode equations \eqref{eq:single_mode} with $k_x=8$, $k_y=0$,  $R_\rho=40$, $Pr=7$, $\tau=0.01$, $Ra_T=10^5$. The first row shows the profiles of (a) $z+\bar{S}_0$, (b) $\widehat{S}$, (c) $\widetilde{u}$ and (d) $\widehat{w}$. The second row shows the reconstructed total salinity using \eqref{eq:normal_mode_S} and \eqref{eq:total_T_S} and the isocontours of the streamfunction for (e) A2, (f) A3.
        }
    \label{fig:profile_R_rho_T2S_40_tau_0p01_Pr_7_A2_A3}
\end{figure}

We also examined secondary bifurcations of the solutions S1, S2, and S3 in figure \ref{fig:bif_diag_low_Ra_S2T}(a), focusing on the bifurcation points closest to their high wavenumber onset. The resulting secondary branches are of two types, corresponding to tilted fingers (TF1) and asymmetric layer spacing (A2 and A3). These states break the left-right reflection and the midplane reflection symmetry, respectively, as revealed by the corresponding solution profiles shown in figures \ref{fig:profile_R_rho_T2S_40_tau_0p01_Pr_7_TF1} and \ref{fig:profile_R_rho_T2S_40_tau_0p01_Pr_7_A2_A3}. The mean salinity profile $z+\bar{S}_0$ for TF1 in figure \ref{fig:profile_R_rho_T2S_40_tau_0p01_Pr_7_TF1}(a) is associated with two mixed regions near $z\in[0.2,0.4]$ and $z\in[0.6,0.8]$, while in the interior $z\in[0.4,0.6]$ the profile is close to linear. The corresponding profile of $\widehat{S}$ in figure \ref{fig:profile_R_rho_T2S_40_tau_0p01_Pr_7_TF1}(b) has both real and imaginary components with even and odd symmetry with respect to the midplane, respectively, indicating that the harmonics are no longer in phase in the vertical. The reconstructed total salinity shown in figure \ref{fig:profile_R_rho_T2S_40_tau_0p01_Pr_7_TF1}(c) reveals that the finger is now tilted; see also the streamfunction shown in figure \ref{fig:profile_R_rho_T2S_40_tau_0p01_Pr_7_TF1}(d). The tilted finger generates a nonzero large-scale shear with $\underset{z}{\text{max}}\;\bar{U}_0(z)=3.16\times 10^{-4}$. The profile of the large-scale shear is shown farther below (figures \ref{fig:profile_R_rho_T2S_2_tau_0p01_TF1_TW1}(a) and \ref{fig:profile_R_rho_T2S_40_tau_0p01_Pr_0p05_TF1_TW1}(b)) for parameter values for which it is much stronger. A similar tilted finger state accompanied by large-scale shear was observed in earlier simulations of the single-mode equations \citep{paparella1999sheared} while observations of tilted fingers are reported in the NATRE \citep[figure 3]{st1999contribution} and C-SALT field measurements \citep[figure 1(b)]{kunze1990evolution} as well as in laboratory experiments on salt-finger convection; see, e.g., \citet[figure 2]{taylor1989laboratory} and \citet[figure 2]{krishnamurti2009heat}. Shear-associated tilting has also been widely reported in experiments on Rayleigh-B\'enard convection \citep{krishnamurti1981large} as well as in direct numerical simulations of 2D Rayleigh-B\'enard convection \citep{goluskin2014convectively}; see the review by \citet{siggia1994high}.

The solution profiles in figures \ref{fig:profile_R_rho_T2S_40_tau_0p01_Pr_7_A2_A3}(a)-(d) indicate that the A2 and A3 solutions spontaneously break the midplane reflection symmetry; states obtained via reflection in the midplane are therefore also solutions. In particular, the profile $z+\bar{S}_0(z)$ no longer passes through $(1/2,1/2)$. However, the reconstructed total salinity profile and the isocontours of the streamfunction in figures \ref{fig:profile_R_rho_T2S_40_tau_0p01_Pr_7_A2_A3}(e)-(f) still resemble, qualitatively, the S2 and S3 profiles that distinguish these two asymmetric solutions. In particular, the A2 streamlines show two counter-rotating but unequal rolls in the vertical while the A3 solution exhibits three counter-rotating rolls, much as in the S2 and S3 states shown in figures \ref{fig:profile_R_rho_T2S_40_tau_0p01}(i)-(j). Solutions that are asymmetric with respect to midplane reflection have been seen in magnetoconvection with a depth-dependent magnetic diffusivity \citep[figure 17]{julien2000nonlinear} but are a consequence of forced symmetry breaking. Here, such asymmetric solutions originate from spontaneous symmetry breaking. 

The stability of these solutions is indicated by thick (stable) and thin (unstable) lines in figure \ref{fig:bif_diag_low_Ra_S2T}. The single-mode S1 solution is stable near both ends but loses stability to TF1 for intermediate wavenumbers $k_x$. However, in either case the stable solution corresponds to the largest $Sh$ among all the solutions shown in the figure, a finding that is broadly consistent with the `relative stability' criterion of \citet{malkus1958finite}. However, this is no longer so for smaller Prandtl numbers, as discussed in \S\ref{sec:results_Pr_0p05}

\begin{table}
    \centering
     \begin{tabular}{c|ccccccccccc}
    \hline
     \backslashbox{I.C.}{$L_x$} &  $2\pi/18$  & $2\pi/16$ & $2\pi/14$ & $2\pi/12$ & $2\pi/10$ & $2\pi/8$ & $2\pi/6$ & $2\pi/4$ & $2\pi/2$ & $2\pi$ & $4\pi$ \\ \hline
     S1 & S1 & S1 & S1 & S1 & S1 & S1 (2) & S1 (2) & S1 (3) & S1 (7) & C & C\\
     S2 & - & S1 & S1 & S1 & S1 & RTF & S1 (2) & S1 (3) & S1 (5) & - & - \\
     S3 & - & - & S1 & S1 & S1 & S1 (2) & S1 (2) & S1 (3) & - & - & -\\
     TF1 & - & - & - & - & - & - & S1 (2) & S1 (3) & S1 (6) & - & -\\
     \hline
    \end{tabular}
    \caption{The flow structures from 2D DNS simulations at $t=3,000$ in domains of size $L_x$ and initial condition (I.C.) constructed from S1, S2, S3 and TF1 solutions using the ansatz \eqref{eq:normal_mode} with $k_x=2\pi/L_x$, $k_y=0$. RTF indicates direction-reversing tilted fingers and C represents chaotic behavior; `-' indicates that a nonzero single-mode solution at $k_x=2\pi/L_x$ is not present based on figure \ref{fig:bif_diag_low_Ra_S2T}(a). The number $n\in \mathbb{Z}$ inside a bracket indicates that the final horizontal wavenumber reached by the solution is $k_x= 2\pi n/L_x$, $n>1$, i.e. that the evolution results in a changed wavenumber.}
    \label{tab:DNS_transition_low_Ra_S2T_Pr_7}
\end{table}

\begin{figure}
    \centering
    \includegraphics[width=0.5\textwidth]{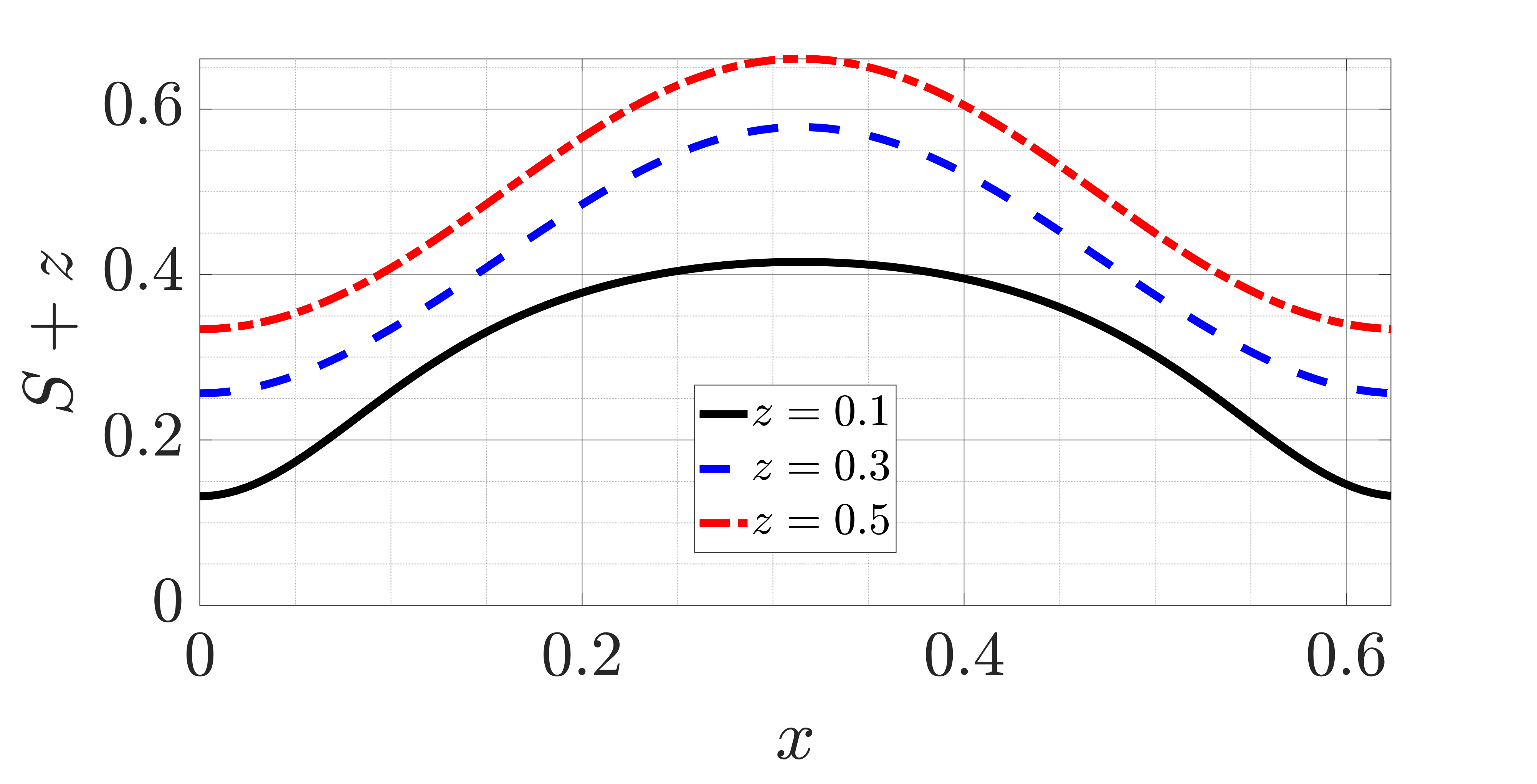}
    \caption{DNS results showing the total salinity $z+S(x,z,t)$ across an S1 finger at $z=0.1$, 0.3 and 0.5 in a $L_x=2\pi/10$ domain with parameters $R_\rho=40$, $Pr=7$, $\tau=0.01$, $Ra_{T}=10^5$.}
    \label{fig:horizontal_variation}
\end{figure}

The dynamics of unstable solutions are typically not easy to isolate and analyze without suitable initial conditions. Here, we use our single-mode solutions as initial conditions for 2D DNS to provide additional insight into their stability. We set the horizontal domain size as $L_x\in [2\pi /18, 4\pi]$ and then use the single-mode solution profile at $k_x=2\pi/L_x$, $k_y=0$, to construct a 2D initial condition using the ansatz in \eqref{eq:normal_mode}. The final state after $t=3,000$ for different $L_x$ and initial conditions based on S1, S2, S3, and TF1 solutions is summarized in table \ref{tab:DNS_transition_low_Ra_S2T_Pr_7}. For small horizontal domains ($L_x\leq 2\pi /10$) the domain constrains the finger to be tall and thin and the DNS results with S1, S2, and S3 initial conditions all transition to a solution resembling the one-layer solution S1. This is consistent with the stability observation in the bifurcation diagram in figure \ref{fig:bif_diag_low_Ra_S2T}. The DNS also provides the final $Sh$ numbers and these values are plotted using black squares in figure \ref{fig:bif_diag_low_Ra_S2T}(a). The predicted $Sh$ from the single-mode solution is close to the DNS results, especially close to the high wavenumber onset of the S1 solution ($k_x=19.251$). The accuracy of the single-mode equations close to the high wavenumber onset is a consequence of the strong damping of modes with wavenumbers $k_x=2\pi n/L_x$ ($n\in \mathbb{Z}$ and $n\geq 2$, e.g., $k_x=4\pi/L_x$, $k_x=6\pi /L_x$) that leaves only the trivial solution at these wavenumbers (see figure \ref{fig:bif_diag_low_Ra_S2T}). As $k_x$ decreases, the DNS results start to deviate from the single-mode predictions (figure \ref{fig:bif_diag_low_Ra_S2T}(a)), a consequence of the departure of the horizontal salinity profile from the assumed sinusoidal form; see, e.g., the total salinity at $z=0.1$ in figure \ref{fig:horizontal_variation}. 

\begin{figure}
    \centering
    (a) $\langle S\rangle_{h}(z,t)$ \hspace{0.45\textwidth} (b) $z+\langle S\rangle_{h,t}$
    
    \includegraphics[width=0.57\textwidth]{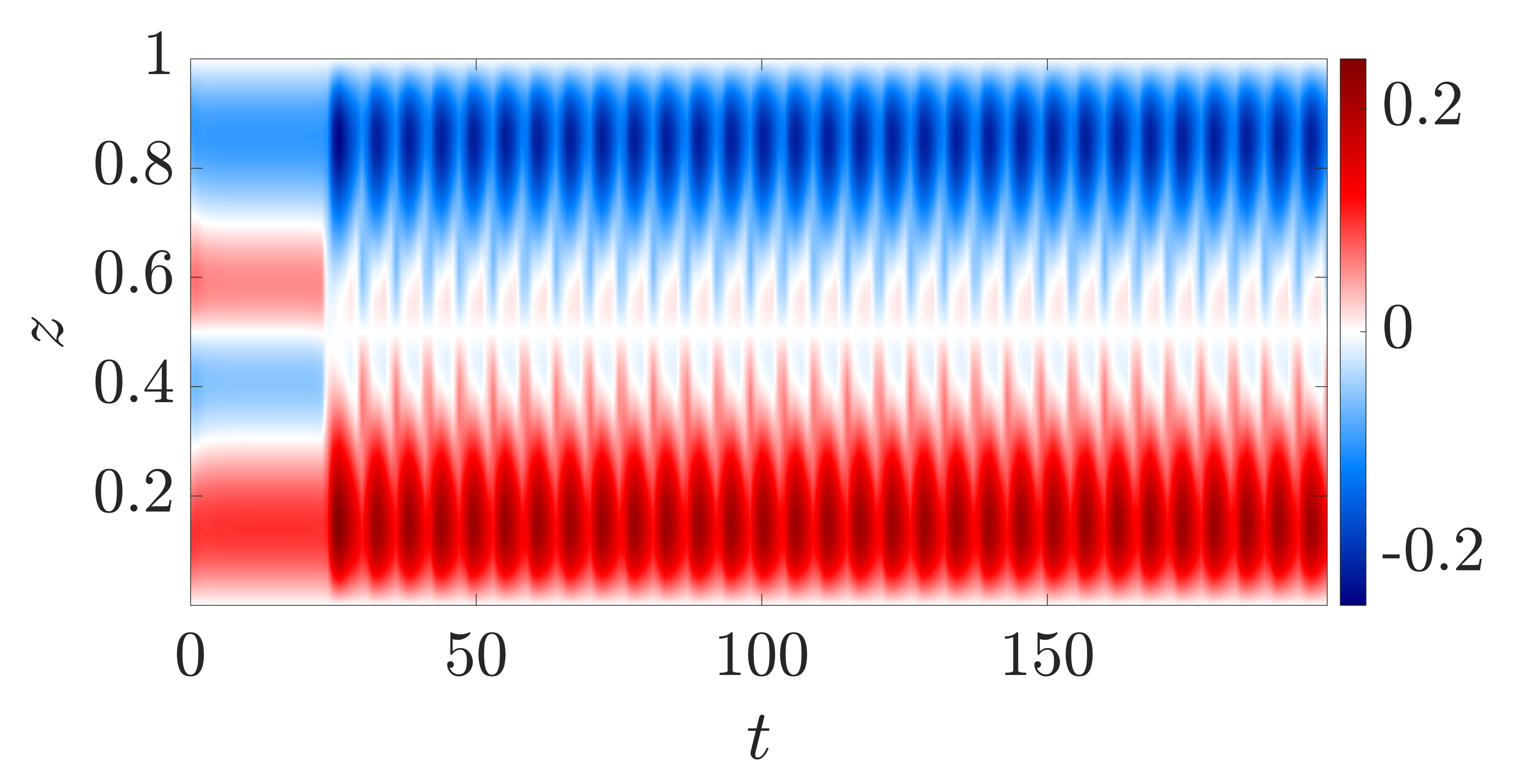}
        \includegraphics[width=0.19\textwidth,trim=-0 -0.8in 0 0]{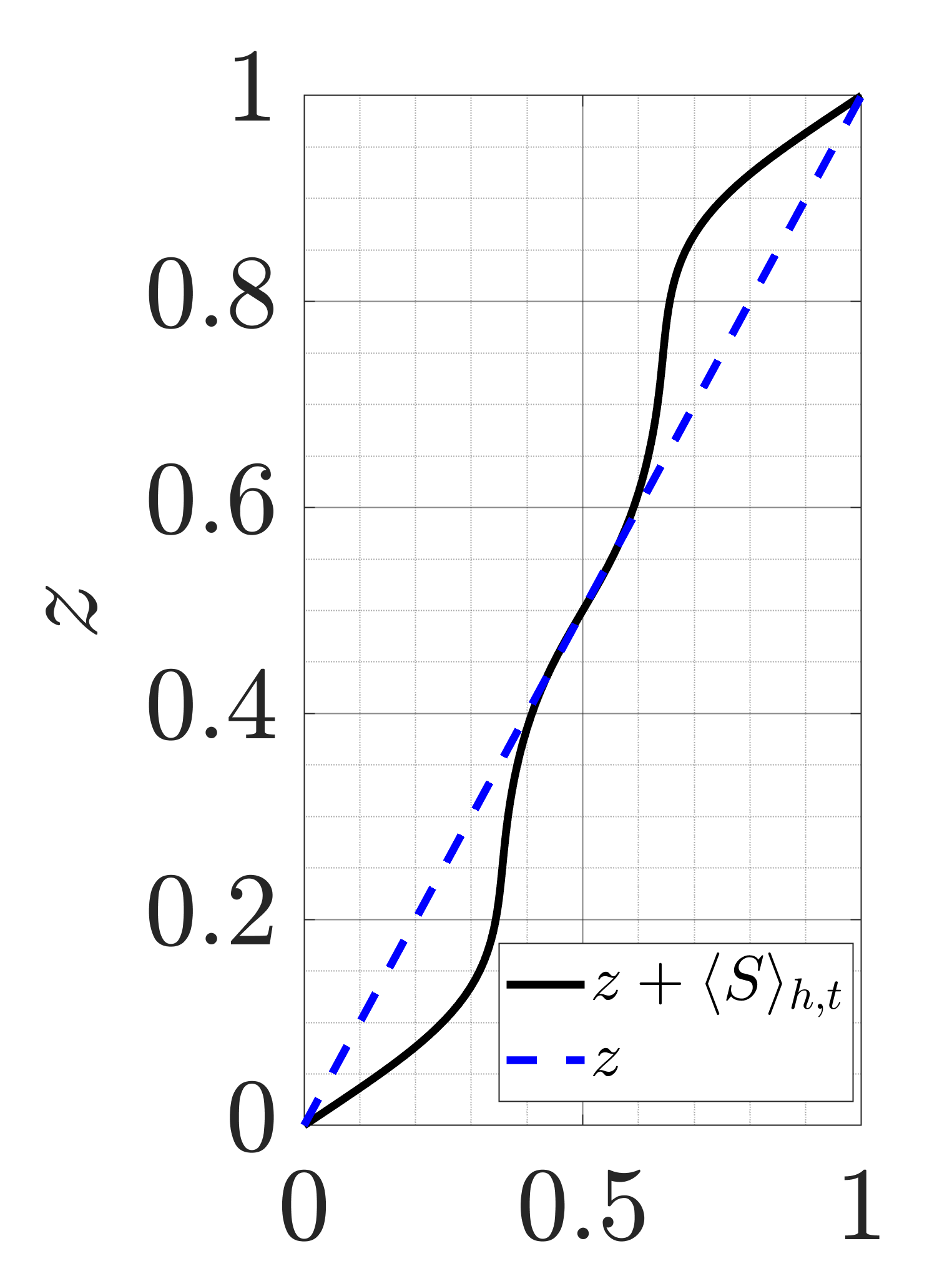}

$z+S(x,z,t)$

     (c) \hspace{0.3\textwidth} (d) \hspace{0.3\textwidth} (e)
    
    \includegraphics[width=0.32\textwidth]{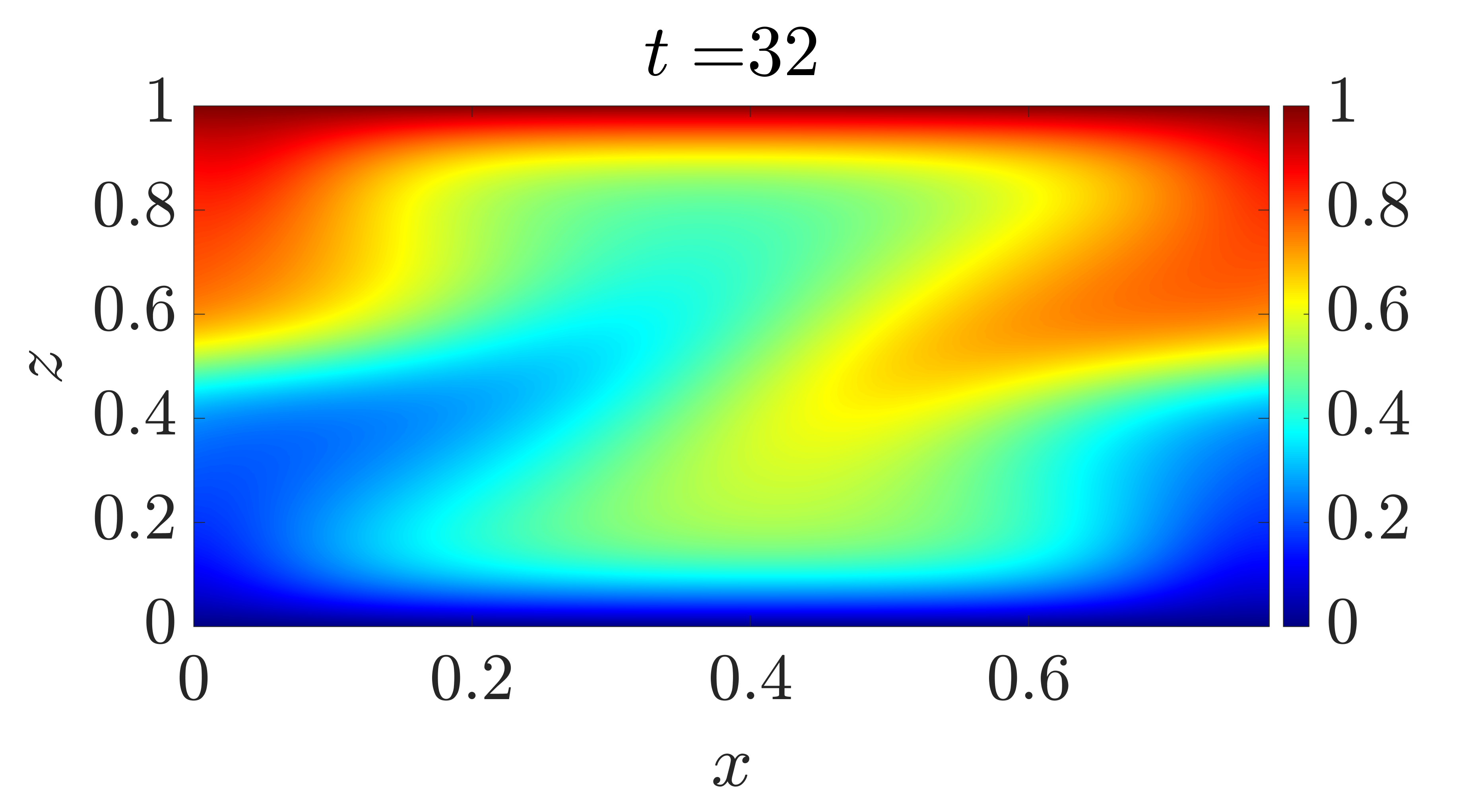}
    \includegraphics[width=0.32\textwidth]{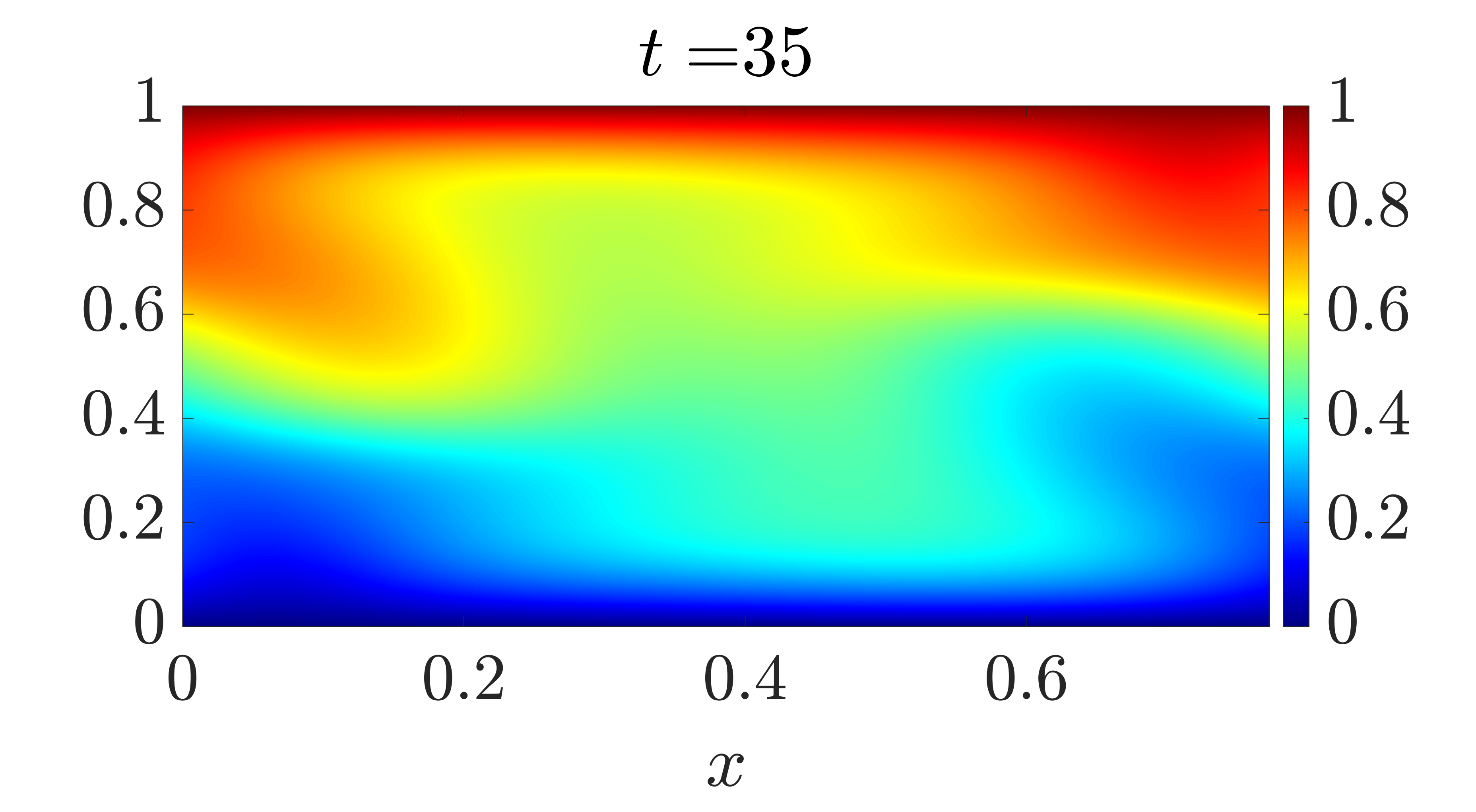}
    \includegraphics[width=0.32\textwidth]{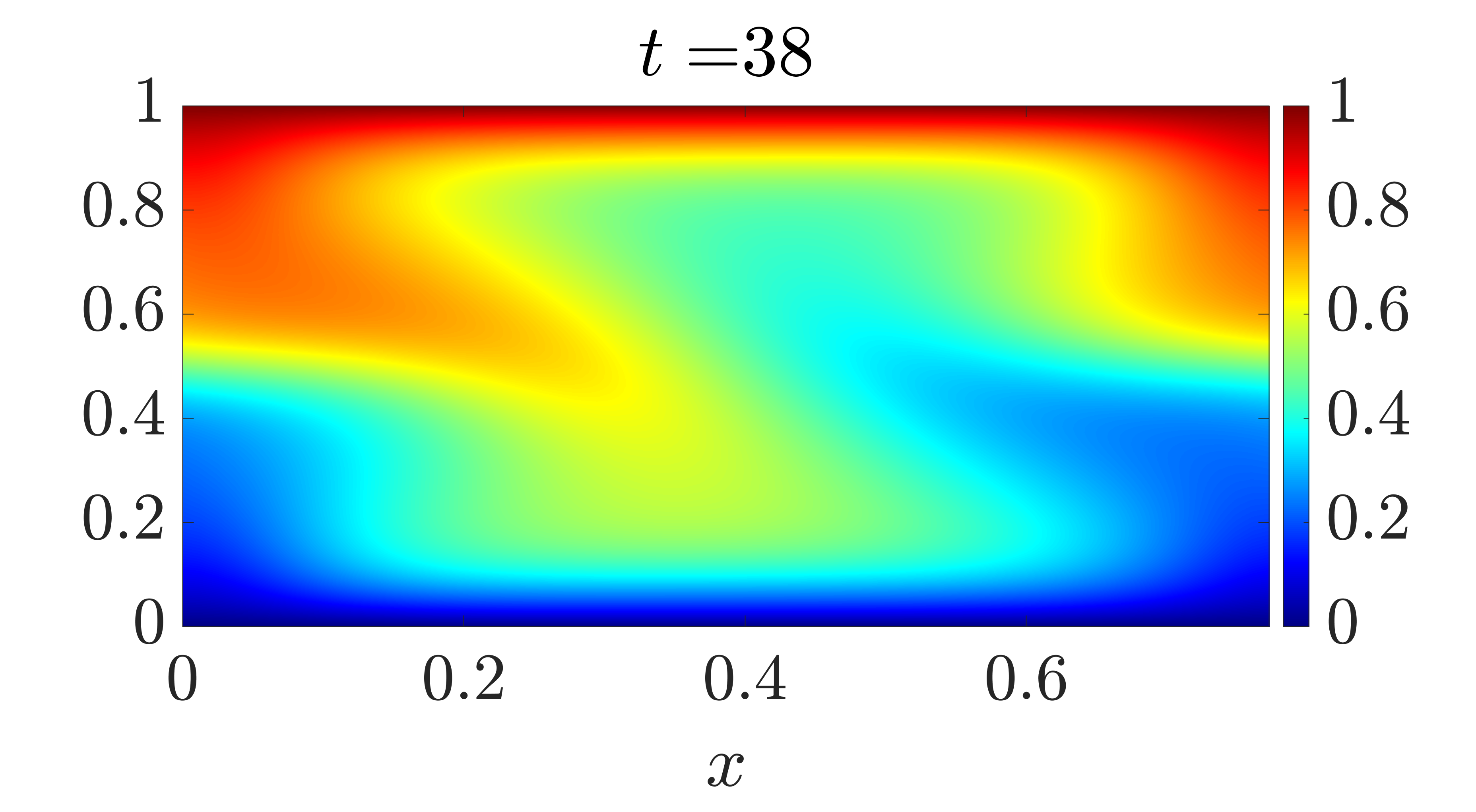}
    \caption{2D DNS showing (a) horizontally averaged salinity $\langle S\rangle_{h}(z,t)$ and (b) time-averaged and horizontally averaged total salinity $z+\langle S\rangle_{h,t}$. The second row shows isocontours of the total salinity $z+S(x,z,t)$ at (c) $t=32$, (d) $t=35$, and (e) $t=38$. The horizontal domain size is $L_x=2\pi/8$ with initial condition in the form of a S2 single-mode solution with $k_x=2\pi/L_x=8$, $k_y=0$. The parameters are $R_\rho=40$, $Pr=7$, $\tau=0.01$, $Ra_{T}=10^5$. See supplementary Movie 1. }
    \label{fig:DNS_Pr_7_S2}
\end{figure}

\begin{figure}
    \centering
    
    (a) $\langle S\rangle_h(z,t)$ \hspace{0.3\textwidth} (b) $z+S(x,z,t)$ at $z=0.1$
    
    \includegraphics[width=0.49\textwidth]{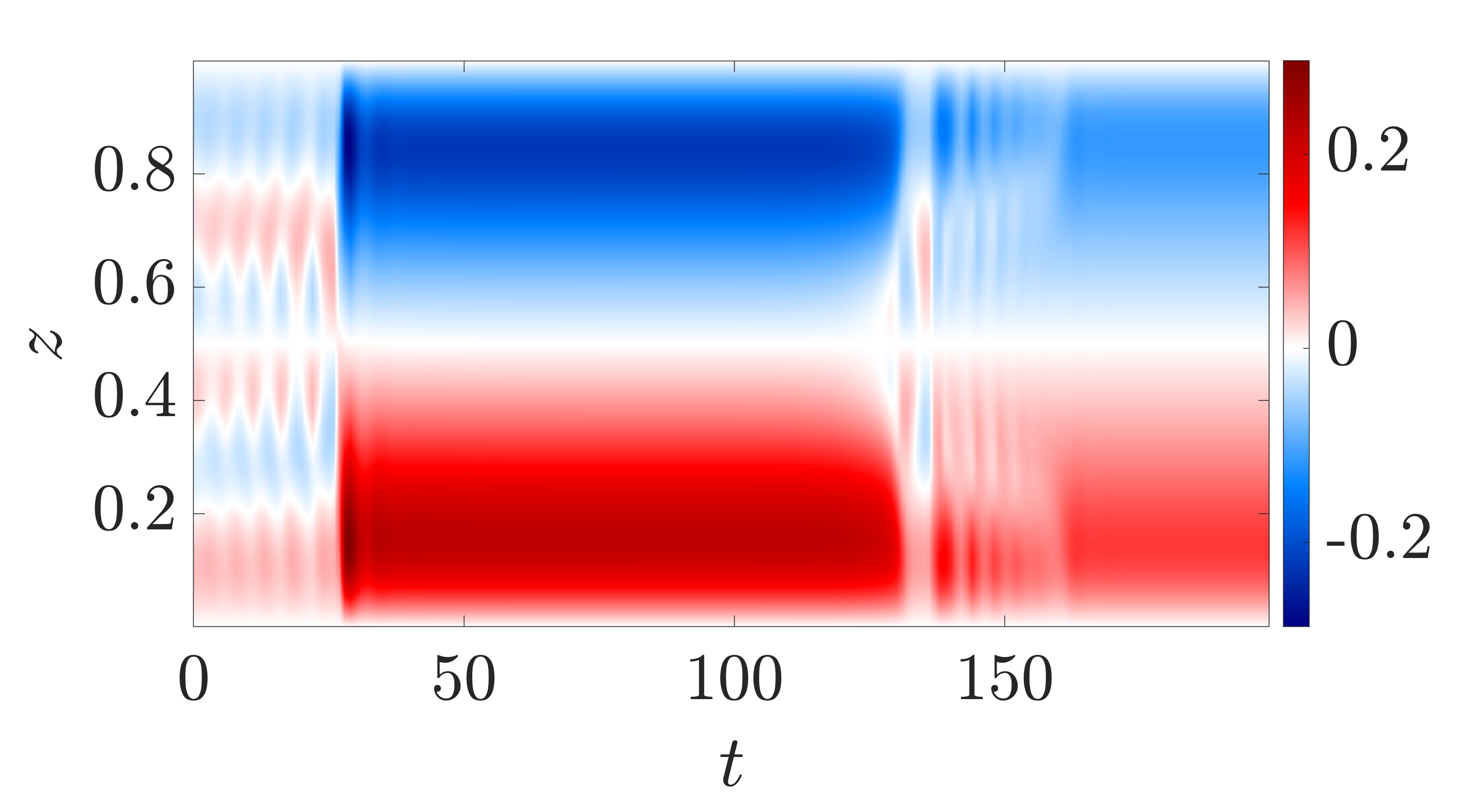}
    \includegraphics[width=0.49\textwidth]{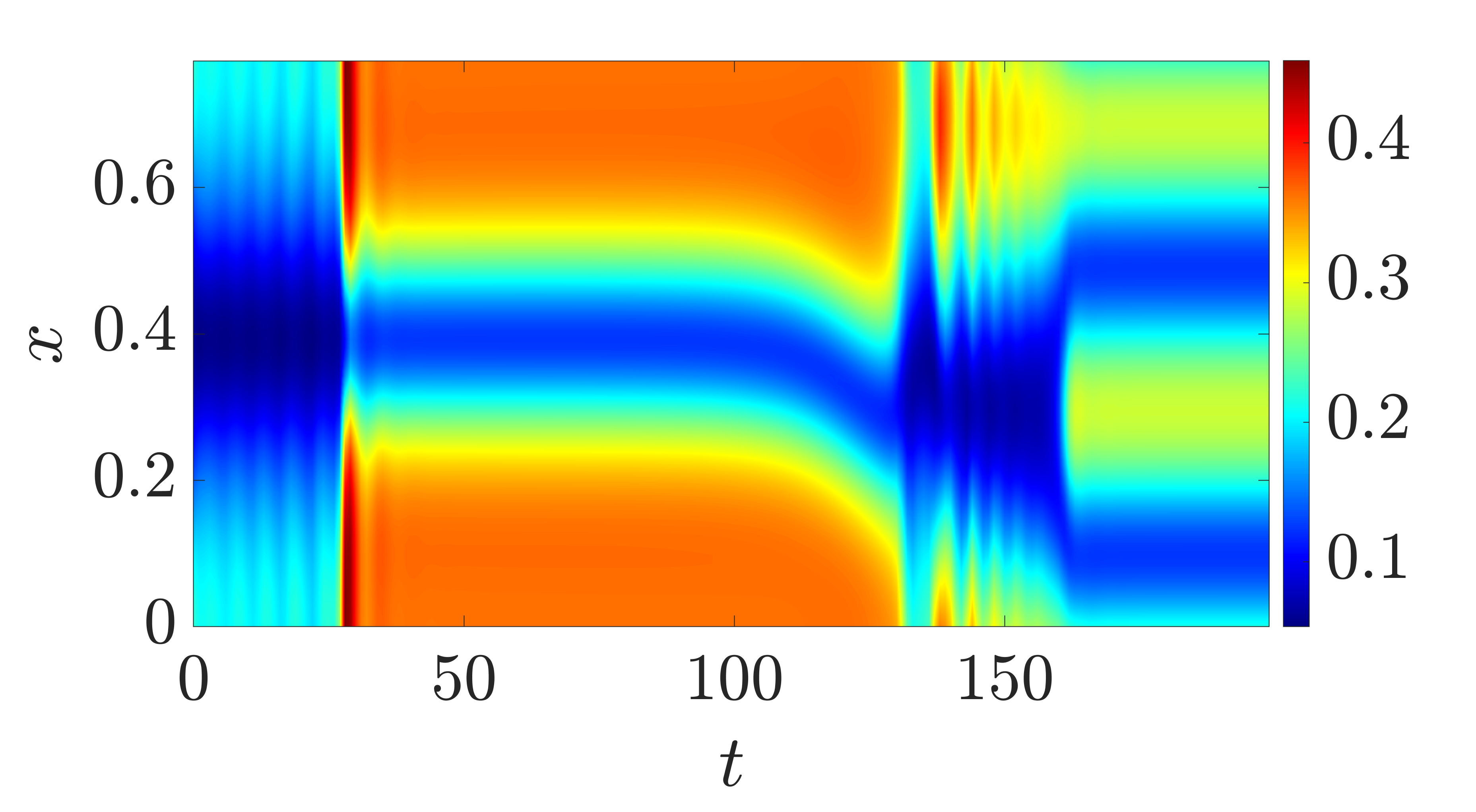}
    \caption{2D DNS results showing (a) the horizontally averaged salinity $\langle S\rangle_h(z,t)$ and (b) the total salinity $ z+S(x,z,t)$ at $z=0.1$. The horizontal domain size is based on $L_x=2\pi/8$ with S3 single-mode initial condition with $k_x=2\pi/L_x=8$, $k_y=0$. The parameters are $R_\rho=40$, $Pr=7$, $\tau=0.01$, $Ra_{T}=10^5$. See supplementary Movie 2.   }
    \label{fig:DNS_Pr_7_S3}
\end{figure}

\begin{figure}
  \centering
  
$z+S(x,z,t)$

(a) I.C.: S1\hspace{0.23\textwidth} (b) I.C.: S2 \hspace{0.23\textwidth} (c) I.C.: TF1
  
    \includegraphics[width=0.32\textwidth]{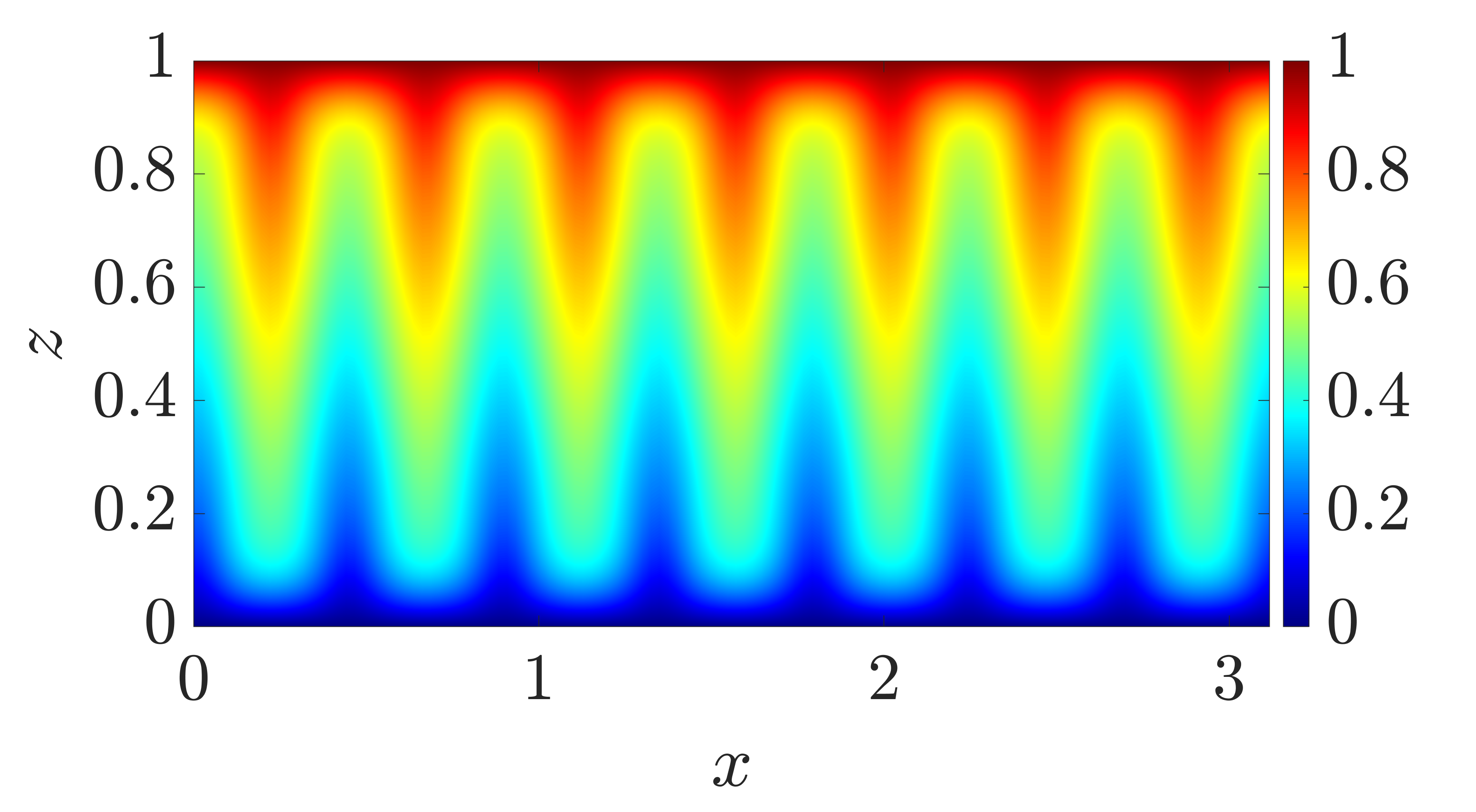}
   \includegraphics[width=0.32\textwidth]{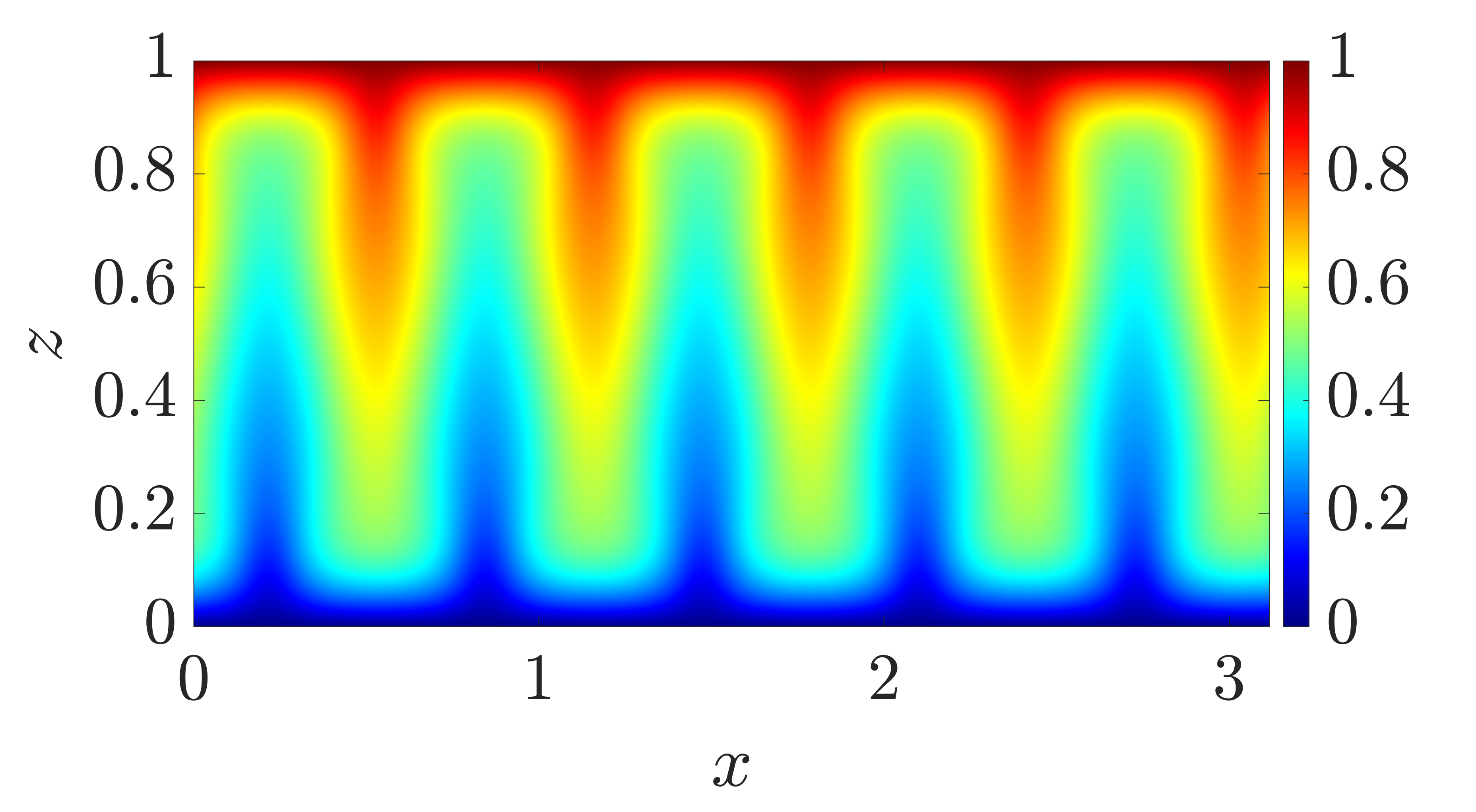}
    \includegraphics[width=0.32\textwidth]{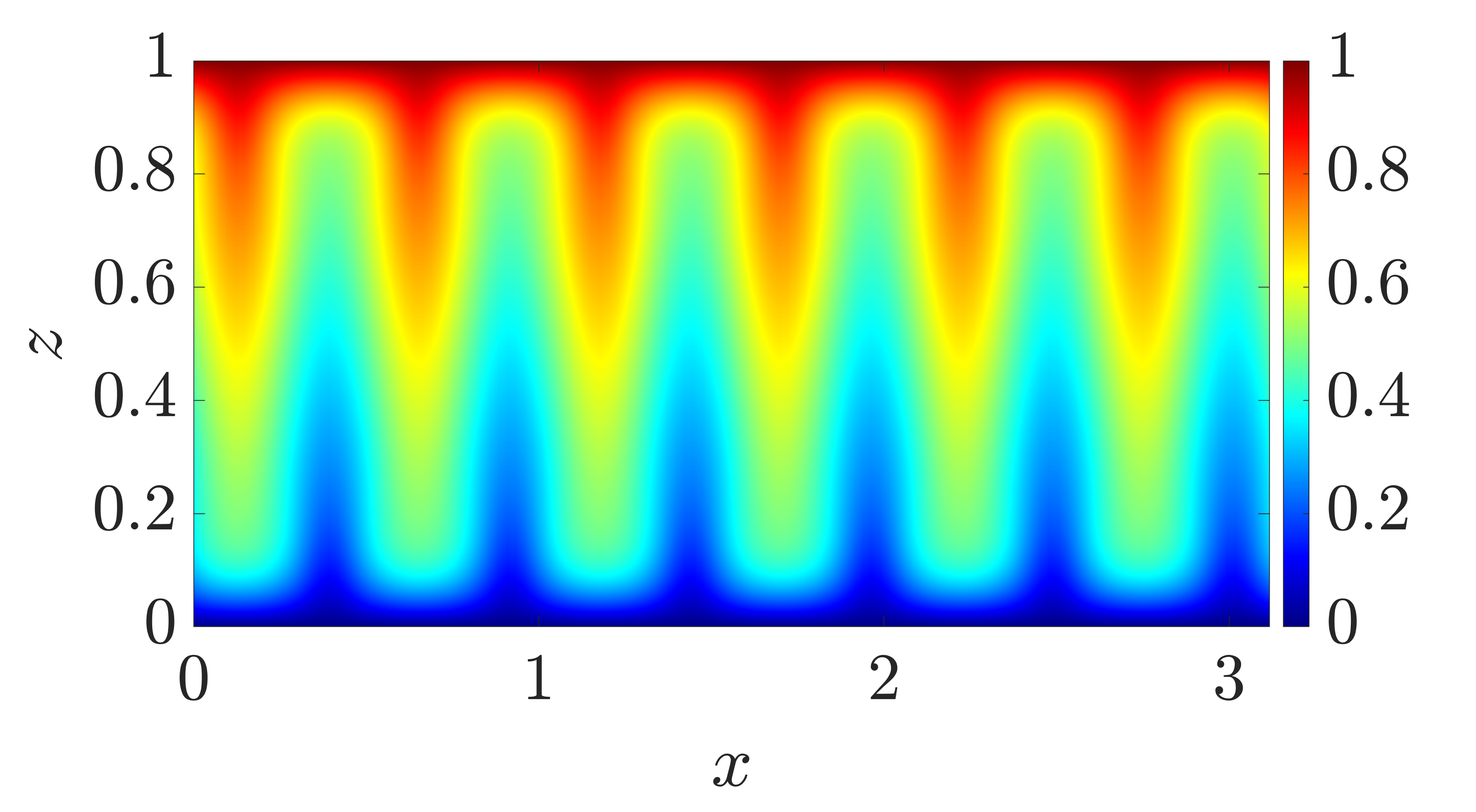}
    \caption{2D DNS results for the total salinity $z+S(x,z,t)$ at $t=3,000$ in a horizontal domain of size $L_x=\pi$ initialized by (a) S1, (b) S2, and (c) TF1 single-mode solutions with $k_x=2\pi/L_x=2$, $k_y=0$.  }
    \label{fig:DNS_R_rho_40_Pr_7_kx_2_S1_S2_TF1}
\end{figure}

\begin{figure}
    \centering

    (a) $z+S(x,z,t)$ at $z=1/2$ \hspace{0.1\textwidth}(b) $\langle |\mathcal{F}_x(S)| \rangle_t(z;k_x)$ \hspace{0.1\textwidth} (c) $\langle |\mathcal{F}_x(u)| \rangle_t(z;k_x)$

    \includegraphics[width=0.4\textwidth]{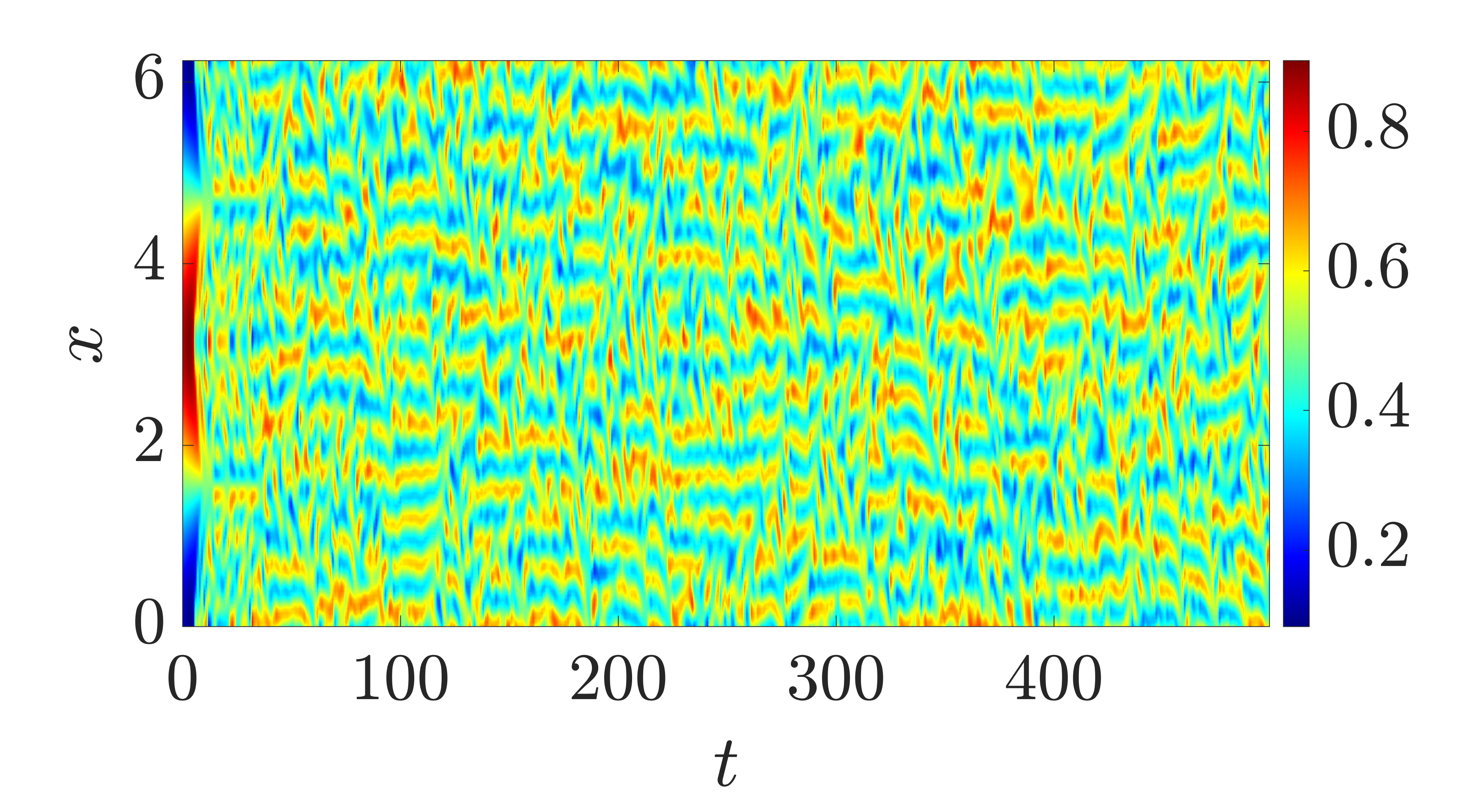}
    \includegraphics[width=0.29\textwidth]{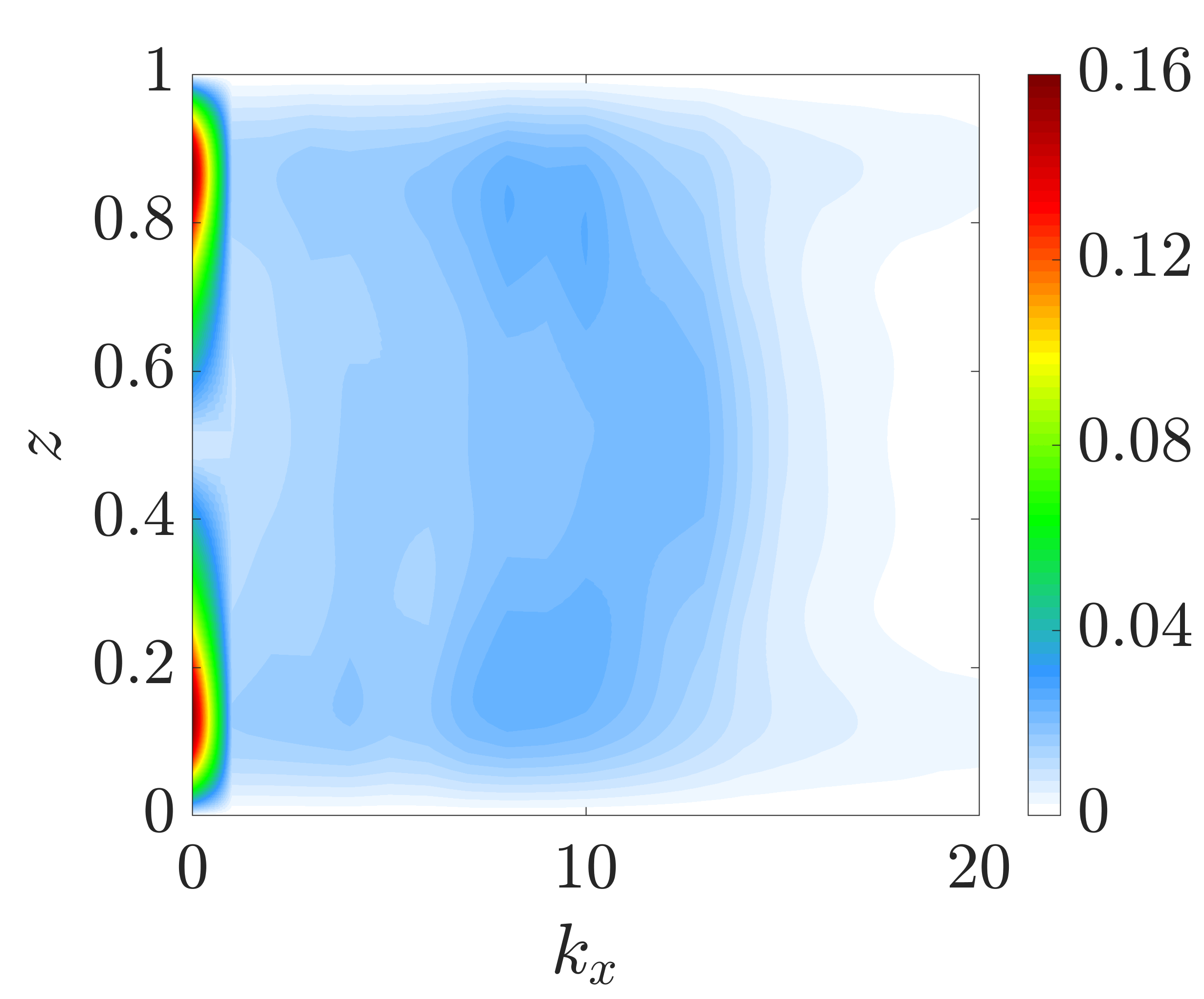}
    \includegraphics[width=0.29\textwidth]{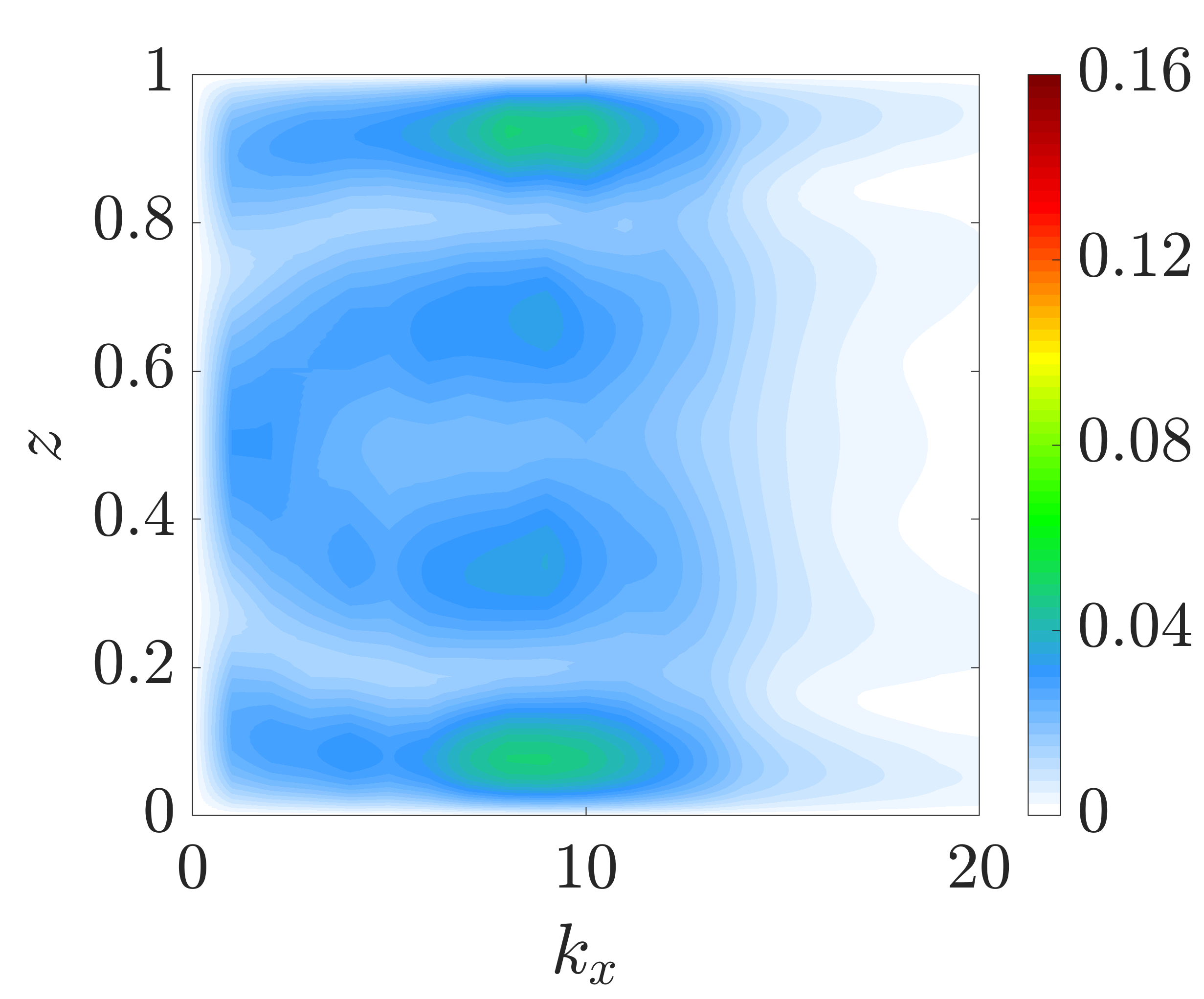}

 \hspace{0.05\textwidth}(d) \hspace{0.13\textwidth}(e) $t=416$ \hspace{0.12\textwidth} (f) $t=437$ \hspace{0.12\textwidth} (g) $t=478$

    \includegraphics[width=0.23\textwidth]{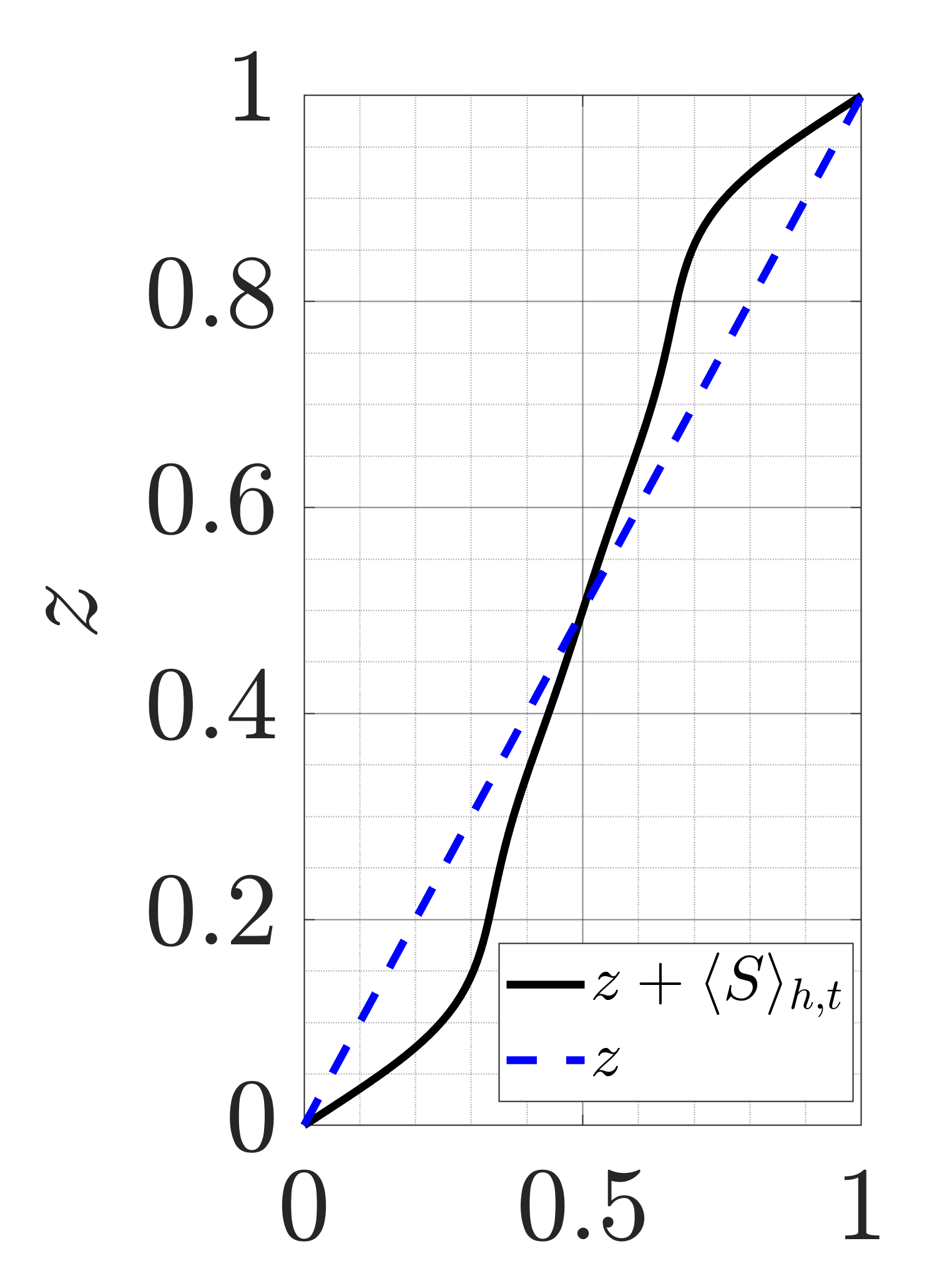}
    \includegraphics[width=0.23\textwidth]{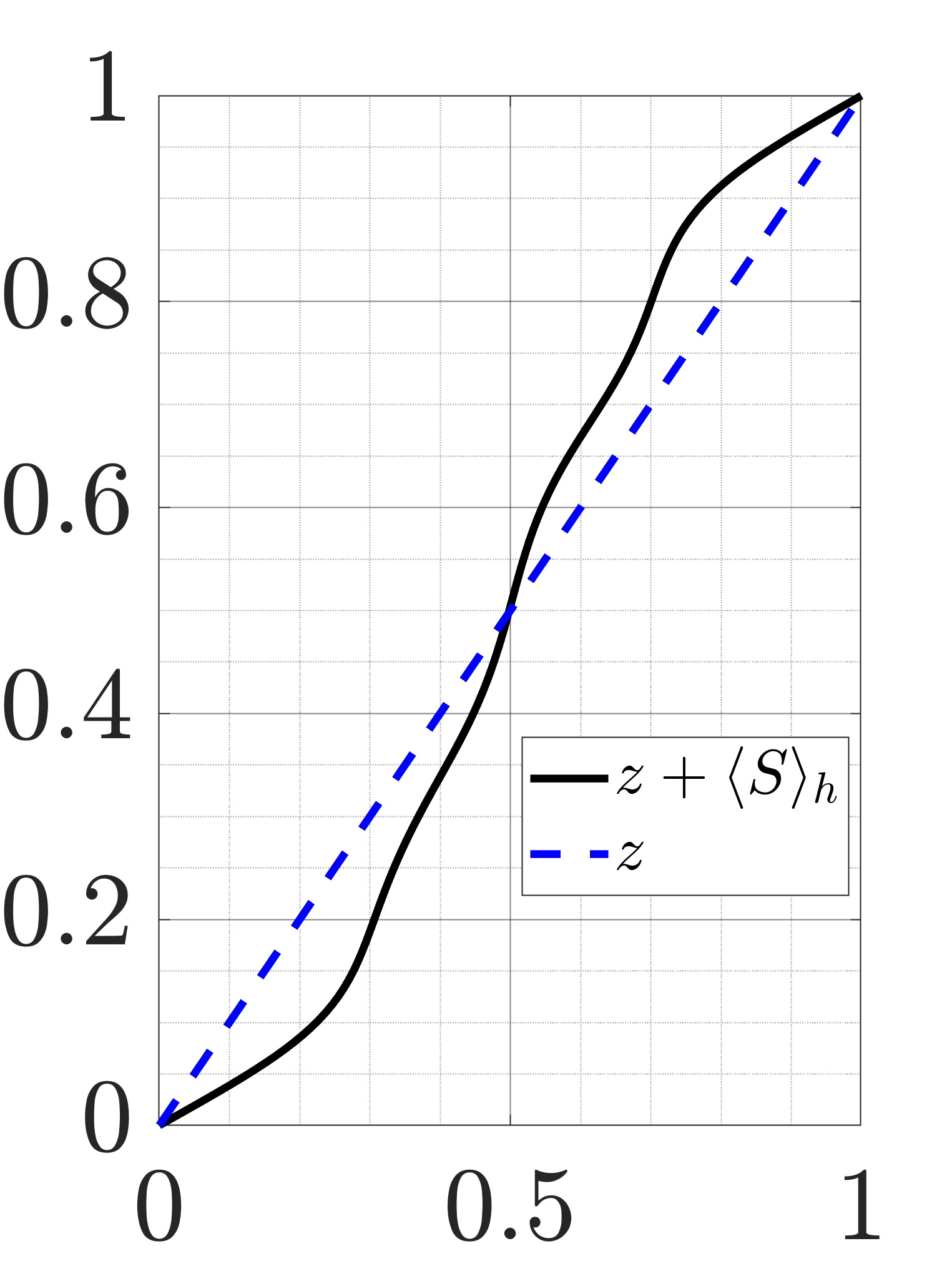}
    \includegraphics[width=0.23\textwidth]{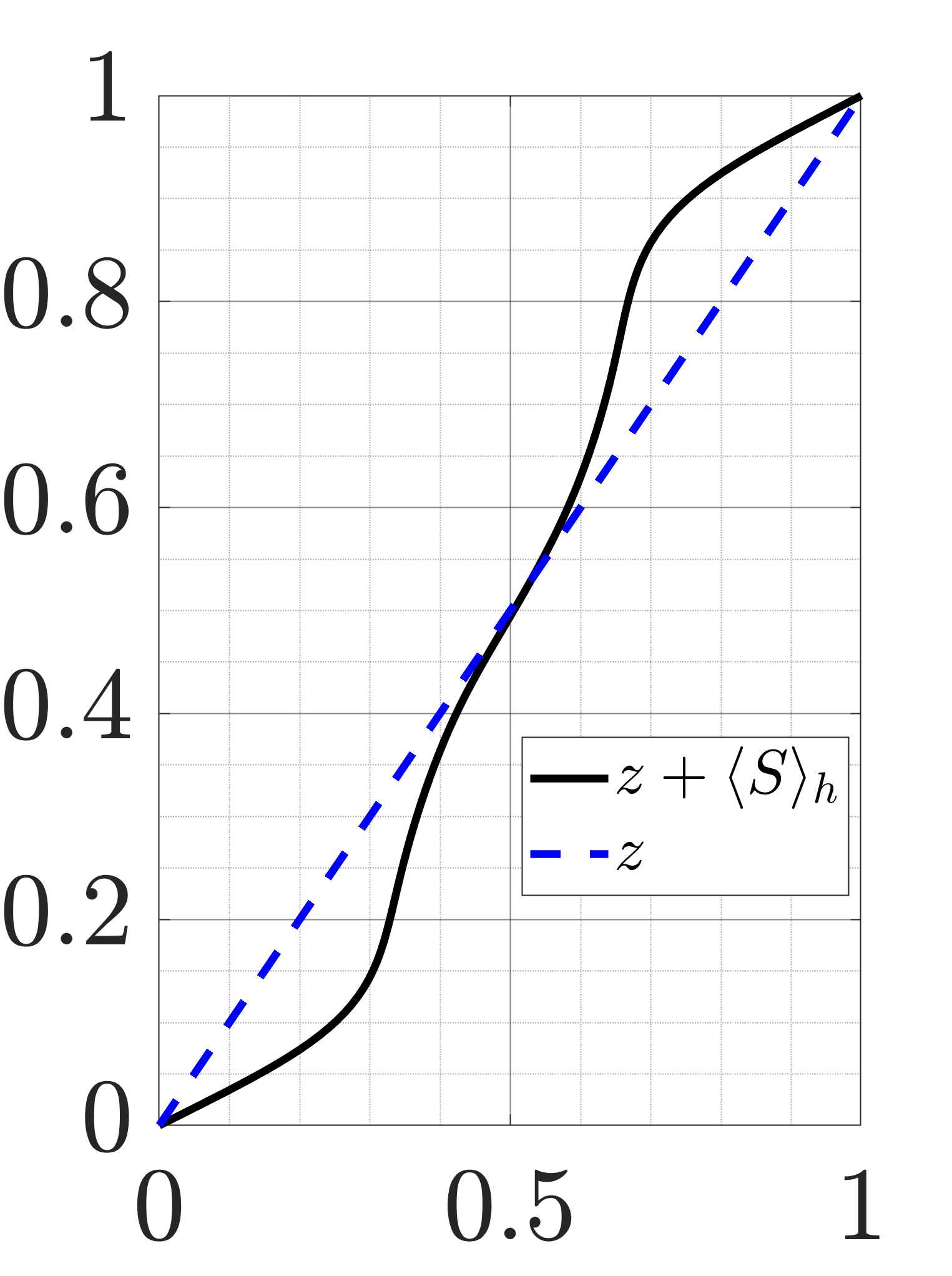}
    \includegraphics[width=0.23\textwidth]{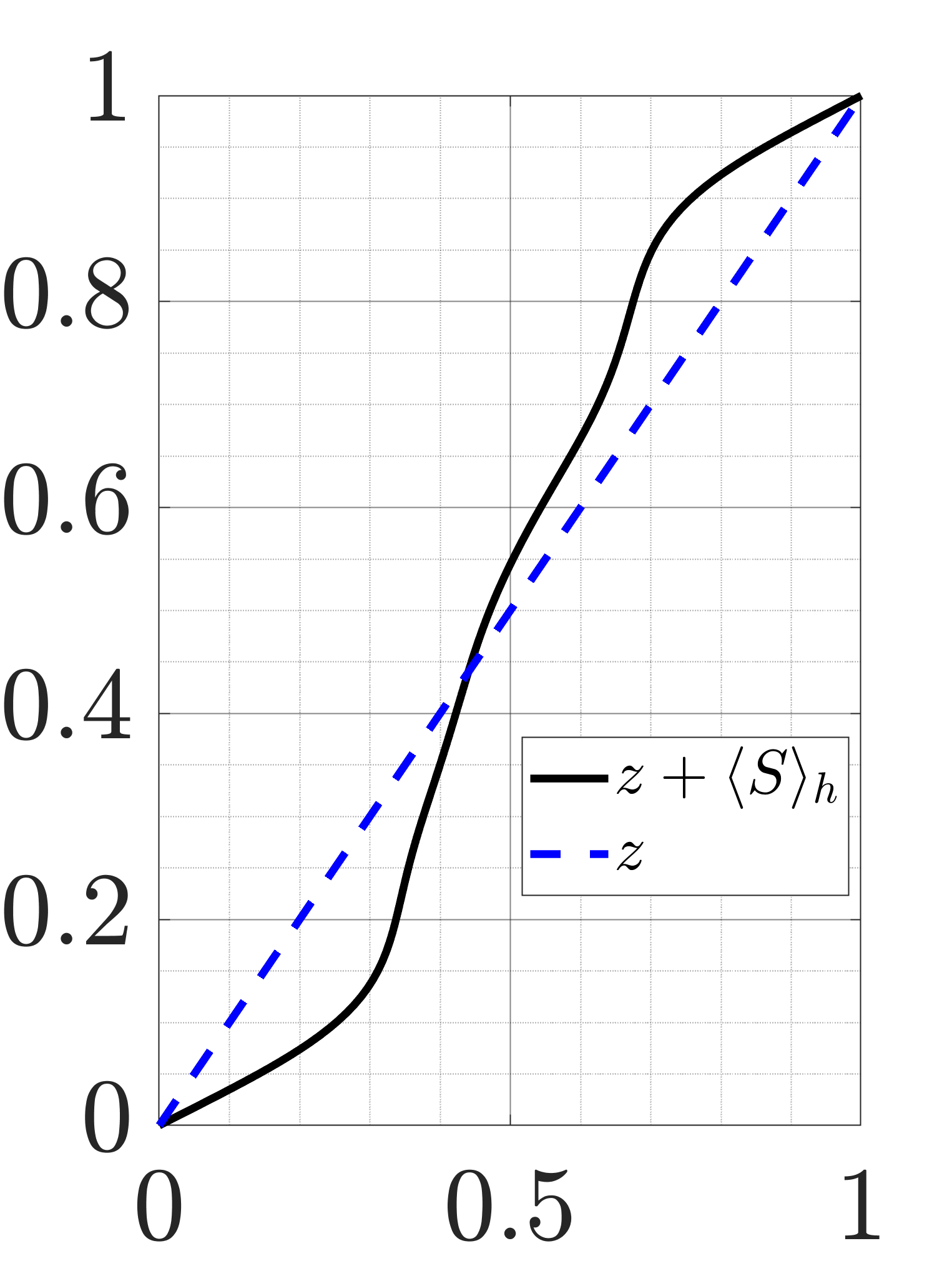}

    \caption{2D DNS results displaying (a) total salinity $z+S(x,z,t)$ at $z=1/2$, (b) $\langle |\mathcal{F}_x(S)| \rangle_t(z;k_x)$, (c) $\langle |\mathcal{F}_x(u)| \rangle_t(z;k_x)$, (d) $z+\langle S\rangle_{h,t}(z)$, and snapshots of the horizontally averaged total salinity $z+\langle S\rangle_h(z,t)$ at (e) $t=416$, (f) $t=437$ and (g) $t=478$. The horizontal domain size is $L_x=2\pi$ initialized by a S1 solution with $k_x=2\pi/L_x=1$, $k_y=0$. The governing parameters are $R_\rho=40$, $Pr=7$, $\tau=0.01$, and $Ra_T=10^5$. }
    \label{fig:DNS_R_rho_40_Pr_7_kx_1_S1}
\end{figure}

The final states in a domain of size $L_x=2\pi/8$ initialized by S2 and S3 show different flow structures. Figure \ref{fig:DNS_Pr_7_S2} shows DNS results initialized with S2 and $L_x=2\pi/8$. The horizontally averaged salinity deviation $\langle S\rangle_{h}(z,t)$ in figure \ref{fig:DNS_Pr_7_S2}(a) takes the requisite two-layer form for short times but then begins to oscillate in time. After averaging over time and in the horizontal direction, the total salinity profile $z+\langle S\rangle_{h,t}$ in figure \ref{fig:DNS_Pr_7_S2}(b) shows two mixed regions, with a broad interior region that is close to a linear profile. In fact, this oscillation and the two associated mixed regions in the mean salinity profile persist to at least $t=30,000$ (not shown). This behavior is similar to the TF1 single-mode solution as shown in figure \ref{fig:profile_R_rho_T2S_40_tau_0p01_Pr_7_TF1}(a). The mean salinity profile in figure \ref{fig:DNS_Pr_7_S2}(b) also resembles the mean temperature profile in Rayleigh-B\'enard convection with induced large-scale shear \citep[figure 5(b)]{goluskin2014convectively}, where the temperature profile shows two mixed regions close to the boundary, and an interior that approaches a linear profile with increasing Rayleigh number. The three snapshots of the total salinity in figure \ref{fig:DNS_Pr_7_S2}(c)-(e) reveal that this behavior is associated with a new state, a direction-reversing tilted finger (RTF). Such behavior is also observed within simulations of the single-mode equations \citep{paparella1999sheared} (termed a layering instability there) and is accompanied by an oscillating large-scale shear. The direction-reversing tilted finger state also resembles (at least phenomenologically) the `pulsating wave' found by \citet[figure 7]{proctor1994nonlinear} and \citet[figure 5]{matthews1993pulsating} in magnetoconvection as well as large-scale flow reversals observed in Rayleigh-B\'enard convection \citep{sugiyama2010flow,chandra2013flow,winchester2021zonal}. We have not explored the origin of this state but note that similar `pulsating waves' in other systems originate from a sequence of local and global bifurcations of a tilted convection roll \citep{matthews1993pulsating,proctor1994nonlinear,rucklidge1996analysis}.  

Figure \ref{fig:DNS_Pr_7_S3} shows the DNS results with an S3 initial condition and $L_x=2\pi/8$ and displays the dynamics that are typically involved in transitioning to a higher wavenumber S1 solution. The profile of $\langle S\rangle_h$ in figure \ref{fig:DNS_Pr_7_S3}(a) initially shows a three-layer structure but starts to oscillate prior to a transition to the one-layer state S1 at $t\approx 30$. After $t\approx160$, this one-layer solution transitions to another one-layer solution but with a smaller magnitude in $\langle S \rangle_{h}$. The total salinity near the lower boundary, $z=0.1$, is shown in figure \ref{fig:DNS_Pr_7_S3}(b) and further explains this transition scenario. The one-layer solution starts to tilt at $t\approx 110$ and begins to show oscillations at $t\approx 130$. Beyond $t\approx 160$, the horizontal wavenumber doubles leading to the S1 (2) state in table \ref{tab:DNS_transition_low_Ra_S2T_Pr_7}. Such a transition to a higher wavenumber S1 solution is also observed with horizontal domains $L_x\in [2\pi/6, 2\pi/2]$, albeit with different possibilities for the final wavenumber. Note that such transitions between different horizontal wavenumbers are not possible within the single-mode approach. 

For $L_x=\pi$, the final state shows different horizontal wavenumbers depending on the three different initial conditions S1, S2, and TF1 as indicated in table \ref{tab:DNS_transition_low_Ra_S2T_Pr_7}. The total salinity at $t=3,000$ in these three final states is shown in figure \ref{fig:DNS_R_rho_40_Pr_7_kx_2_S1_S2_TF1}. Although the horizontal domain size and the governing parameters are the same, different initial conditions lead to different final horizontal wavenumbers and the DNS results suggest that these persist up to at least $t=3,000$. This observation suggests that for $L_x=\pi$ the three initial conditions S1, S2, and TF1 all lie in the basin of attraction of S1 solutions but associated with different and larger horizontal wavenumbers. This is of course a consequence of the preferred horizontal scale of the fingers. Note that the S3 solution does not exist for this domain size; see figure \ref{fig:bif_diag_low_Ra_S2T}(a) and table \ref{tab:DNS_transition_low_Ra_S2T_Pr_7}. Rolls with different horizontal wavenumber obtained from different initial conditions are a familiar phenomenon in Rayleigh-B\'enard convection \citep{wang2020multiple} and in spanwise rotating plane Couette flow \citep{xia2018multiple,yang2021bifurcation}.

Solutions that do not reach a well-organized structure will be referred to as chaotic. Such behavior is observed in large $L_x$ domains. For example, single-mode S1 solutions with $k_x=1$ ($L_x=2\pi$) and $k_x=0.5$ ($L_x=4\pi$) are stable as shown in figure \ref{fig:bif_diag_low_Ra_S2T}(a) but DNS initialized by these solutions ultimately exhibit chaotic behavior. This is a consequence of the harmonics of the fundamental wavenumber $k_x$ included in the DNS. The total salinity in the midplane for $L_x=2\pi$ is shown in figure \ref{fig:DNS_R_rho_40_Pr_7_kx_1_S1}(a), and displays chaotic behavior with multiple downward and upward plumes in the horizontal. In order to further characterize the horizontal length scale, we computed $\langle |\mathcal{F}_x(S)|\rangle_t(z;k_x)$ and $\langle |\mathcal{F}_x(u)|\rangle_t(z;k_x)$, where $\mathcal{F}_x(\cdot)$ is the Fourier transform in the $x$-direction and $|\cdot|$ represents the modulus of the obtained complex Fourier coefficients. The results are shown in figures \ref{fig:DNS_R_rho_40_Pr_7_kx_1_S1}(b)-(c). The $k_x=0$ component of $\langle |\mathcal{F}_x(\cdot)|\rangle_t(z;k_x)$ is related to the amplitude of the horizontal mean $\bar{(\cdot)}_0$ while the $k_x\neq 0$ contributions to $\langle |\mathcal{F}_x(\cdot)|\rangle_t(z;k_x)$ are associated with the harmonic components $\widehat{(\cdot )}$ in the single-mode ansatz \eqref{eq:normal_mode}. Figure \ref{fig:DNS_R_rho_40_Pr_7_kx_1_S1}(b) displays a peak at $k_x=0$ that corresponds to the deviation of the mean salinity from its linear profile. A second peak around $k_x\approx 9$ in figure \ref{fig:DNS_R_rho_40_Pr_7_kx_1_S1}(b) corresponds to a wavenumber that is close to that providing the largest $Sh$ in the S1 solution shown in figure \ref{fig:bif_diag_low_Ra_S2T}(a). However, in the midplane, the peak wavenumber shifts to $k_x= 13$.

The Fourier transform of the horizontal velocity enables us to isolate the role of the S1, S2 and S3 solutions. The plot of $\langle |\mathcal{F}_x(u)| \rangle_t(z;k_x)$ in figure 10(c) reveals a stronger peak at $k_x=9$ but now at multiple locations dominated by regions near the boundaries, with a lower amplitude near $z=1/3$ and $2/3$, the peak region of the horizontal velocity $\widetilde{u}$ in the three-layer solution S3 shown in figure 3(c). Note that the horizontal velocity in the S1 and S2 solutions almost vanishes at these locations as shown in figure 3(c). Furthermore, $\langle |\mathcal{F}_x(u)| \rangle_t(z;k_x)$ also shows a local peak at $k_x=1$ near the midplane resembling the horizontal velocity $\widetilde{u}$ of the two-layer solution S2, while the horizontal velocity in the S1 and S3 single-mode solutions vanishes in the midplane as shown in figure 3(c). These observations suggest that the unstable S2 and S3 solutions both play a role in the chaotic behavior and that their properties may manifest themselves in the statistics of the chaotic state, cf. \citet{kawahara2001periodic} and the reviews by \citet{kawahara2012significance} and \citet{graham2021exact}.

In order to further quantify the role of unstable solutions, we plot in figures \ref{fig:DNS_R_rho_40_Pr_7_kx_1_S1}(e)-(g) the horizontally averaged profiles at three different times for comparison with the horizontally averaged time-averaged state in figure \ref{fig:DNS_R_rho_40_Pr_7_kx_1_S1}(d). Although the salinity profile in the latter state does not show a significant staircase, the horizontally averaged salinity profiles at the three times shown in figures \ref{fig:DNS_R_rho_40_Pr_7_kx_1_S1}(e)-(f) do in fact show a tendency towards three and two steps, respectively, while the snapshot in figure \ref{fig:DNS_R_rho_40_Pr_7_kx_1_S1}(g) shows instead a salinity profile that temporarily breaks the midplane reflection symmetry. These results provide evidence that the chaotic state visits the neighborhood of not only the unstable staircase-like solutions S2 and S3 but also of the asymmetric solution. The time-averaged $Sh$ within $t\in [100,500]$ is $Sh\approx 2.8693$, a value lower than the maximal $Sh$ for S1 but higher than those for S2 and S3 at the suggested peak wavenumber $k_x=9$ as shown in figure \ref{fig:bif_diag_low_Ra_S2T}(a), and suggests that these states may play a role in determining the time-averaged salinity transport in this chaotic state.
\subsection{No-slip versus stress-free boundary conditions}
\label{subsec:results_Pr_7_stress_free}

\begin{figure}
    \centering
    
    (a) 2D: $k_y=0$ \hspace{0.35\textwidth} (b) 3D: $k_y=k_x$
    
    \includegraphics[width=0.49\textwidth]{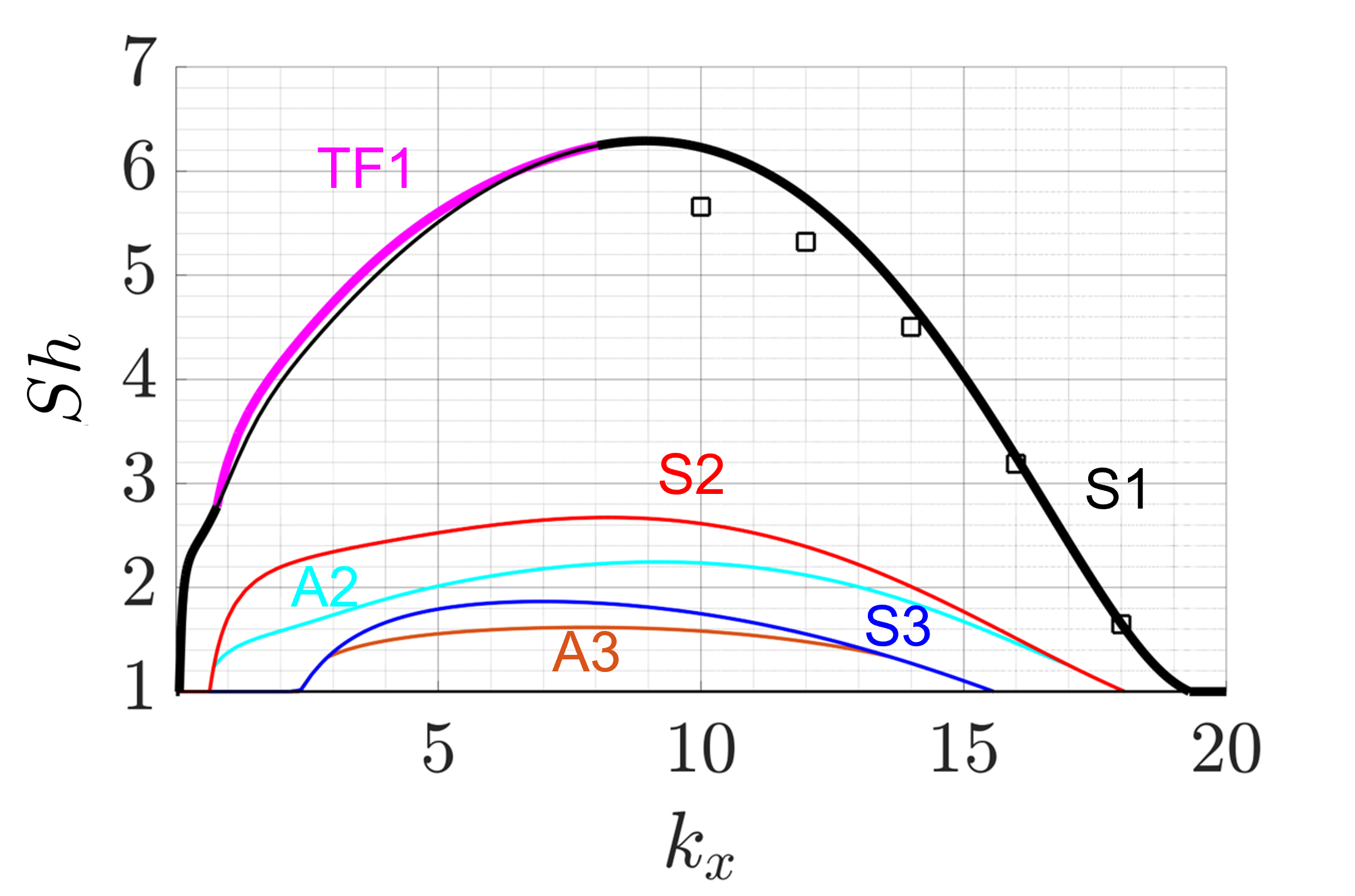}
    \includegraphics[width=0.49\textwidth]{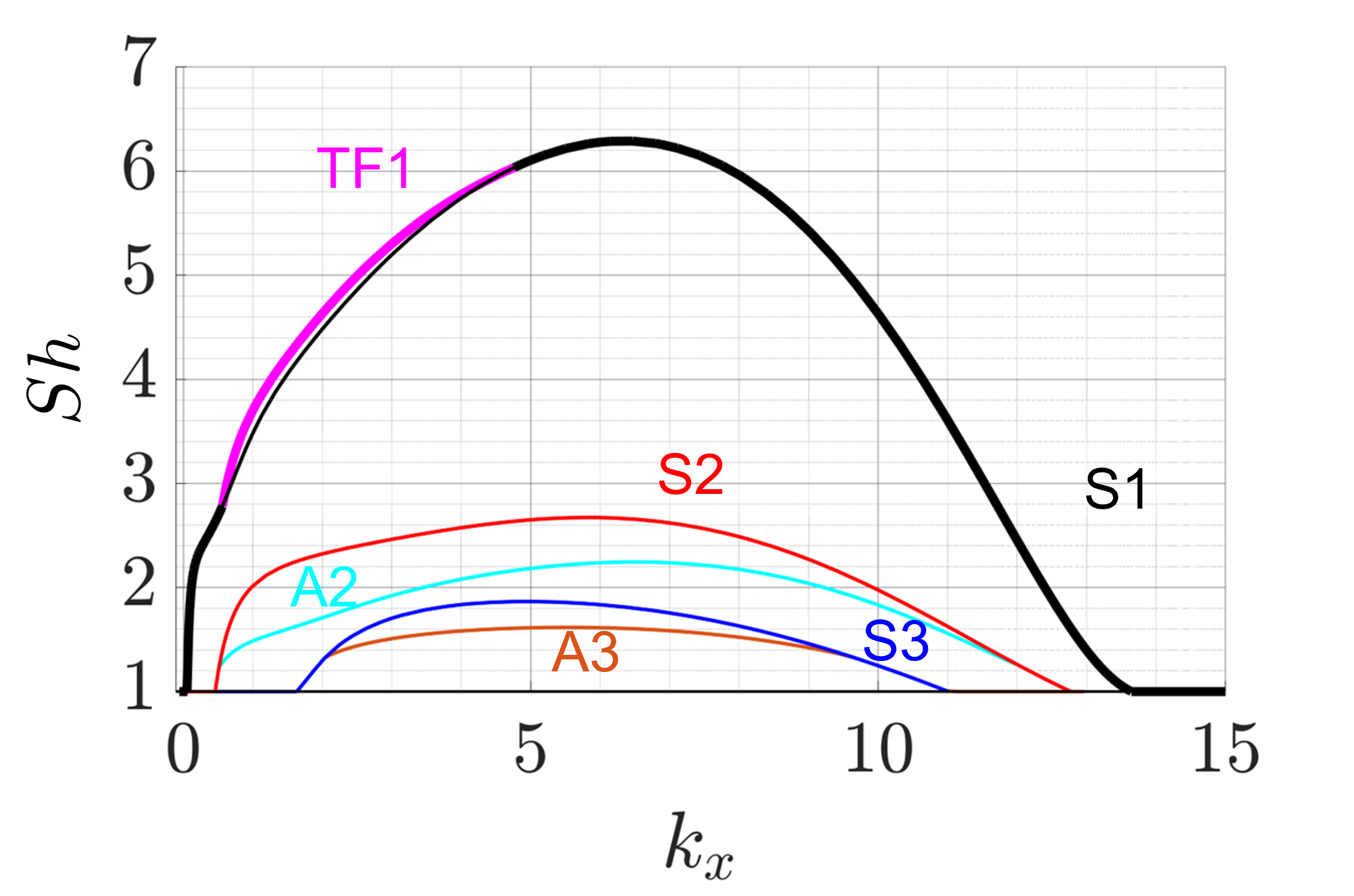}
    
    (c) \hspace{0.49\textwidth} (d)
    \includegraphics[width=0.49\textwidth]{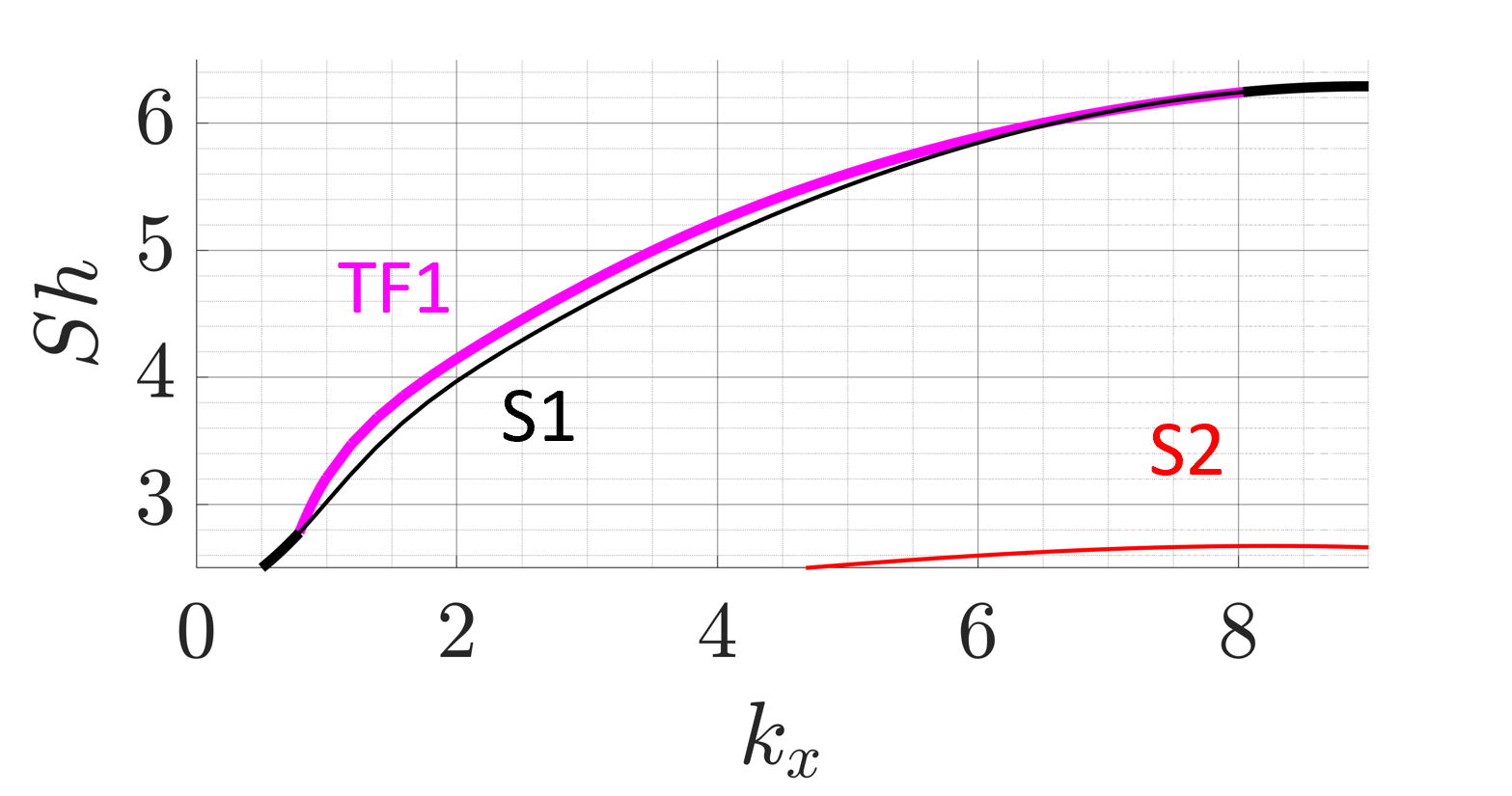}
    \includegraphics[width=0.49\textwidth]{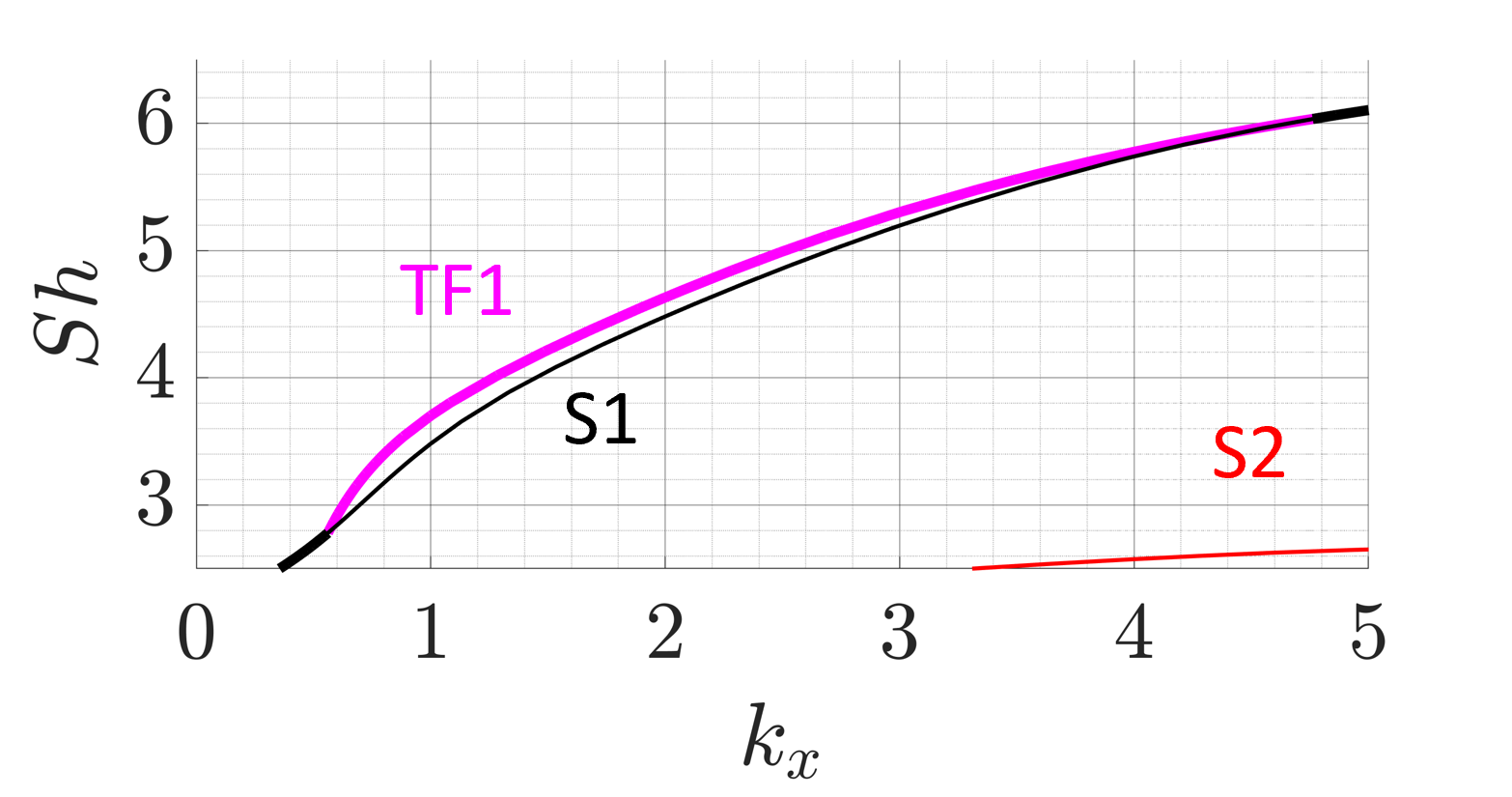}
    \caption{Bifurcation diagrams computed from the single-mode equations \eqref{eq:single_mode} with the same parameter values as in figure \ref{fig:bif_diag_low_Ra_S2T} but stress-free velocity boundary conditions at top and bottom. The black squares show the corresponding steady state $Sh$ reached using 2D DNS in domains of size $L_x=2\pi /k_x$. Panels (c) and (d) show zooms of the 2D and 3D results near the TF1 branch, respectively.}
    \label{fig:bif_diag_low_Ra_S2T_stress_free}
\end{figure}

\begin{figure}
\centering
    (a) \hspace{0.2\textwidth} (b) \hspace{0.2\textwidth} (c) \hspace{0.2\textwidth} (d)
    
    \includegraphics[width=0.24\textwidth]{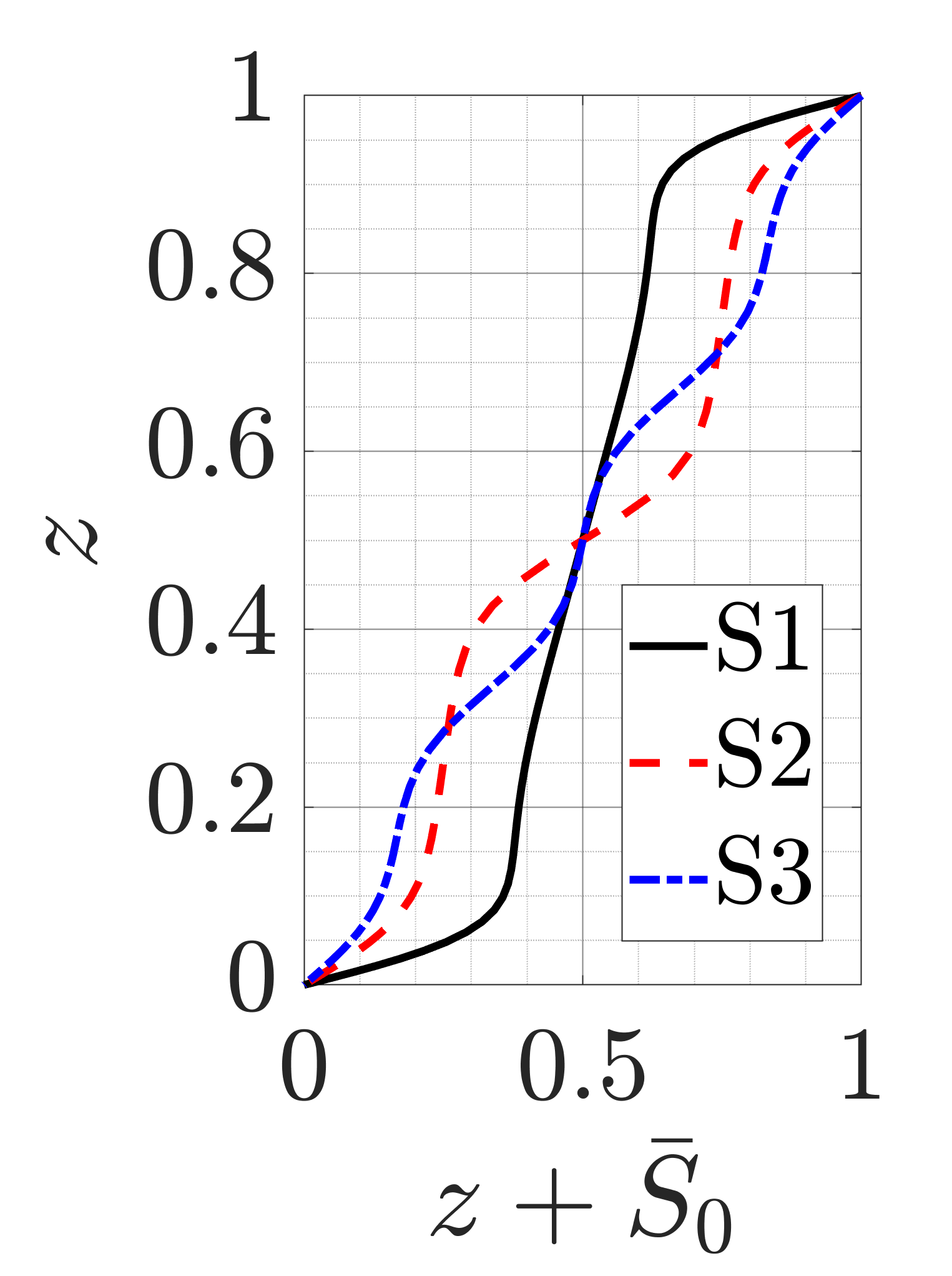}
    \includegraphics[width=0.24\textwidth]{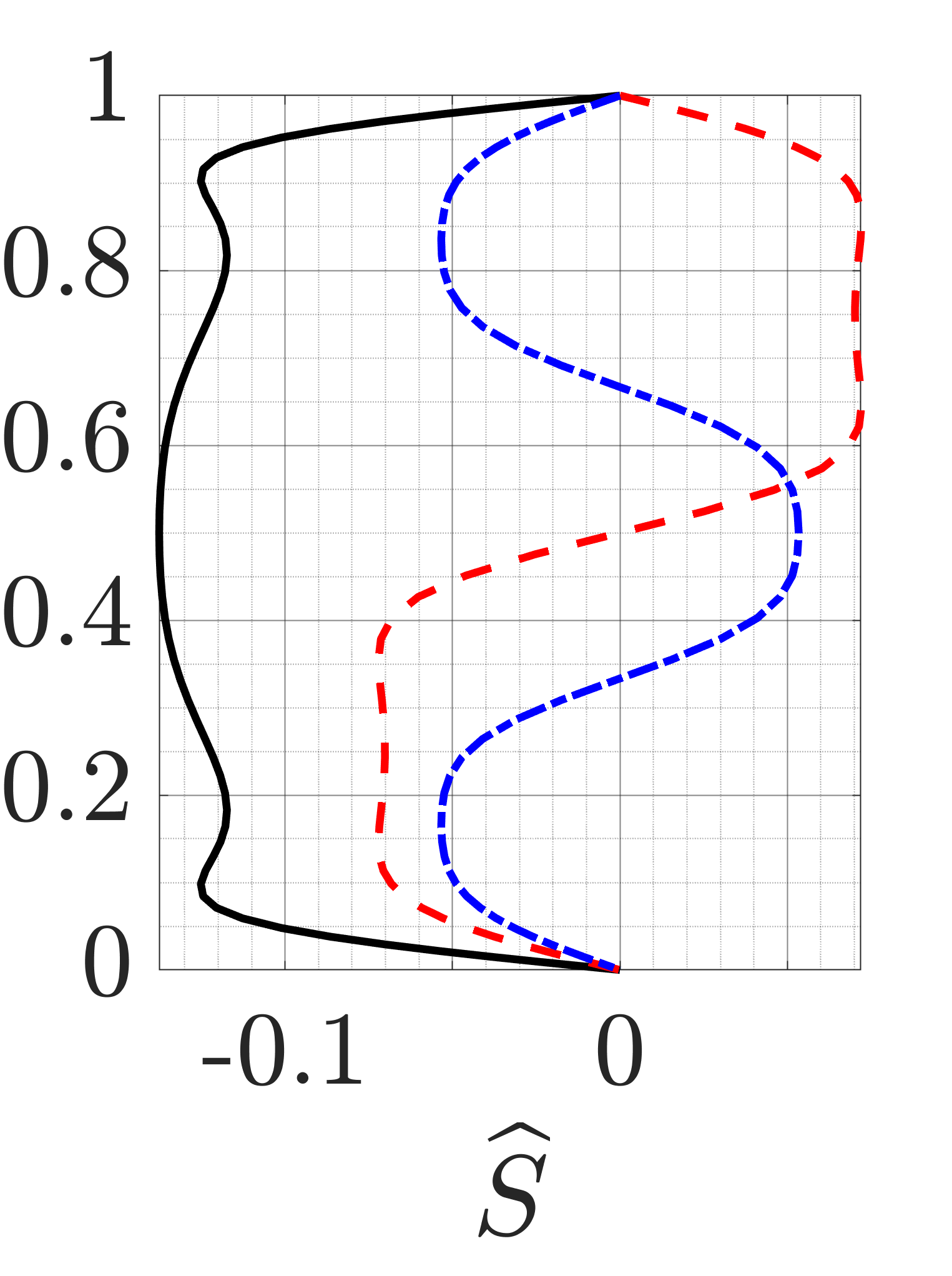}
    \includegraphics[width=0.24\textwidth]{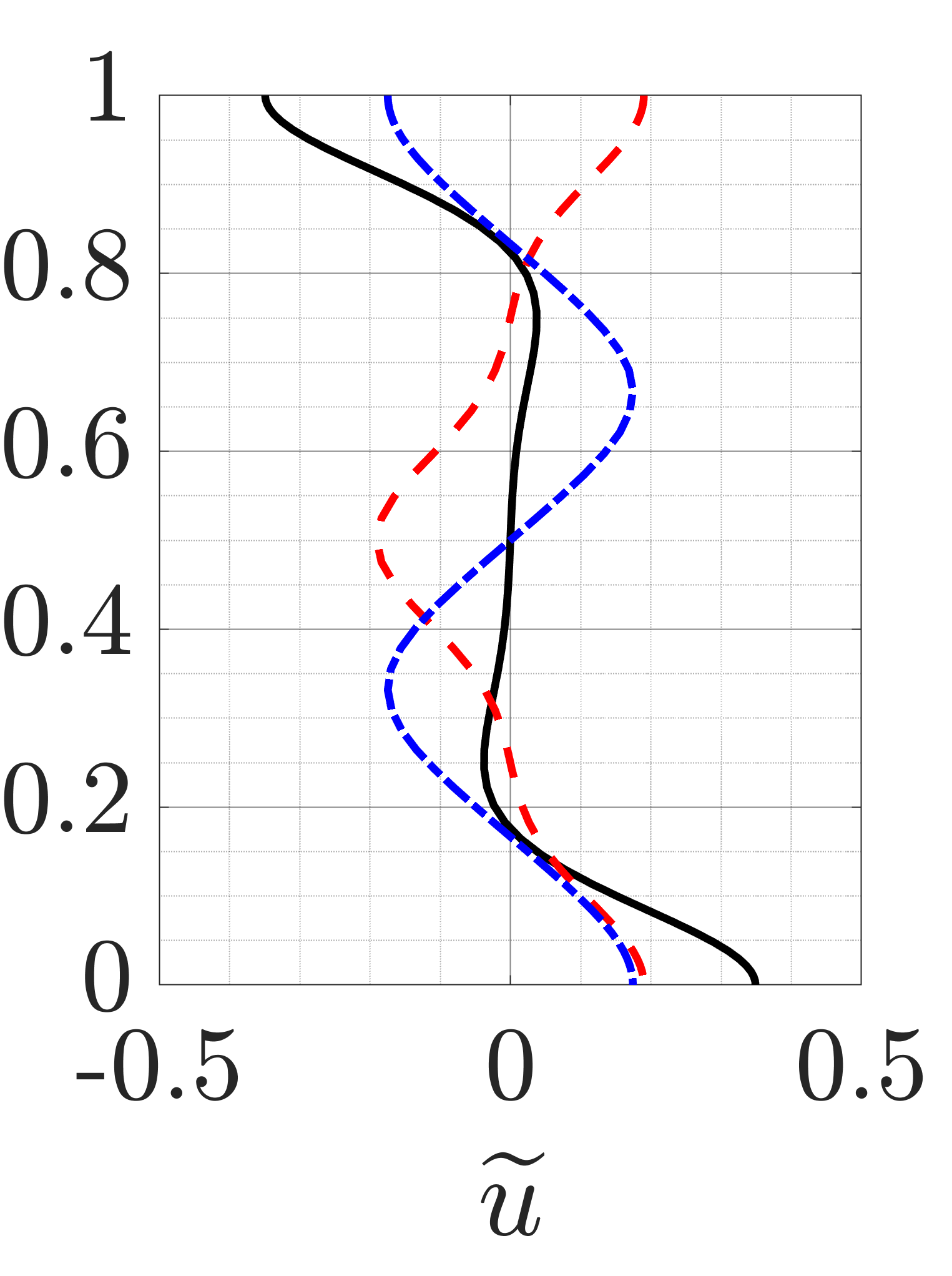}
    \includegraphics[width=0.24\textwidth]{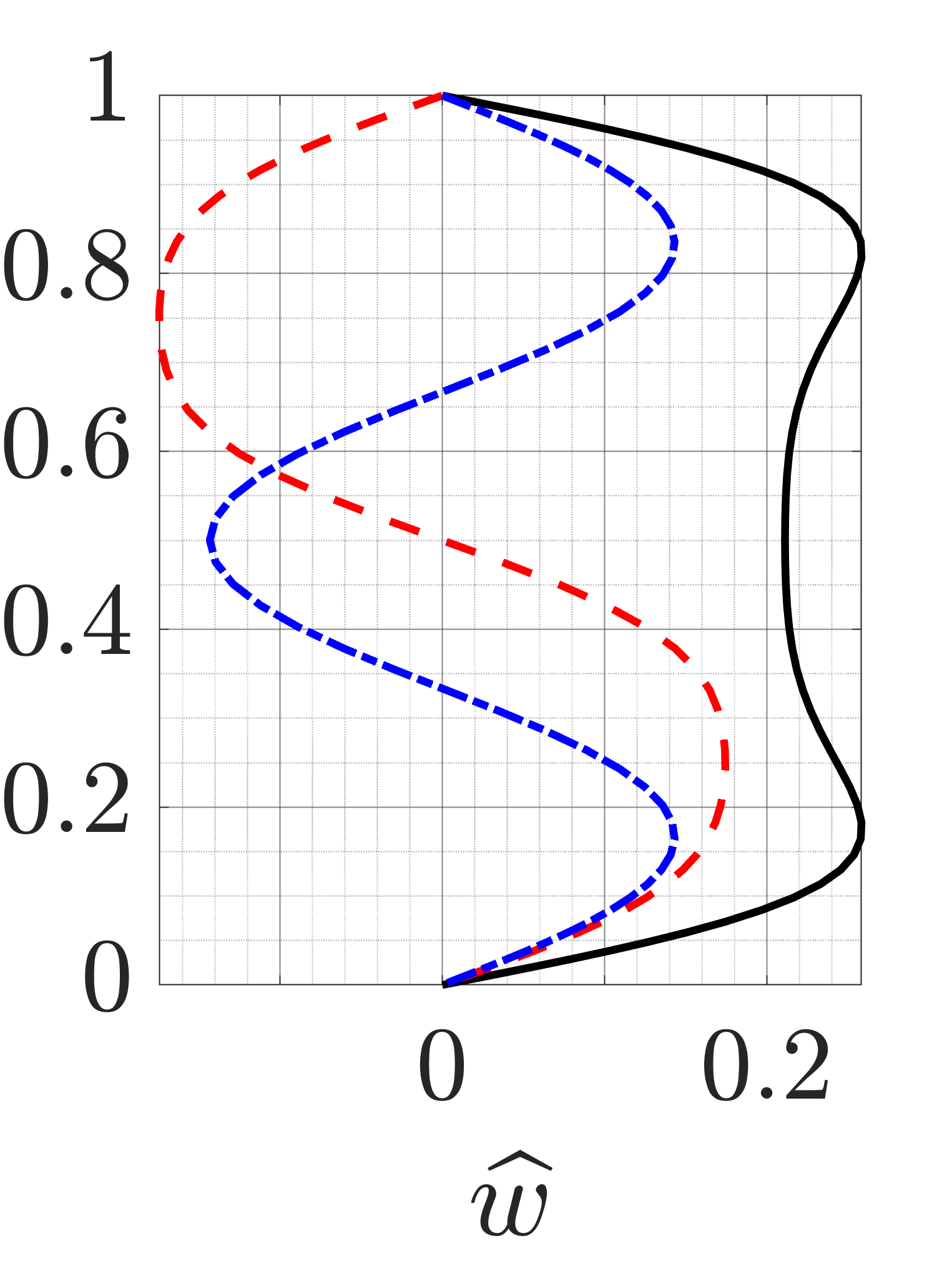}

    (e) S1 \hspace{0.24\textwidth} (f) S2 \hspace{0.24\textwidth} (g) S3
        
    \includegraphics[width=0.31\textwidth]{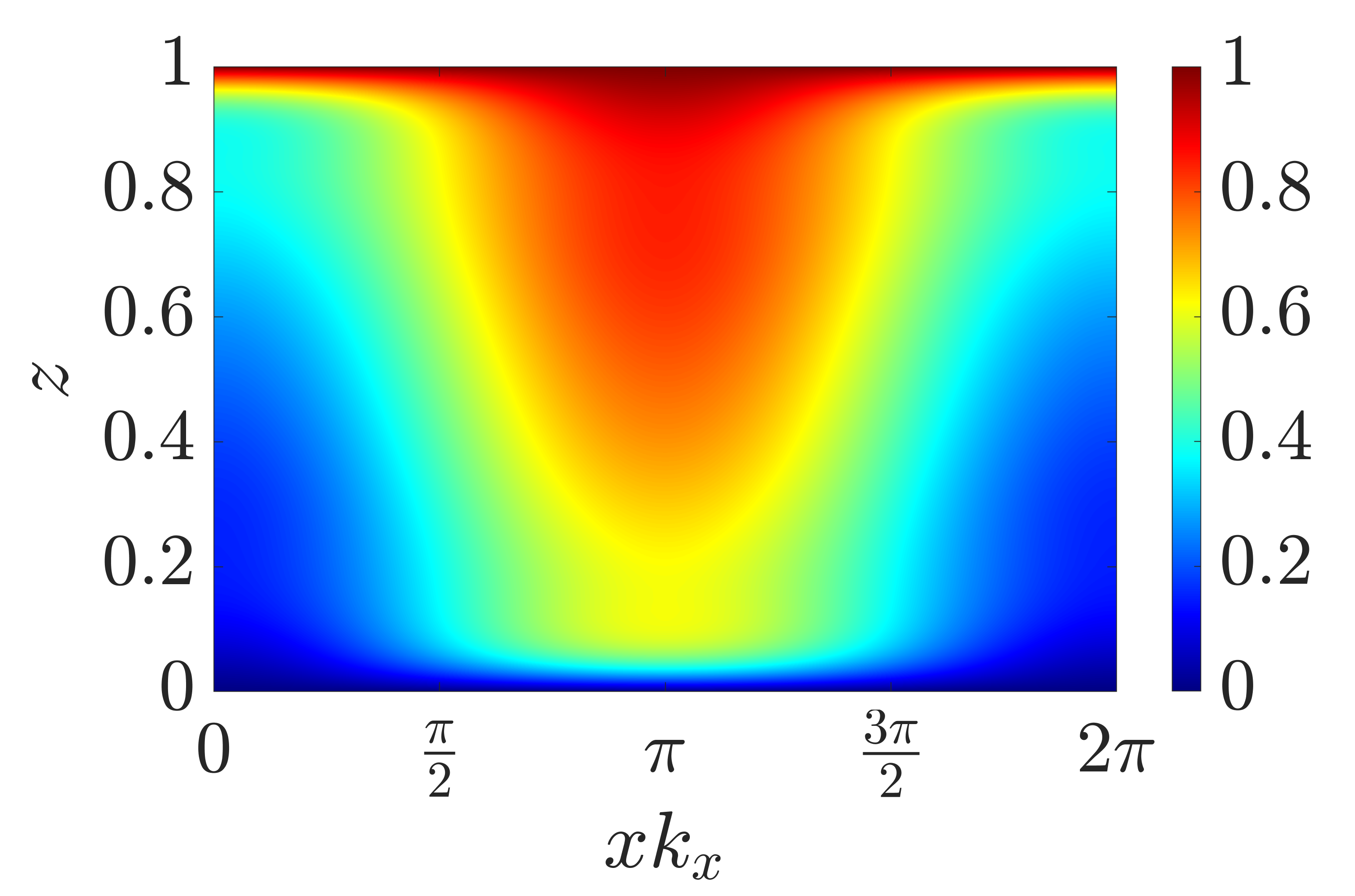}
    \includegraphics[width=0.31\textwidth]{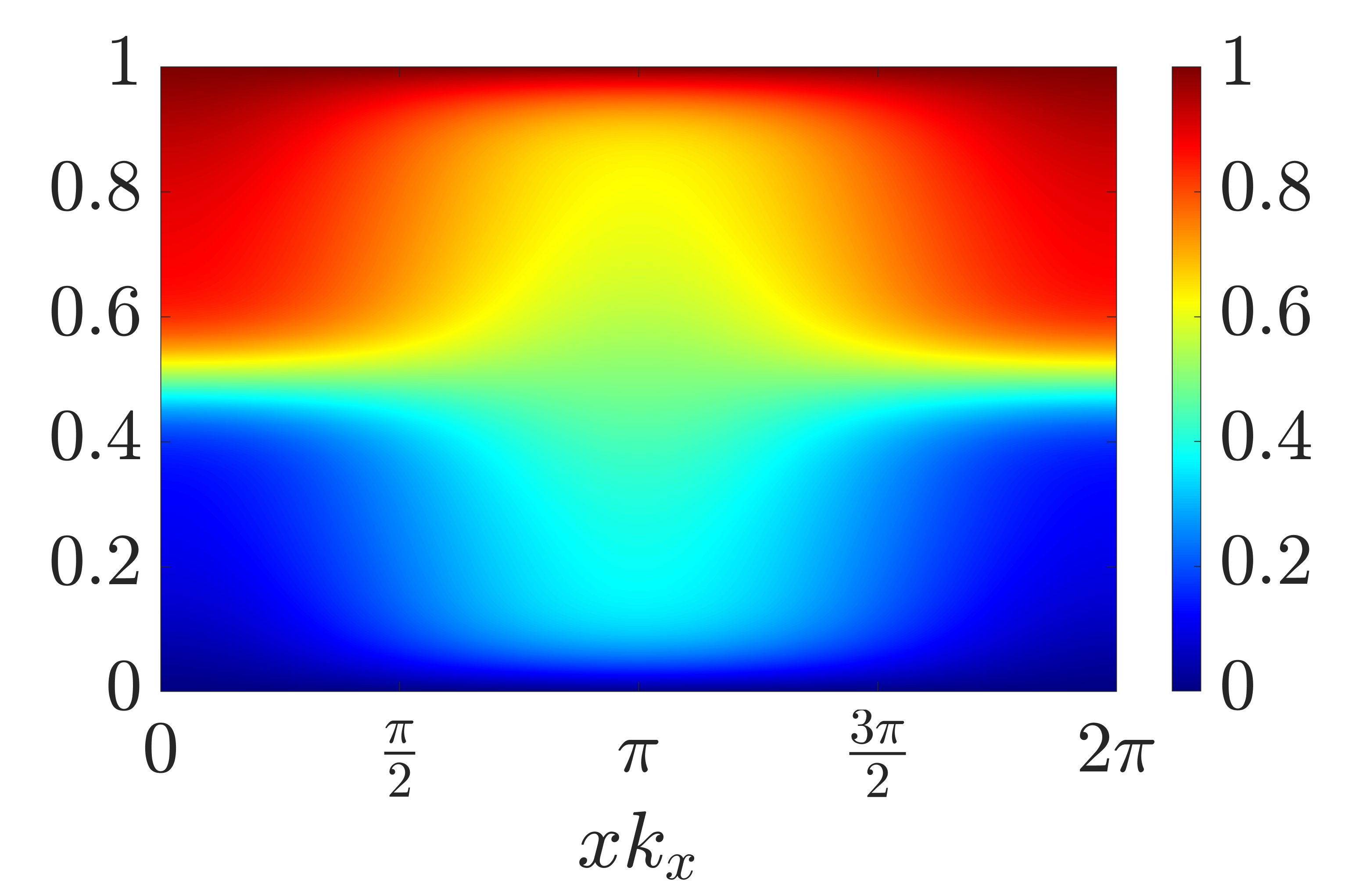}
    \includegraphics[width=0.31\textwidth]{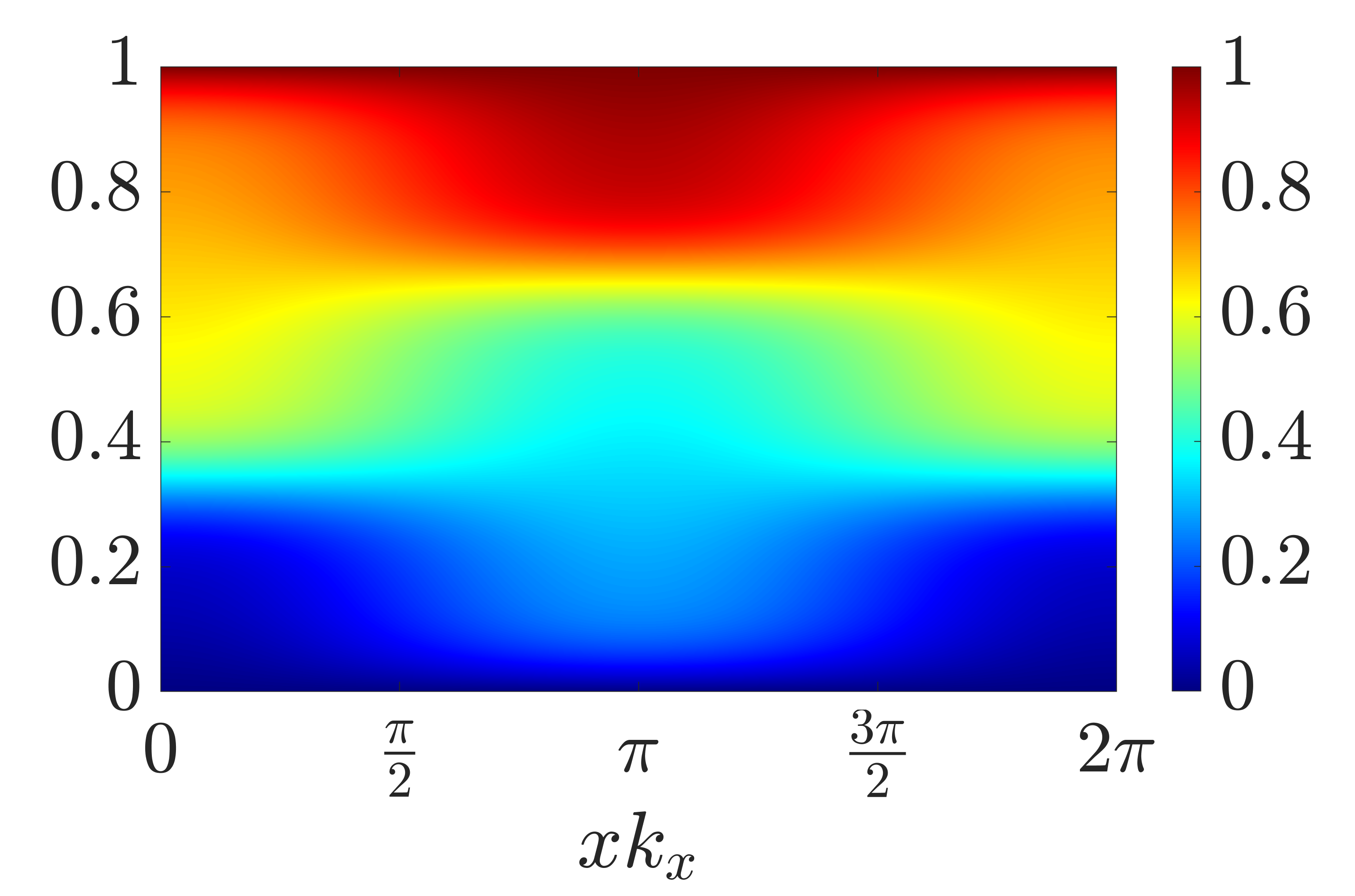}

    \caption{Solution profiles from the single-mode equations \eqref{eq:single_mode} in the same parameter regime as figure \ref{fig:profile_R_rho_T2S_40_tau_0p01} but with stress-free velocity boundary conditions at top and bottom. The first row shows the profiles of (a) $z+\bar{S}_0$, (b) $\widehat{S}$, (c) $\widetilde{u}$, and (d) $\widehat{w}$. The second row shows the reconstructed total salinity using \eqref{eq:normal_mode_S} and \eqref{eq:total_T_S} for (e) S1, (f) S2, and (g) S3 solutions.}
        \label{fig:profile_R_rho_T2S_40_tau_0p01_stress_free}
\end{figure}

\begin{figure}
(a) \hspace{0.22\textwidth} (b) \hspace{0.22\textwidth} (c) \hspace{0.22\textwidth} (d)

    \centering
    \includegraphics[width=0.24\textwidth]{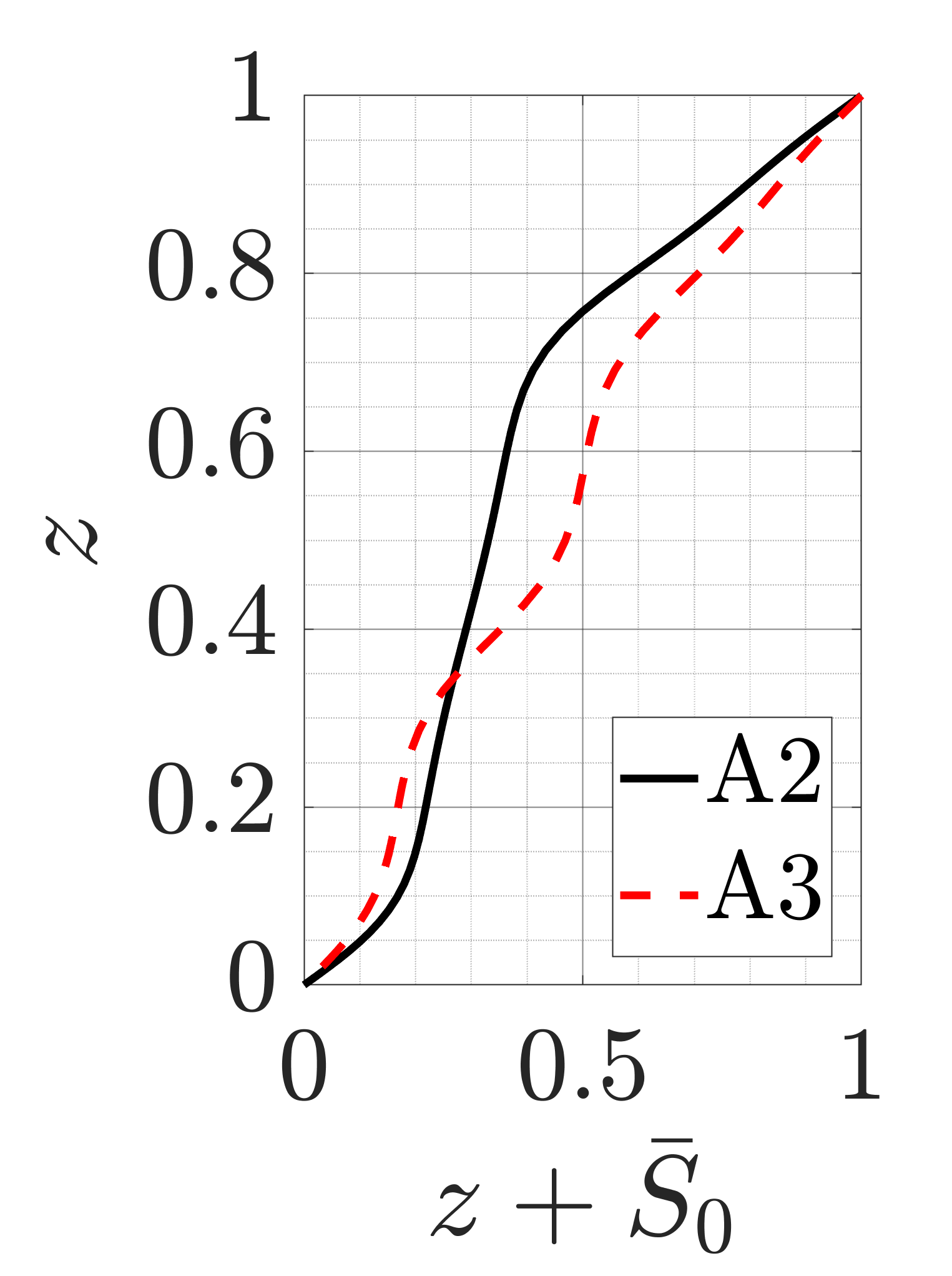}
    \includegraphics[width=0.225\textwidth,trim=-0 -0.5in 0 0]{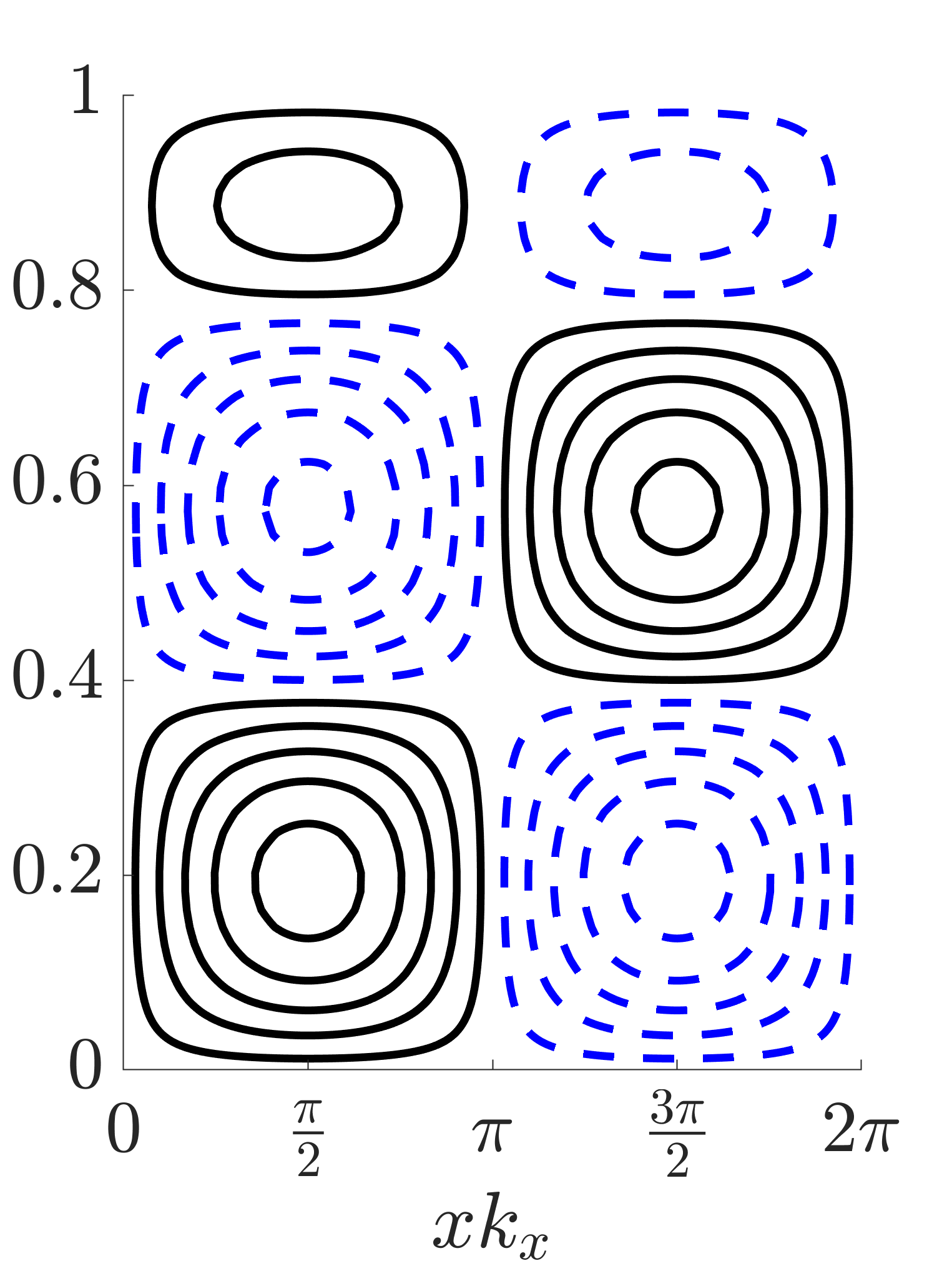}
    \includegraphics[width=0.24\textwidth]{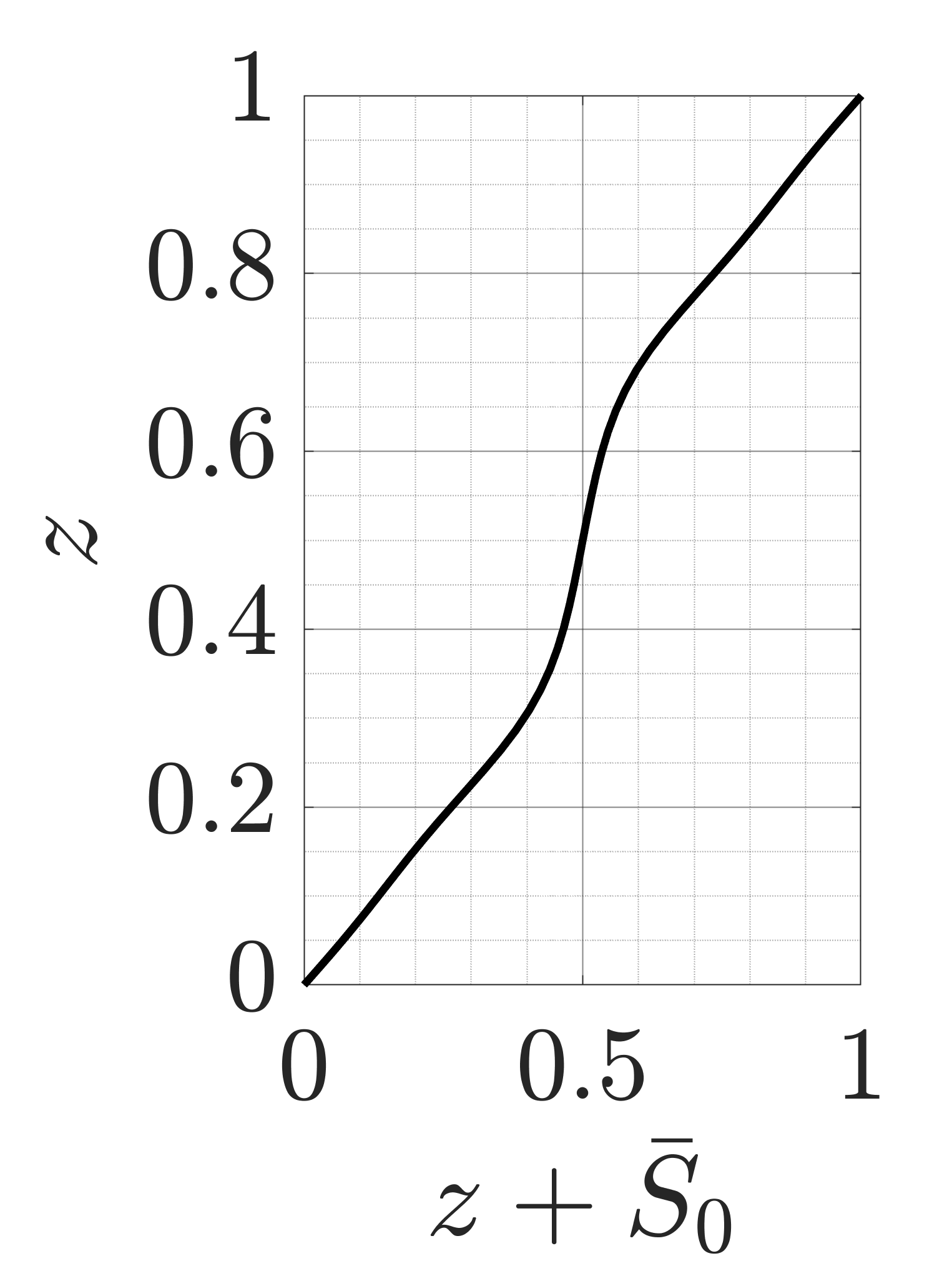}
    \includegraphics[width=0.225\textwidth,trim=-0 -0.5in 0 0]{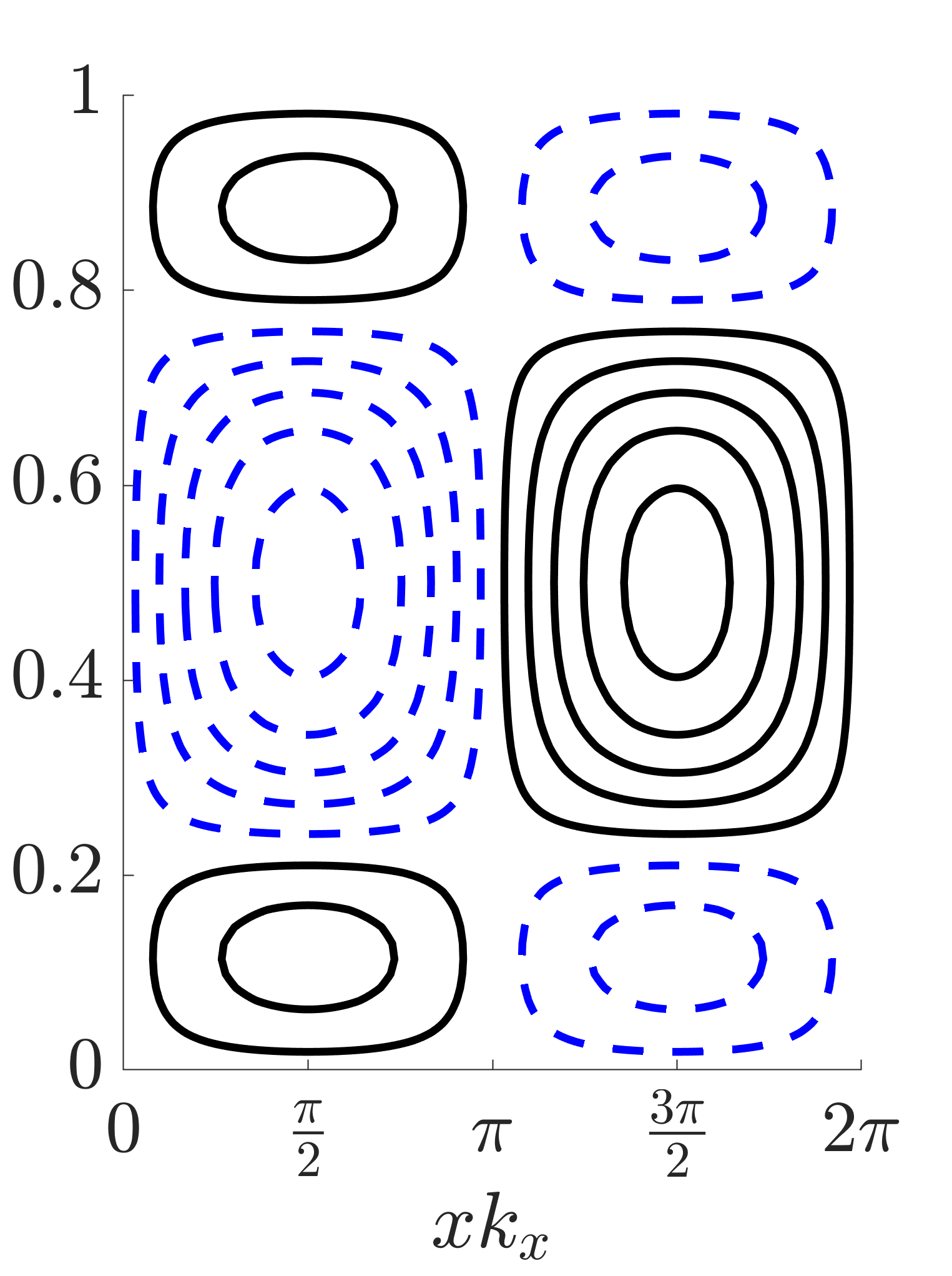}
    \caption{Solution profiles on the secondary branches computed from the single-mode equations \eqref{eq:single_mode} in the same parameter regime as figure \ref{fig:profile_R_rho_T2S_40_tau_0p01_Pr_7_A2_A3} but with stress-free velocity boundary conditions at the top and bottom. Panel (a) shows the profiles of $z+\bar{S}_0$ of the A2 and A3 solutions, while panel (b) displays the isocontours of the streamfunction of the A3 solution. Panels (c) and (d) show $z+\bar{S}_0$ and the isocontours of the streamfunction for a steady solution preserving midplane reflection symmetry generated in a secondary bifurcation of the S3 solution. }
    \label{fig:profile_R_rho_T2S_40_tau_0p01_Pr_7_A2_A3_stress_free}
\end{figure}

Although no-slip boundary conditions are relevant for experiments \citep{hage2010high}, stress-free boundary conditions are more appropriate for oceanographic applications. Here we compare the bifurcation diagram in the no-slip case with the corresponding diagram with stress-free boundary conditions. For this purpose we replace the original no-slip boundary conditions in \eqref{eq:BC_no_slip} by the stress-free boundary conditions
\begin{subequations}
    \label{eq:BC_vel_stress_free}
\begin{align}
    &w(x,y,z=0,t)=w(x,y,z=1,t)=0,\\
    &\partial_z u(x,y,z=0,t)=\partial_z u(x,y,z=1,t)=0,\;\;\\\
    &\partial_z v(x,y,z=0,t)=\partial_z v(x,y,z=1,t)=0.
\end{align}
\end{subequations}
Within the single-mode equations, the corresponding velocity boundary conditions in \eqref{eq:BC_no_slip_single_mode} are thus replaced as:
\begin{subequations}
\label{eq:BC_stress_free}
\begin{align}
    &\widehat{w}(z=0,t)=\widehat{w}(z=1,t)=\partial_{z}^2\widehat{w}(z=0,t)=\partial_{z}^2\widehat{w}(z=1,t)\\
    =\,&\partial_z\widehat{\zeta}(z=0,t)=\partial_z\widehat{\zeta}(z=1,t)=\partial_z\bar{U}_0(z=0,t)=\partial_z\bar{U}_0(z=1,t)\\
    =\,&0.
\end{align}
\end{subequations}
Note that the governing equations with stress-free boundary conditions admit an additional Galilean symmetry: the large-scale shear can be shifted by an arbitrary constant, $\bar{U}_0\rightarrow \bar{U}_0+C$. To obtain unique solutions we therefore impose the additional constraint $\int_{0}^{1}\bar{U}_0(z,t)dz=0$ following the procedure in \citet[\S 6.9]{uecker2021numerical}.

Figure \ref{fig:bif_diag_low_Ra_S2T_stress_free} shows the resulting bifurcation diagrams for comparison with figure \ref{fig:bif_diag_low_Ra_S2T}. The diagrams are very similar: all of the previously discussed solution branches are still present. The main difference is that the stress-free boundary conditions typically lead to a larger $Sh$ than the no-slip boundary conditions at the same horizontal wavenumber.  This behavior is also consistent with the observation by \citet{yang2016vertically}, where 3D DNS results indicate that the flow morphology is qualitatively similar but stress-free boundary conditions display a larger $Sh$. Figure \ref{fig:bif_diag_low_Ra_S2T_stress_free}(a) compares the single-mode results with the corresponding steady state $Sh$ reached using 2D DNS in domains of size $L_x=2\pi /k_x$ (black squares) showing good agreement provided the domain width $L_x$ is sufficiently narrow, i.e., for $k_x$ sufficiently close to the stress-free onset wavenumber $k_x=19.298$.  

Figure \ref{fig:profile_R_rho_T2S_40_tau_0p01_stress_free} displays solution profiles and total salinity for the S1, S2 and S3 solutions at the same parameter values as in figure \ref{fig:profile_R_rho_T2S_40_tau_0p01} but with stress-free instead of no-slip boundary conditions. Here, the horizontally averaged total salinity $z+\bar{S}_0$ still displays the one-layer, two-layer and three-layer solution profiles in vertical. The horizontal structure of these states is qualitatively similar to the no-slip results in figure \ref{fig:profile_R_rho_T2S_40_tau_0p01}. The main difference is that the horizontal velocity $\widetilde{u}$ and the gradient of vertical velocity $\partial_z \widehat{w}$ no longer vanish at the boundaries.

Figure \ref{fig:profile_R_rho_T2S_40_tau_0p01_Pr_7_A2_A3_stress_free}(a) displays the A2 and A3 solution profiles with stress-free boundary conditions. The A2 solution exhibits a similar profile as in the no-slip boundary condition case shown in figure \ref{fig:profile_R_rho_T2S_40_tau_0p01_Pr_7_A2_A3}(a), while the horizontally averaged salinity in A3 differs from that found with the no-slip boundary conditions. Figure \ref{fig:profile_R_rho_T2S_40_tau_0p01_Pr_7_A2_A3_stress_free}(b) shows the isocontours of the streamfunction in A3, and shows two large and one small recirculation cell in the vertical direction, while figure \ref{fig:profile_R_rho_T2S_40_tau_0p01_Pr_7_A2_A3}(f) shows one large and two small cells in the vertical. Here, the A3 solution in the stress-free boundary condition case originates from a secondary bifurcation at $k_x=13.665$ on the S3 branch, while there is another secondary bifurcation at $k_x=13.703$ (not shown in figure \ref{fig:bif_diag_low_Ra_S2T_stress_free}) that is closer to the high wavenumber onset of the S3 state $(k_x=15.573)$ and that preserves midplane reflection symmetry. The mean salinity of the state shown in figure \ref{fig:profile_R_rho_T2S_40_tau_0p01_Pr_7_A2_A3_stress_free}(c) resembles the mean temperature profile in magnetoconvection \citep[figure 7(b)]{julien2000nonlinear}, and the isocontours of the streamfunction in figure \ref{fig:profile_R_rho_T2S_40_tau_0p01_Pr_7_A2_A3_stress_free}(d) show one large and two small closed streamlines that are similar to those in the A3 solution with no-slip boundary conditions in figure \ref{fig:profile_R_rho_T2S_40_tau_0p01_Pr_7_A2_A3}(f). The fact that this additional secondary bifurcation appears closer to the high wavenumber onset of S3 than the bifurcation to the A3 branch likely contributes to the difference in the solution profiles of the A3 state with stress-free and no-slip boundary conditions. 
\subsection{Dependence on density ratio}
\label{subsec:results_Pr_7_R_rho_dependence}

Oceanographic conditions display a wide range of density ratios, and this subsection therefore explores the dependence of our results on the density ratio. In particular, the staircases observed in oceanographic observations, laboratory measurements, and DNS are typically associated with the parameter regime $R_\rho\sim O(1)$ \citep{schmitt1987c,krishnamurti2003double,krishnamurti2009heat,radko2003mechanism} that is far away from the onset of the salt-finger instability $R_{\rho,{\rm crit}}=1/\tau$ with $\tau=0.01$. 

\begin{figure}
    \centering
    
    2D: $k_y=0$ \hspace{0.38\textwidth} 3D: $k_y=k_x$
    
    (a) \hspace{0.49\textwidth} (b)
    
    \includegraphics[width=0.49\textwidth]{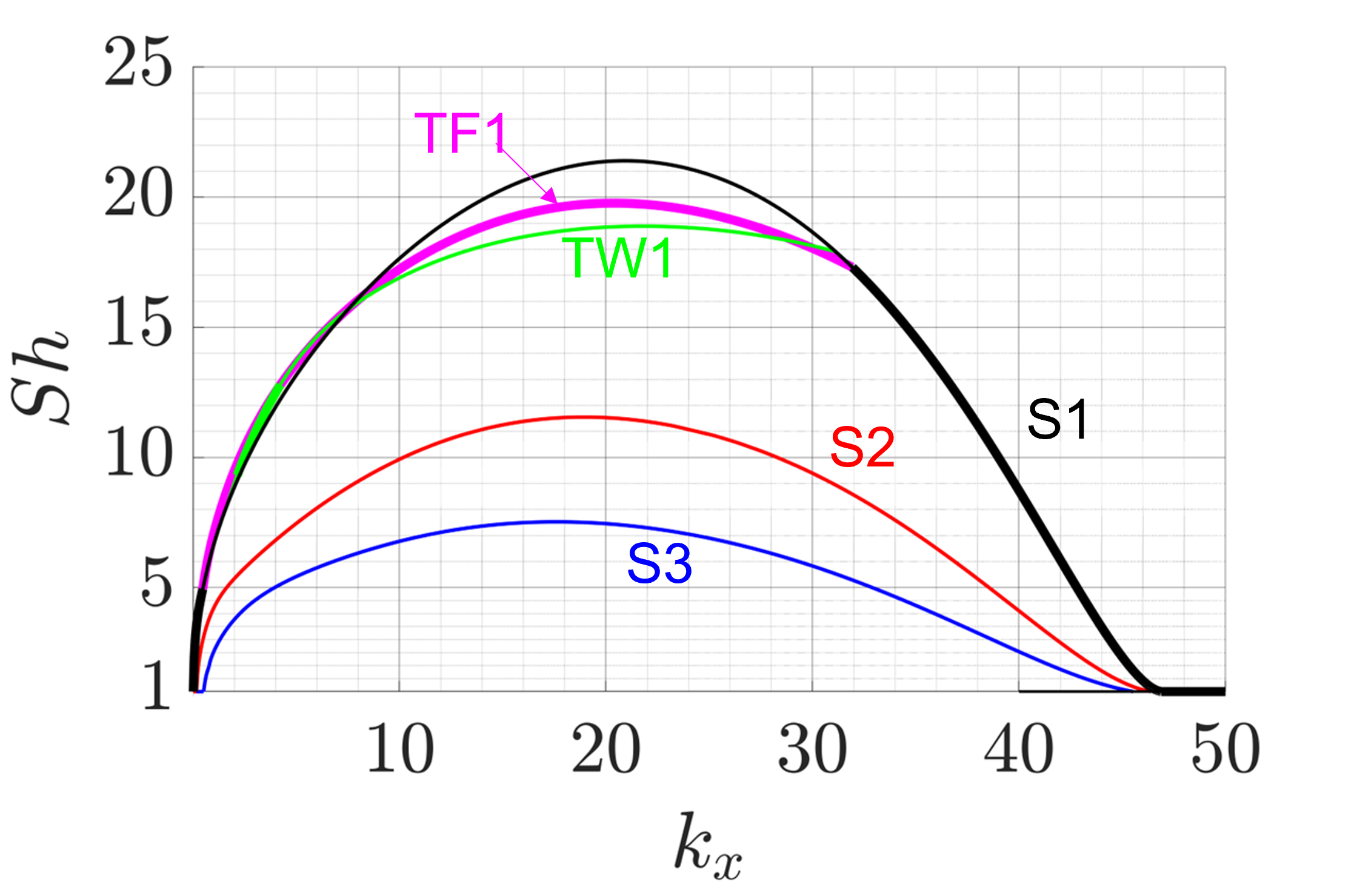}
    \includegraphics[width=0.49\textwidth]{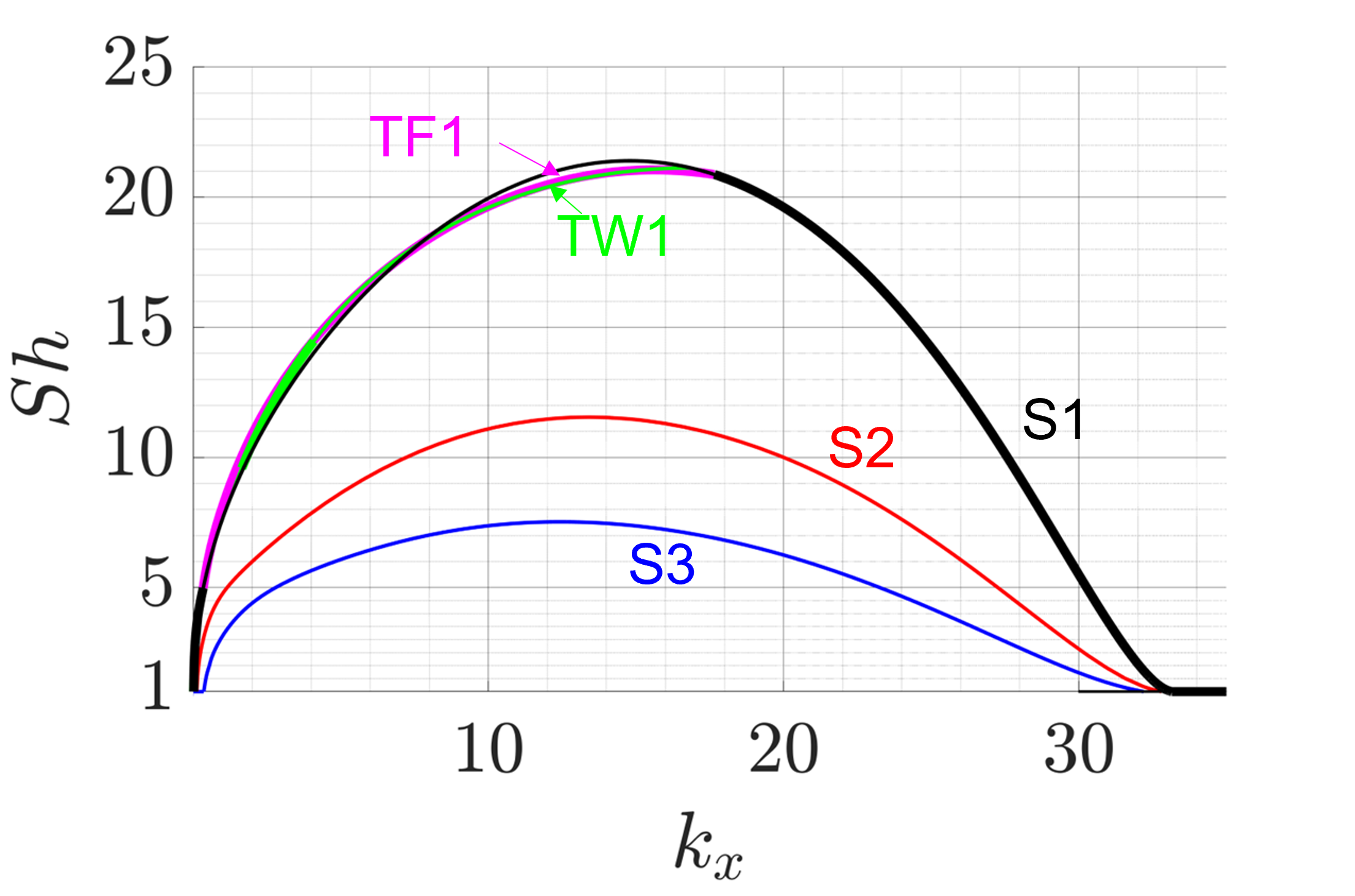}
    
    (c) \hspace{0.2\textwidth} (d) \hspace{0.2\textwidth} (e) \hspace{0.2\textwidth} (f)

    \includegraphics[width=0.245\textwidth]{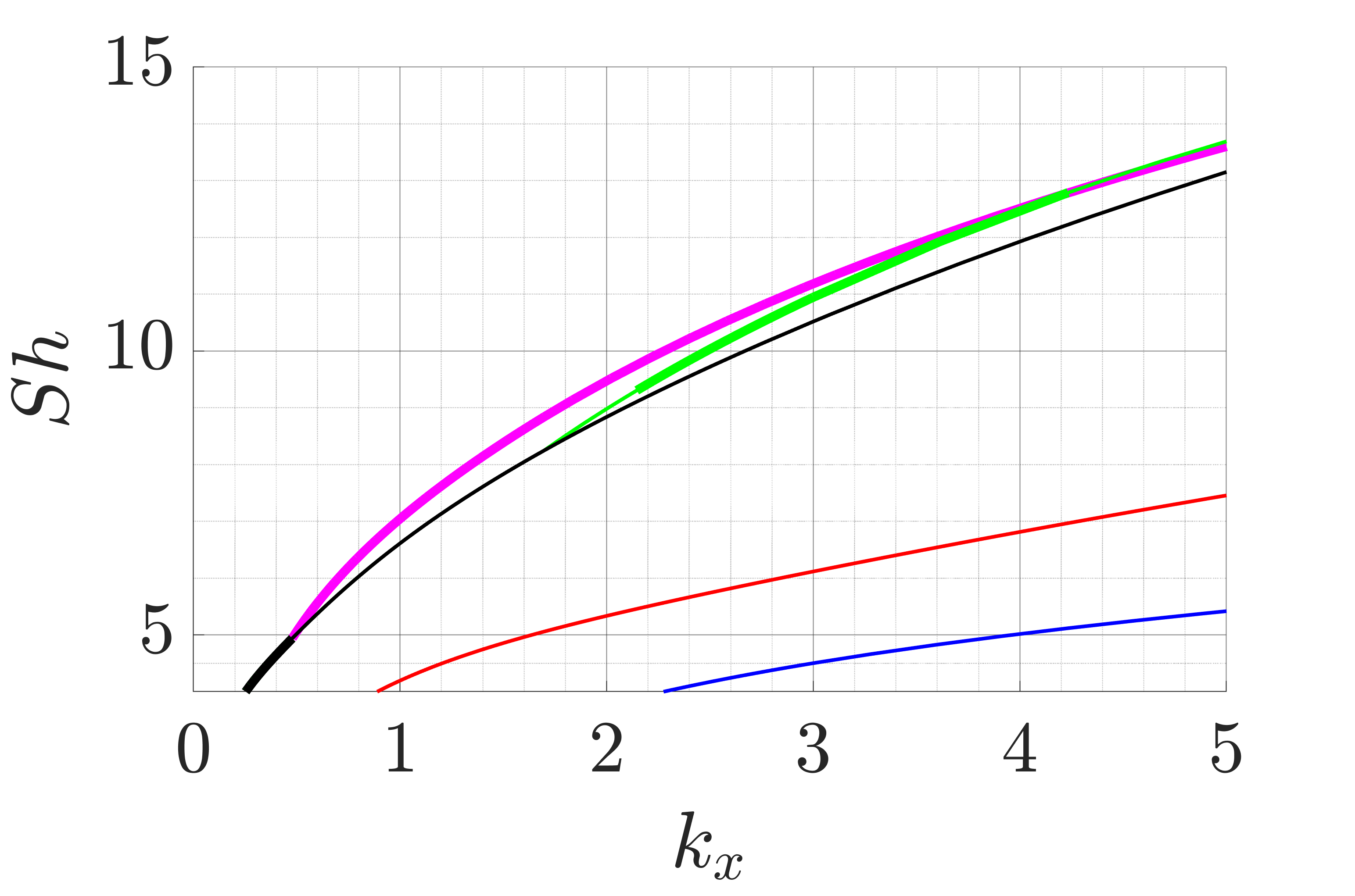}
    \includegraphics[width=0.245\textwidth]{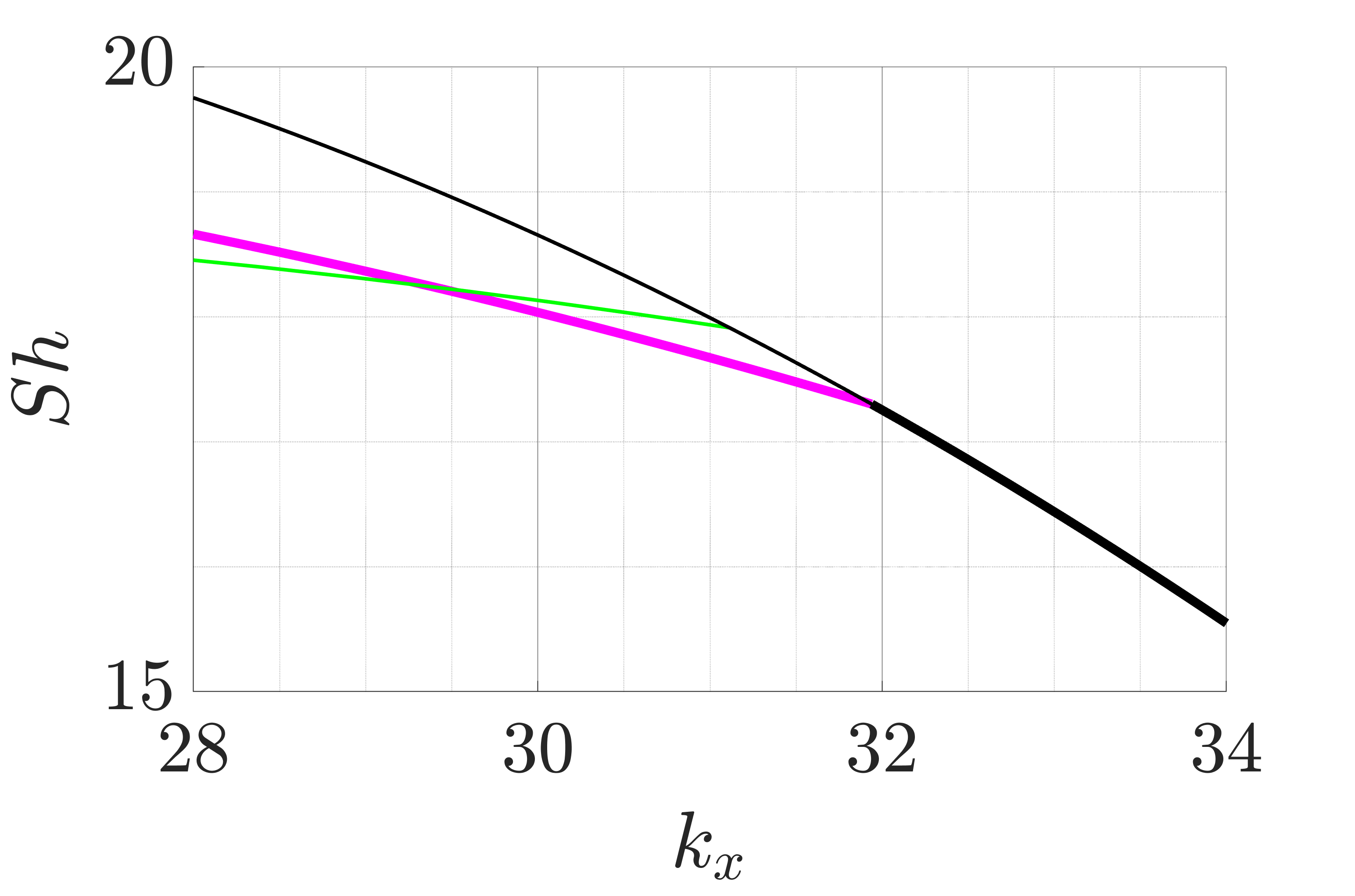}
    \includegraphics[width=0.245\textwidth]{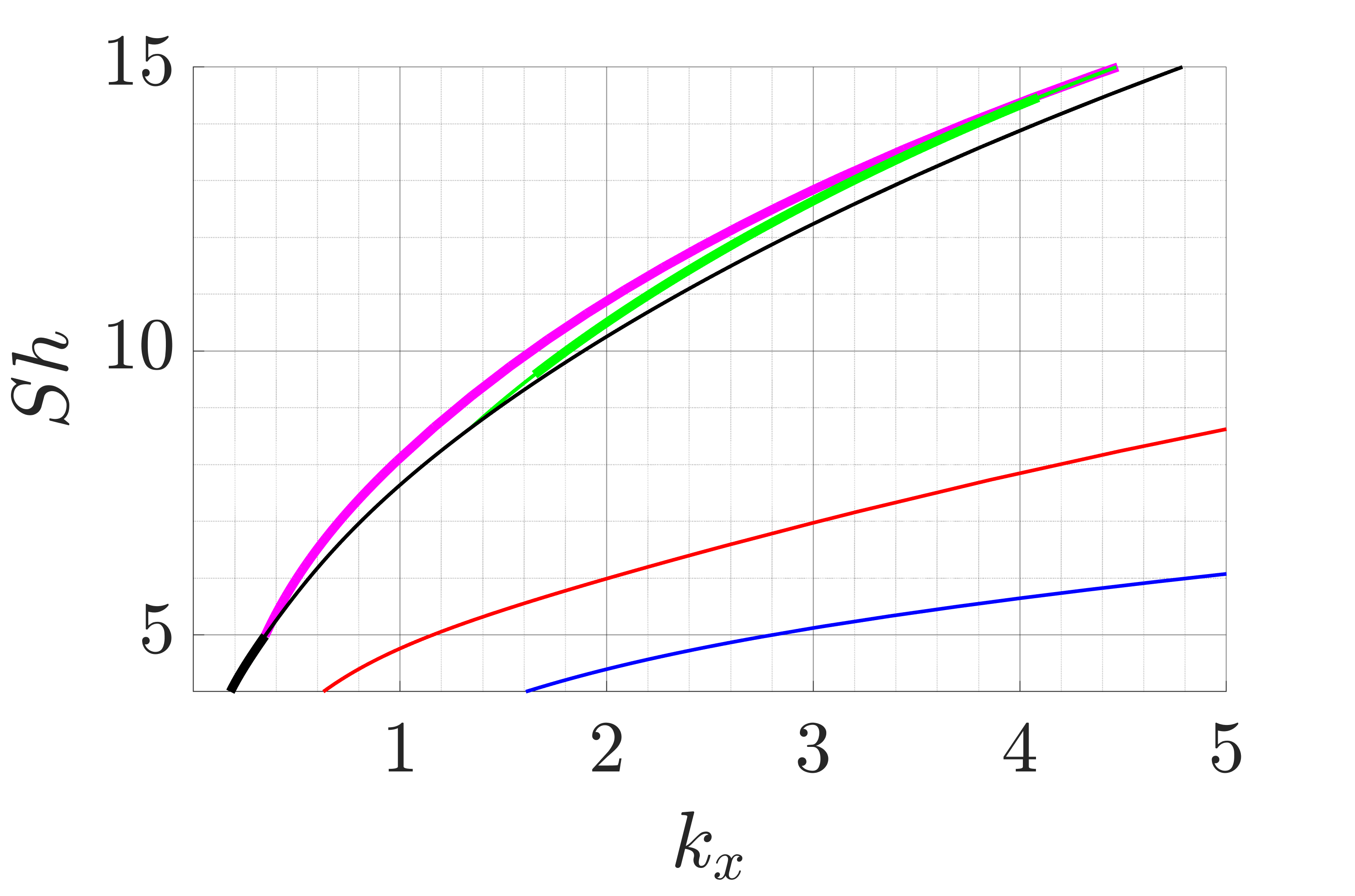}
    \includegraphics[width=0.245\textwidth]{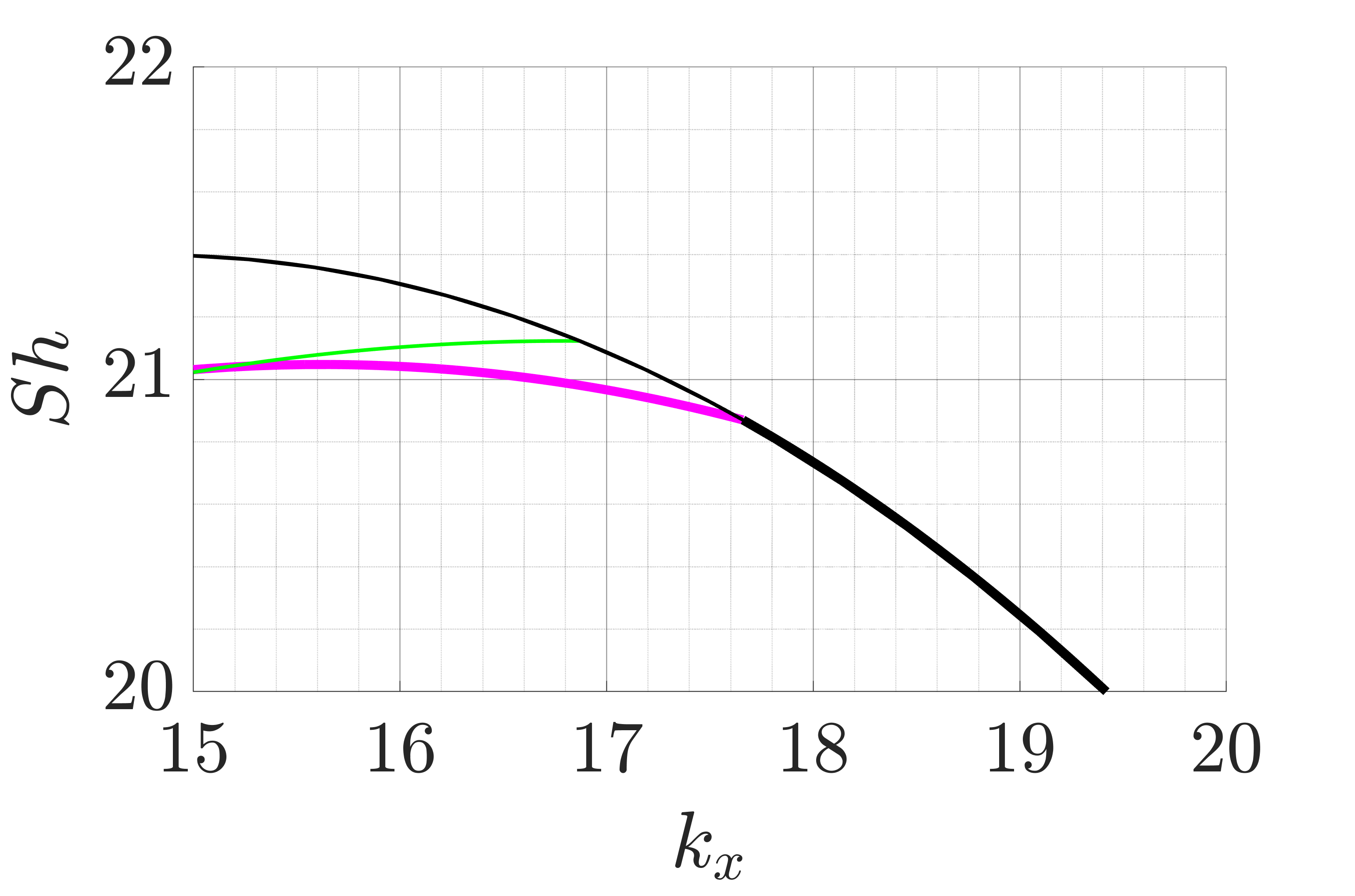}

    \caption{Bifurcation diagrams as a function of the wavenumber $k_x$ from the single-mode equations \eqref{eq:single_mode} with (a) 2D: $k_y=0$ and (b) 3D: $k_y=k_x$. The governing parameters are $R_\rho=2$, $Pr=7$, $\tau=0.01$, and $Ra_T=10^5$. Panels (c)-(d) are a zoom of the 2D results in panel (a) and panels (e)-(f) are a zoom of the 3D results in panel (b). }
    \label{fig:bif_diag_mid_Ra_S2T}
\end{figure}

Figure \ref{fig:bif_diag_mid_Ra_S2T} shows the bifurcation diagram with density ratio $R_\rho=2$ and no-slip boundary conditions for (a) 2D results with $k_y=0$ and (b) 3D results with $k_y=k_x$. Here, we focus on the S1, S2, S3, TF1 solution branches, as well as a new traveling wave (TW1) solution branch that does not exist when $R_\rho=40$. Note that in this case nontrivial solutions exist over a wider range of horizontal wavenumbers than for $R_\rho=40$ (figure \ref{fig:bif_diag_low_Ra_S2T}), suggesting that the finger width depends not only on $Ra_T$ but also on the density ratio $R_\rho$. Here, the $Sh$ of the S1, S2, and S3 solutions in the 2D and 3D configurations are still the same after converting the results by rescaling the horizontal wavenumber according to \eqref{eq:kx_2D}. However, the stability of the S1 solutions is now different and the secondary bifurcation points on S1 that lead to TF1 and TW1 also differ. This can be seen in the zoom of regions close to the appearance of TF1 and TW1 in figures \ref{fig:bif_diag_mid_Ra_S2T}(c)-(f). Specifically, the bifurcation points leading to the TF1 and TW1 branches in 2D are closer to the high wavenumber onset $k_x=46.884$ of the S1 solution than in 3D, suggesting that the 2D configuration forms tilted fingers or traveling wave solutions more readily. This is consistent with the suggestion that large-scale shear is generated more easily in a 2D configuration than in 3D \citep{goluskin2014convectively}.

\begin{figure}
\centering
    (a) \hspace{0.2\textwidth} (b) \hspace{0.2\textwidth} (c) \hspace{0.2\textwidth} (d)
    
    \includegraphics[width=0.24\textwidth]{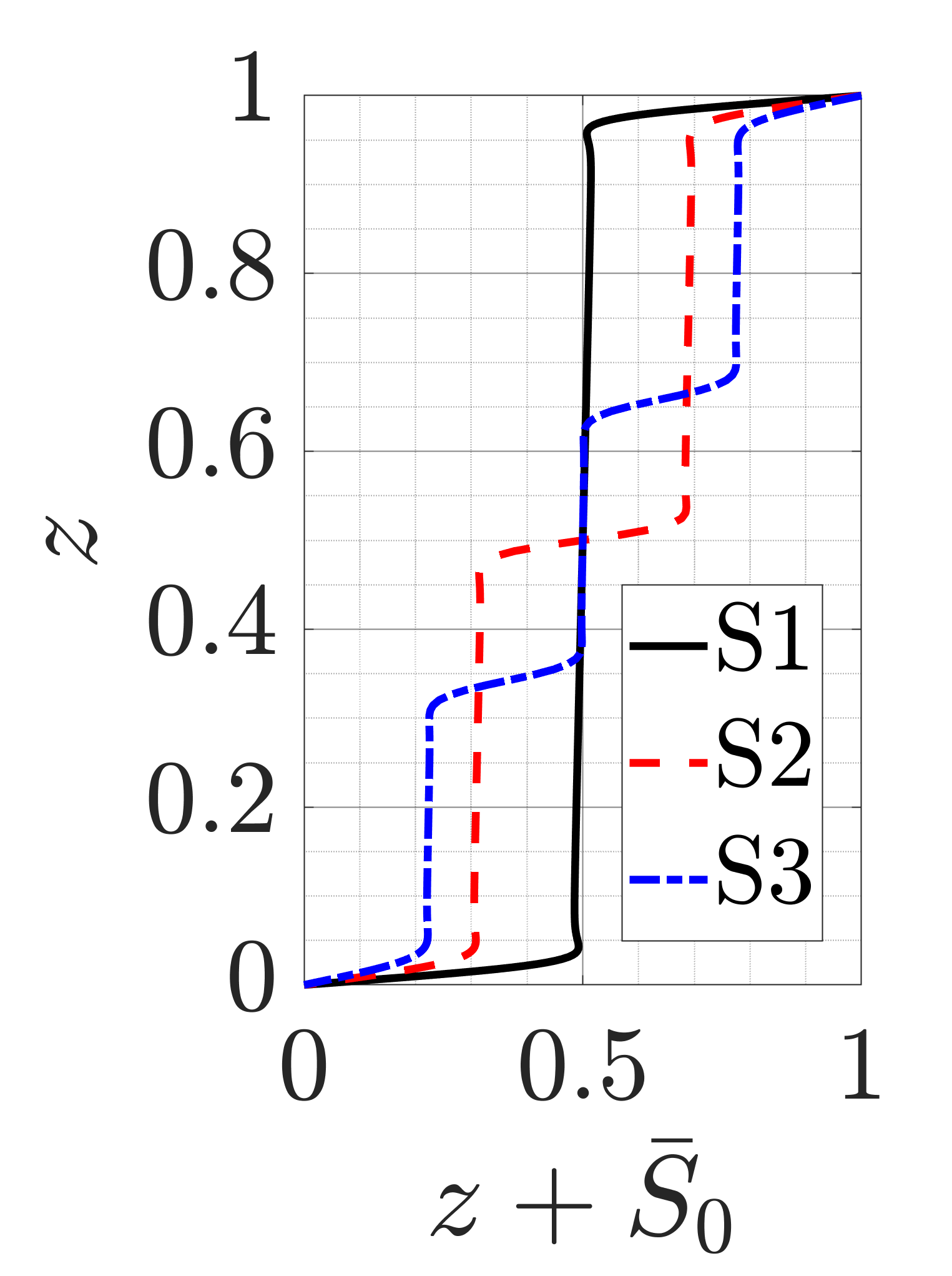}
    \includegraphics[width=0.24\textwidth]{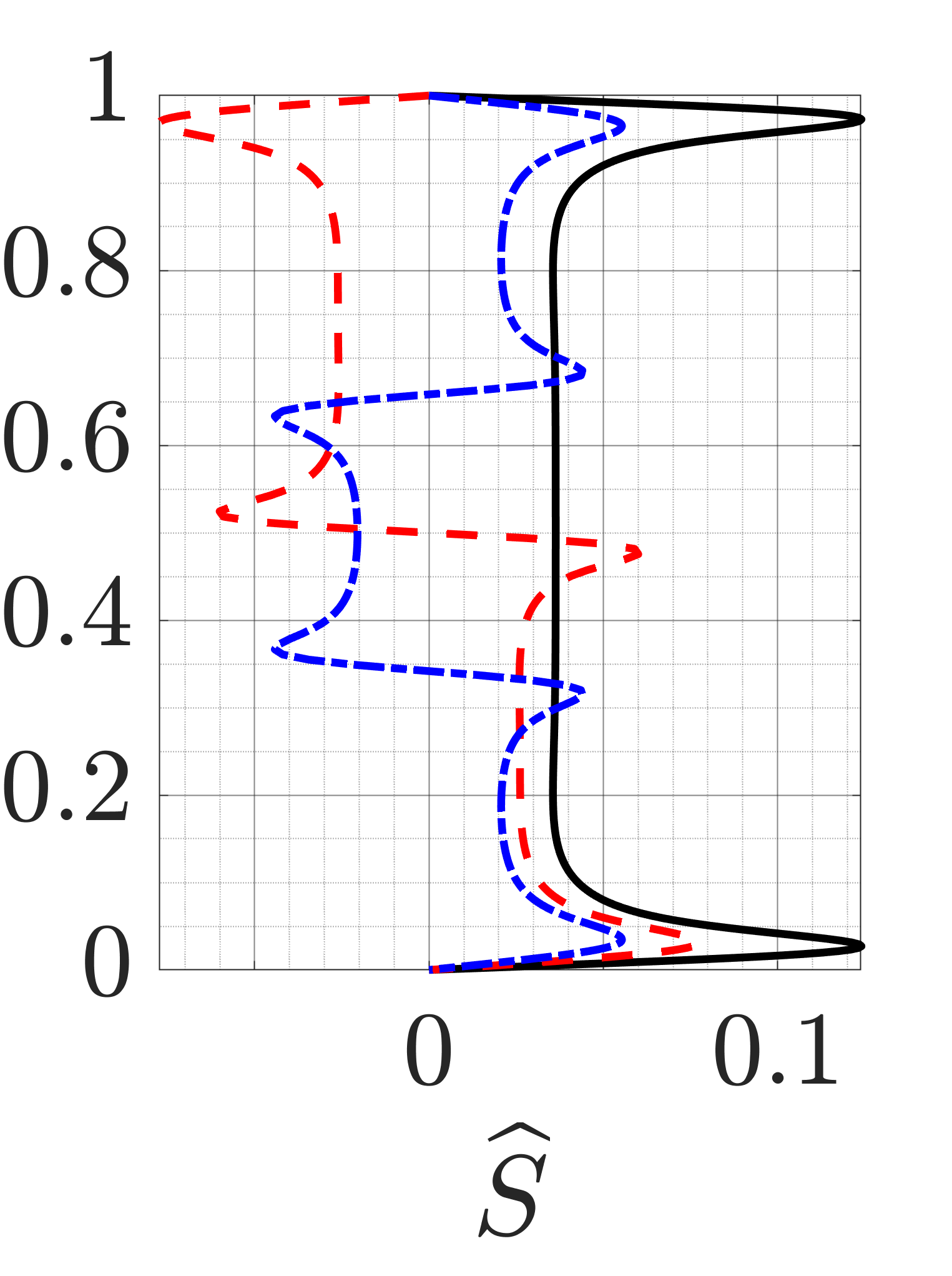}
    \includegraphics[width=0.24\textwidth]{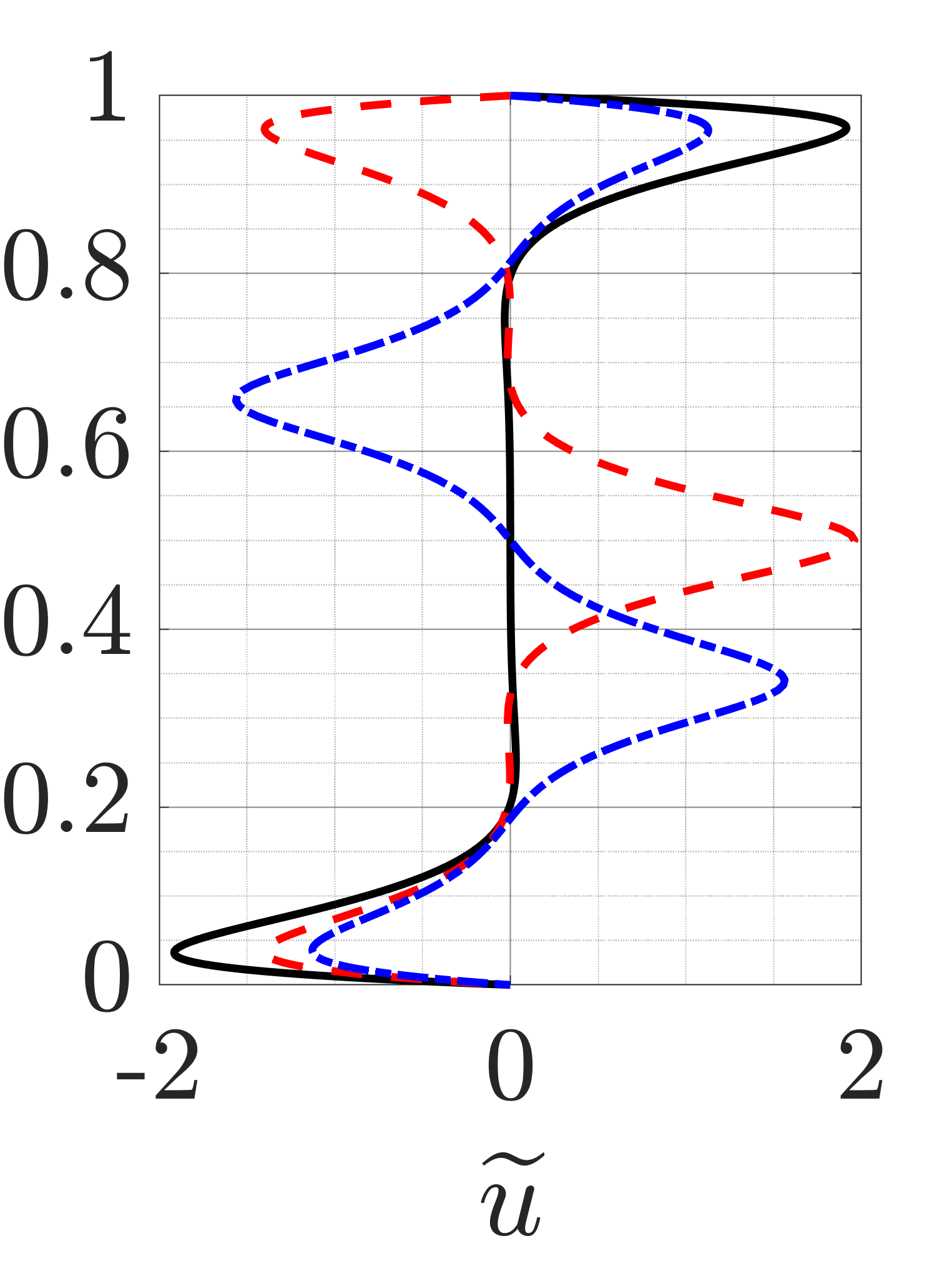}
    \includegraphics[width=0.24\textwidth]{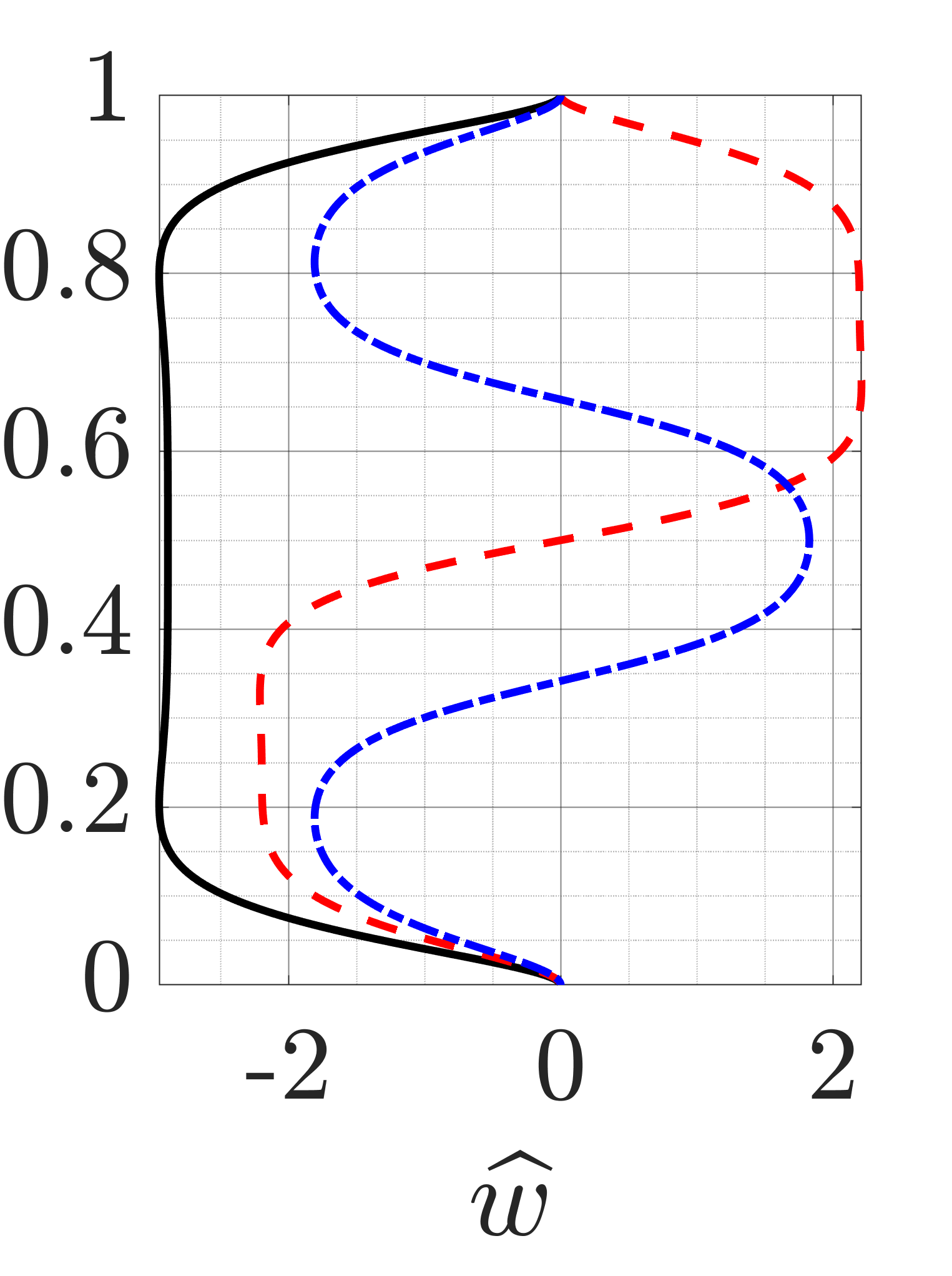}

    (e) S1 \hspace{0.24\textwidth} (f) S2 \hspace{0.24\textwidth} (g) S3
        
    \includegraphics[width=0.31\textwidth]{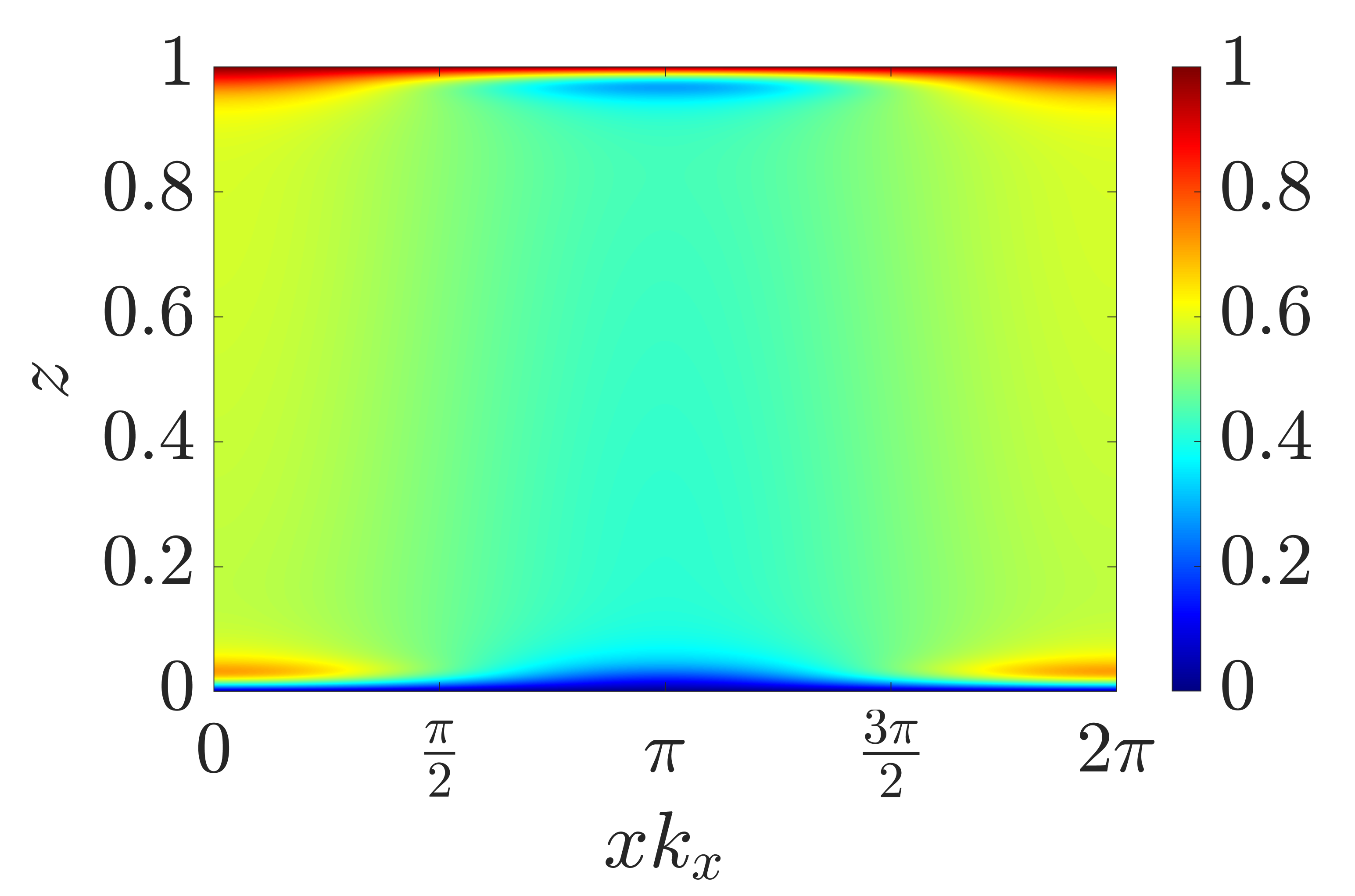}
    \includegraphics[width=0.31\textwidth]{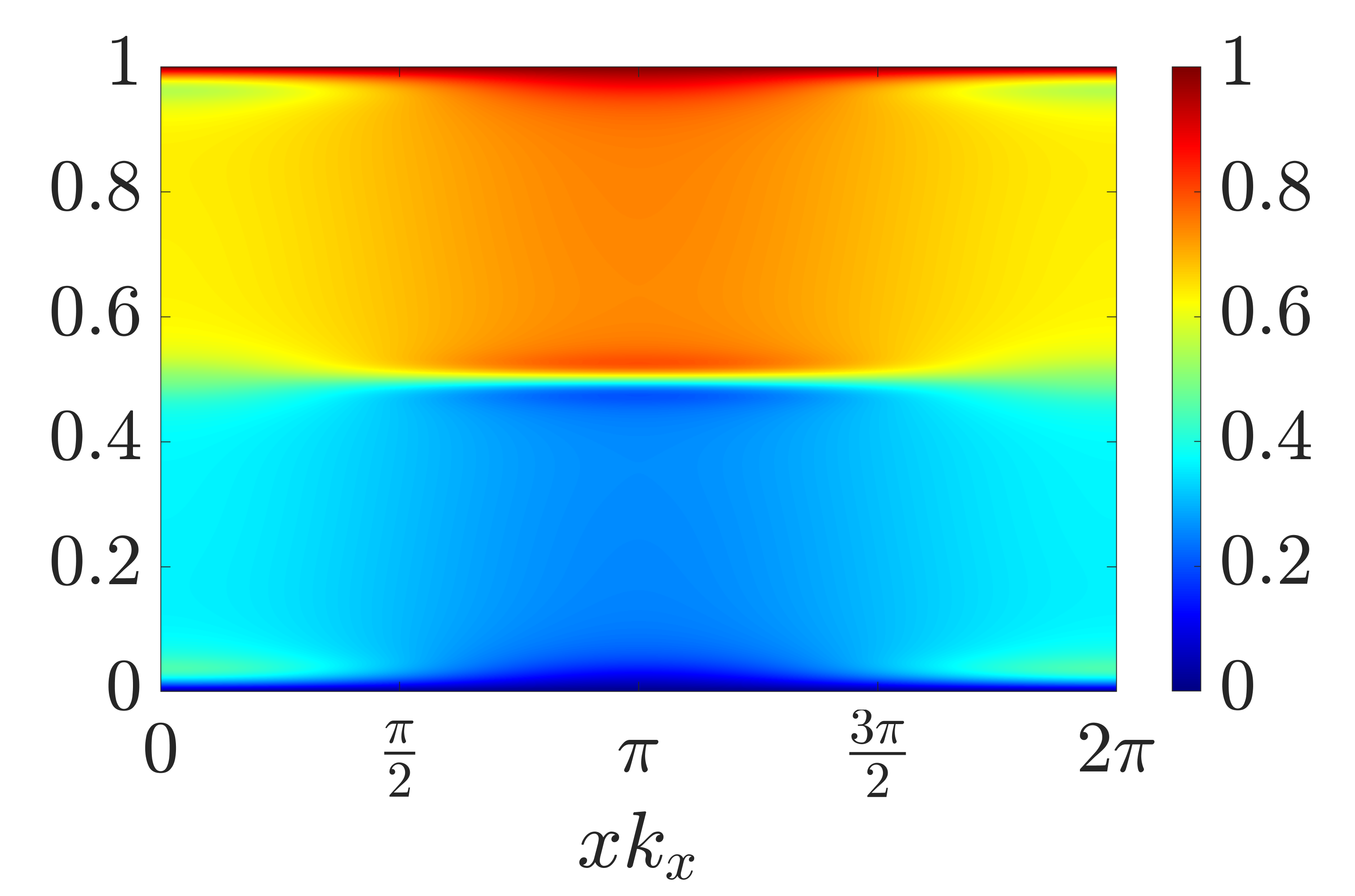}
    \includegraphics[width=0.31\textwidth]{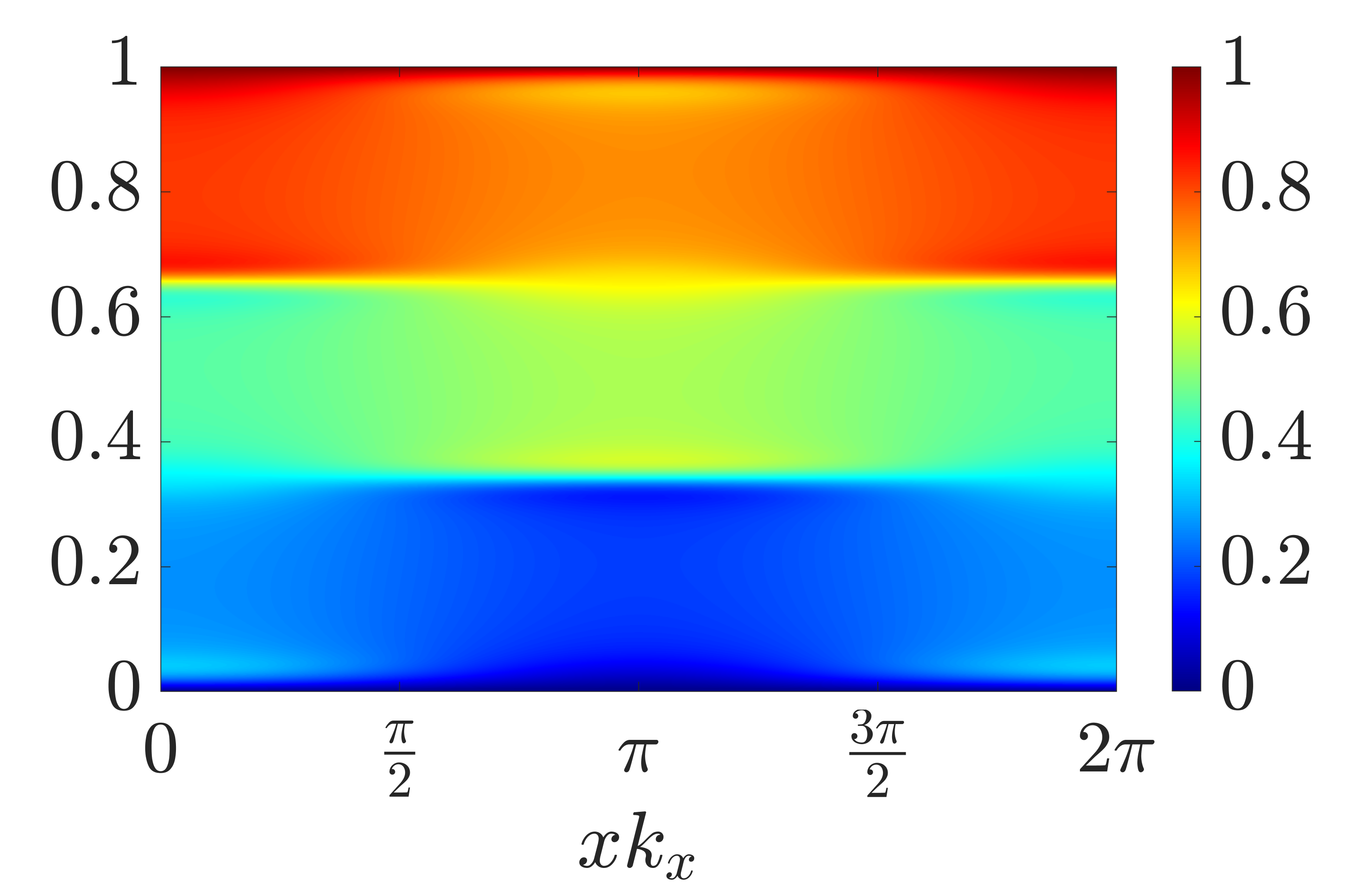}

    \caption{Solution profiles from the single-mode equations \eqref{eq:single_mode}. The first row shows the profiles of (a) $z+\bar{S}_0$, (b) $\widehat{S}$, (c) $\widetilde{u}$ and (d) $\widehat{w}$. The second row shows the reconstructed total salinity using \eqref{eq:normal_mode_S} and \eqref{eq:total_T_S} for (e) S1, (f) S2 and (g) S3 solutions. The parameter values are $k_x=17.5$, $k_y=0$, $R_\rho=2$, $Pr=7$, $\tau =0.01$ and $Ra_T=10^5$. }
    \label{fig:profile_R_rho_T2S_2_tau_0p01}
\end{figure}

The S1, S2 and S3 solution branches at $R_\rho=2$ in figure \ref{fig:bif_diag_mid_Ra_S2T} exhibit much larger $Sh$ than those at $R_\rho=40$ in figure \ref{fig:bif_diag_low_Ra_S2T}. This is expected as a lower density ratio indicates stronger destabilization by the salinity gradient. Figure \ref{fig:profile_R_rho_T2S_2_tau_0p01} shows profiles of the S1, S2 and S3 solutions at $R_\rho=2$ and $k_x=17.5$, $k_y=0$. Here, the mixed region in the horizontally averaged total salinity $z+\bar{S}_0(z)$ in figure \ref{fig:profile_R_rho_T2S_2_tau_0p01}(a) corresponds to almost constant salinity compared with the solution profile at $R_\rho=40$ in figure \ref{fig:profile_R_rho_T2S_40_tau_0p01}(a). Outside of this region, both the profile $z+\bar{S}_0(z)$ and the isocontours of total salinity in figures \ref{fig:profile_R_rho_T2S_2_tau_0p01}(e)-(g) exhibit sharper interfaces than those at $R_\rho=40$ in figure \ref{fig:profile_R_rho_T2S_40_tau_0p01}. This is consistent with the observation of well-defined staircases in flow regimes with $R_\rho \sim O(1)$ in field measurements \citep{schmitt1987c} and laboratory experiments \citep{krishnamurti2003double,krishnamurti2009heat} as well as in DNS \citep{radko2003mechanism}. The vertical profiles of the salinity harmonic $\widehat{S}$ in the S1, S2 and S3 states shown in figure \ref{fig:profile_R_rho_T2S_2_tau_0p01}(b) all show a local peak at the region associated with sharp interfaces. Both the horizontal velocity and the vertical velocity in figures \ref{fig:profile_R_rho_T2S_2_tau_0p01}(c)-(d) show a larger magnitude compared with the $R_\rho=40$ results in figures \ref{fig:profile_R_rho_T2S_40_tau_0p01}(c)-(d).

\begin{figure}
    \centering
    
    \hspace{0.02\textwidth} (a) $\bar{S}_0=0$ \hspace{0.12\textwidth} (b) S1 \hspace{0.12\textwidth} (c) S2 \hspace{0.12\textwidth} (d) S3
    
    \includegraphics[width=0.9\textwidth]{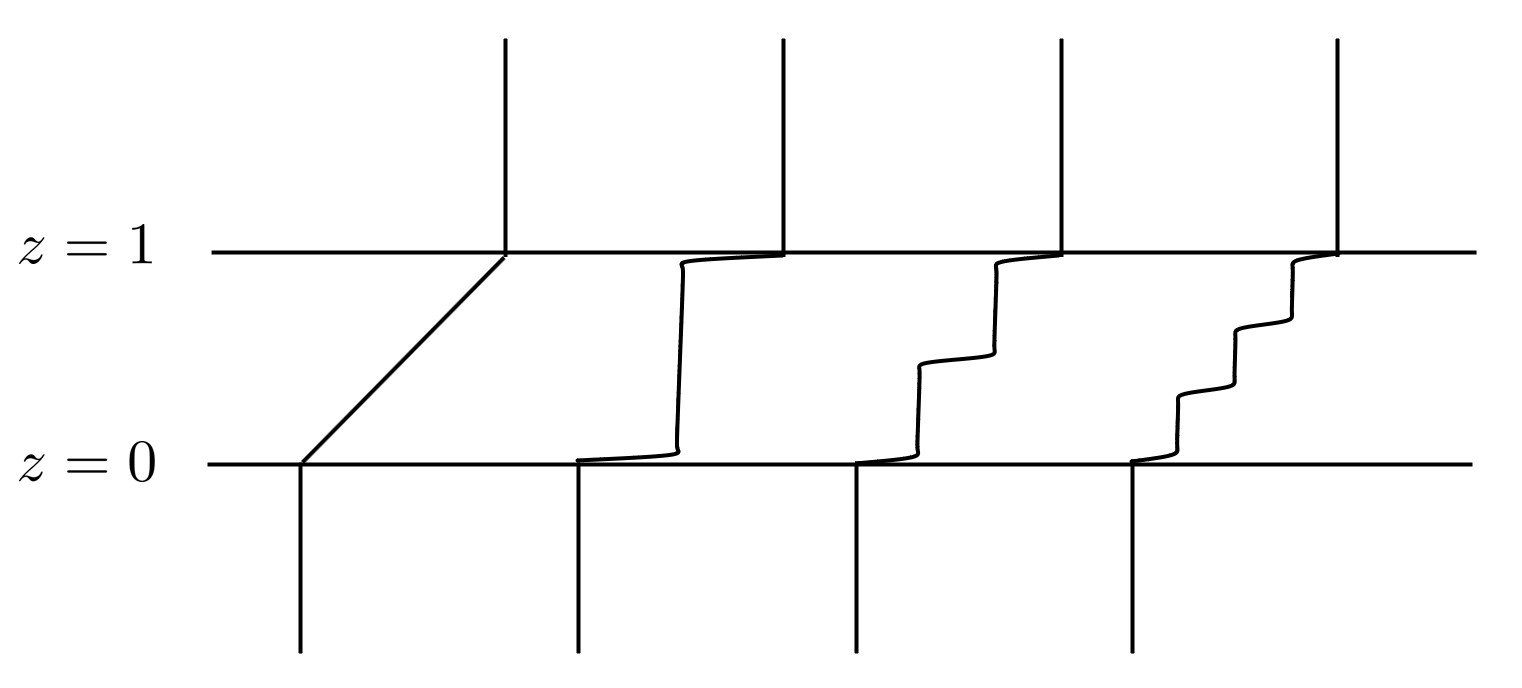}
    \caption{Sketch of the mean total salinity profile $z+\bar{S}_0(z)$ between two well-mixed regions, where the interior $z\in [0,1]$ is associated with (a) a linear base state, (b) S1 solution, (c) S2 solution and (d) S3 solution. The S1, S2 and S3 single-mode solutions are obtained from figure \ref{fig:profile_R_rho_T2S_2_tau_0p01}(a) associated with $k_x=17.5$, $k_y=0$, $R_\rho=2$, $Pr=7$, $\tau =0.01$ and $Ra_T=10^5$.}
    \label{fig:staircase_ocean}
\end{figure}

Figure \ref{fig:staircase_ocean} combines the mean salinity profiles of the S1, S2 and S3 single-mode solutions in figure \ref{fig:profile_R_rho_T2S_2_tau_0p01}(a) with constant salinity outside of the region $z\in [0,1]$ on the assumption that each layer is well mixed. The figure shows that salt-finger convection can distort a linear mean salinity profile into a staircase-like profile in $z\in [0,1]$, particularly when $R_{\rho}$ is relatively small. While other mechanisms may be involved in the ocean leading to layer formation, merger and migration, including the presence of a diffusive regime (cold fresh water on top of warm salty water) \citep{timmermans2008ice,radko2016thermohaline,yang2022layering} or stratified shear flow \citep{oglethorpe2013spontaneous,taylor2017multi,lucas2017layer,lucas2019layer}, our work suggests that in horizontally extended domains shearing instabilities may disrupt the finger zones that form the S1, S2 and S3 profiles without destroying the associated staircase structure, leading to well-mixed layers of the type observed in the oceans.

\begin{figure}
    \centering
    (a) \hspace{0.2\textwidth} (b)TF1 \hspace{0.27\textwidth} (c)TW1\hspace{0.05\textwidth}

    \includegraphics[width=0.186\textwidth]{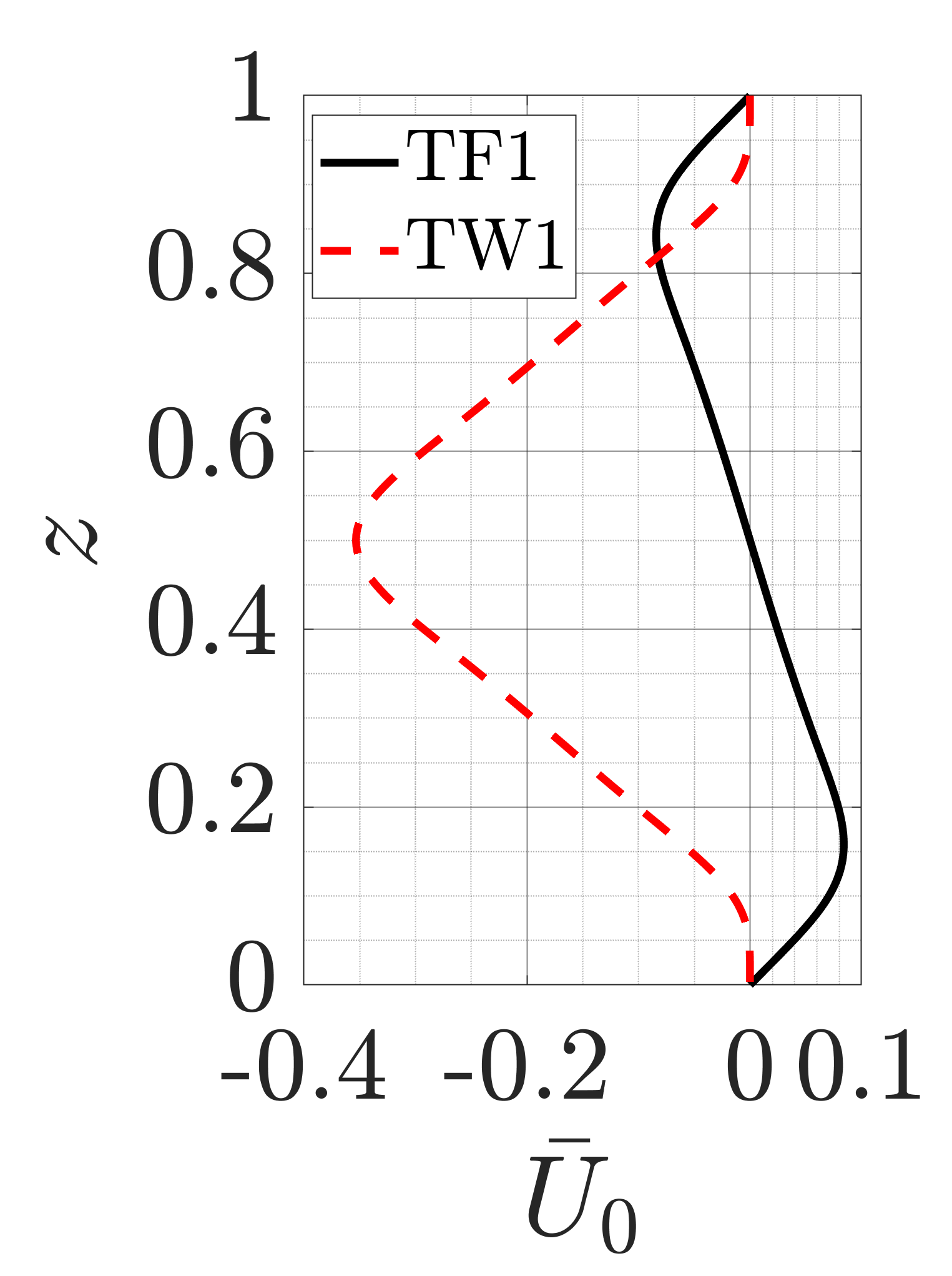}
    \includegraphics[width=0.365\textwidth,trim=-0 -0.25in 0 0]{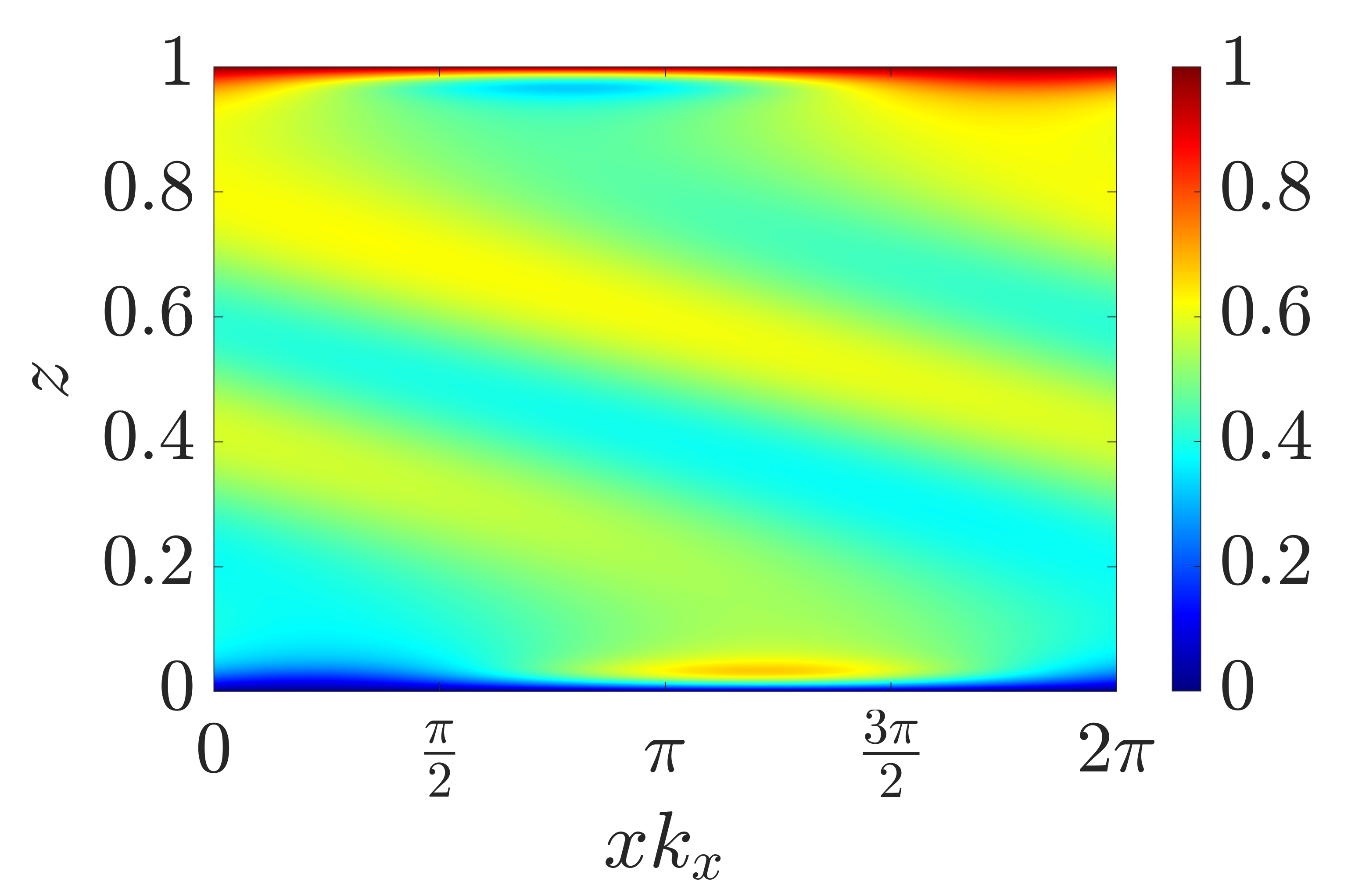}
    \includegraphics[width=0.365\textwidth,trim=-0 -0.25in 0 0]{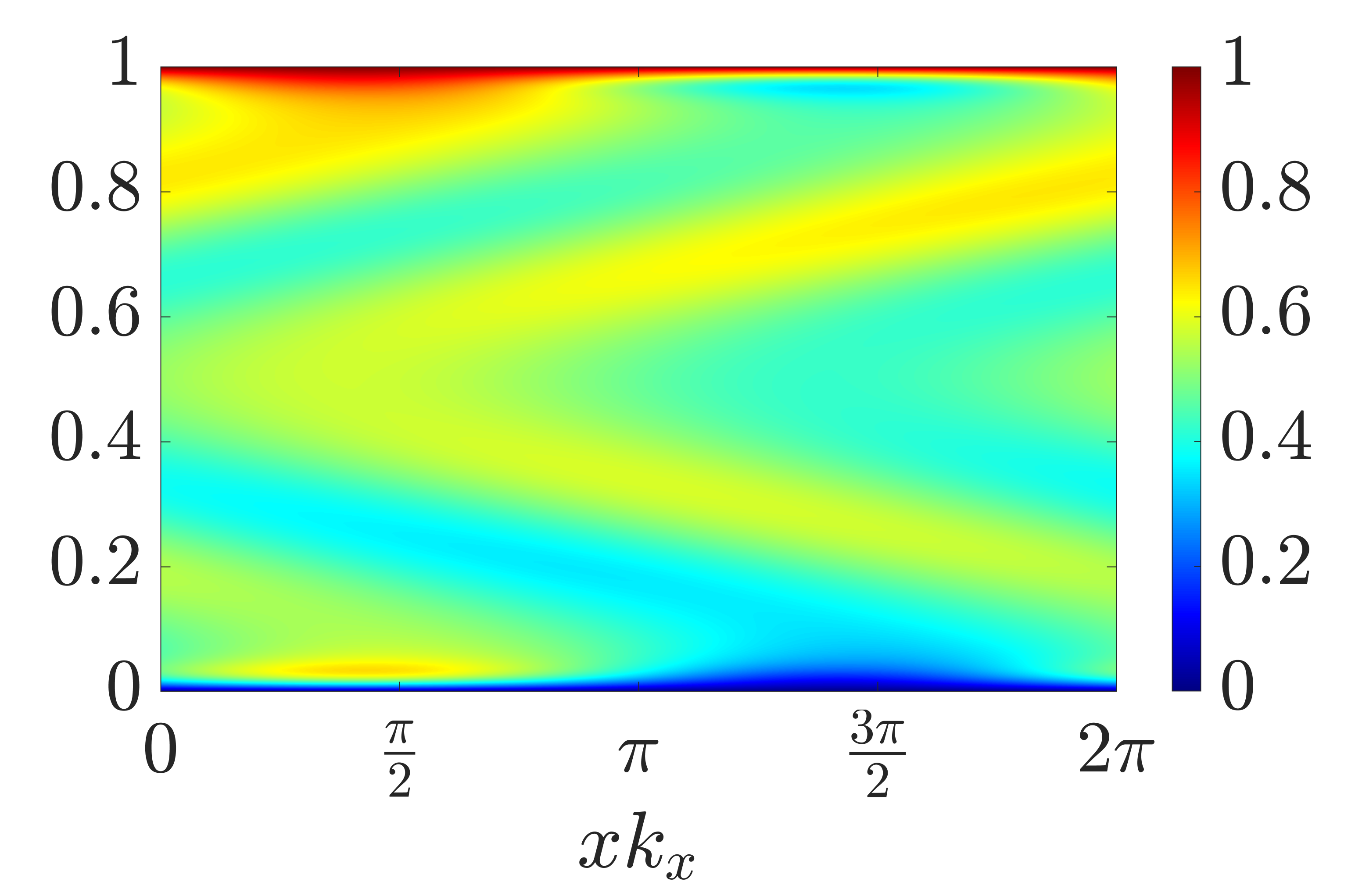}
    \caption{Solution profiles from the single-mode equations \eqref{eq:single_mode} showing (a) the large-scale shear $\bar{U}_0$ and the isocontours of total salinity for (b) TF1 and (c) TW1. The governing parameters are $k_x=17.5$, $k_y=0$, $R_\rho=2$, $Pr=7$, $\tau =0.01$ and $Ra_T=10^5$.}
    \label{fig:profile_R_rho_T2S_2_tau_0p01_TF1_TW1}
\end{figure}

We now focus on the solutions TF1 and TW1 originating from secondary bifurcations of the S1 state. Figure \ref{fig:profile_R_rho_T2S_2_tau_0p01_TF1_TW1}(a) shows the associated large-scale shear $\bar{U}_0(z)$ of TF1 and TW1. This shear is antisymmetric with respect to the midplane for TF1 but symmetric for TW1. Figure \ref{fig:profile_R_rho_T2S_2_tau_0p01_TF1_TW1}(b) shows the total salinity profile for a tilted finger with a larger tilt angle compared with the tilt at $R_\rho=40$ shown in figure \ref{fig:profile_R_rho_T2S_40_tau_0p01_Pr_7_TF1}(e). Figure \ref{fig:profile_R_rho_T2S_2_tau_0p01_TF1_TW1}(c) shows the total salinity of TW1 in the comoving frame, with structures that tilt in opposite directions above and below the midplane. This alternating tilt direction resembles the `wavy fingers' tilted either left or right observed in experiments \citep[figure 7]{krishnamurti2003double}, although whether such  `wavy fingers' travel depends on the boundary conditions in the horizontal.

The stability properties of these solutions are also indicated in figure \ref{fig:bif_diag_mid_Ra_S2T} and are similar to those for high density ratios (figure \ref{fig:bif_diag_low_Ra_S2T}). Here, the S1 solution is stable near the onset, but loses stability to tilted fingers. The solutions S2 and S3 are still unstable within all of the current parameter regime. The traveling wave appears to gain stability within a certain wavenumber regime close to the low wavenumber onset, as indicated in the zoom in figures \ref{fig:bif_diag_mid_Ra_S2T}(c) and \ref{fig:bif_diag_mid_Ra_S2T}(e). Since the single-mode solutions are expected to be more accurate in the high wavenumber regime, cf. figure \ref{fig:bif_diag_low_Ra_S2T}(a), we postpone a study of the dynamics of TF1 and TW1 to the lower Prandtl number considered in \S \ref{sec:results_Pr_0p05} where these states form at higher wavenumbers.

\begin{figure}
    \centering
    \includegraphics[width=0.49\textwidth]{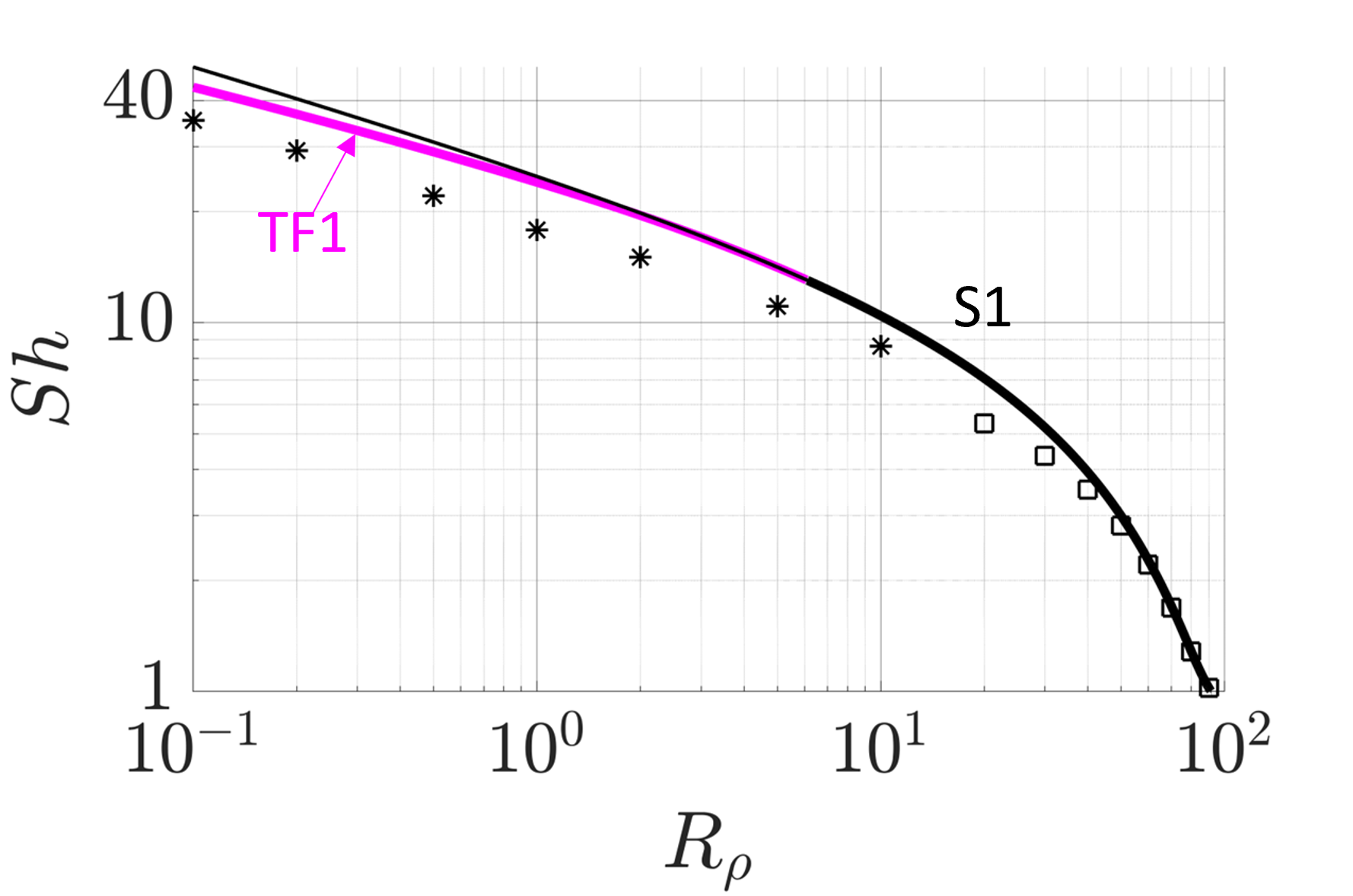}
        \includegraphics[width=0.49\textwidth]{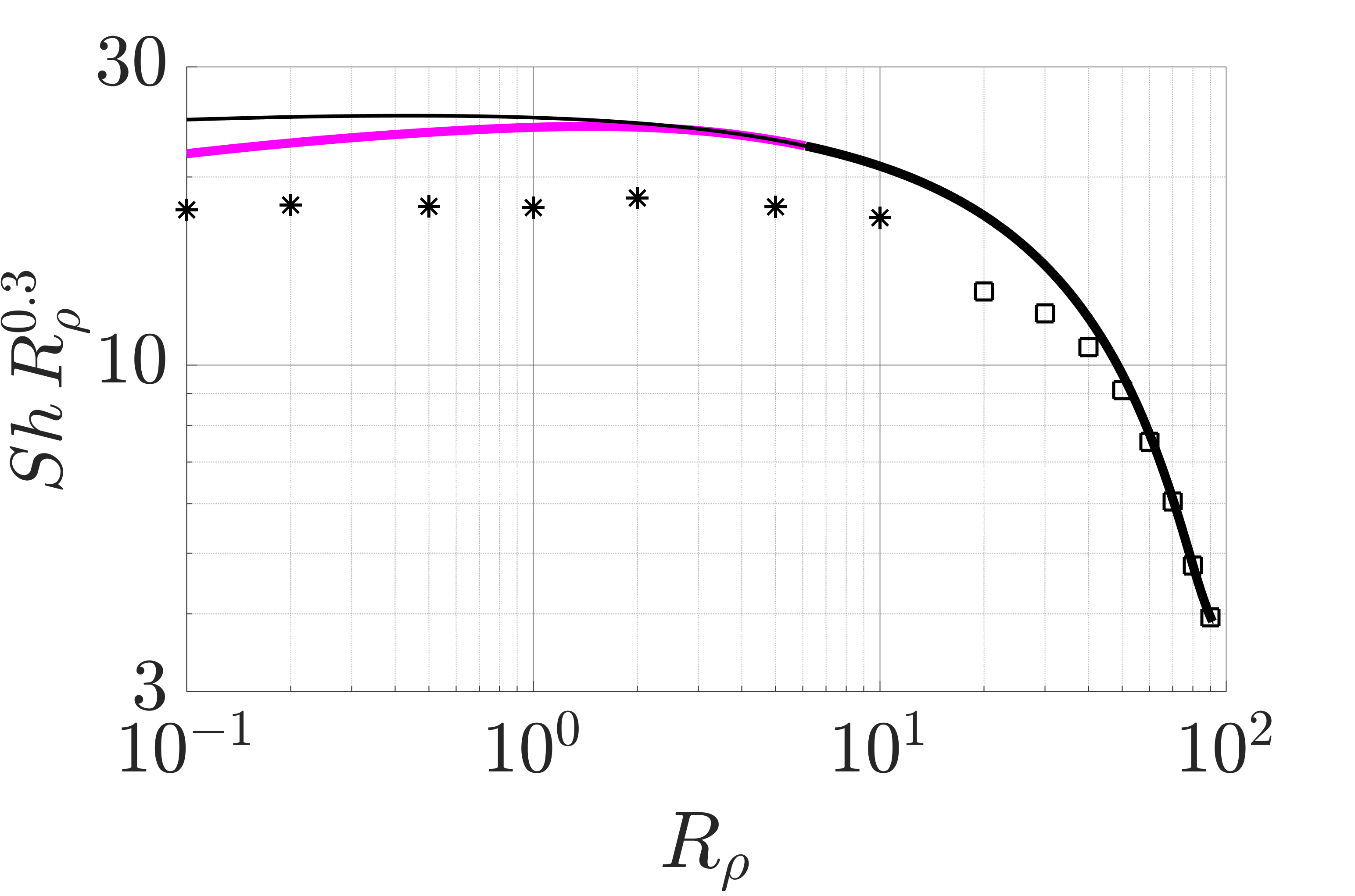}

    \caption{ (a) Bifurcation diagrams displaying $Sh$ as a function of the density ratio $R_\rho$ from the single-mode equations \eqref{eq:single_mode} with wavenumbers $k_x$ and $k_y$ given by the scaling law \eqref{eq:kx_scaling_yang}. The other parameters are $Pr=7$, $\tau =0.01$ and $Ra_T=10^5$. The black star $(*)$ is obtained from 3D DNS \citep[table 1]{yang2015salinity}. The black square $(\square)$ is obtained from 2D DNS results in a horizontal domain of size $L_x=2\pi /k_{x,2D}=2\pi/(\sqrt{2}k_x)$ with $k_x$ given by \eqref{eq:kx_scaling_yang}. Panel (b) plots the results in (a) in compensated form to exhibit the scaling exponent. }
    \label{fig:bif_diag_R_rho_T2S_tau_0p01}
\end{figure}

\begin{figure}
    \centering
    \includegraphics[width=0.8\textwidth]{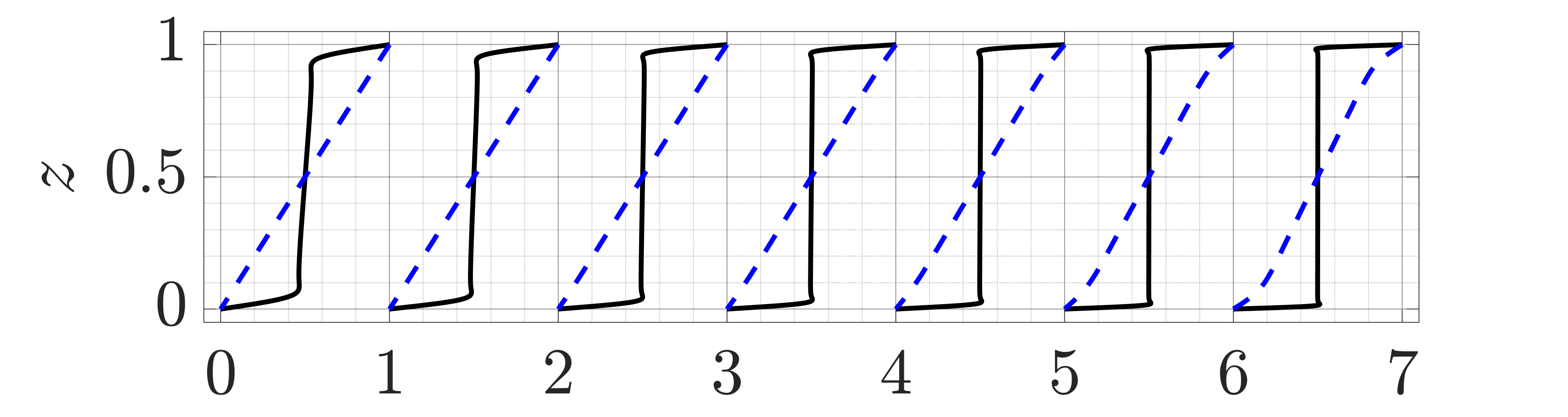}
    \caption{Solution profiles from the single-mode equations \eqref{eq:single_mode} displaying horizontally averaged total salinity (black solid lines) and horizontally averaged total temperature (blue dashed lines). Solution profiles are labeled by integers and correspond to $R_\rho=10,5,2,1,0.5,0.2$, and $0.1$ from left to right. Other parameter values are $Pr=7$, $\tau =0.01$, $Ra_T=10^5$, with $k_x=k_y$ given by the scaling law in \eqref{eq:kx_scaling_yang}.}
    \label{fig:profile_R_rho_T2S_range_mean_S_T}
\end{figure}

Figure \ref{fig:bif_diag_R_rho_T2S_tau_0p01} displays the dependence of the $Sh$ on the density ratio $R_\rho$ for $\tau =0.01$ for 3D results with $k_y=k_x$ based on the scaling law 
\begin{subequations}
\begin{align}
    k_x=&k_y=R_\rho^{0.125}\pi/(14.8211Ra_S^{-0.2428}),
    \label{eq:kx_scaling_yang}\\
    Ra_S:=&\frac{ g \beta\Delta S h^3}{\kappa_S\nu }=\frac{Ra_T}{R_\rho \tau},
    \label{eq:Ra_S_def}
\end{align}
\end{subequations}
obtained through a least-squares fit of the finger width computed from 3D DNS \citep[figure 13(b)]{yang2016scaling}. Here, we can see that the S1 solution remains stable near onset, and then bifurcates to stable tilted fingers (TF1) as $R_\rho$ decreases. The figure also shows the $Sh$ from 3D DNS data \citep[table 1]{yang2015salinity} using black stars. We also perform 2D direct numerical simulations in domains of size $L_x=2\pi/k_{x,2D}=2\pi/(\sqrt{2}k_x)$ with $k_x$ given by \eqref{eq:kx_scaling_yang}. Note that our simulation results do not display a smooth transition to the $Sh$ from the 3D results \citep{yang2015salinity}, which is likely due to the intrinsic difference between 2D and 3D simulations and possibly different domain sizes.  However, the $Sh$ obtained from the S1 single-mode solutions agree well with the DNS results near the high density ratio onset. At lower density ratios, the single-mode solution S1 and TF1 both overpredict the DNS results. This may be because both are steady-state solutions, while the DNS results exhibit time-dependent behavior which likely reduces the time-averaged $Sh$ number. 

We find that $Sh\sim R_\rho ^{-0.30}$ for the single-mode solutions within the range $R_\rho\in [0.1,2]$ and fixed $Ra_T=10^5$ and $\tau=0.01$, indicating that for these values of $Ra_T$ and $\tau$, $Sh\sim Ra_S^{0.30}$ as a function of the salinity Rayleigh number defined in \eqref{eq:Ra_S_def}. This result agrees well with the scaling law of Nusselt number $Nu\sim Ra_T^{0.30}(\text{ln} Ra_T)^{0.20}$ obtained from single-mode solutions for Rayleigh-B\'enard convection and wavenumber scaling as $k_x\sim Ra_T^{1/4}$ \citep[p. 713]{gough1975modal}. The scaling law in \eqref{eq:kx_scaling_yang} employed here with a fixed $R_\rho$ is also compatible with $k_x\sim Ra_T^{1/4}$ employed in Rayleigh-B\'enard convection \citep{gough1975modal}. Figure \ref{fig:bif_diag_R_rho_T2S_tau_0p01}(b) displays the compensated Sherwood number $Sh\,R_\rho^{0.30}$, showing that the S1 solutions from both the single-mode approach and DNS \citep{yang2015salinity} follow the same $Sh\sim R_\rho^{-0.3}$ scaling law. Note that the DNS results for salt-finger convection in the asymptotic regime of high $Ra_S$ instead suggest $Sh\sim Ra_S^{1/3}$ \citep[figure 7(a)]{yang2015salinity} while experimental results suggest $Sh\sim Ra_S^{4/9}$ \citep[figure 8]{hage2010high}. The parameter regime considered here is within the regime $Ra_S\leq 10^8$, a value that may be insufficient to reliably establish the asymptotic scaling; see e.g.,~\citep[figure 7(a)]{yang2015salinity} and \citep[figure 3(a)]{yang2016vertically}. The TF1 solutions suggest the scaling law $Sh\sim R_\rho^{-0.27}$ or $Sh\sim Ra_S^{0.27}$ for fixed $Ra_T$ and $\tau$. This is consistent with the observation in Rayleigh-B\'enard convection that the formation of the large-scale shear decreases the scaling exponent $\eta$ of $Nu\sim Ra_T^{\eta}$; see e.g., \citet[figure 4]{goluskin2014convectively}. 

Figure \ref{fig:profile_R_rho_T2S_range_mean_S_T} shows the corresponding mean salinity and temperature profiles at the same parameter values as those used for the 3D DNS results in \citet[figure 2(a)]{yang2015salinity}. The solution profiles from the single-mode equations closely match the qualitative trend of the mean salinity and temperature profiles as a function of the density ratio in the  DNS results \citep[figure 2(a)]{yang2015salinity}. In particular, both the single-mode and DNS results show that the mean salinity profile displays a mixed region in the layer interior, and that both show a small overshoot (a thin stably stratified region) near the top and bottom boundaries. The mean total temperature remains close to a linear profile, however, and shows a visible deviation from a linear profile only at $R_\rho=0.1$, a fact also consistent with DNS observation \citep[figure 2(a)]{yang2015salinity}.

\section{Tilted salt fingers and traveling waves at $Pr=0.05$}
\label{sec:results_Pr_0p05}

The bifurcation diagram for a low density ratio, $R_\rho=2$, and $Pr=7$ displays a branch of steady tilted fingers (TF1) and a branch of traveling waves (TW1), both of which bifurcate from the symmetric one-layer solution S1 at intermediate wavenumbers. Here, we study these states in the low Prandtl number regime that pushes the bifurcations to TF1 and TW1 to higher wavenumbers and hence closer to the high wavenumber onset, where we expect the single-mode solutions to be accurate. Low Prandtl number salt-finger convection is of interest in astrophysical applications, where molecular weight gradients compete with thermal buoyancy but heat transport is dominated by photon diffusion \citep{garaud2018double}. However, to keep the notation consistent with the previous sections, we continue to use the symbol $S$ to represent the higher molecular weight component and refer to it as salinity. The results presented here for $Pr=0.05$ parallel those for $Pr=7$ in \S \ref{subsec:results_Pr_7_R_rho_40} in order to highlight the effect of a low Prandtl number.

\begin{figure}
    \centering

    (a) 2D: $k_y=0$ \hspace{0.35\textwidth} (b) 3D: $k_y=k_x$

    \includegraphics[width=0.49\textwidth]{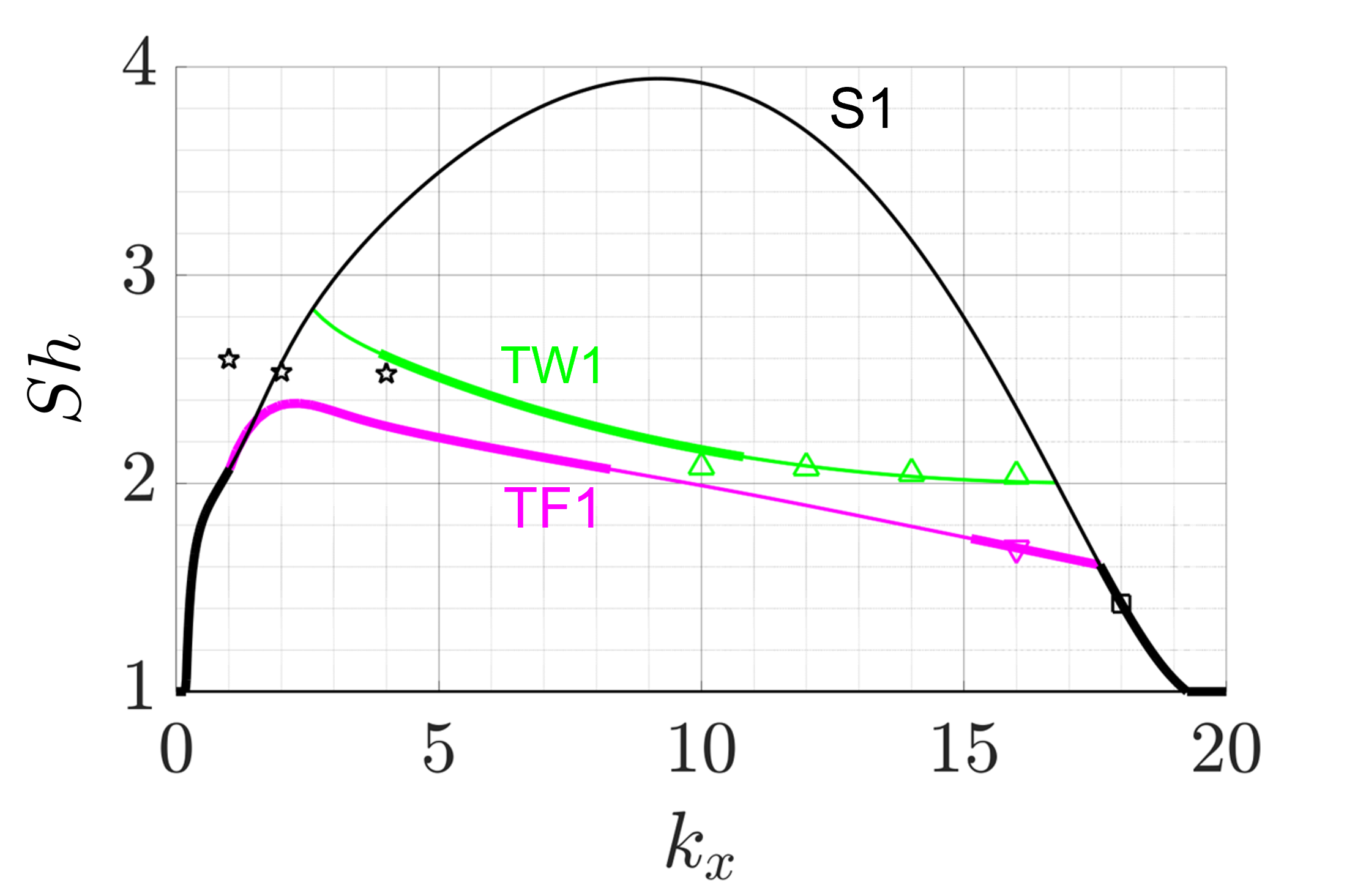}
    \includegraphics[width=0.49\textwidth]{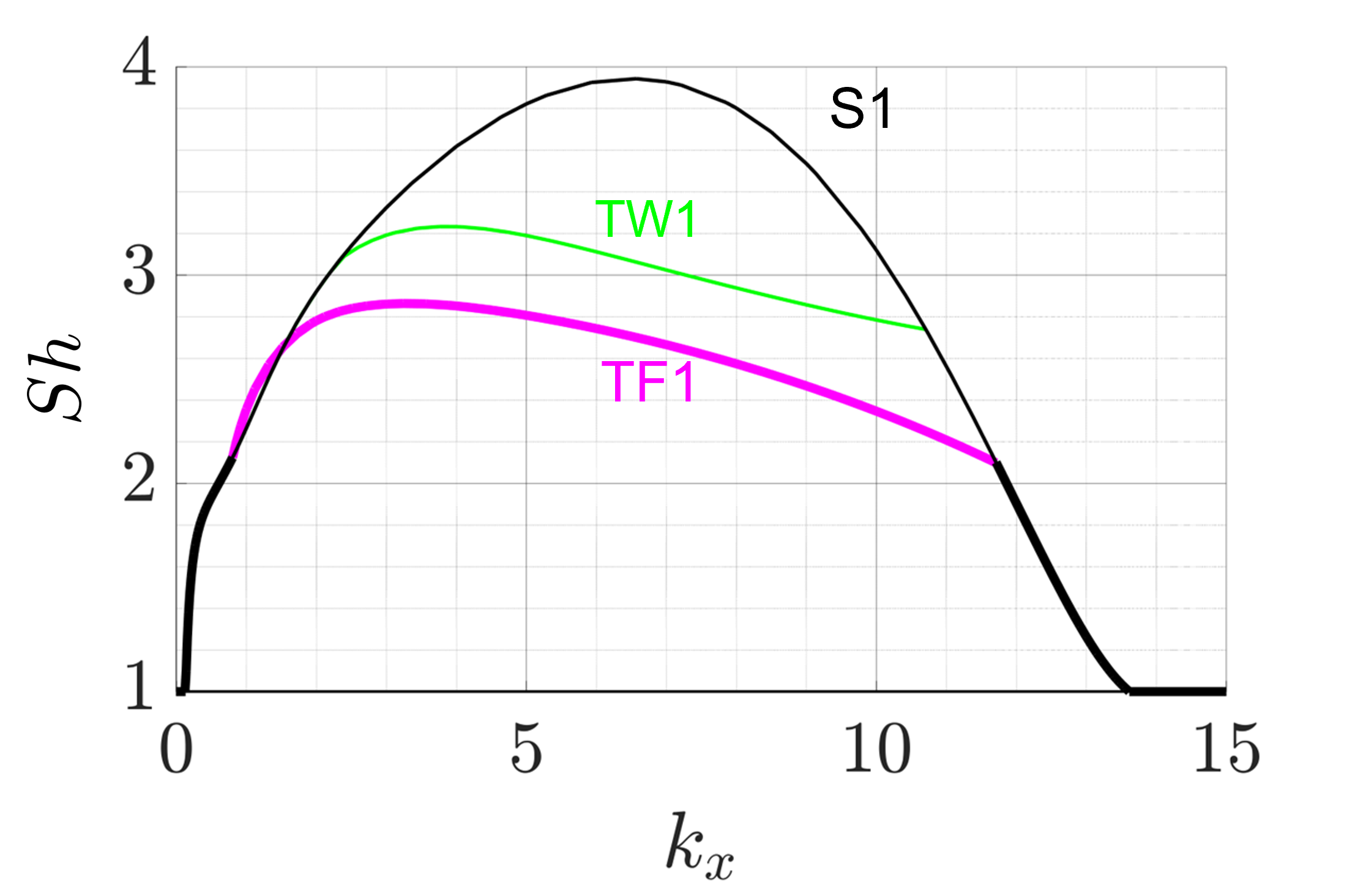}
    
    \caption{Bifurcation diagrams as a function of the wavenumber $k_x$ from the single-mode equations \eqref{eq:single_mode} for (a) 2D: $k_y=0$ and (b) 3D: $k_y=k_x$. Other parameters are $R_\rho=40$, $Pr=0.05$, $\tau=0.01$, and $Ra_T=10^5$.  The markers correspond to the $Sh$ associated with steady states resembling S1 ($\square$), TF1 ({\color{magenta} $\triangledown $}) and TW1 ({\color{green}$\triangle$}) solutions from 2D DNS in domains of size $L_x=2\pi/k_x$ (table \ref{tab:DNS_transition_low_Ra_S2T_Pr_0p05}). The black pentagrams corresponding to the $Sh$ at $L_x=2\pi/k_x$ ($k_x=1$, 2, 4) indicate chaotic behavior. }
    \label{fig:bif_diag_low_Ra_S2T_low_Pr}
\end{figure}

Figure \ref{fig:bif_diag_low_Ra_S2T_low_Pr} shows the bifurcation diagram for the parameter values used in figure \ref{fig:bif_diag_low_Ra_S2T} but with $Pr=0.05$. Here, we focus on the solution branches S1, TF1 and TW1; the S2 and S3 branches are omitted since neither their Sherwood number nor their overall stability changes when the Prandtl number changes from $Pr=7$ to $Pr=0.05$. Compared with the $Pr=7$ results in figure \ref{fig:bif_diag_low_Ra_S2T}, we see that the $Sh$ associated with the S1 solution remains the same, but the bifurcation to TF1 ($k_x= 17.593$) now occurs much closer to the high wavenumber onset of the S1 solution ($k_x=19.251$). Moreover, the traveling wave branch TW1 that is present here does not appear at $R_\rho=40$, $Pr=7$ in figure \ref{fig:bif_diag_low_Ra_S2T}. Evidently low Prandtl numbers favor spontaneous formation of large-scale shear, as found in DNS by \citet[figures 1-2]{radko2010equilibration} and \citet{garaud20152d} as well as in a reduced model valid in the asymptotic limit of low $\tau$ and low $Pr$ \citep{xie2019jet}. A related phenomenon is found in Rayleigh-B\'enard convection, where at low $Pr$ a steady convection roll becomes immediately unstable to a large-scale (zonal) mode \citep{winchester2022onset}.

\begin{figure}
    \centering
    (a) \hspace{0.2\textwidth} (b) \hspace{0.2\textwidth} (c)TF1 \hspace{0.2\textwidth} (d)TW1 
    \includegraphics[width=0.24\textwidth]{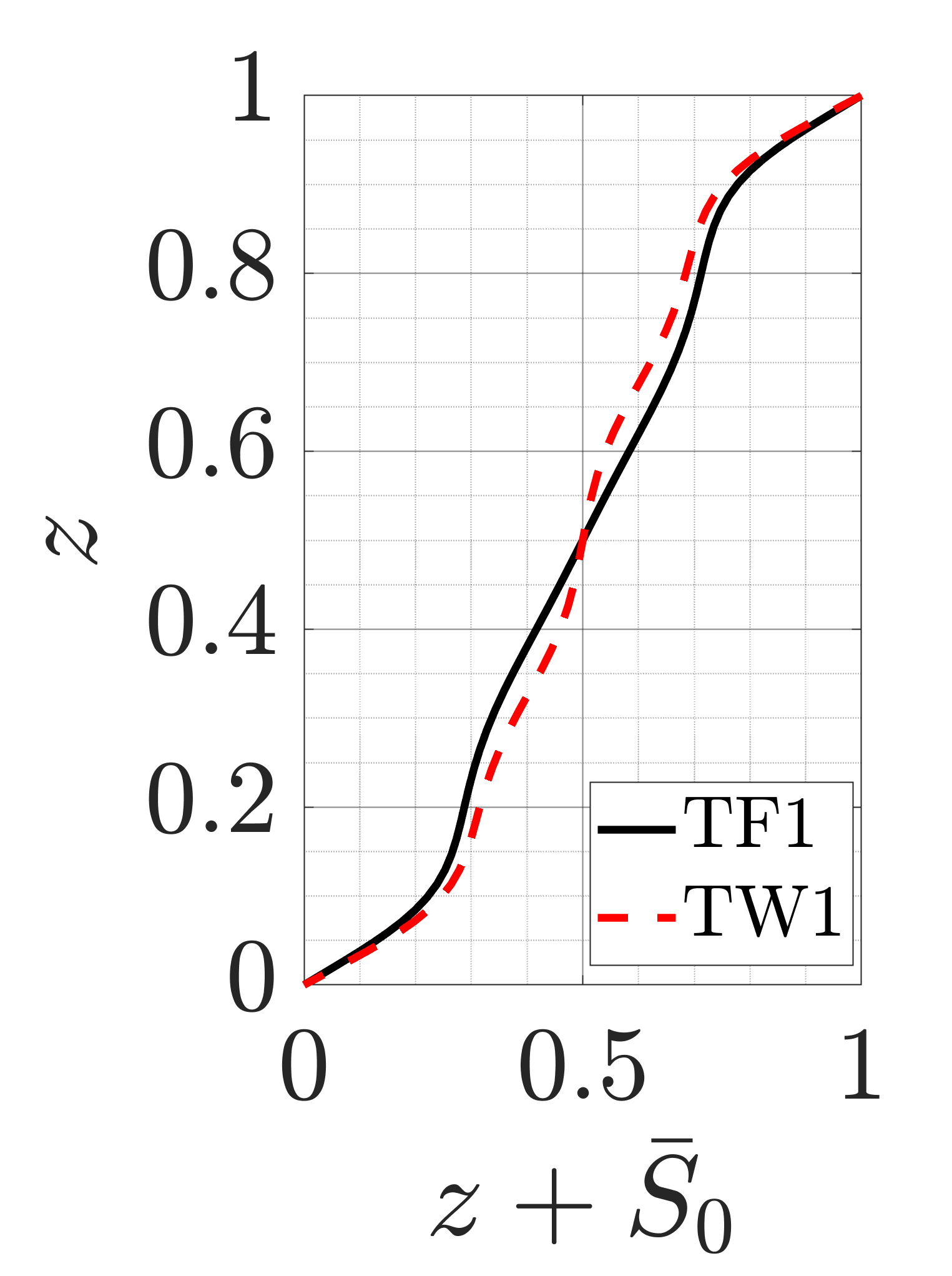}
    \includegraphics[width=0.24\textwidth]{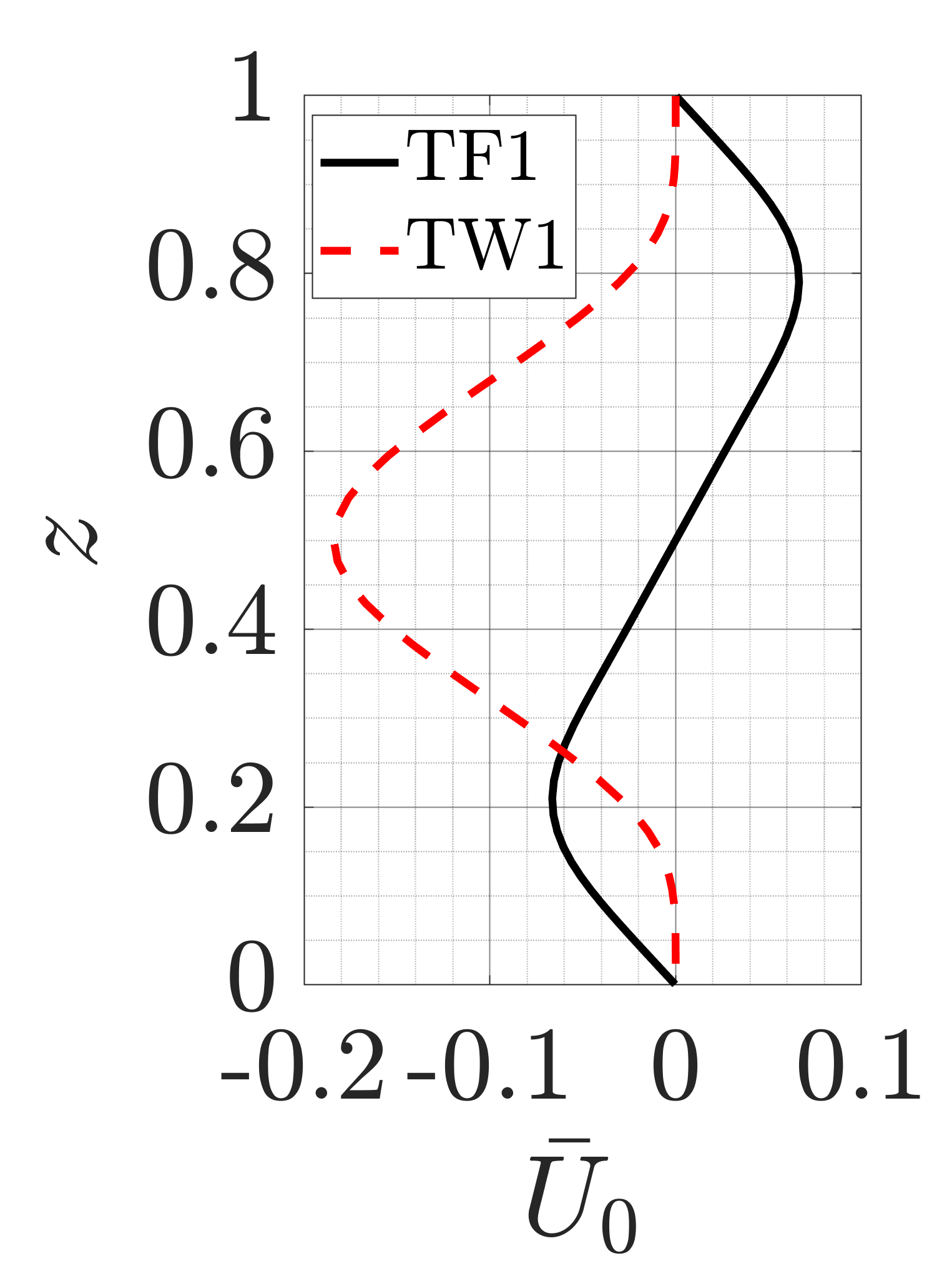}
    \includegraphics[width=0.24\textwidth]{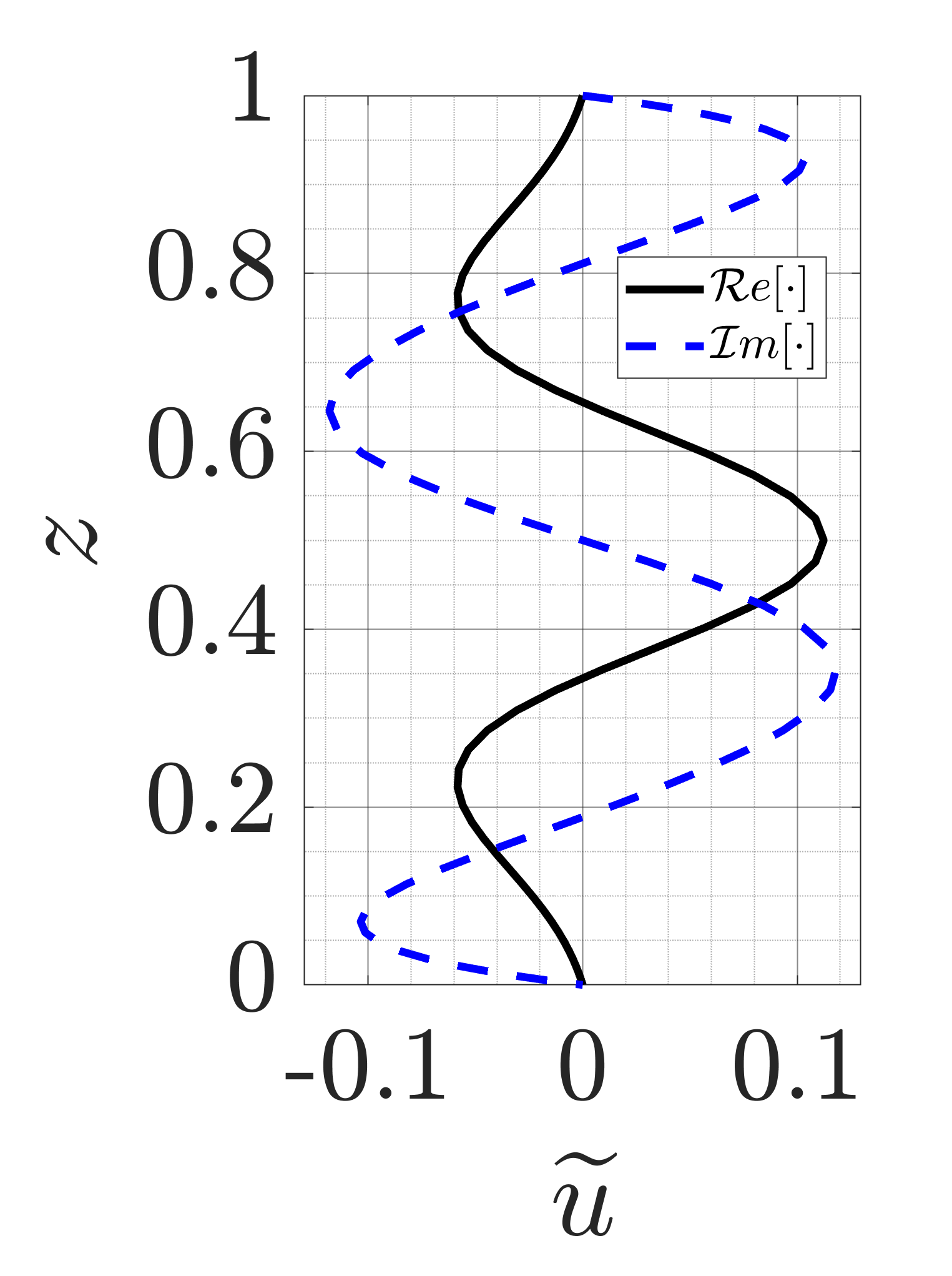}
    \includegraphics[width=0.24\textwidth]{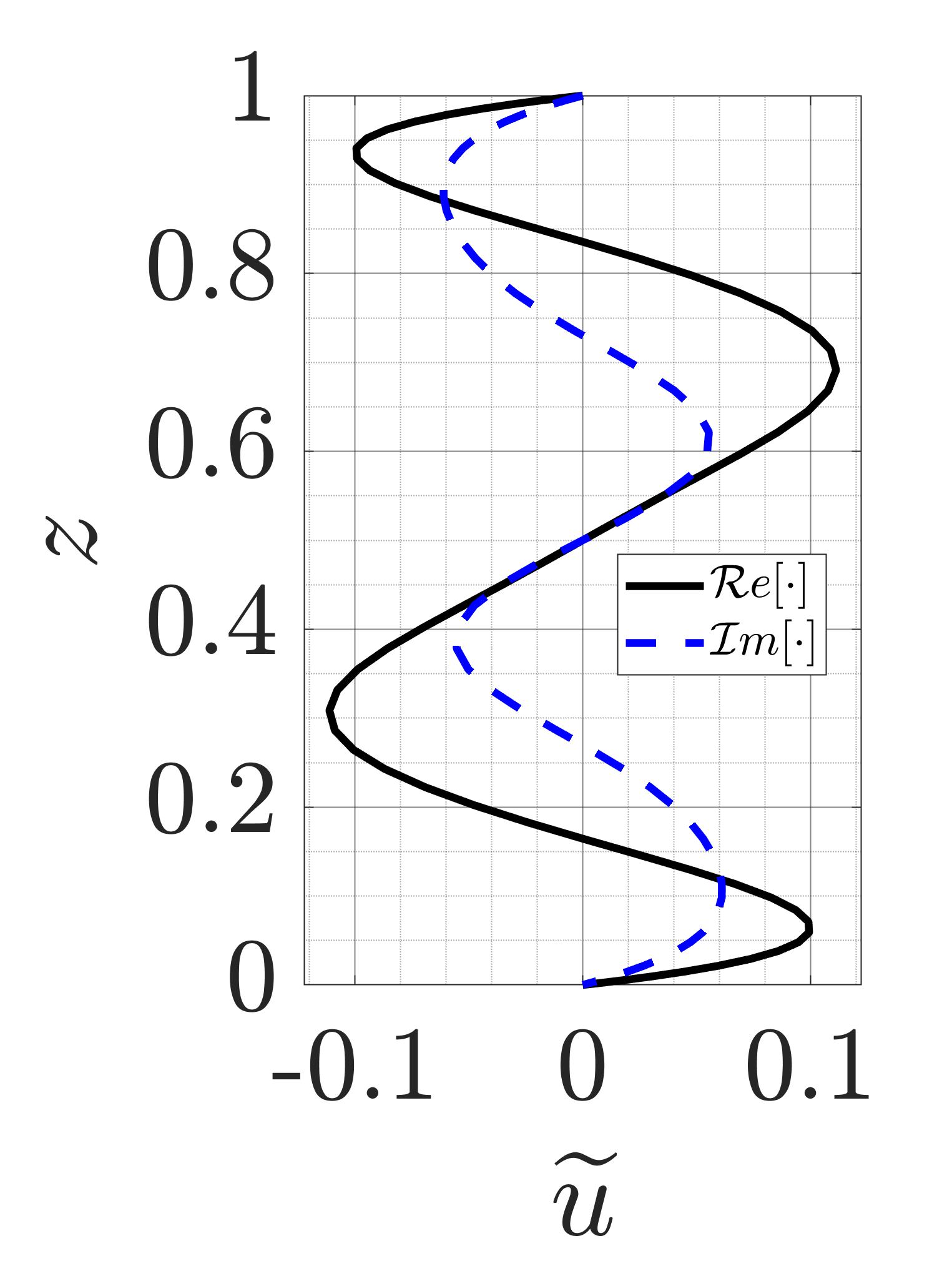}

    (e)TF1 \hspace{0.1\textwidth} (f)TW1
     \hspace{0.16\textwidth}   (g)TF1 \hspace{0.22\textwidth} (h)TW1\hspace{0.05\textwidth}

     \includegraphics[width=0.16\textwidth]{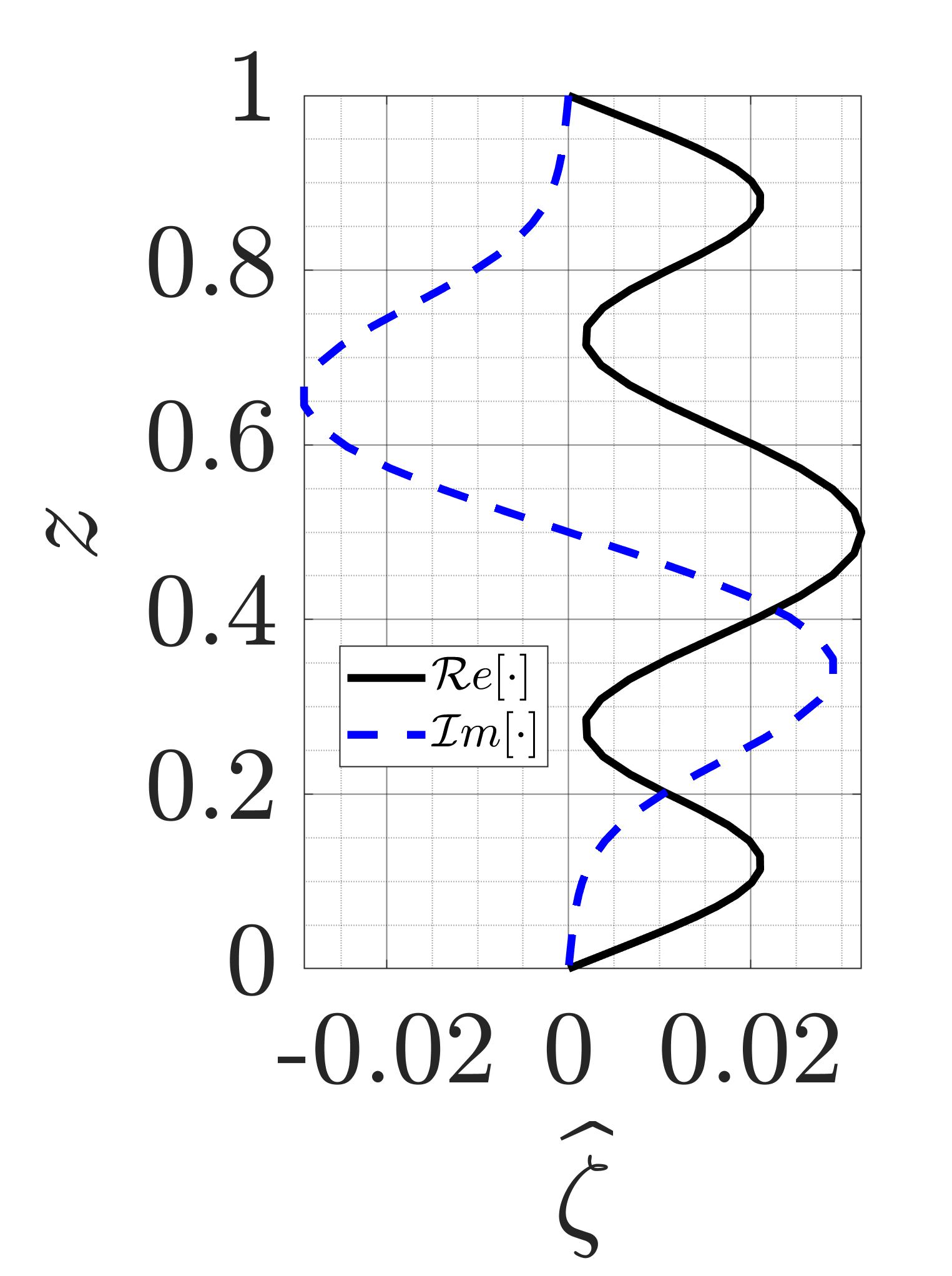}
    \includegraphics[width=0.16\textwidth]{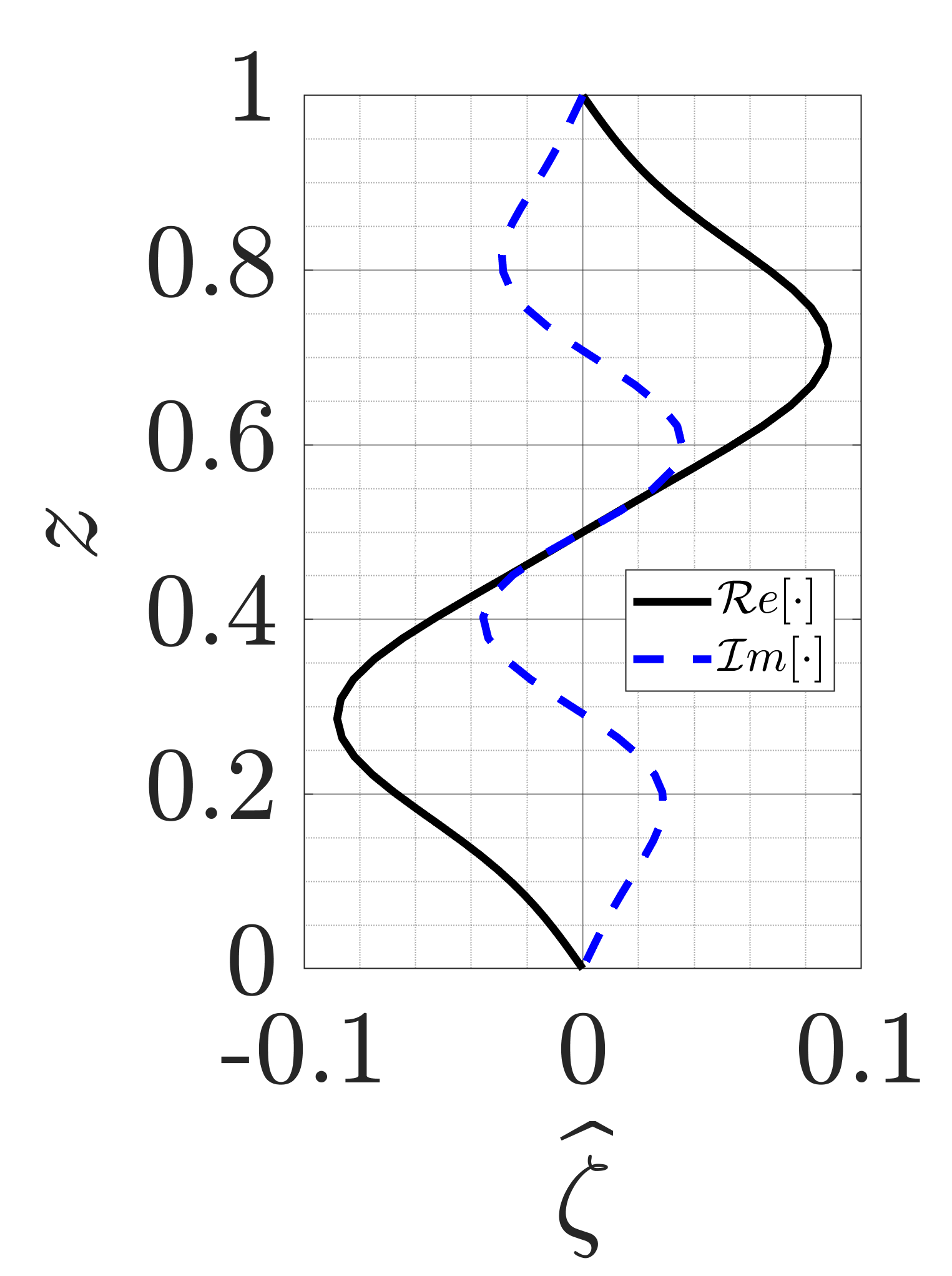}
    \includegraphics[width=0.32\textwidth,trim=-0 -0in 0 0]{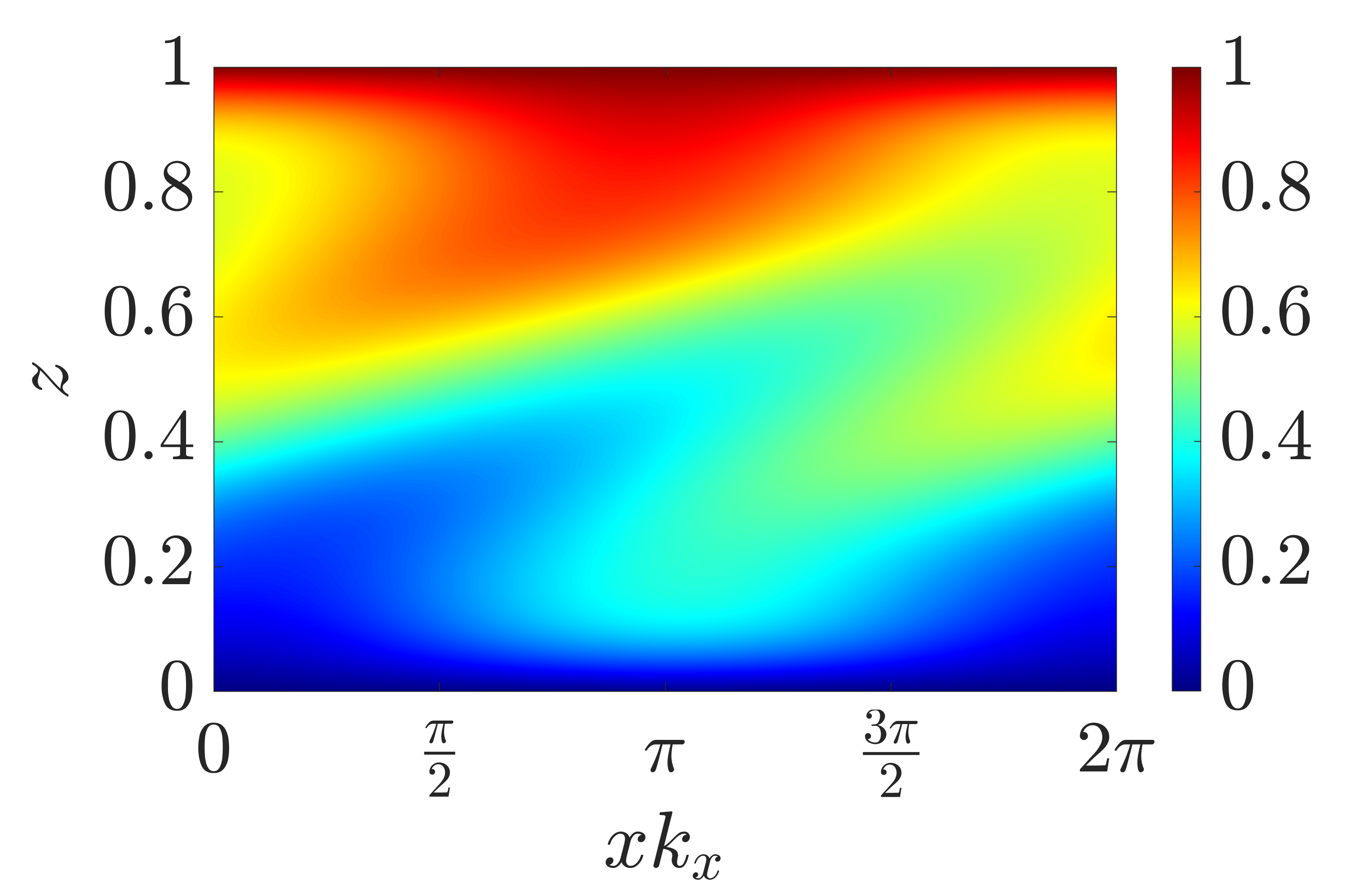}
    \includegraphics[width=0.32\textwidth,trim=-0 -0in 0 0]{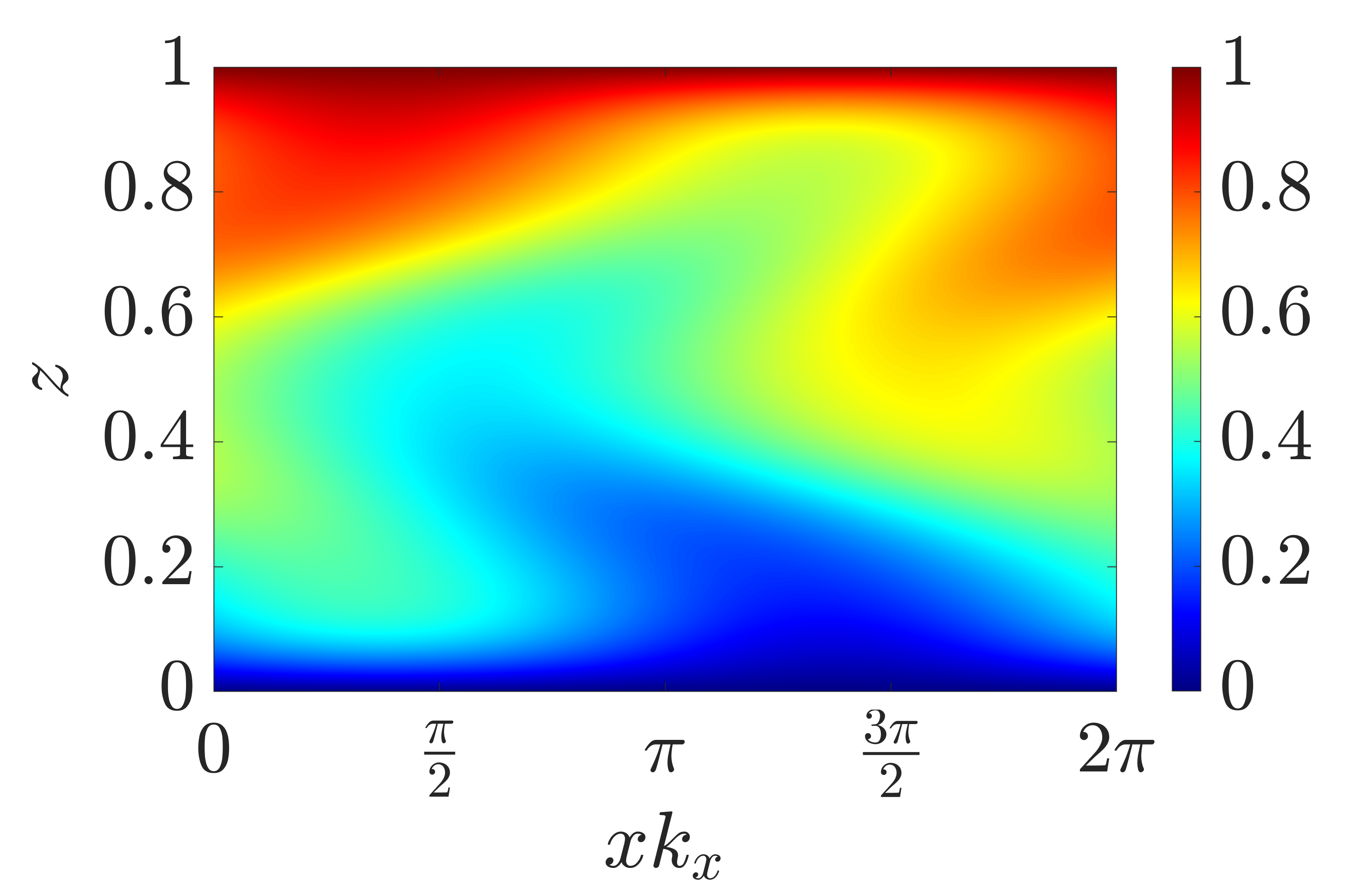}    
    \caption{Solution profiles from the single-mode equations \eqref{eq:single_mode} for the TF1 and TW1 solutions displaying (a) $z+\bar{S}_0$, (b) $\bar{U}_0$ and the horizontal velocity $ \widetilde{u}$ for (c) TF1 and (d) TW1. The second row shows the vertical vorticity mode for (e) TF1 and (f) TW1 and the isocontours of the total salinity for (g) TF1 and (h) TW1 using \eqref{eq:normal_mode_S} and \eqref{eq:total_T_S}. Other parameter values are $k_x=k_y=6.873$, $R_\rho=40$, $Pr=0.05$, $\tau =0.01$ and $Ra_T=10^5$.  }
    \label{fig:profile_R_rho_T2S_40_tau_0p01_Pr_0p05_TF1_TW1}
\end{figure}

A comparison between 2D and 3D results at $Pr=0.05$ in figure \ref{fig:bif_diag_low_Ra_S2T_low_Pr} shows that the bifurcation point of TF1 and TW1 is closer to the high wavenumber onset of S1 in the 2D configuration. The difference between 2D and 3D at $Pr=0.05$ in figure \ref{fig:bif_diag_low_Ra_S2T_low_Pr} is more evident than that at $Pr=7$ in figure \ref{fig:bif_diag_low_Ra_S2T}, as is the case in Rayleigh-B\'enard convection \citep{van2013comparison}. Recalling that both TF1 and TW1 are associated with the formation of large-scale shear (see table \ref{tab:summary_case}, and figures \ref{fig:profile_R_rho_T2S_2_tau_0p01_TF1_TW1}(a) and \ref{fig:profile_R_rho_T2S_40_tau_0p01_Pr_0p05_TF1_TW1}(b)), this difference suggests that the 2D configuration favors the formation of large-scale shear, cf. \citep{garaud20152d}.

\begin{table}
    \centering
    \begin{tabular}{c|ccccccccccc}\hline
         \backslashbox{I.C.}{$L_x$} & $2\pi/18$ & $2\pi/16$ & $2\pi/14$ & $2\pi/12$ & $2\pi/10$ & $2\pi/8$ & $2\pi/6$ & $2\pi/4$ & $2\pi/2$ & $2\pi $ & $4\pi$  \\
         \hline
        S1 & S1 & TF1 & RTF & RTF & RTF & C & RTF (2) & C & C & C & C \\
        TF1 & - & TF1 & RTF & RTF & RTF & C & RTF (2) & C & C & - & - \\
        TW1 & - & TF1 &RTF & RTF & RTF & C & RTF (2) & C & - & - & -\\ \hline
    \end{tabular}
    \caption{The flow structures from 2D DNS at $t=3,000$ in domains of different sizes $L_x$ and initial conditions (I.C.) constructed from the single-mode approximations to the S1, TF1 and TW1 states using the ansatz \eqref{eq:normal_mode} with $k_x=2\pi/L_x$, $k_y=0$. RTF indicates direction-reversing tilted fingers and C represents chaotic behavior; `-' indicates that a nonzero single-mode solution at $k_x=2\pi/L_x$ is not present based on figure \ref{fig:bif_diag_low_Ra_S2T_low_Pr}(a). The number $n\in \mathbb{Z}$ inside a bracket indicates the horizontal wavenumber $k_x=2\pi n /L_x$ reached at $t=3,000$ if different from the initial wavenumber $n=1$.}
    \label{tab:DNS_transition_low_Ra_S2T_Pr_0p05}
\end{table}

Figure \ref{fig:profile_R_rho_T2S_40_tau_0p01_Pr_0p05_TF1_TW1} shows the solution profiles in the TF1 and TW1 states. Here we choose a 3D wavenumber $k_y=k_x=6.873$ close to the wavenumber that maximizes the $Sh$ in the S1 solution. Similar to the earlier observation of tilted fingers in figure \ref{fig:profile_R_rho_T2S_40_tau_0p01_Pr_7_TF1}, the mean total salinity profile $z+\bar{S}_0(z)$ of TF1 shows two mixed regions with a linear profile in the middle of the layer. In contrast, the TW1 shows instead a three-layer structure in $z+\bar{S}_0(z)$. Panel (b) in figure \ref{fig:profile_R_rho_T2S_40_tau_0p01_Pr_0p05_TF1_TW1} displays the associated large-scale shear $\bar{U}_0(z)$, and exhibits a similar structure to that observed at $R_\rho=2$, $Pr=7$ in figure \ref{fig:profile_R_rho_T2S_2_tau_0p01_TF1_TW1}(a). Figures \ref{fig:profile_R_rho_T2S_40_tau_0p01_Pr_0p05_TF1_TW1}(c)-(d) display the vertical profile of the first harmonic of the horizontal velocity $\widetilde{u}$ in the TF1 and TW1 states showing local peaks close to $z=1/3$, $1/2$, and $2/3$, in contrast to the corresponding S1 solution that displays peaks near the boundaries. These properties are in turn reflected in the staircase profile of the mean salinity shown in figure \ref{fig:profile_R_rho_T2S_40_tau_0p01_Pr_0p05_TF1_TW1}(a). 

Furthermore, in 3D the TF1 and TW1 solutions are also associated with nonzero vertical vorticity as shown in figures \ref{fig:profile_R_rho_T2S_40_tau_0p01_Pr_0p05_TF1_TW1}(e)-(f). Here, the generation of the vertical vorticity originates from the large-scale horizontal shear $\bar{U}_0$, which provides a source term for the vertical vorticity equation \eqref{eq:single_mode_b}. Vertical vorticity is also generated in bifurcations of steady convection rolls (resembling S1) in single-mode solutions of Rayleigh-B\'enard convection on a hexagonal lattice \citep{lopez1983time,murphy1984influence,massaguer1988instability,massaguer1990nonlinear} with a source term provided by self-interaction on this lattice (e.g. \citet[equation (2.2b)]{massaguer1990nonlinear}). As the current single-mode formulation is limited to roll or square planforms, this source term is not present. Instead, it is the associated large-scale shear $\bar{U}_0$ that provides the source term for vertical vorticity, a possibility not included in the previous work \citep{lopez1983time,murphy1984influence,massaguer1988instability,massaguer1990nonlinear}.

The isocontours of total salinity of TF1 and TW1 are shown in figures \ref{fig:profile_R_rho_T2S_40_tau_0p01_Pr_0p05_TF1_TW1}(g)-(h). These panels show qualitative similarity with the TF1 and TW1 shown in figures \ref{fig:profile_R_rho_T2S_2_tau_0p01_TF1_TW1}(b)-(c), where the TF1 solution is tilted in one direction, while the TW1 is tilted in opposite directions above and below the midplane. However, the TF1 and TW1 at $R_\rho=2$, $Pr=7$ in figure \ref{fig:profile_R_rho_T2S_2_tau_0p01_TF1_TW1} show a relatively well-mixed region in the interior, $z\approx 1/2$, compared with the TF1 and TW1 at the higher density ratio $R_\rho=40$ but small Prandtl number, $Pr=0.05$, shown in figure \ref{fig:profile_R_rho_T2S_40_tau_0p01_Pr_0p05_TF1_TW1}. 

The stability of these solutions is indicated in figure \ref{fig:bif_diag_low_Ra_S2T_low_Pr} by thick (stable) and thin (unstable) lines. In 2D TF1 is stable near the onset of this branch, but becomes unstable at an intermediate wavenumber. The traveling wave is instead unstable near the onset, but then gains stability over an interval of intermediate wavenumbers before losing it again, cf. the TW1 branch at $R_\rho=2$ and $Pr=7$ in figure \ref{fig:bif_diag_mid_Ra_S2T}. In 3D the TF1 branch is always stable, as in the high Prandtl number regime, cf. figures \ref{fig:bif_diag_low_Ra_S2T} and \ref{fig:bif_diag_mid_Ra_S2T}, while the TW1 branch is always unstable, cf. figure \ref{fig:bif_diag_low_Ra_S2T_low_Pr}(b).

We next compare these results with those from 2D DNS following our analysis of the $Pr=7$ results in table \ref{tab:DNS_transition_low_Ra_S2T_Pr_7}. The domain length $L_x$ in the horizontal is taken from $[2\pi/18, 4\pi]$ and the initial conditions are, respectively, constructed using the S1, TF1, and TW1 single-mode solutions with $k_x=2\pi/L_x$ and $k_y=0$. In this $L_x$ interval single-mode theory predicts that S1, TF1 or TW1 may be stable. In table \ref{tab:DNS_transition_low_Ra_S2T_Pr_0p05} we record the state reached in each case at $t=3,000$. We find that the DNS returns S1 or TF1 depending on initial conditions but also supports direction-reversing tilted fingers (RTF) and chaotic states (C). In particular, with horizontal domain $L_x=2\pi/18$, the final state exhibits S1 behavior consistent with the single-mode prediction and the $Sh$ from DNS overlaps with the prediction from single-mode equations as shown in figure \ref{fig:bif_diag_low_Ra_S2T_low_Pr}(a). For $L_x=2\pi/16$, the final state in table \ref{tab:DNS_transition_low_Ra_S2T_Pr_0p05} shows TF1, which is the only stable solution in single-mode theory at $k_x=16$. The corresponding $Sh$ at $L_x=2\pi/16$ from DNS also overlaps with the TF1 single-mode solution as shown in figure \ref{fig:bif_diag_low_Ra_S2T_low_Pr}(a). 

\begin{figure}
    \centering
(a) $\langle u\rangle_h(z,t)$ \hspace{0.33\textwidth} (b)  \hspace{0.15\textwidth} (c)

    \includegraphics[width=0.49\textwidth]{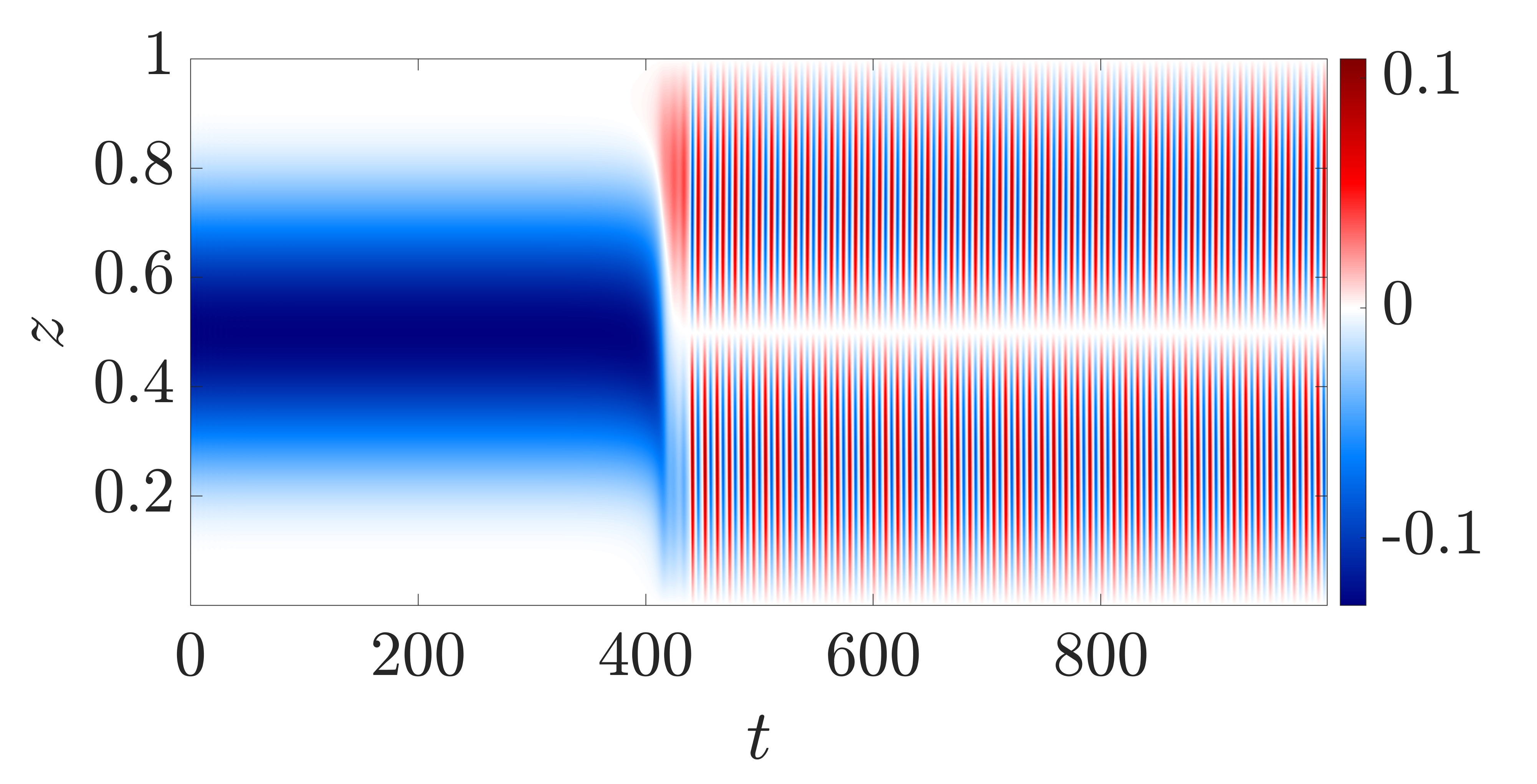}
    \includegraphics[width=0.183\textwidth]{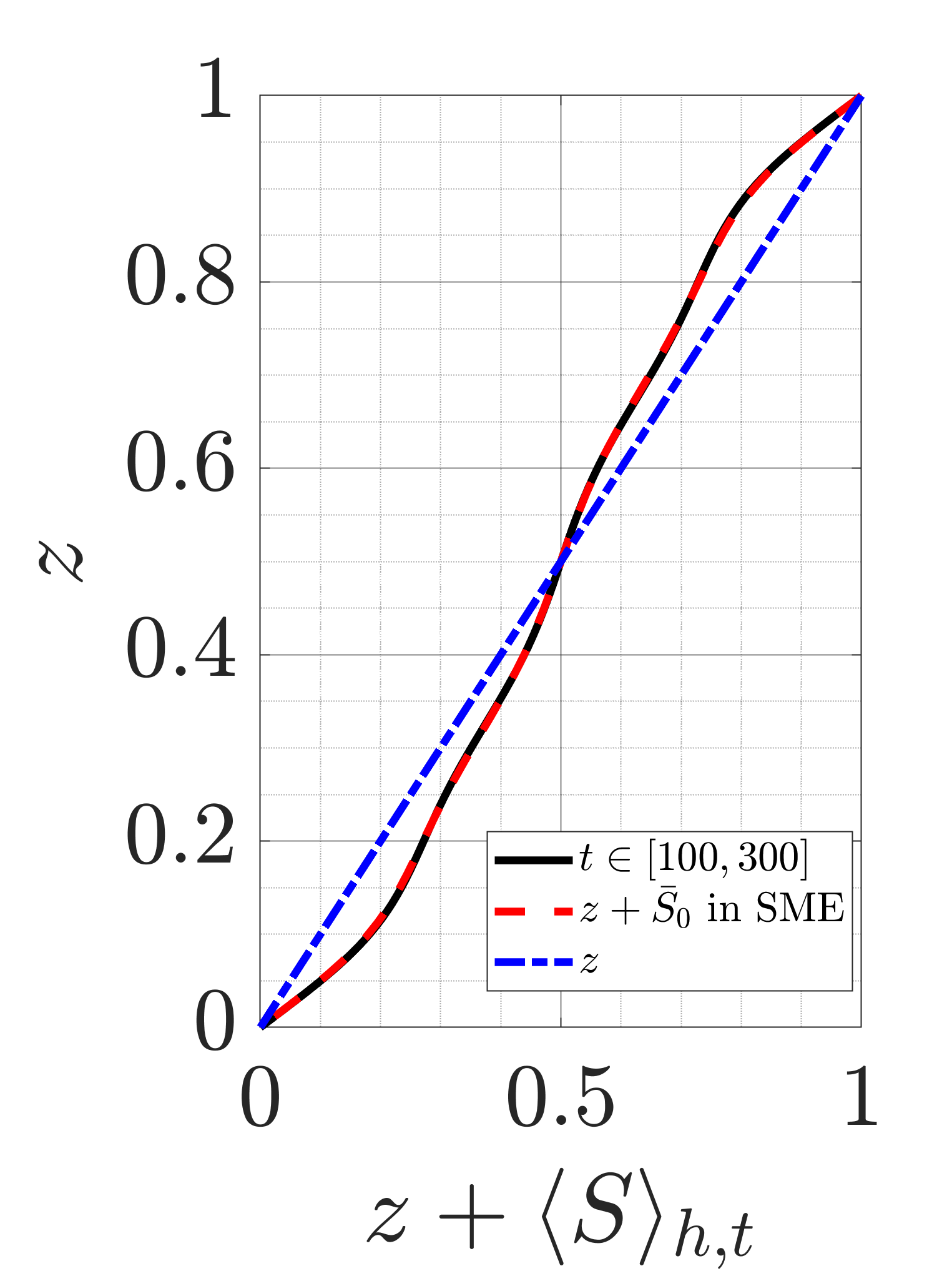}
    \includegraphics[width=0.183\textwidth]{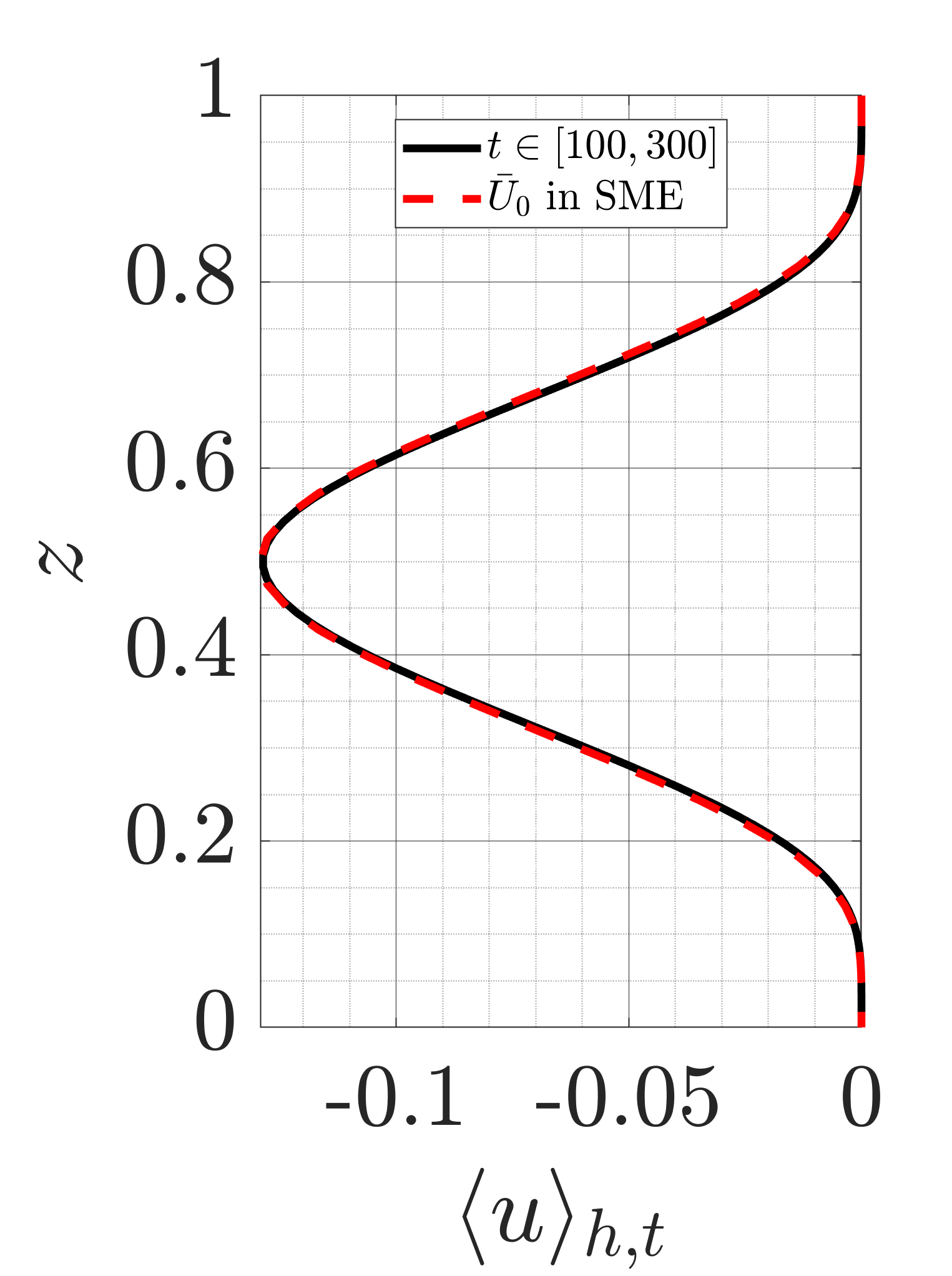}
    \caption{2D DNS of a traveling wave showing (a) $\langle u\rangle_h(z,t)$, (b) $z+\langle S\rangle_{h,t}(z)$ and (c) $\langle u\rangle_{h,t}(z)$ averaged over $t\in[100,300]$ (black lines) for comparison with $z+\bar{S}_0$ and $\bar{U}_0$ from the single-mode TW1 branch (red dashed lines). The parameters are $R_\rho=40$, $Pr=0.05$, $\tau=0.01$, and $Ra_T=10^5$. The initial condition is a TW1 solution with $L_x=2\pi/14$ for the DNS; the single-mode solutions are computed for $k_x=2\pi/L_x=14$, $k_y=0$. }
    \label{fig:DNS_Pr_0p05_kx_14_TW1}
\end{figure}

In order to demonstrate the fidelity of the single-mode TW1 solutions near the high wavenumber onset, figure \ref{fig:DNS_Pr_0p05_kx_14_TW1} compares one DNS run with $L_x=2\pi/14$ and a TW1 single-mode initial condition. The solution takes the form of a traveling wave up to $t\approx 400$, and consequently we compare the time average of this state over $t\in [100,300]$ with the corresponding single-mode results. Specifically, figure \ref{fig:DNS_Pr_0p05_kx_14_TW1}(a) shows the horizontally averaged horizontal velocity $\langle u\rangle_h(z,t)$ corresponding to the large-scale shear mode $\bar{U}_0(z,t)$ in the single-mode equations. Figures \ref{fig:DNS_Pr_0p05_kx_14_TW1}(b)-(c) demonstrate an essentially perfect agreement between the $z+\langle S\rangle_{h,t}$ and $\langle u\rangle_{h,t}$ profiles computed from the DNS in this time interval and their counterparts $z+\bar{S}_0$ and $\bar{U}_0$ computed from the TW1 branch of the single-mode solutions at $k_x=2\pi/L_x=14$ and the same parameters. The $Sh$ also agrees with the single-mode prediction as indicated in figure \ref{fig:bif_diag_low_Ra_S2T_low_Pr}(a). At $t\approx 400$, the TW1 state undergoes an abrupt transition to an oscillatory state with an antisymmetric mean shear profile suggesting that this state is a tilted finger state whose tilt direction reverses periodically in time, i.e., it is direction-reversing tilted finger (RTF). Similar behavior was also found at $Pr=7$ with $L_x=2\pi/8$ initialized by an S2 solution; see table \ref{tab:DNS_transition_low_Ra_S2T_Pr_7} and figure \ref{fig:DNS_Pr_7_S2}. All RTF states in tables \ref{tab:DNS_transition_low_Ra_S2T_Pr_7} and \ref{tab:DNS_transition_low_Ra_S2T_Pr_0p05} were identified based on the direction-reversing behavior of the associated large-scale shear $\langle u\rangle_h(z,t)$. 

\begin{figure}
    \centering
    (a) $z+S(x,z,t)$ at $z=1/2$ \hspace{0.27\textwidth}(b) $\langle u\rangle_h(z,t)$
    
    \includegraphics[width=0.48\textwidth]{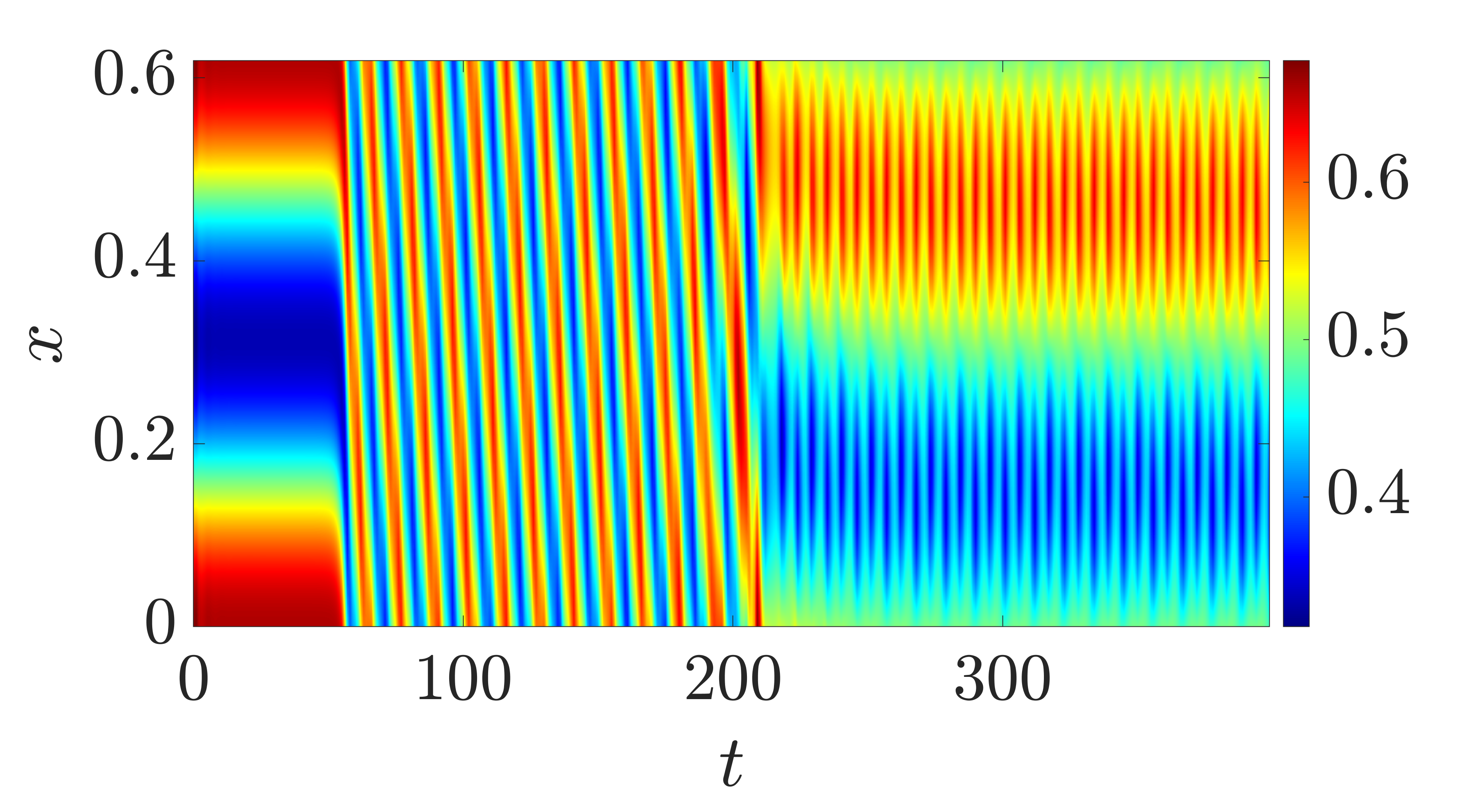}
    \includegraphics[width=0.48\textwidth]{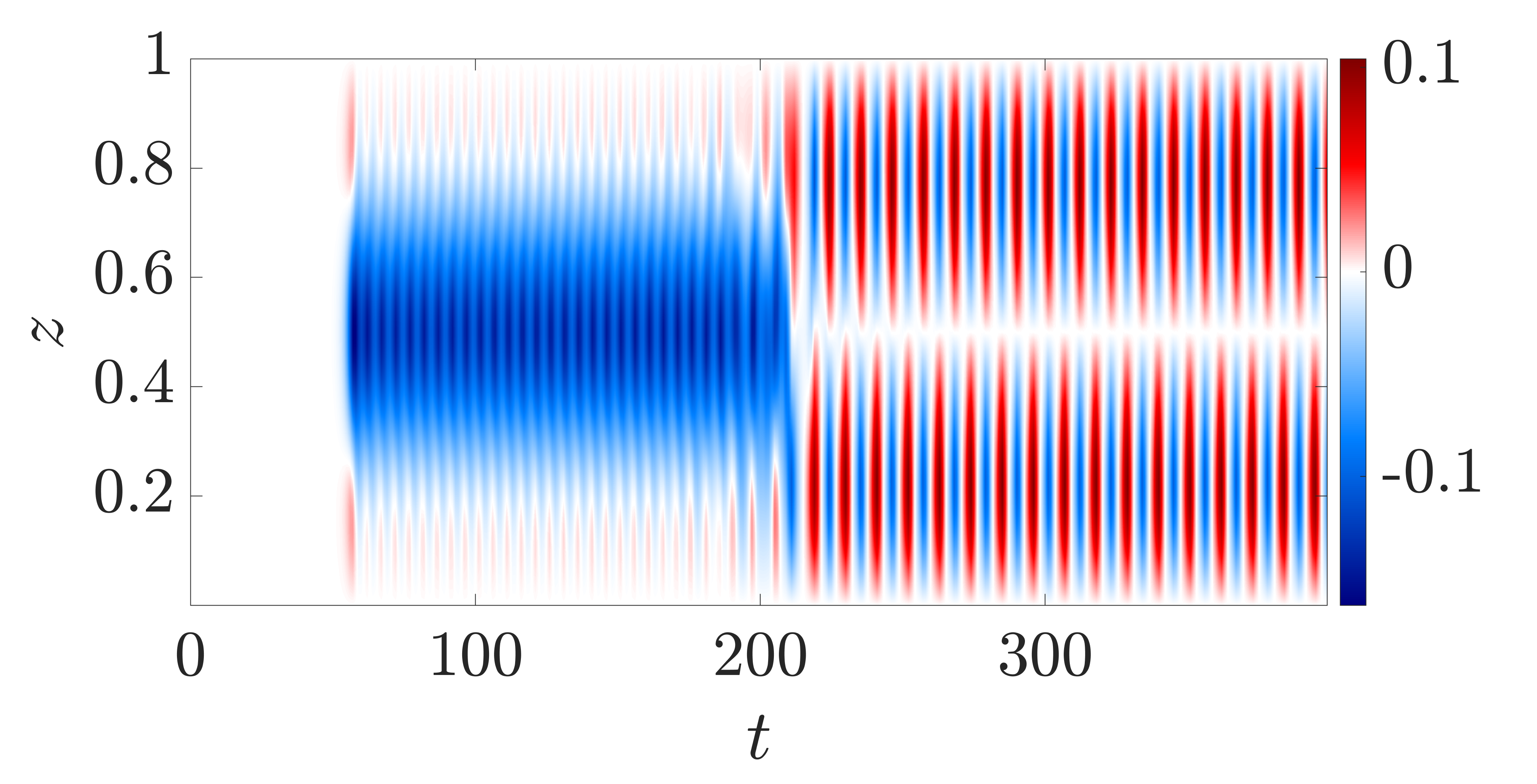}

(c) \hspace{0.2\textwidth}(d) \hspace{0.2\textwidth}(e)\hspace{0.2\textwidth}(f)

\includegraphics[width=0.24\textwidth]{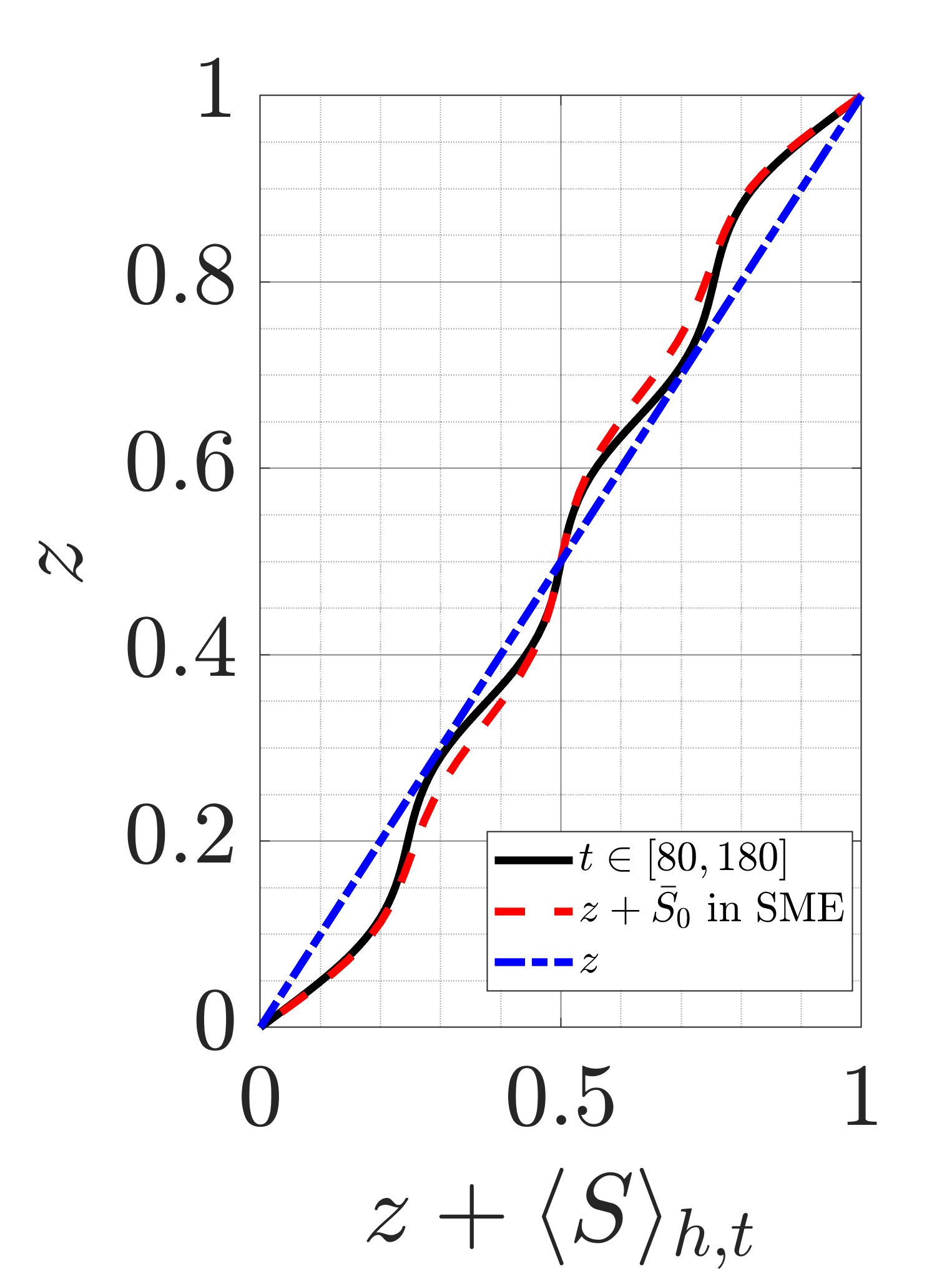}
\includegraphics[width=0.24\textwidth]{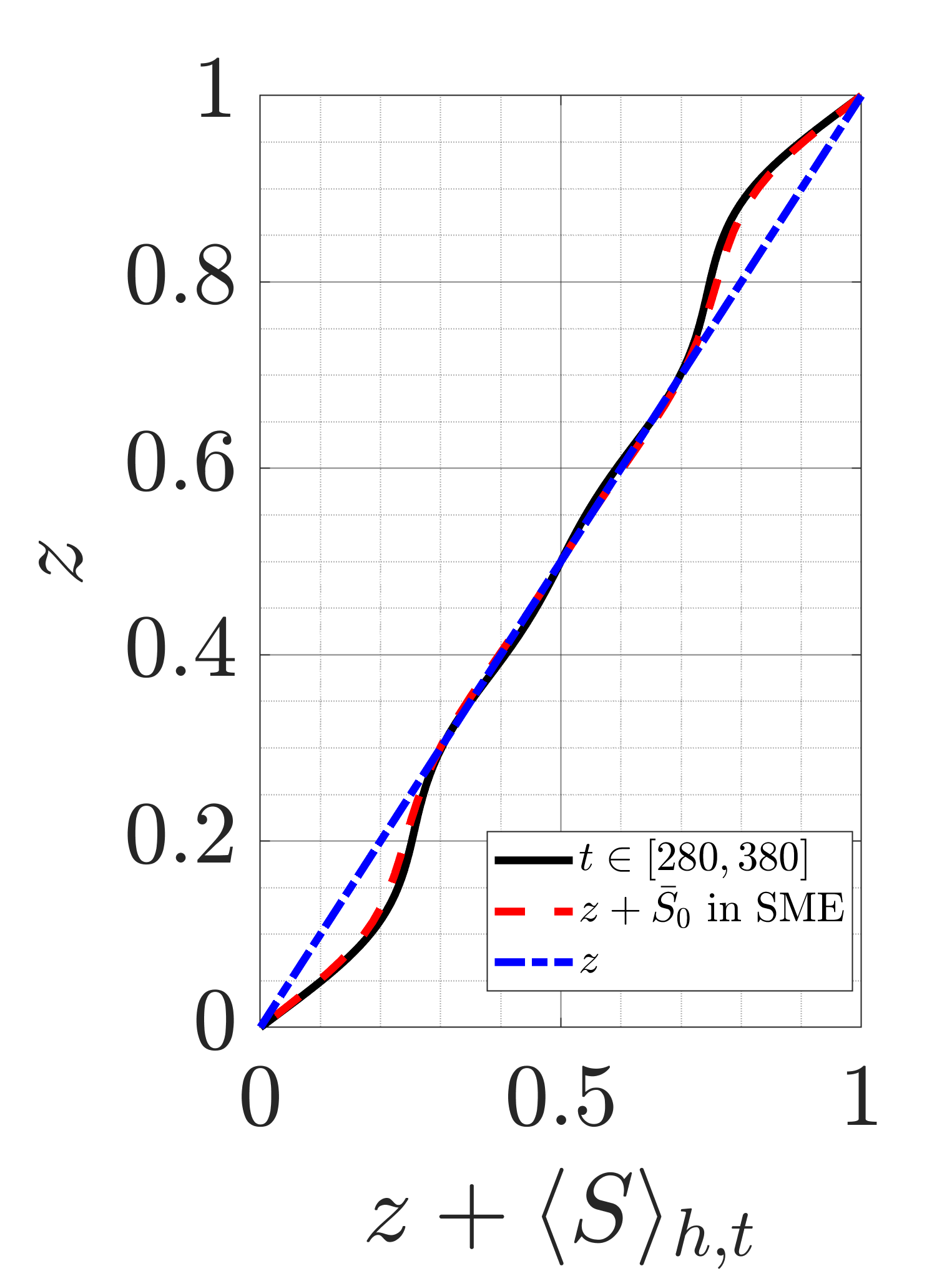}
\includegraphics[width=0.24\textwidth]{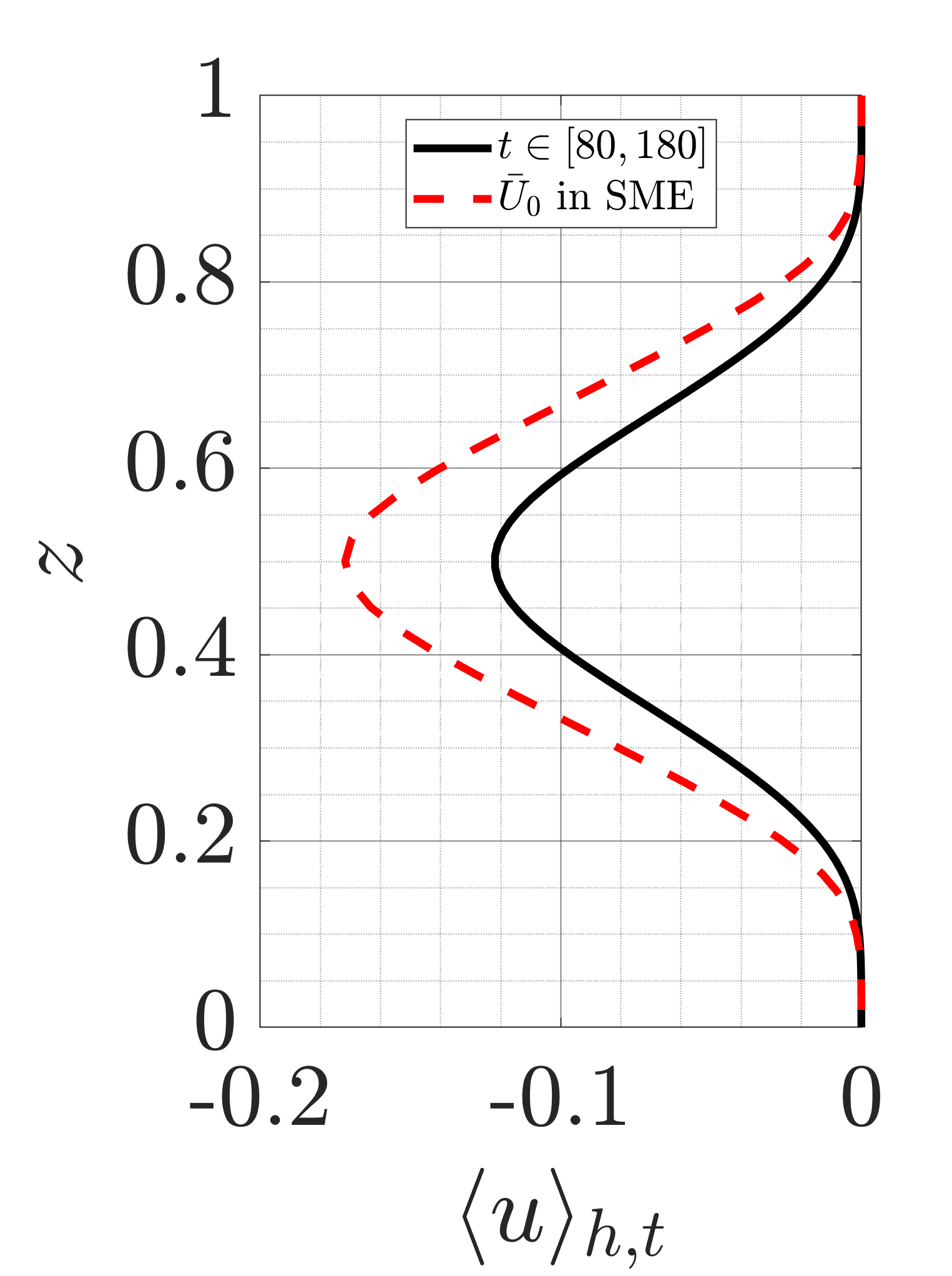}
\includegraphics[width=0.24\textwidth]{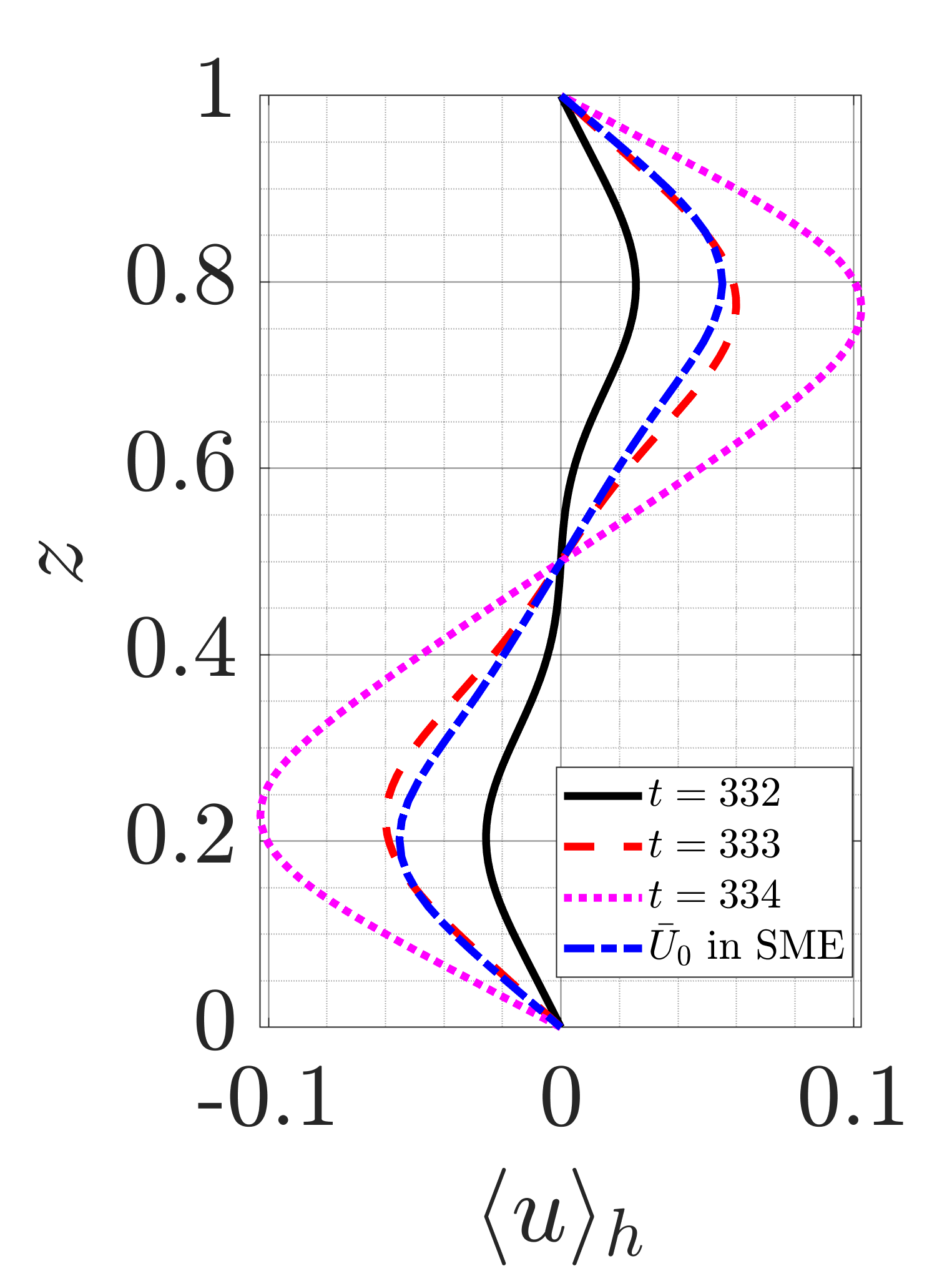}

    \caption{2D DNS showing space-time plots of (a) $z+S(x,z,t)$ at $z=1/2$ and (b) $\langle u\rangle_h(z,t)$. Panels (c) and (d) show  the profiles of $z+\langle S\rangle_{h,t}(z)$ averaged over $t\in[80,180]$ for the traveling wave episode, and $t\in [280,380]$ for direction-reversing tilted fingers, and compared with $z+\bar{S}_0$ from the single-mode TW1 and TF1 solutions. Panel (e) shows $\langle u\rangle_{h,t}$ averaged over $t\in [80,180]$ and compared with $\bar{U}_0$ for the single-mode TW1 branch, while (f) shows $\langle u\rangle_h(z,t)$ at $t=332$, 333 and 334 for comparison with $\bar{U}_0$ from the single-mode TF1 solution. The initial condition is the S1 solution with $L_x=2\pi/10$, and other parameters are $R_\rho=40$, $Pr=0.05$, $\tau=0.01$, $Ra_T=10^5$. The single-mode solutions are calculated for $k_x=2\pi/L_x=10$, $k_y=0$. See supplementary Movie 3. }
    \label{fig:DNS_Pr_0p05_kx_10_S1}
\end{figure}

\begin{figure}
\centering

    (a) $z+S(x,z,t)$ at $z=1/2$ \hspace{0.1\textwidth}(b) $\langle |\mathcal{F}_x(S)| \rangle_t(z;k_x)$ \hspace{0.1\textwidth} (c) $\langle |\mathcal{F}_x(u)| \rangle_t(z;k_x)$

    \centering
    \includegraphics[width=0.4\textwidth]{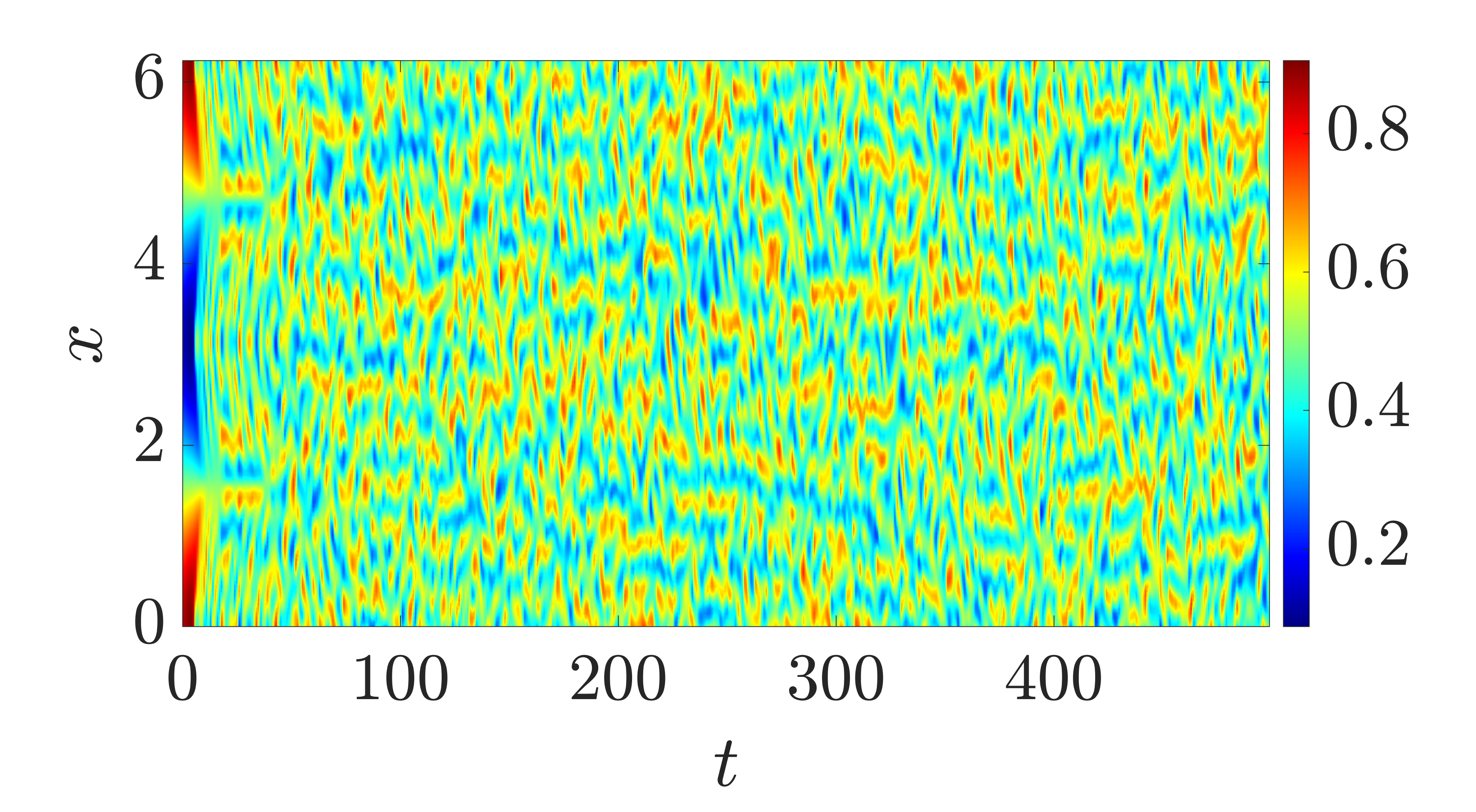}
    \includegraphics[width=0.29\textwidth]{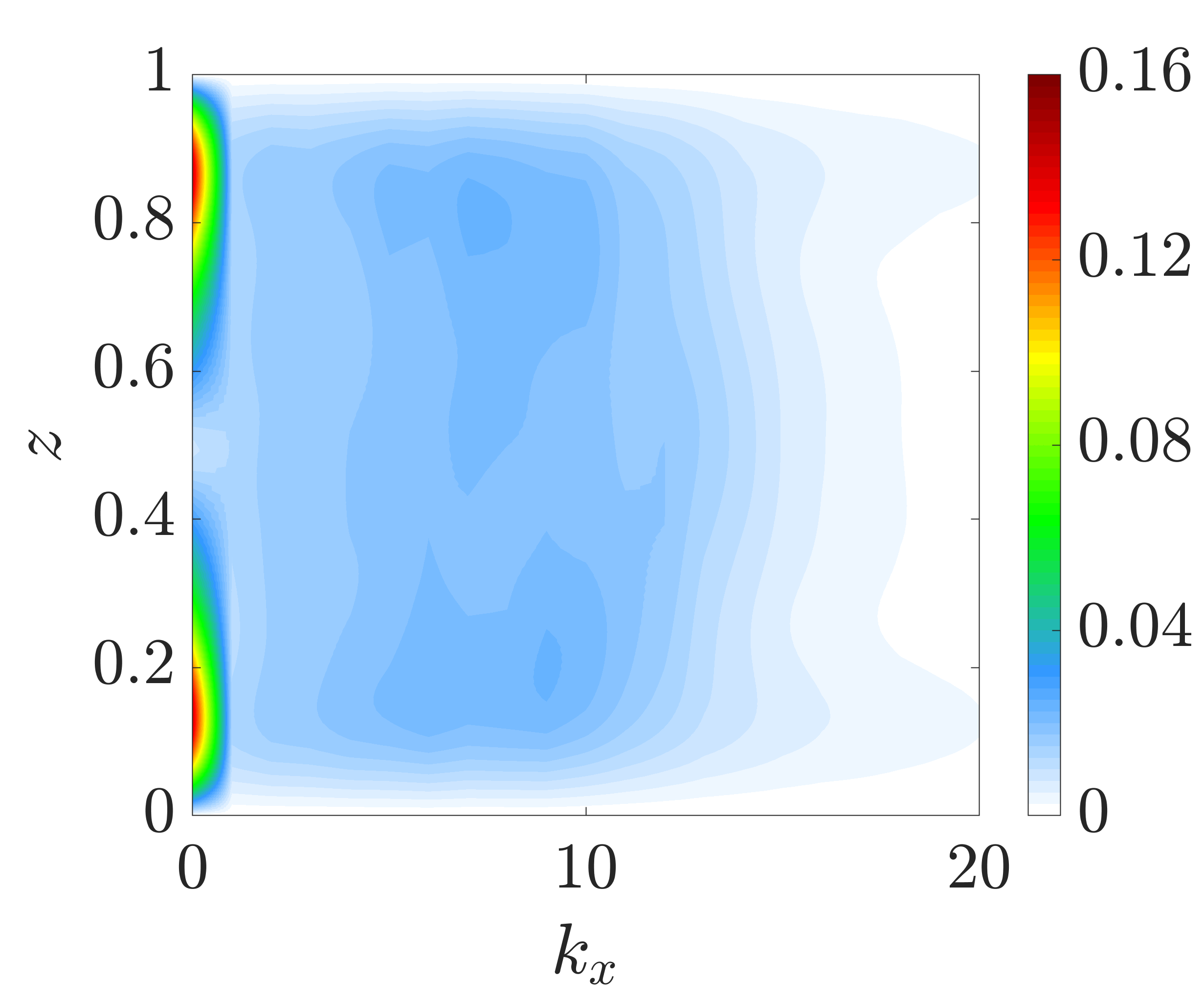}
    \includegraphics[width=0.29\textwidth]{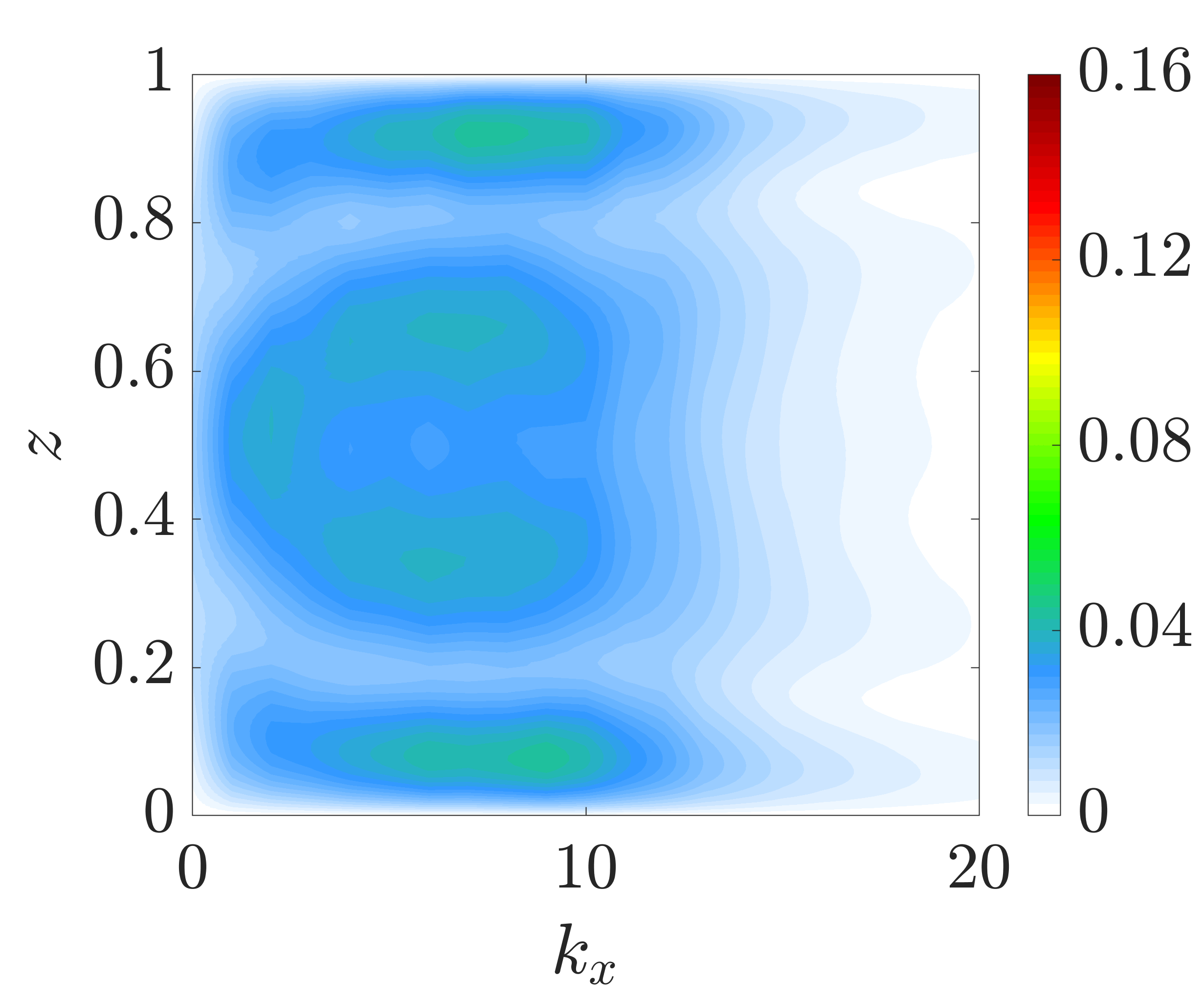}
    
    (d) $\langle u \rangle_h(z,t)$ 

    \includegraphics[width=0.4\textwidth]{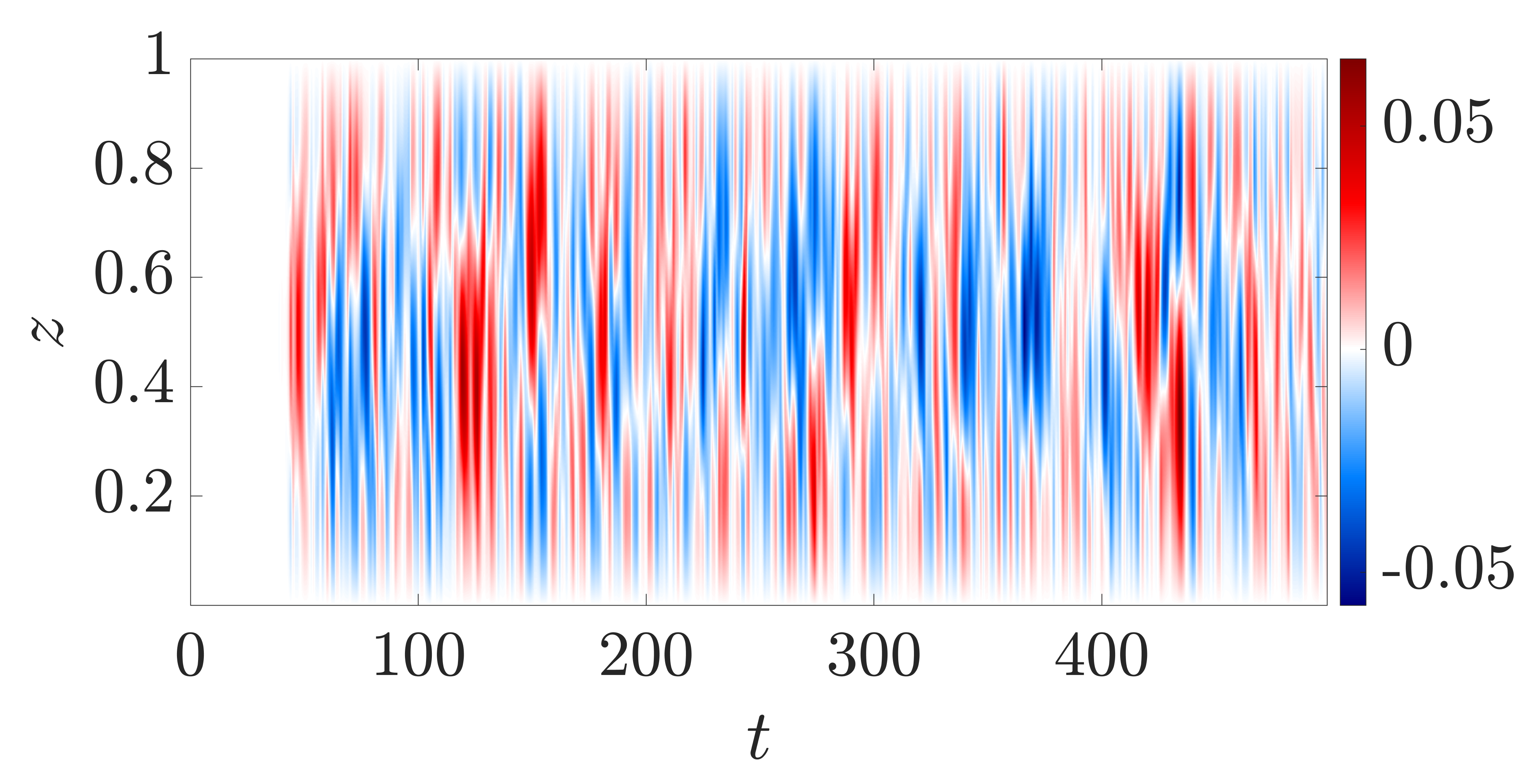}
    \caption{2D DNS results showing (a) total salinity at $z=1/2$, (b) $\langle |\mathcal{F}_x(S)| \rangle_t(z;k_x)$, (c) $\langle |\mathcal{F}_x(u)| \rangle_t(z;k_x)$ and (d) $\langle u\rangle_h(z,t)$. The horizontal domain size is $L_x=2\pi$ initialized by the corresponding S1 solution. The parameters are $R_\rho=40$, $Pr=0.05$, $\tau=0.01$, and $Ra_T=10^5$.  }
    \label{fig:DNS_R_rho_40_Pr_0p05_kx_1_S1}
\end{figure}

Direction-reversing tilted fingers are also observed as the final state in larger horizontal domains with $L_x=2\pi/12$, $2\pi/10$ and $2\pi/6$ as shown in table \ref{tab:DNS_transition_low_Ra_S2T_Pr_0p05}. Figure \ref{fig:DNS_Pr_0p05_kx_10_S1} shows the DNS results with $L_x=2\pi/10$ and initial condition S1. The total salinity $z+S$ at the midplane $z=1/2$ in figure \ref{fig:DNS_Pr_0p05_kx_10_S1}(a) shows that the initial state is sinusoidal in space and steady, corresponding to the prescribed S1 solution. After $t\approx 80$, the S1 state transitions into a traveling wave and subsequently, at $t\approx 220$, into an RTF state. Figure \ref{fig:DNS_Pr_0p05_kx_10_S1}(b) shows that $\langle u\rangle_h(z,t)=0$ at $t=0$ but then this quantity becomes nonzero and midplane-symmetric at the transition to the TW1 state with superposed periodic oscillations. Figures \ref{fig:DNS_Pr_0p05_kx_10_S1}(c)-(d) present the horizontally averaged total salinity profiles, respectively averaged over $t\in [80,180]$ and $t\in [280,380]$, corresponding to the traveling wave and the direction-reversing tilted fingers, and compared with $z+\bar{S}_0$ for the single-mode TW1 and TF1 solutions with $k_x=2\pi/L_x$ and the same parameter values. Here, both the DNS results and the single-mode solutions show a modest three-layer staircase characteristic of the traveling wave state, while both show a two-layer staircase with a linear profile in the middle that is associated with direction-reversing tilted fingers in DNS and the tilted finger state of the single-mode equations.

Figure \ref{fig:DNS_Pr_0p05_kx_10_S1}(e) shows $\langle u \rangle_{h,t} $ from the DNS averaged over $t\in [80,180]$ corresponding to the traveling wave, and compares it with $\bar{U}_0$ from the corresponding single-mode TW1 solution. Here, the large-scale shear from the single-mode theory overpredicts the amplitude compared to that observed in DNS. This is expected as figure \ref{fig:DNS_Pr_0p05_kx_10_S1}(b) indicates that the state is in fact a modulated traveling wave with superposed oscillations in $\langle u \rangle_{h} $ in $t\in [80,180]$. Moreover, as shown previously in figures \ref{fig:bif_diag_low_Ra_S2T}(a) and \ref{fig:bif_diag_low_Ra_S2T_low_Pr}(a), the accuracy of the single-mode equation is expected to decrease with decreasing wavenumber (here $k_x=10$), a fact that may account for the difference in the solution profiles between DNS and single-mode theory. This difference in the large-scale shear may also contribute to the difference in the mean salinity profiles in figure \ref{fig:DNS_Pr_0p05_kx_10_S1}(c) and the associated $Sh$ in figure \ref{fig:bif_diag_low_Ra_S2T_low_Pr}(a). As the TF1 solution in the single-mode equation is steady, we compare the large-scale shear $\bar{U}_0$ with three different snapshots of $\langle u\rangle_h(z,t)$ at $t=332$, 333, 334 in figure \ref{fig:DNS_Pr_0p05_kx_10_S1}(f). The shape and magnitude of $\langle u\rangle_h$ at $t=333$ is close to a TF1 single-mode solution, but the maximum amplitude of the large-scale shear of RTF observed in 2D DNS is larger than $\bar{U}_0$ from the TF1 single-mode solution.

For larger $L_x$, the solution may transition to a double wavenumber RTF, see $L_x=2\pi/6$ in table \ref{tab:DNS_transition_low_Ra_S2T_Pr_0p05}, and such a transition to a higher wavenumber flow structure was already described for $Pr=7$ as summarized in table \ref{tab:DNS_transition_low_Ra_S2T_Pr_7} and figures \ref{fig:DNS_Pr_7_S3} and \ref{fig:DNS_R_rho_40_Pr_7_kx_2_S1_S2_TF1}. Since the final state does not exhibit well-organized flow structures when $L_x\in [2\pi/4, 4\pi]$ we refer to it as chaotic. Figure \ref{fig:DNS_R_rho_40_Pr_0p05_kx_1_S1} presents (a) the total salinity at the midplane, (b) the Fourier-transformed salinity deviation $\langle |\mathcal{F}_x(S)|\rangle_t$, and (c) the Fourier-transformed horizontal velocity $\langle |\mathcal{F}_x(u)| \rangle_t$. Here, we select $L_x=2\pi$ and the initial condition is the S1 solution, so the only difference between the results in figures \ref{fig:DNS_R_rho_40_Pr_0p05_kx_1_S1} and \ref{fig:DNS_R_rho_40_Pr_7_kx_1_S1} is the value of the Prandtl number. The total salinity at the midplane $z=1/2$ in figure \ref{fig:DNS_R_rho_40_Pr_0p05_kx_1_S1}(a) at $Pr=0.05$ does not show a very clear difference from figure \ref{fig:DNS_R_rho_40_Pr_7_kx_1_S1}(a) at $Pr=7$, while the Fourier spectrum of the salinity at $Pr=0.05$ in figure \ref{fig:DNS_R_rho_40_Pr_0p05_kx_1_S1}(b) is broader than the same quantity at $Pr=7$ in figure \ref{fig:DNS_R_rho_40_Pr_7_kx_1_S1}(b). The Fourier-transformed horizontal velocity in figure \ref{fig:DNS_R_rho_40_Pr_0p05_kx_1_S1}(c) shows a greater difference from the corresponding result at $Pr=7$ in figure \ref{fig:DNS_R_rho_40_Pr_7_kx_1_S1}(c), where the peaks near $z=1/3$ and $z=2/3$ and the peak at $z=1/2$ at $k_x\approx 1$ have magnitudes similar to those near the boundary. This suggests that both TF1 and TW1 are involved in the chaotic behavior. Moreover, $\langle |\mathcal{F}_x(u)|\rangle_t$ shows a nonzero value at $k_x=0$ corresponding to the large-scale shear in figure \ref{fig:DNS_R_rho_40_Pr_0p05_kx_1_S1}(c), while the white region in $\langle |\mathcal{F}_x(u)|\rangle_t$ at $k_x=0$ in figure \ref{fig:DNS_R_rho_40_Pr_7_kx_1_S1}(c) for $Pr=7$ indicates a value close to zero. This is consistent with the fact that low Prandtl numbers favor the formation of large-scale shear. Indeed, $\langle u\rangle_h(z,t)$ in figure \ref{fig:DNS_R_rho_40_Pr_0p05_kx_1_S1}(d) reaches a maximum value of 0.065, while the same quantity has the maximum amplitude $5.37\times 10^{-4}$ when $Pr=7$. 

At other times, the large-scale shear $\langle u\rangle_h(z,t)$ in figure \ref{fig:DNS_R_rho_40_Pr_0p05_kx_1_S1}(d) resembles the large-scale shear $\bar{U}(z)$ associated with either TF1 or TW1 in figure \ref{fig:profile_R_rho_T2S_40_tau_0p01_Pr_0p05_TF1_TW1}(b) based on the parity of $\langle u\rangle_h(z,t)$ with respect to $z=1/2$. The observed time series suggests that the solution visits the neighborhoods of both the traveling waves and the tilted fingers in a chaotic manner. The mean $Sh$ associated with this state in domains of size $L_x=\pi/2$, $\pi$ and $2\pi$, and initialized using the S1 single-mode solution are shown in figure \ref{fig:bif_diag_low_Ra_S2T_low_Pr}(a), and fall within the range predicted by S1, TF1, and TW1 single-mode states. Note that for these relatively large domain sizes, transition to a higher wavenumber mode is also possible but cannot be predicted within the single-mode approach. The comparison in figure \ref{fig:bif_diag_low_Ra_S2T_low_Pr}(a) supports the role of the (unstable) TF1 and TW1 solutions in determining the salinity transport in this chaotic state. 

 \section{Conclusions and future work}
\label{sec:conclusion}

This work performed bifurcation analysis of vertically confined salt-finger convection using single-mode equations obtained from a severely truncated Fourier expansion in the horizontal. The resulting equations were solved for the vertical structure of the solutions as a function of the density ratio, the Prandtl number, and the assumed horizontal wavenumber. We fixed the diffusivity ratio and thermal Rayleigh number and focused almost exclusively on the case of no-slip velocity boundary conditions. We computed strongly nonlinear staircase-like solutions having one (S1), two (S2) and three (S3) well-mixed mean salinity regions in the vertical direction. These bifurcate in successive bifurcations from the trivial solution. In each case, salinity gradients are expelled from regions of closed streamlines resembling the mechanism described by \citet{rhines1983rapidly}, leading to a well-mixed interior. Secondary bifurcations of S1 lead to tilted fingers (TF1) or traveling waves (TW1) both of which spontaneously break reflection symmetry in the horizontal owing to the spontaneous generation of large-scale shear. Secondary bifurcations from S2 and S3 lead to asymmetric solutions (A2 and A3) that spontaneously break symmetry with respect to the horizontal midplane. 

The stability and relevance of the single-mode solutions we obtained were further analyzed with the assistance of 2D DNS. Near onset, the one-layer solution S1 is stable and corresponds to maximum salinity transport among the solutions, an observation consistent with the prediction of `relative stability' criterion \citep{malkus1958finite}. However, when a secondary bifurcation destabilizes S1 the superseding stable states (TF1 or TW1) may result in reduced salinity transport as found in the low Prandtl number regime. The associated Sherwood number near the high wavenumber end of the S1 solution is in excellent agreement with DNS in small horizontal domains.

In larger domains DNS reveals that the final state reverts to a higher wavenumber S1 state closer to the natural finger scale or exhibits chaotic behavior, a process that begins to set in once the domain size is comparable to approximately twice the finger scale. In general the final state exhibits a strong dependence on initial conditions. The S2, S3, A2 and A3 solutions are all unstable within the parameter regime explored, although the chaotic solutions we found appear to visit neighborhoods of these unstable solutions at different instants in time. With stress-free velocity boundary conditions the S1, S2, S3, TF1, A2 and A3 solutions persist whenever the density ratio and the Prandtl number are high enough, but exhibit larger Sherwood numbers.

At lower density ratios, the S1, S2 and S3 solutions exhibit sharper staircase structures, while the TF1 solution displays stronger large-scale shear and tilt angle. In addition S1 also bifurcates to the TW1 state, a bifurcation that is not present at high density ratio. The scaling of the S1 Sherwood number with the density ratio closely matches the DNS results within the currently explored parameter regime, while the corresponding change in the mean salinity and temperature profiles resembles that seen in DNS.

The dynamics of the secondary TF1 and TW1 states in the low Prandtl number regime also show excellent agreement with the 2D DNS results in small domains, likely because they appear, in this regime, quite close to the high wavenumber onset of S1. These states are typically associated with two- and three-layer mean salinity profiles, respectively. The final state seen in low Prandtl number 2D DNS is a prominent direction-reversing tilted finger (RTF) in which the tilt and the associated large-scale shear reverse with time in a manner that resembles the `pulsating wave' state identified in magnetoconvection \citep{matthews1993pulsating,proctor1994nonlinear} as well as the large-scale flow reversals observed in Rayleigh-B\'enard convection \citep{sugiyama2010flow,chandra2013flow,winchester2021zonal}. In smaller domains the time-averaged salinity and shear profiles computed from 2D DNS resemble the two-layer and three-layer mean salinity and shear profiles corresponding to RTF and TW1, respectively, while in larger domains such profiles are still evident but only episodically.

The bifurcation diagrams of single-mode solutions shown here also shed light on the difference between 2D and 3D. While the corresponding S1, S2, S3, A2 and A3 states in 3D can be obtained from the 2D results by a simple wavenumber rescaling, this is no longer so for states associated with large-scale shear, i.e. TF1 and TW1. The resulting difference is most prominent at low density ratio or low Prandtl numbers. A comparison shows that 2D favors the generation of large-scale shears, a fact consistent with DNS observations \citep{garaud20152d}.

The bifurcation diagram presented here is far from complete. Whether other primary or secondary bifurcations possess the potential to provide stable staircase solutions is an interesting question for future exploration. The secondary Hopf and global bifurcations are also of interest since they provide additional insight into the origin of the direction-reversing tilted finger state and of chaotic salt-finger convection, respectively. Extension of the present framework to 3D states with hexagonal coordination in the horizontal is also of interest, as is a study of the doubly- or triply-periodic configuration with periodic boundary conditions in the vertical as well as in the horizontal, a formulation that is likely to have greater relevance to oceanography. Finally, density staircases have also been widely observed in the diffusive regime of double-diffusive convection (cold fresh water on top of warm salty water) \citep{timmermans2008ice,radko2016thermohaline,yang2022layering} as well as in stratified shear flow \citep{oglethorpe2013spontaneous,taylor2017multi,lucas2017layer,lucas2019layer} and elsewhere. It is of interest to explore the applicability of the present bifurcation-theoretic approach to the study of staircase states in these flow regimes.

\section*{Acknowledgment}
This work was supported by the National Science Foundation under Grant Nos. OCE 2023541 (C.L. and E.K.) and OCE-2023499 (K.J.). 

\section*{Declaration of Interests}
The authors report no conflict of interest.

\bibliography{ref_bounded_salt_finger_clean_v7}
\bibliographystyle{jfm}

\end{document}